\documentclass{aa}
\usepackage{natbib,graphicx,ulem}
\usepackage{txfonts}
\usepackage{longtable,lscape}
\usepackage{hyperref}
\begin{document}

\title{The evolution of the cluster optical galaxy luminosity function
  between z=0.4 and 0.9 in the DAFT/FADA survey.~\thanks{ Based on
    observations made with the FORS2 multi-object spectrograph mounted
    on the Antu VLT telescope at ESO-Paranal Observatory (programs
    085.A-0016, 089A-0666, 191.A-0268; PI: C. Adami). Also based on
    observations made with the Italian Telescopio Nazionale Galileo
    (TNG) operated on the island of La Palma by the Fundaci\'on
    Galileo Galilei of the INAF (Istituto Nazionale di Astrofisica) at
    the Spanish Observatorio del Roque de los Muchachos of the
    Instituto de Astrof\'isica de Canarias.  Based on observations
    obtained with MegaPrime/MegaCam, a joint project of CFHT and
    CEA/IRFU, at the Canada-France-Hawaii Telescope (CFHT) which is
    operated by the National Research Council (NRC) of Canada, the
    Institut National des Science de l'Univers of the Centre National
    de la Recherche Scientifique (CNRS) of France, and the University
    of Hawaii. This work is based in part on data products produced at
    Terapix available at the Canadian Astronomy Data Centre as part of
    the Canada-France-Hawaii Telescope Legacy Survey, a collaborative
    project of NRC and CNRS.  Also based on observations obtained at
    the WIYN telescope (KNPO). The WIYN Observatory is a joint
    facility of the University of Wisconsin-Madison, Indiana
    University, Yale University, and the National Optical Astronomy
    Observatory. Kitt Peak National Observatory, National Optical
    Astronomy Observatory, is operated by the Association of
    Universities for Research in Astronomy (AURA) under cooperative
    agreement with the National Science Foundation.  Also based on
    observations obtained at the MDM observatory (2.4m telescope).
    MDM consortium partners are Columbia University Department of
    Astronomy and Astrophysics, Dartmouth College Department of
    Physics and Astronomy, University of Michigan Astronomy
    Department, The Ohio State University Astronomy Department, Ohio
    University Dept. of Physics and Astronomy.  Also based on
    observations obtained at the Southern Astrophysical Research
    (SOAR) telescope, which is a joint project of the Minist\'{e}rio
    da Ci\^{e}ncia, Tecnologia, e Inova\c{c}\~{a}o (MCTI) da
    Rep\'{u}blica Federativa do Brasil, the U.S. National Optical
    Astronomy Observatory (NOAO), the University of North Carolina at
    Chapel Hill (UNC), and Michigan State University (MSU).  Also
    based on observations obtained at the Cerro Tololo Inter-American
    Observatory, National Optical Astronomy Observatory, which are
    operated by the Association of Universities for Research in
    Astronomy, under contract with the National Science Foundation.
    Also based on observations made with the Gran Telescopio Canarias
    (GTC), installed in the Spanish Observatorio del Roque de los
    Muchachos of the Instituto de Astrof\'isica de Canarias, in the
    island of La Palma.  Also based on archive data collected at the
    Subaru Telescope, which is operated by the National Astronomical
    Observatory of Japan.  Finally, this research has made use of the
    VizieR catalogue access tool, CDS, Strasbourg, France.  }}

\titlerunning{Evolution of the optical cluster galaxy
  luminosity function} \authorrunning{Martinet et al.}

\author{Nicolas Martinet\inst{1}, Florence Durret\inst{1}, Loic
  Guennou\inst{2,3}, Christophe Adami\inst{3}, Andrea
  Biviano\inst{4,1}, M. P. Ulmer\inst{5}, Douglas Clowe\inst{6}, Claire
  Halliday\inst{7}, Olivier Ilbert\inst{3}, Isabel Marquez\inst{8},
  Mischa Schirmer\inst{9,10}} \offprints{Nicolas Martinet,
  \email{martinet@iap.fr}}

\institute{UPMC Universit\'e Paris 06, UMR~7095, Institut
  d'Astrophysique de Paris, 98bis Bd Arago, F-75014, Paris, France
  \and Astrophysics and Cosmology Research Unit, University of
  KwaZulu-Natal, Durban, 4041, SA \and LAM, OAMP, Universit\'e
  Aix-Marseille \& CNRS, P\^ole de l'Etoile, Site de Ch\^ateau
  Gombert, 38 rue Fr\'ed\'eric Joliot-Curie, 13388 Marseille 13 Cedex,
  France \and INAF/Osservatorio Astronomico di Trieste, via Tiepolo
  11, 34143 Trieste, Italy \and Dept of Physics and Astronomy \&
  Center for Interdisciplinary Exploration and Research in
  Astrophysics (CIERA), Northwestern University, Evanston, IL
  60208-2900, USA \and Department of Physics and Astronomy, Ohio
  University, 251B Clippinger Lab, Athens, OH 45701, USA \and 23, rue
  d'Yerres, 91230 Montgeron, France \and Instituto de Astrof\'isica de
  Andaluc\'ia, CSIC, Glorieta de la Astronom\'ia s/n, 18008, Granada,
  Spain \and Gemini Observatory, Casilla 603, La Serena, Chile \and
  Argelander-Institut f\"ur Astronomie, Universit\"et Bonn, Auf dem
  H\"ugel 71, 53121, Bonn, Germany }

\setcounter{page}{1}

\abstract {There are some disagreement about the abundance of faint
  galaxies in high-redshift clusters, with contradictory results in
  the literature arising from studies of the optical galaxy luminosity
  function (GLF) for small cluster samples.} {We compute GLFs for one
  of the largest medium-to-high-redshift ($0.4\leq z<0.9$) cluster
  samples to date in order to probe the abundance of faint galaxies in
  clusters. We also study how the GLF depends on cluster redshift,
  mass, and substructure and compare the GLFs of clusters with those
  of the field. We separately investigate the GLFs of blue and 
  red-sequence (RS) galaxies to understand the evolution of different cluster
  populations.} {We calculate GLFs for 31 clusters taken from the
  DAFT/FADA survey in the B, V, R, and I rest-frame bands. We use
  photometric redshifts computed from BVRIZJ images to constrain
  galaxy cluster membership. We carry out a detailed estimate of the
  completeness of our data. We distinguish the red-sequence and blue
  galaxies using a V-I versus I colour magnitude diagram. We study the
  evolution of these two populations with redshift. We fit Schechter
  functions to our stacked GLFs to determine average cluster
  characteristics.}  {We find that the shapes of our GLFs are
    similar for the B, V, R, and I bands with a drop at the red GLF
    faint ends that is more pronounced at high-redshift: $\alpha_{red}
    \sim -0.5$ at $0.40\leq z<0.65$ and $\alpha_{red} > 0.1$ at
    $0.65\leq z<0.90$. The blue GLFs have a steeper faint end
    ($\alpha_{blue} \sim -1.6$) than the red GLFs, that appears to be
    independent of redshift. For the full cluster sample, blue and red
    GLFs meet at $M_V=-20$, $M_R=-20.5$, and $M_I=-20.3$. A study of
    how galaxy types evolve with redshift shows that late type
    galaxies appear to become early types between $z\sim 0.9$ and
    today. Finally, the faint ends of the red GLFs of more massive
    clusters appear to be richer than less massive clusters, which is
    more typical of the lower redshift behaviour.} {Our results
    indicate that these clusters form at redshifts higher than $z=0.9$
    from galaxy structures that already have an established red
    sequence. Late type galaxies then appear to evolve into early
    types, enriching the red-sequence between this redshift and
    today. This effect is consistent with the evolution of the faint
    end slope of the red-sequence and the galaxy type evolution that
    we find. Finally, faint galaxies accreted from the field
    environment at all redshifts might have replaced the blue late
    type galaxies that converted into early types, explaining the lack
    of evolution in the faint end slopes of the blue GLFs.}


\maketitle

\section{Introduction }
\label{sec:intro}

It is widely accepted that cluster elliptical galaxies are a passively
evolving population formed at high-redshift ($z>1$) in a short
duration event \citep[e.g.][]{deprop99, delucia04, Andreon06,
  deprop07, delucia07, Muzzin08, Mancone10, Mancone12, deprop13}. This
scenario is strongly supported by the lack of evolution in the colour
magnitude relation for the bright galaxies in clusters from $z=1$ to
$z=0$ \citep[e.g.][]{delucia04}. However, there is still a strong
debate about whether cluster galaxies migrated from the field to
clusters at lower redshift ($z \sim 0.8$) \citep[e.g.][]{delucia04,
  Poggianti06, delucia07}, or if they joined clusters at yet higher
redshift, or even originally formed in clusters. This debate arises
from the different behaviours of the faint end slope of galaxy
luminosity functions (hereafter GLFs) observed at high $z$. At low
$z$, cluster GLFs mainly have flat faint ends populated by
low mass galaxies \citep[e.g.][]{Secker97, Rud09}. We note that
\citet{Popesso06} find an upturn of the very faint population of the
GLF for $M^*_g>-18$ in nearby clusters, but our data are not deep
enough to investigate this population of dwarf galaxies at high
redshift.

The faint end of GLFs is found to either decrease with increasing
redshift  \citep[e.g.][]{Smail98, delucia04, Tanaka+05, delucia07,
    Stott+07, Gilbank+08, Rud09, Vulcani+11} or remain constant with
redshift \citep[e.g.][]{deprop03, Andreon06, deprop07, deprop13}. The
first type of behaviour is the most commonly observed, but the cold
dark matter scenario predicts a larger number of low mass galaxies
\citep{Andreon+06, Rud09}. Hence, additional processes are often
invoked within clusters, such as ram pressure stripping \citep{Gunn72}
and harassment \citep[e.g.][]{Moore96, Moore98}. These processes are
found to weakly affect the results in the simulation of
\citet{Lanzoni05} and are likely to depend on mass \citep{Muzzin08}.
The study of the abundance of faint galaxies at high-redshift is
the main objective of this paper. 
\\

We also investigate whether GLFs are universal or depend on
environment. This can help us determine whether the red cluster galaxy
population originates from the field at higher redshift. Many studies
find a universal GLF that does not depend on environment
\citep{Lugger86, Lugger89, Colless89, Gaidos97, Rauzy98, Trentham98,
  Paolillo01, Yagi02, Andreon04}, while others \citep{Dressler78,
  Loper97, Lumsden97, Valotto97, Driver+98a, Garilli+99, Goto+02,
  deprop03, Christlein03, Popesso06, Muzzin08, Rud09} observe
differences between clusters and field GLFs. The most widely observed
trend is a flattening of the GLF as the environment becomes less dense
(see e.g. \citet{deprop03} for observations and \citet{Lanzoni05} for
simulations). This behaviour could be explained by either star
formation being inhibited in dense environments \citep{Tully02,
  deprop03, Muzzin08} or merging processes being more common in the
field where the relative velocities of galaxies are lower
\citep{Menci+02}. This last explanation does not apply to single
objects falling onto groups of galaxies, which can trigger large amounts
of star formation \citep[e.g.][]{Adami09}. In addition, we note that
\citet{Il05} find a steepening of the faint end of the field GLF with
increasing redshift such that they do not see the usual flattening of
the field GLF at high-redshift.

Additional support for the GLF dependence on the environment is the
perturbation of the GLF caused by cluster merging
\citep[e.g.][]{Durret10}. Finally, \citet{deprop03} and \citet{Boue08}
find different GLFs for cluster cores and outskirts. The first authors
find an excess of bright galaxies in cluster cores and the second a
steeper faint end in the outskirts.
\\

The Dark energy American French Team (DAFT, in French FADA) survey is
ideal for investigating the faint end of the GLF and field to cluster
differences at relatively high-redshift. The DAFT/FADA survey
encompasses $\sim 90$ high-redshift ($0.4 \leq z \leq 0.9$) massive
(M$\geq 2\times 10^{14}$~M$_\odot$) clusters of galaxies with Hubble
Space Telescope (HST) imaging available, and multi-band optical and
near infrared ground based imaging, using 4m class telescopes, that is
now almost complete.  The main goals of the survey are to form a
comprehensive database to study clusters and their evolution, and to
test cosmological constraints geometrically by means of weak lensing
tomography. In addition to the first DAFT/FADA paper establishing the
reference basis for the photometric redshift (hereafter photo$-z$)
determination \citep[][ hereafter G10]{Gue10}, results concerning
several topics have been obtained by the DAFT/FADA team. The current
status of the survey, with a list of refereed publications, can be
found at \url{http://cencosw.oamp.fr/DAFT/index.php}.
\\

An outline of the paper is as follows; we first present the
photo-$z$ measurements and the improvements that we have
made since G10. We then describe our method for computing GLFs. We
present the optical GLFs for 31 clusters of the DAFT/FADA survey in
the $0.4\leq z<0.9$ redshift range for the B, V, R, and I rest-frame
bands. The cluster membership of galaxies is based on photo$-z$s
computed with U or B, V, R, I, Z, and J or Ks band data and a field
subtraction. We take special care to estimate the completeness of our
data, and we show that the GLFs are strongly correlated to the 90\%
completeness limit. We investigate average cluster behaviours by
stacking them and discuss the dependence of GLFs on cluster redshifts,
masses, and substructures. We compare the GLF behaviour in the cluster
core and outskirts. We also separate blue and red-sequence galaxies to
investigate the evolution of cluster different galaxy populations.
Finally, we compare our cluster GLFs to the field GLFs computed with COSMOS data \citep{Il+09} made in the same redshift
intervals.  We discuss our results in light of the
literature. Throughout the paper, we use the standard cosmological
model with $\Omega_M=0.3$, $\Omega_\Lambda=0.7$, and
$H_0=70$~km~Mpc$^{-1}$~s$^{-1}$.

\section{Photometric redshifts}

\subsection{Context}
\label{sec:context}

We measure our photo-$z$s as in G10, with the LePhare package
\citep[e.g.][]{Arnouts+99, Il06}. We refer the reader to these papers
for details but we provide here the salient points of the technique.
The aim of the method is to compare observed magnitudes with predicted
ones created by templates, in order to estimate the redshift and other
parameters such as the photometric type. This type varies between 1 and 31
with the chosen templates (see below). Numbers 1--7 correspond to
early type galaxies, numbers 8--12 to early spiral galaxies, numbers
13--19 to late spiral galaxies, and numbers 20--31 to very blue
galaxies. The last category corresponds to very blue templates which have
been generated to compensate for the lack of very blue templates in
\citet{Polletta+06, Polletta+07}.

In a similar way to G10, we select spectral energy distributions
(hereafter SEDs) with emission lines from \citet{Polletta+06,Polletta+07}, with a Calzetti et al. extinction law
\citep[e.g.][]{Calzetti99} applied to different galaxy classes (see
below).

The available spectroscopic redshift catalogues are another important
ingredient (as in G10) of our calculations. As LePhare is able to
estimate possible shifts in photometric zero-points by comparing
photometric and spectroscopic redshifts (used as training sets), this
allows us to compensate for the various origins of our ground-based
images. We collected spectroscopic catalogues for all clusters in the
present paper.

\subsubsection{Input magnitudes}

The first difference from G10 is the photometric bands that we
used. As already demonstrated, having near infrared bands is mandatory
to obtain a robust estimate of photo-$z$s at $z \geq 1$. In G10, we used space based IRAC data in the infrared. We did not
do so in the present paper for the following reasons:

- These data are unavailable for the entire sample presented here.

- The angular resolution of IRAC is very poor compared to regular
ground based data (typically 4 times worse) and this forced us in G10
to estimate correcting factors in order not to be biased. These
factors were typically up to 1.5 magnitudes for small objects.

- Another problem is the small angular extent of our
clusters, for which typical galaxy-galaxy separations are often smaller
than the IRAC spatial resolution, leading to considerable confusion
in the central parts of clusters.

- The IRAC bands are very red (3.6 and 4.5~$\mu$m) compared to the
reddest optical ground based images at our disposal (typically the z'
band at 0.9~$\mu$m), leading to a large wavelength gap, and
making constraints on redshifts rather poor.

Here, we choose to use J and/or K band data instead of IRAC data. The
typical seeings vary between 0.7 and 1.2 arcsec. Simulations similar
to the ones we performed in G10 (see their Fig. 9) show that the
shifts induced by the different spatial resolutions will be of the
order of 0.1 to 0.2 magnitudes, which can be easily compensated for by the
capacity of LePhare to adapt the photometric zero-points when
spectroscopic redshift catalogues are available.

\subsubsection{Image registration}

Our data reduction procedure uses the Scamp and Swarp packages
\citep{Bertin02, Bertin06} and is identical to that in G10. We
produce calibrated median images with cosmic rays and other image
defects removed.

The second difference from G10 resides in the image registration
between different bands. In G10, we considered data acquired by only
three different instruments (IRAC, ESO/FORS2, and CTIO/MOSAIC). It was
therefore possible to  align precisely all the images and to extract
magnitude catalogues with SExtractor \citep{sex} in double-image
mode.

\begin{figure}[ht!]
 \centering
 \includegraphics[angle=270,width=0.48\textwidth,clip]{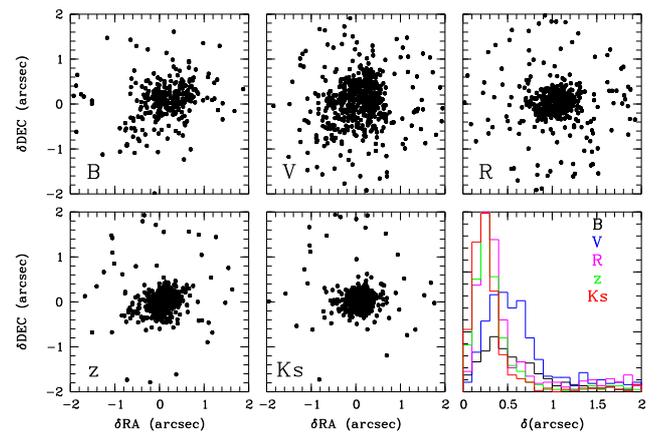}%
 \caption{Astrometric shifts ($\alpha$, $\delta$) for CL~J0152.7-1357
   between the objects detected in the i' band and the B, V, R, z', and Ks
   bands. The figure at the lower right shows the histograms of the shifts
   between the i' band and the other bands.}
\label{fig:C1}
\end{figure}

In the present paper, we consider more than three times as many
clusters (31 clusters, compared to 10 in G10) for which
we collected near infrared data, in addition to optical
band data. Owing to the relatively deep nature of our catalogues (with a
typical 90\% completeness limit of I$\sim$24 for stellar objects),
this represents a large survey, gathering about 350 hours of
observations in both hemispheres on 4m class telescopes. Given also
that we used all possible  images available in public databases to
minimize the amount of new data to be acquired, our project involves
a wide range of very different ground-based data: we use data obtained
with about 10 different telescopes and more than 12 different cameras
(see Table~\ref{tab:listamas}). Although we reduced all the data from its
raw form to ensure that the final imaging products are as homogeneous
as possible, it is impossible to always have image astrometry more
precise than 0.5~arcsec everywhere in the fields. Our final images are
sometimes still plagued by high frequency astrometric differences of
this order. As an example, we show in Fig.~\ref{fig:C1} the
astrometric diagrams for CL~J0152.7-1357, for which we collected B, V,
R, i', z', and Ks data at SOAR (SOI), Subaru (Suprime), and ESO
(HawkI). We see that for a non negligible number of objects, the
astrometric shift is larger than 0.5 arcsec. However, the
astrometry of sources in the data of CL~J0152.7-1357 is among the
poorest of all our collated data.

\subsubsection{Magnitude homogenization}

As already mentioned, we have very different sources for our images,
in contrast to G10.  It is obviously uninteresting to compute galaxy
luminosity functions in several bands that vary from one cluster to
another. We therefore choose to take advantage of our spectroscopic
catalogues and the ability of LePhare to compute magnitude 
zero-point shifts. This allows us to convert our various magnitudes homogeneously into a
single system (the $common~system$ in
the following) that we choose to be VLT/FORS2 B, V, R, I, z' (AB
system), and VLT/HawkI J or Ks band (Vega system). These filter
  shifts are applied after the completion of the photo-$z$
  calculations.

We therefore classify our images into several general classes. All U-
and B-like magnitudes are translated into VLT/FORS2 B, all V- and
g'-like magnitudes into VLT/FORS2 V, all R- and r'-like magnitudes
into VLT/FORS2 R, all I- and i'-like magnitudes into VLT/FORS2 I, all
z'-like magnitudes into VLT/FORS2 z', all J-like magnitudes into
VLT/HawkI J band, and all Ks-like magnitudes into VLT/HawkI Ks band.
In the particular case of Abell~851, we also consider the
CFHT/WIRCAM Y and H bands directly.

Since we cannot use SExtractor in double image mode here, we choose to
apply it in single image mode, computing total magnitudes (MAG\_AUTO)
in each of the considered bands. We then cross-correlate the different
catalogues to generate a final catalogue including all magnitudes for
all objects, with an identification distance of 2~arcsec and a
  minimization of this distance when several objects are within the same
  radius. This is almost twice the maximum astrometric difference
observed between the different bands. We checked that the results obtained with this
  correlation method do not differ considerably from those of a
  double image mode detection by comparing the results of both methods for clusters
  with imaging of good astrometry acquired with the same camera in the 5
  optical bands. Both methods agree well except for very faint
  galaxies, which are detected in larger numbers in double image mode owing to the use
  of the deepest band (the i band) as the reference detection image for the
  double image mode. However, this only concerns objects far below
  the completeness limit of our images. In some cases, there is also a
  small difference at the bright end of the magnitude distribution because
  of foreground objects larger than our 2 arcsec criterion.

  We also varied the MAG\_AUTO minimum aperture radius from 3.5 to 1.5
  pixels to verify that we were not missing light in faint objects, as
  explained in \citet{Rud09}. We did not find any significant
  variation in the magnitude distribution between these two radii.

\subsection{Optimization and estimate of the LePhare performances}

\subsubsection{Zero-point shifts}

\begin{figure}[ht!]
 \centering
 \includegraphics[angle=270,width=0.48\textwidth,clip]{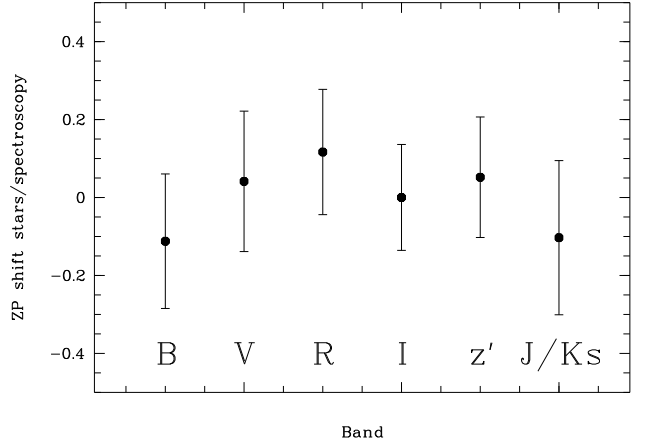}    
 \caption{Mean and uncertainty in the difference between the star and
   galaxy spectroscopic redshift-based shifts for the different
   magnitude bands considered over our 31 fields.}
  \label{fig:C2}
\end{figure}

\begin{figure}[h!]
 \centering
 \includegraphics[angle=270,width=0.48\textwidth,clip]{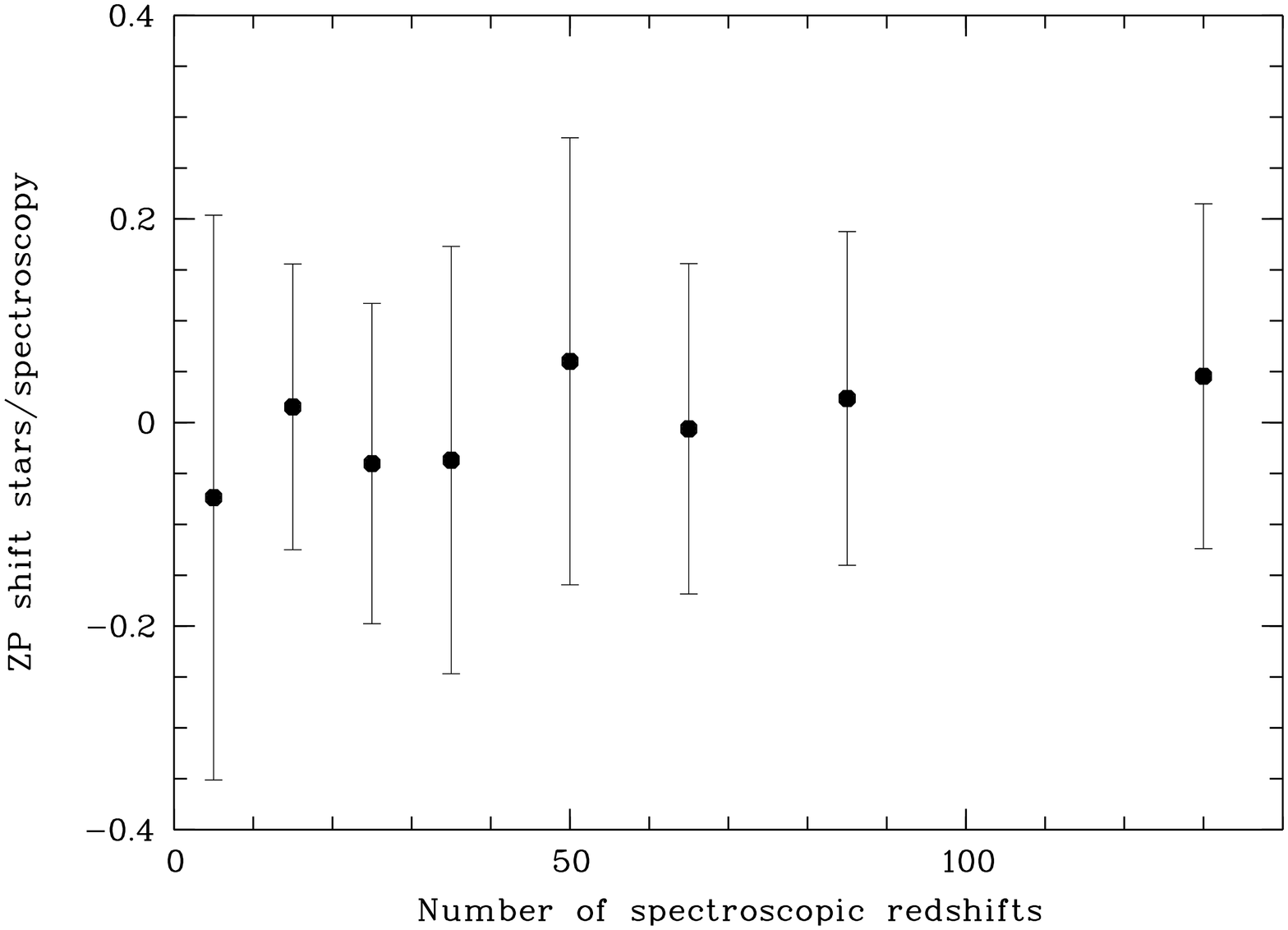}  
 \includegraphics[angle=270,width=0.48\textwidth,clip]{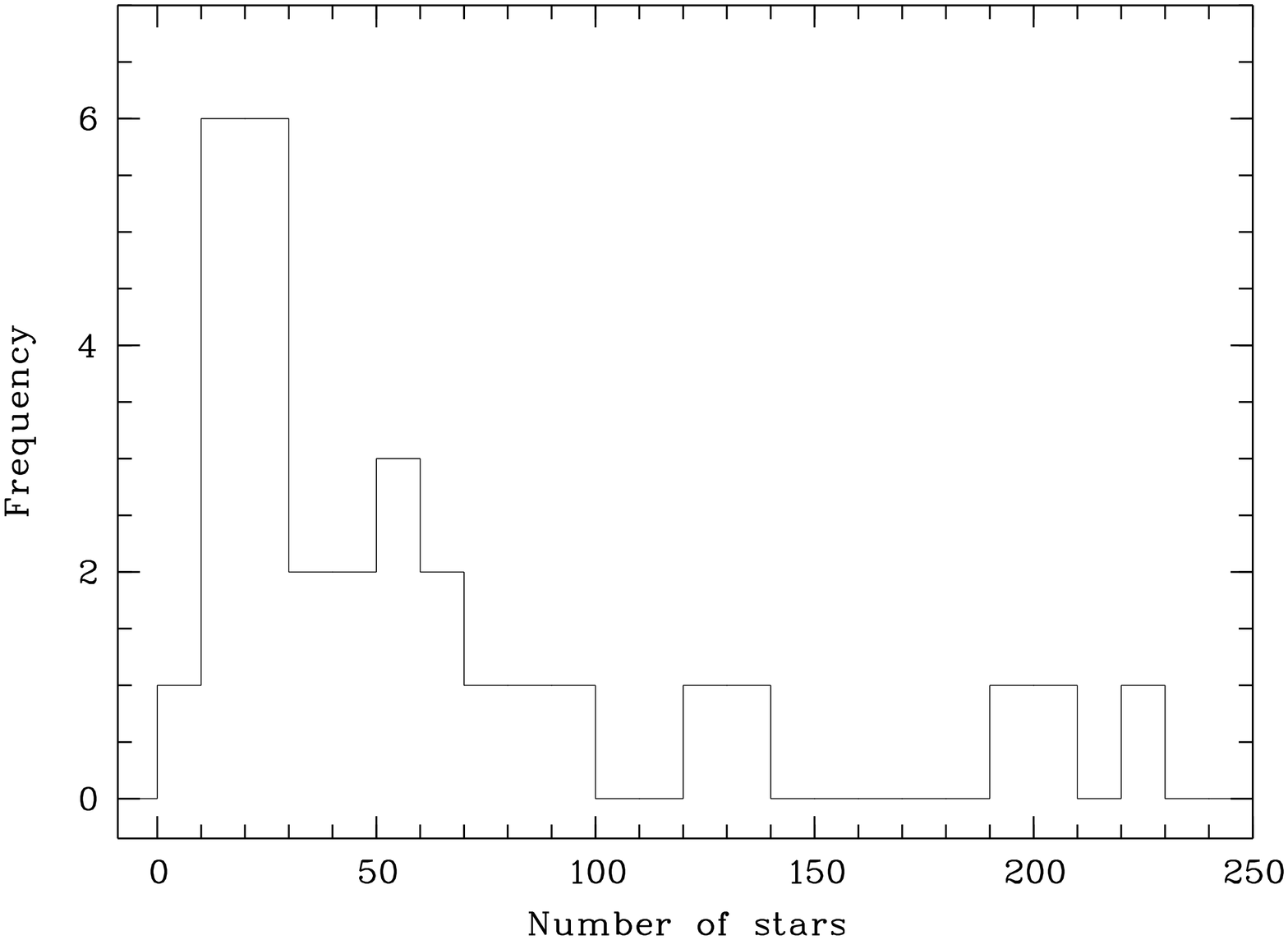}  
 \caption{Top: Mean and uncertainty on the difference between the star
   and galaxy spectroscopic redshift-based shifts in the I band versus
   the number of spectroscopic redshifts available over our 31
     fields. Bottom: Histogram of the number of stars used for the
   calibration of each cluster.}
  \label{fig:C3}
\end{figure}

We entered catalogues of galaxy spectroscopic redshifts into LePhare
to correct for small zero-point variations and compute zero-point
shifts when computing magnitudes in our $common~system$. We obviously
need to estimate the typical errors induced by this process.

The zero-point shift computation in LePhare is a complex interplay
between the selected templates, the considered redshifts,
and the selection function of the spectroscopic catalogue,
which is almost always impossible to compute precisely, owing to the
wide range of origins of our spectroscopic redshifts. As a
consequence, the only possible way to estimate the errors induced by
the LePhare zero-point shift computation is to consider several
catalogues of objects for which the redshifts are known.

\begin{landscape}
\begin{table}
  \caption{Data used in our present study. (O) represents observed data and (A) data taken from
    archives. A * indicates that the cluster was only partially observed in the
    field. The last column states whether we were able to calculate a luminosity
    function. The LCDCS clusters come from \citet{Gonzalez+01}.
    For clarity, we display an abbreviated name of cameras used,
    such that WIYN/M corresponds to WIYN/MiniMo, SOAR/S to SOAR/SOI,
    CFHT/M to CFHT/Megacam, CFHT/W to CFHT/WIRCAM, CFHT/C to CFHT/CFH12K,
    VLT/F1 and VLT/F2 to VLT/FORS1 and VLT/FORS2, Subaru/S to
    Subaru/Suprime, Subaru/M to Subaru/MOIRCS, CTIO/M to CTIO/MOSAIC,
    GTC/O to GTC/Osiris, ESO/H to ESO/HawkI TNG/N to TNG/NICS and MDM/R to
    MDM/Red4K. In addition, clusters XDCS\_cm\_J032903.1+025640 and BMW-HRI\_J122657.3+333253 
    are respectively abbreviated to XDCS\_cm\_J032903.1 and BMW-HRI\_J122657.3.}

\centering
\begin{tabular}{lllllllllll}
  \hline
  \hline
  Cluster & RA  & Dec. & z & U/B & V & R & I & Z & J/Ks & LF\\ 
 \hline 

CL\_0016+1609 	           & 00 18 33.33 & +16 26 35.84 & 0.5455 & WIYN/M B (O)    & WIYN/M V (O)     & WIYN/M r' (O)    & WIYN/M i' (O)   & WIYN/M z' (O)    & Subaru/M Ks (A) & Y \\
CL\_J0152.7-1357           & 01 52 40.99 & -13 57 45.00& 0.8310  & SOAR/S B (O)       & Subaru/S V (A)  & Subaru/S R (A)  & SOAR/S I (O)       & Subaru/S z' (A) & ESO/H Ks (A)     & N \\
PDCS\_018 	           & 02 27 25.50 & +00 40 04.00 & 0.4000 & CFHT/M u (A)   & CFHT/M g' (A)  	& SOAR/S r' (O)  	& CFHT/M i' (A)  & CFHT/M z' (O)   & CFHT/W J (O)    & Y \\
XDCS\_cm\_J032903.1       & 03 29 02.81 & +02 56 25.18 & 0.4122 & CFHT/M u (A)   & CFHT/M g' (A) 	& CFHT/M r' (A) 	& SOAR/S i'  (O)     & SOAR/S z'  (O)      & CFHT/W J (O)    & Y \\
F1557.19TC                & 04 12 54.69 & -65 50 57.58 & 0.5100  & VLT/F2 B (A)      & VLT/F2 V (A) 	& VLT/F2 R (A)  	& VLT/F2 I (A)      & SOAR/S z'  (O)      & ESO/H Ks (A)     & N \\
MACS-J0454.1-0300          & 04 54 10.92 & -03 01 07.14 & 0.5377 & VLT/F2 U (A)      & VLT/F2 V (A)  	& VLT/F2 R (A)  	& VLT/F2 I (A)      & Subaru/S z' (A) & ESO/H Ks (A)     & Y \\
MACS\_J0647.7+7015         & 06 47 45.89 & +70 15 02.98 & 0.5907 & WIYN/M B (O)    & WIYN/M V (O) 	& WIYN/M R (O) 	& WIYN/M I (O)    & WIYN/M z' (O)    & TNG/N J (O)       & N \\
MACS\_J0744.9+3927         & 07 44 51.79 & +39 27 33.01 & 0.6860 & WIYN/M B (O)    & WIYN/M V (O) 	& WIYN/M R (O) 	& WIYN/M I (O)    & WIYN/M z' (O)    & Subaru/M J (A)  & N \\
RX\_J0848.8+4455           & 08 48 49.30 & +44 55 45.98 & 0.5430 & Subaru/S B (A) & Subaru/S V (A) 	& Subaru/S R (A) 	& Subaru/S I (A) & Subaru/S z' (A) & MDM/R Y (O)      & Y \\
ABELL\_0851                & 09 42 56.64 & +46 59 21.91 & 0.4069 & CFHT/M u (A)   & CFHT/M g' (A)  	& CFHT/M r' (A)  	& CFHT/M i' (A)  & CFHT/M z' (A)   & CFHT/W Y (A)    & Y \\
LCDCS\_0130                & 10 40 41.59 & -11 55 50.98 & 0.7043 & CTIO/M B (O) 	& VLT/F2 V (A)       & VLT/F2 R (A)       & VLT/F2 I (A)      & VLT/F2 z' (A)      & ESO/H Ks (A)     & Y \\
SEXCLAS\_12                & 10 52 38.20 & +57 30 49.28 & 0.6100 & WIYN/M B (O)*   & WIYN/M V (O)*    & WIYN/M R (O)*    & WIYN/M I (O)*   & CFHT/M z' (O)   & MDM/R Y (O)      & N \\
LCDCS\_0173                & 10 54 43.50 & -12 45 50.00 & 0.7498 & CTIO/M B (O) 	& VLT/F2 V (A)       & VLT/F2 R (A)       & VLT/F2 I (A)      & VLT/F2 z' (A)      & ESO/H Ks (A)     & Y \\
MS\_1054-03                & 10 57 00.22 & -03 37 27.40 & 0.8231 & VLT/F1 B (A)  	& VLT/F1 V (A) 	& VLT/F2 R (A)  	& VLT/F2 I (A)      & GTC/O z'(O)      & ESO/H J (A)      & Y \\
RXC\_J1206.2-0848          & 12 06 11.97 & -08 48 00.03 & 0.4400 & Subaru/S B (A)	& VLT/F1 V (A)  	& CFHT/M r' (A)  	& Subaru/S I (A) & VLT/F2 z' (O)      & TNG/N J (O)       & Y \\
LCDCS\_0504               & 12 16 45.10 & -12 01 17.00  & 0.7943 & CTIO/M B (O) 	& VLT/F2 V (A)       & VLT/F2 R (A)       & VLT/F2 I (A)      & VLT/F2 z' (A)      & ESO/H J (A)      & Y \\
BMW-HRI\_J122657.3  & 12 26 58.00 & +33 32 54.09 & 0.8900 & Subaru/S B (A) & Subaru/S V (A)  & Subaru/S R (A)  & Subaru/S I (A) & Subaru/S z' (A) & Subaru/M J (A)  & Y \\
LCDCS\_0531                & 12 27 53.89 & -11 38 20.00 & 0.6355 & CTIO/M B (O) 	& VLT/F2 V (A)       & VLT/F2 R (A)       & VLT/F2 I (A)      & VLT/F2 z' (A)      & ESO/H Ks (A)     & Y \\
HDF:ClG\_J1236+6215        & 12 37 59.99 & +62 15 54.00 & 0.8500 & CFHT/M u (A)   & CFHT/C V (A)  	& Subaru/S R (A) 	& CFHT/C I (A)    & Subaru/S z' (A) & CFHT/W J (A)    & Y \\
MJM98\_034                 & 13 35 13.78 & +37 48 56.30 & 0.5950 & CFHT/M u (A)  	& CFHT/M g' (A)	& Subaru/S R (A) 	& CFHT/M i' (A)  & Subaru/S z' (A) & CFHT/W J (A)    & Y \\
LCDCS\_0829                & 13 47 31.99 & -11 45 42.01 & 0.4510 & VLT/F1 B (A) 	& VLT/F1 V (A)  	& VLT/F1 R (A)  	& VLT/F1 I (A)      & CFHT/M z' (A)   & CFHT/W J (A)    & Y \\
LCDCS\_0853                & 13 54 09.49 & -12 30 59.00 & 0.7627 & CTIO/M B (O) 	& VLT/F2 V (A)       & VLT/F2 R (A)       & VLT/F2 I (A)      & VLT/F2 z' (A)      & ESO/H J (A)      & N \\
3C\_295\_CLUSTER           & 14 11 20.15 & +52 12 09.03 & 0.4600 & CFHT/M u (A)  	& CFHT/M g' (A)  	& CFHT/M r' (A) 	& CFHT/M i' (A)  & CFHT/M z' (A)   & CFHT/W J (O)    & Y \\
MACS\_J1423.8+2404         & 14 23 48.29 & +24 04 46.99 & 0.5450 & Subaru/S B (A) & Subaru/S V (A) 	& Subaru/S R (A) 	& Subaru/S I (A) & Subaru/S z' (A) & CFHT/W Ks (A)   & Y \\
GHO\_1601+4253             & 16 03 13.82 & +42 45 36.17 & 0.5391 & Subaru/S B (A) & Subaru/S V (A) 	& Subaru/S R (A) 	& Subaru/S I (A) & CFHT/M z' (O)   & CFHT/W J (O)    & Y \\
GHO\_1602+4312             & 16 04 25.15 & +43 04 52.71 & 0.8950 & Subaru/S B (A) & Subaru/S V (A) 	& Subaru/S R (A)  & Subaru/S I (A) & CFHT/M z' (O)   & Subaru/M J (A)  & N \\
MACS\_J1621.4+3810         & 16 21 23.99 & +38 10 01.99 & 0.4650 & CFHT/M u (A)	& Subaru/S V (A)	& Subaru/S R (A) 	& Subaru/S I (A) & CFHT/M z' (O)   & CFHT/W J (O)    & Y \\
MACS\_J1621.6+3810         & 16 21 35.99 & +38 10 00.01 & 0.4610 & CFHT/M u (A) 	& Subaru/S V (A)	& Subaru/S R (A) 	& Subaru/S I (A) & CFHT/M z' (O)   & CFHT/W J (O)    & Y \\
MACS\_J2129.4-0741         & 21 29 25.99 & -07 41 27.99 & 0.5889 & Subaru/S B (A)	& Subaru/S V (A)  & SOAR/S r' (O)       &  SOAR/S i' (O)     & SOAR/S z' (O)       & CFHT/W Ks (A)   & Y \\
GHO\_2143+0408             & 21 46 04.79 & +04 23 18.99 & 0.5310 & WIYN/M B (O)	& WIYN/M V (O)	& WIYN/M R (O)	& SOAR/S i' (O)      & VLT/F2 z' (O)      & CFHT/W J (O)    & Y \\
GHO\_2155+0334             & 21 57 55.37 & +03 47 51.53 & 0.4500 & VLT/F2 B (O)	& VLT/F2 V (O)	& VLT/F2 R (O)	& VLT/F2 I (O)      & SOAR/S z' (O)  & CFHT/W J (O)  & N \\
  \hline 
  \hline

\end{tabular}
\label{tab:listamas}
\end{table}
\end{landscape}

We could have divided our galaxy spectroscopic catalogues into several
sub-samples and checked the robustness of the resulting magnitude zero
point shifts. However, this would only have been possible for a few
clusters for which we have a sufficient number of spectroscopic
redshifts. We therefore choose another approach, considering the only
other object class for which the redshifts are known: the stars in our
fields.

We select stars with both ground-based and space-based HST
data. This is done by plotting all the detected objects in diagrams of
central surface brightness versus total magnitude. Space-based data
allow us to detect very faint stars albeit in rather limited sky areas,
while ground-based data allow us to detect relatively bright stars across
larger areas of the sky. By applying LePhare to these star catalogues and fixing
the redshifts to 0, we can compute zero-point shifts for these star
catalogues. The same is done for the catalogues of galaxies with a
known spectroscopic redshift, giving us a second estimate of the zero
point shifts.

Both shifts have no reason to be identical, as we consider in
one case only stellar templates (which are not adapted to our galaxy
catalogues) and in the other case galaxy templates at various
redshifts. However, we expect not to obtain dramatically different
values, since galaxy templates are theoretically nothing but
combinations of stellar templates.

In Fig.~\ref{fig:C2}, we show the mean value and the uncertainty in the
difference between these two shifts for the various magnitude bands
considered. As expected, the mean differences are always smaller than 0.2~magnitudes.
 Similarly, typical uncertainties in the mean differences
are of the order of 0.2~magnitude. The numbers of spectroscopic
redshifts and stars used in the calibration are given in
Fig.~\ref{fig:C3}. On the one hand, this shows that we cannot exclude
the hypothesis that the two shifts are the same, whatever the
photometric band. On the other hand, this also means that given the
uncertainties in the differences, it would be incorrect to use only
star catalogues to estimate the zero-point shifts of our galaxy
catalogues. A 0.2~magnitude shift is indeed large enough to induce
significant errors in our photo-$z$ estimates (see e.g. G10). 
  Hence, star-based zero-point shifts were only used to roughly assess the
  spectroscopic redshift based shifts.

We also test how varying the number of spectroscopic redshifts affects
our photo-$z$ estimates. In Fig.~\ref{fig:C3}, we plot the mean and
uncertainty in the difference of the two shifts in the I band for
different numbers of spectroscopic redshifts available in the
considered catalogues. Except for the very sparse spectroscopic
catalogues ($\leq$10 redshifts) with uncertainties of about
0.3~magnitudes, all shift differences are consistent with zero and all
uncertainties are smaller than 0.2~magnitudes. We therefore choose to
consider only clusters for which we have at least 10 spectroscopic
redshifts along the line of sight.

To conclude, our method
of translating all our magnitudes to a $common~system$ is robust, but
cannot be efficiently applied without spectroscopic catalogues of at
least 10 galaxies per cluster.

\subsubsection{Extinction and photometric redshifts}

One of the main results of G10 was that the precision of our photometric
redshifts was sometimes degraded by a factor of two when considering
cluster galaxies \citep[see also][]{Adami11}. This is probably due to
a lack of galaxy templates typical of high density regions. Even the
reddest galaxy templates are sometimes not red enough. Indeed, the
mean type of cluster galaxies is 15 when taking all galaxies into
account and 22 when taking only galaxies with photo-$z$s differing
from spectroscopic redshifts by more than $1\sigma$. These numbers
highlight the lack of red templates (templates get bluer from type 1
to 31), and force LePhare to increase the galaxy redshifts.

In this framework, we note that in G10, we allowed LePhare to include
extinction only in spiral galaxies. This was in good agreement with
the galaxy properties generally observed, even though
early type galaxies are not always unobscured
\citep[e.g.][]{Martini+13}. Here, we allow LePhare to artificially
introduce extinction in early type galaxies, permitting
galaxy templates to become redder. We are aware that this artificial
  extinction might not be physical but we choose to apply it
  nonetheless because it significantly improves the accuracy of our
  photo-$z$s.

\begin{figure}[h!]
 \centering
 \includegraphics[angle=270,width=0.48\textwidth,clip]{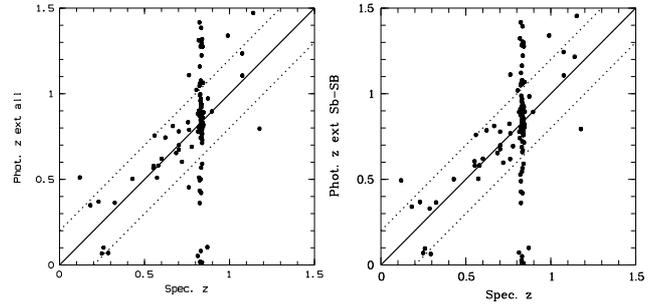}   
 \caption{Photometric versus spectroscopic redshifts when allowing
   extinction in all galaxies (left) or only in spiral galaxies
   (right) for the MS1054-03 field. The full lines show the lines of 
     equality of the photometric and spectroscopic redshifts and the
     dotted lines show the scatter at $\pm0.2$. See text for details.}
  \label{fig:C4}
\end{figure}

To illustrate the effect of this approach, we show the
example of MS~1054-03 at redshift $z=0.8231$. The extinction in the I
band is $0.3\pm0.3$ magnitude for early types and $0.4\pm0.3$
magnitude for late types.  Fig.~\ref{fig:C4} shows the photometric
versus spectroscopic redshifts when allowing extinction only in spiral
galaxies and in all galaxies. We achieve a higher
  photometric redshift accuracy when allowing extinction also
  in early type galaxies. More quantitatively, the dispersion in the
  mean difference between the photometric and spectroscopic redshifts
  of cluster members is $\sigma_{|z_{phot}-z_{spec}|}=0.20$ when
  allowing extinction in early type galaxies, while it is 0.30 when
  allowing extinction only in late type galaxies. We therefore
improve the quality of our photometric redshift estimates in the
cluster by $\sim 50\%$ when allowing extinction in early type
galaxies. Outside the cluster, the effect is clearly less evident
because the dispersion in the mean difference between photometric and
spectroscopic redshifts is 0.22 when permitting extinction in early type
galaxies and 0.21 when allowing extinction only in late type galaxies.

Permitting extinction in early type galaxies therefore does not drastically change
photometric redshifts outside the cluster but increases their accuracy by
50$\%$ within the cluster, enabling us to reach the same
precision inside and outside the cluster. Furthermore, the number of
catastrophic errors inside the cluster is reduced by more than 25\%
when extinction is allowed in early type galaxies.

\subsubsection{Photometric redshift quality}

\begin{figure}
 \centering
 \includegraphics[angle=270,width=0.48\textwidth,clip]{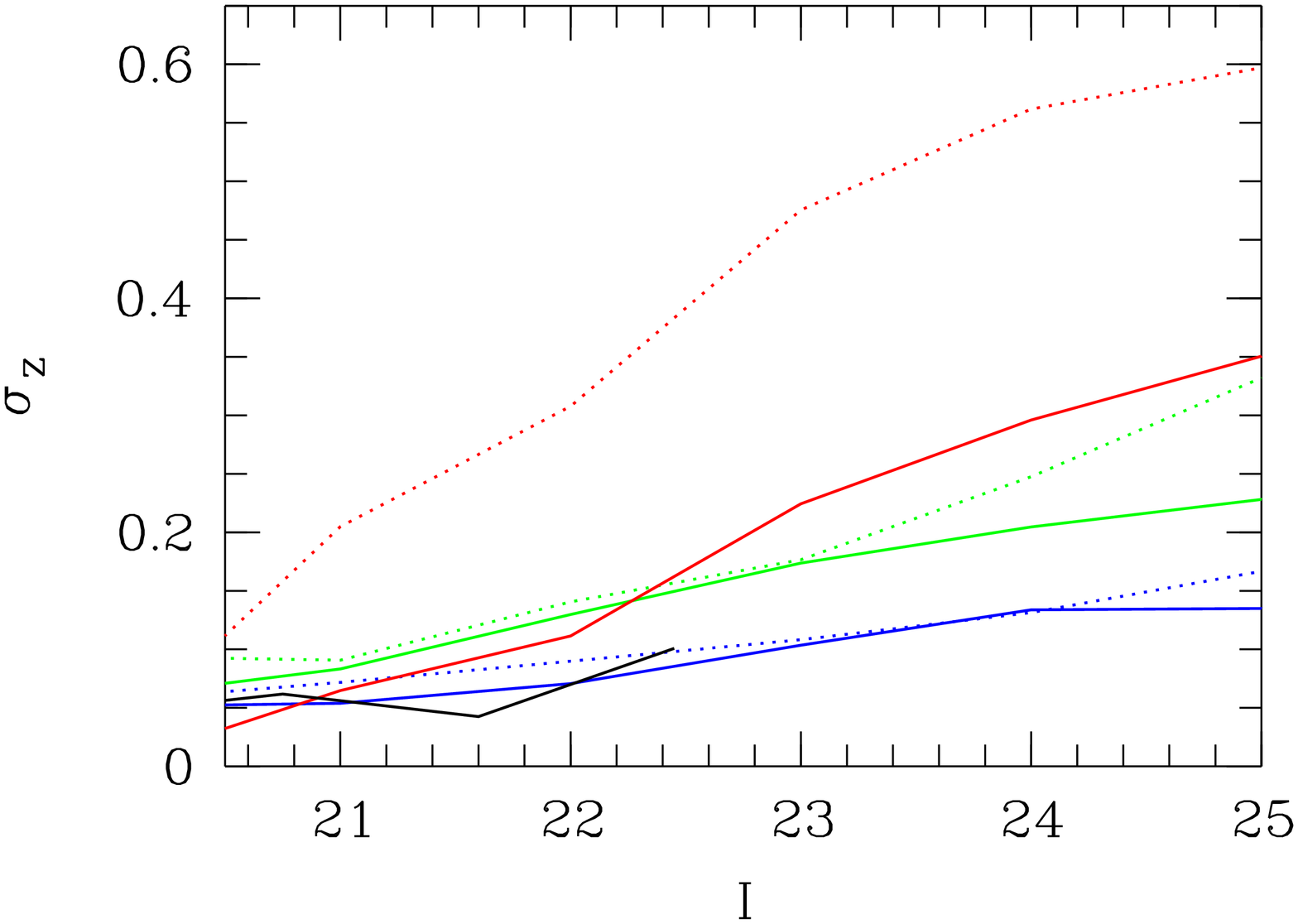} 
 \includegraphics[angle=270,width=0.48\textwidth,clip]{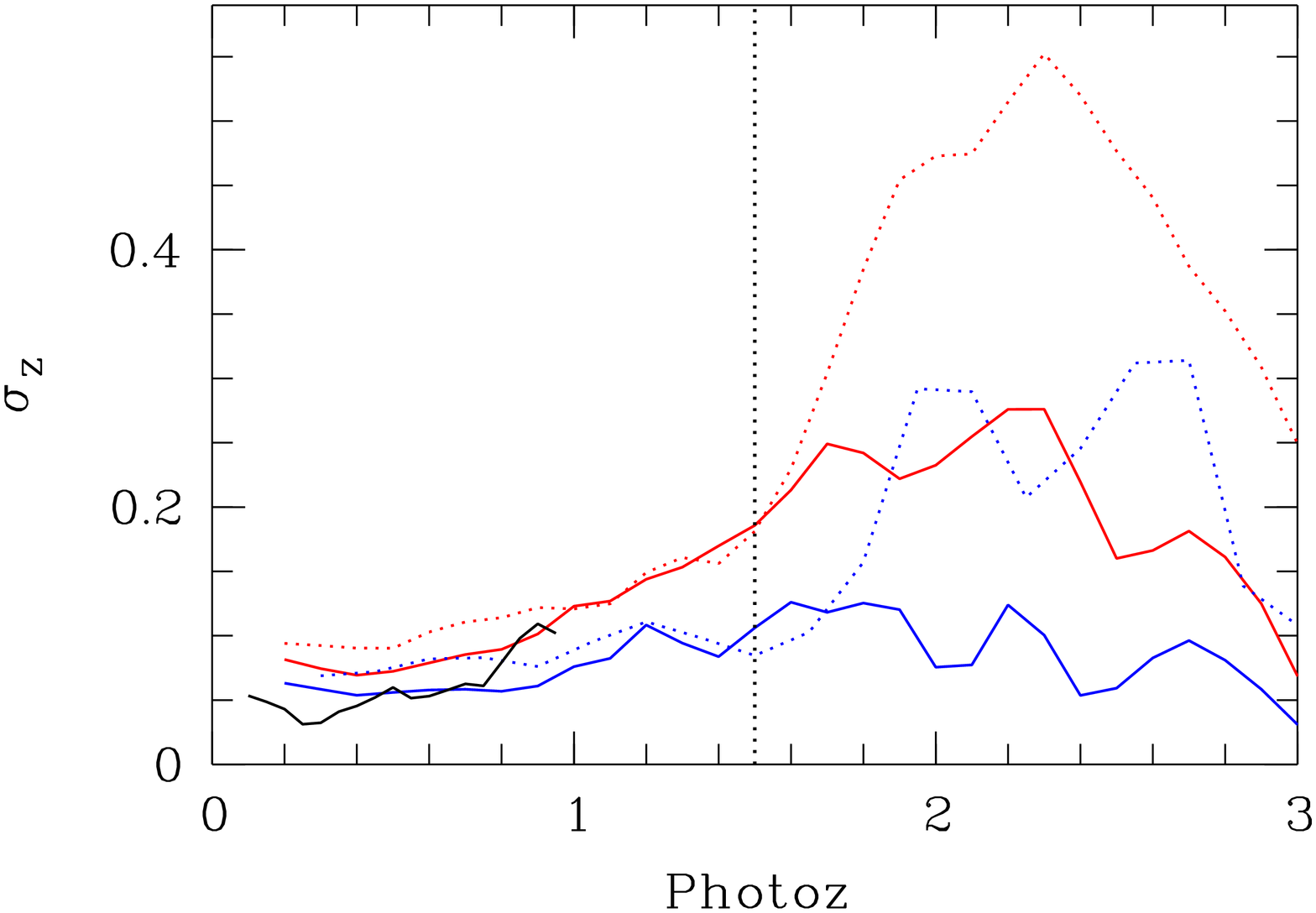}  
 \caption{Top: mean individual photo$-z$ uncertainties as a function
   of I~magnitude in three redshift intervals, colour coded as: blue:
   z = [0.; 1.05], green: z = [1.05; 2.0], red: z = [2.0;
   2.5]. Bottom: mean individual photo$-z$ uncertainties as a function
   of photo$-z$ for various I magnitude intervals, colour coded as:
   blue: I = [19.5; 22.5], red: F814W = [19.5; 24.5]. The vertical
   dotted line shows the z $\leq$ 1.5 limit we suggested to adopt in
   G10. In both plots, dotted lines correspond to G10 and continuous
   lines to the present work. The black curves correspond to the
     CFHTLS with $I<22.5$ and $z<1.05$ and should be compared
     with the blue curves (see text for details).}
  \label{fig:C5}
\end{figure}

We now discuss the quality of our photometric redshifts and compare
our results with those of G10.  The dotted lines in Fig.~\ref{fig:C5}
come from G10 with slightly different redshift and magnitude
intervals.  At z$\leq$1.5 and 
I$\leq$22.5, our photometric redshifts have slightly smaller uncertainties than those of 
G10. The improvement is much more significant at I$\geq$22.5 and
z$\geq$1.5.

 We also compare our photo-$z$s to those of the CFHTLS
  \citep{Coupon09}. To do so, we select cluster galaxies in the
  XXM-LSS survey \citep[e.g.][]{Adami11} with spectroscopic redshifts
  and check their corresponding photo-$z$s in the CFHTLS. We then
  calculate the photo-$z$ uncertainties for this sample ($I<22.5$ and
  $z<1.05$) using LePhare and plot them in Fig.~\ref{fig:C5}. The precision of our photo-$z$s is comparable to that of the
  CFHTLS, and becomes higher for redshifts higher than $z=0.8$ owing
  to the use of near infrared data.

\begin{figure*}
 \centering
 \includegraphics[angle=270,width=0.48\textwidth,clip]{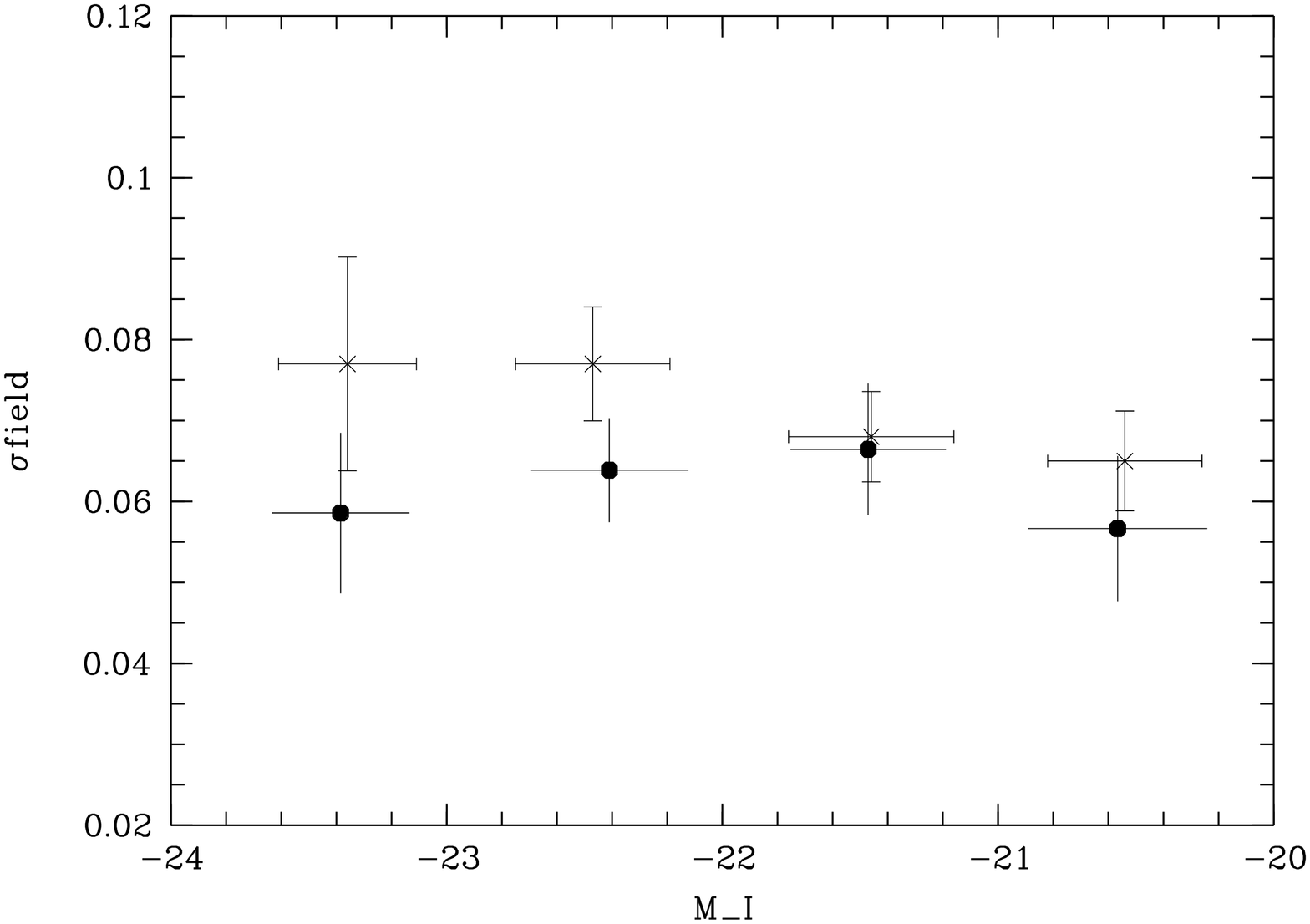}  
 \includegraphics[angle=270,width=0.48\textwidth,clip]{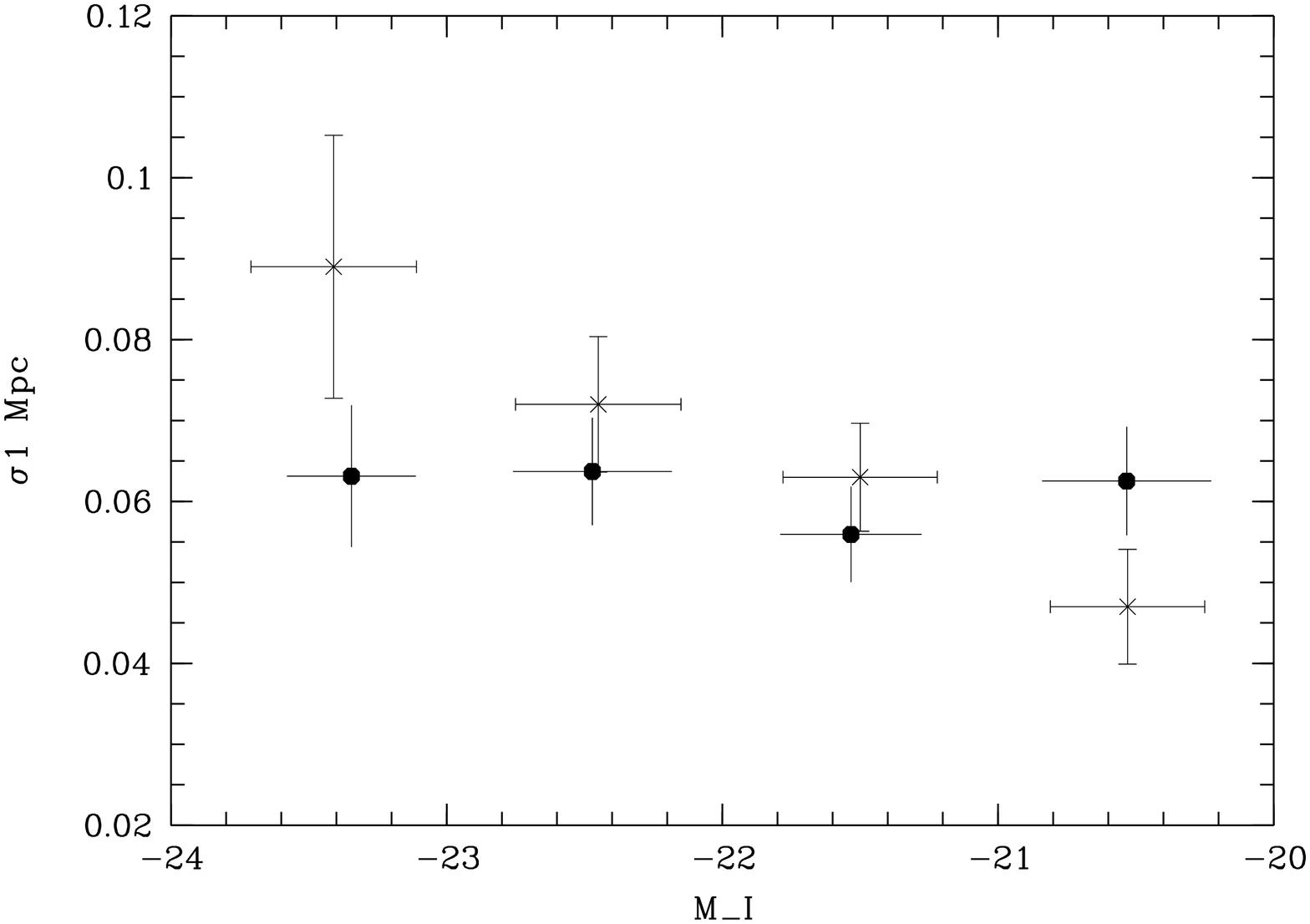} \\ 
 \includegraphics[angle=270,width=0.48\textwidth,clip]{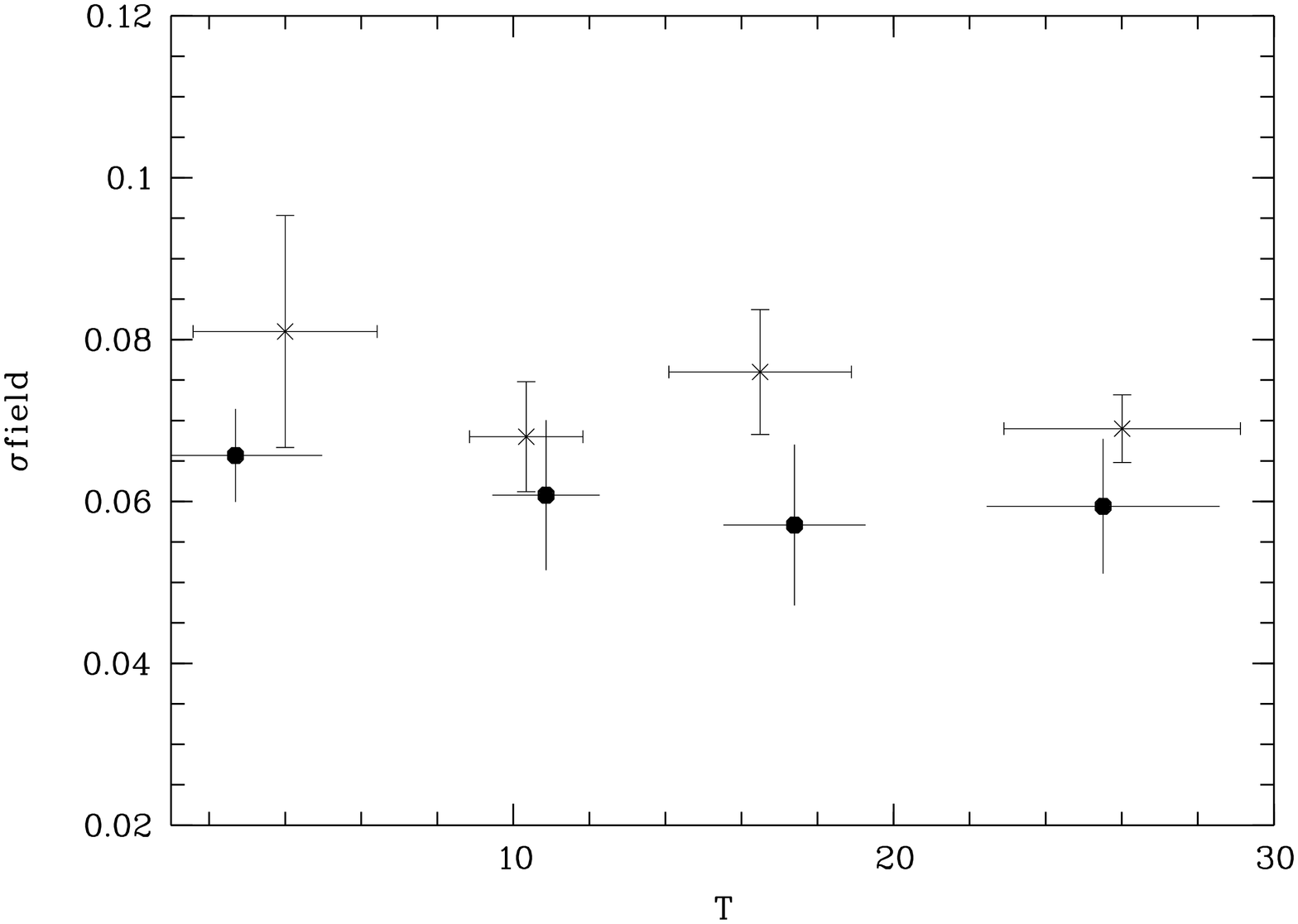}  
 \includegraphics[angle=270,width=0.48\textwidth,clip]{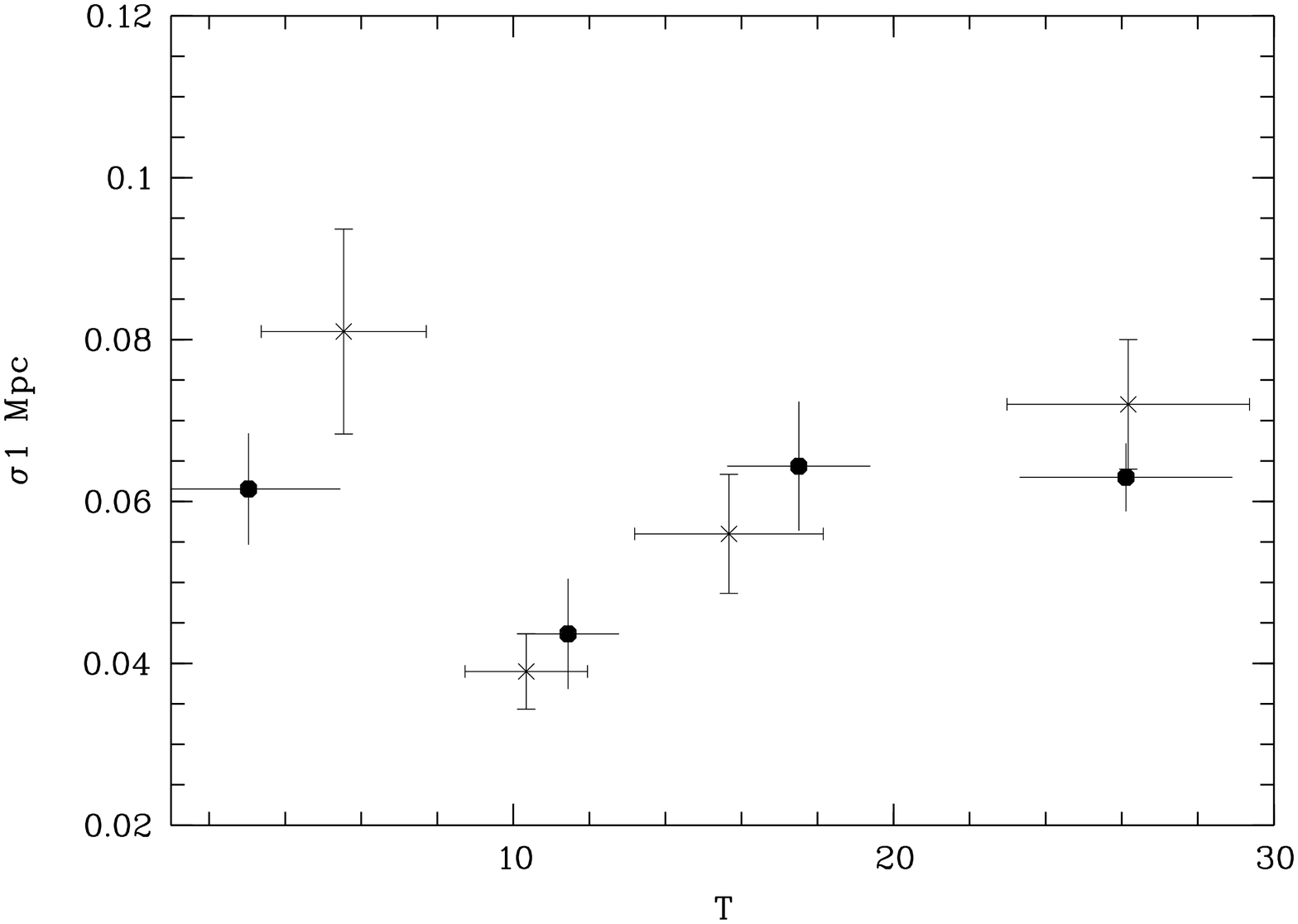}  
 \caption{Reduced $\sigma$ of photo$-z$s versus I band absolute
   magnitude (top) and versus galaxy photometric type T
   (bottom). Left: field galaxies, and right: cluster galaxies inside
   a 1 Mpc radius. Error bars for the reduced $\sigma$ are Poissonian
   and therefore directly proportional to the inverse square root
   of the number of galaxies within the considered bin. Error bars for
   each type are simply the second order momentum of the galaxy type
   distributions in the selected type bins ([1; 7], [8; 12], [13; 19],
   [20; 31]). Crosses correspond to the G10 values, and circles to the
   present values. Shifts in types and magnitudes arise from
     plotting average values for two different samples. Refer to
     section~\ref{sec:context} for details of the photometric types.}
  \label{fig:C6}
\end{figure*}

Similarly, we reproduce Figs.~14 and 15 of G10 in Fig.~\ref{fig:C6},
using data drawn only from spectroscopic catalogues. We find that our present 
photometric redshift computations lead to a modest improvement in the
photometric redshift quality for field galaxies. However, cluster
galaxy photometric redshifts are systematically improved for early
type galaxies, as expected, thanks to the extinction artificially
allowed for such types.

 As a conclusion, the present photometric redshift computations
  allow us to achieve a constant photometric redshift
  precision of $\sigma_z\sim0.06$ for all galaxy environments (field
  or cluster), magnitudes, and galaxy types.

\subsubsection{Improvement achieved by the use of more than one near infrared band}

While the interest of having several near infrared bands is evident at
z$\geq$1, the effect at z$\leq$1 (where all our clusters are) is not
so clear. We test this with Abell~851. This cluster has a comprehensive
range of data, with Y, J, H, and Ks near infrared bands available. Our
spectroscopic redshift catalogue typically extends from
z$\sim$0.2 to z$\sim$0.8, so is perfectly suited to testing the photometric
redshift quality over the entire redshift range covered by our cluster
sample. We therefore compute photometric redshifts for this cluster
by considering, in addition to the optical bands, the z' band, then the
z' and Y bands, the z', Y, and J bands, the z', Y, J, and H bands, and
finally the z', Y, J, H, and Ks bands.

The mean photometric redshift precision (with catastrophic errors removed)
between z=0.2 and 0.8 does not depend significantly on the number of
near infrared bands included. However, the number of completely wrong
photometric redshifts (i.e. for which the difference between the
spectroscopic and photometric redshift is greater than 0.3) tends to
increase when the number of bands decreases. In the case of Abell~851,
this percentage is close to 25$\%$ when at least one band is used among
Y, J, H, or Ks, in addition to B, V, R, I, and z', while it suddenly jumps
to 38$\%$ when only B, V, R, I, and z' are used.

This shows that within the redshift range considered (typically
z$\leq$1), the photo-$z$ accuracy is not significantly improved by collecting data for more than one band
among Y, J, H, or Ks. However, including data for at least one of
these bands will notably diminish the number of catastrophic errors.

\section{Galaxy Luminosity Functions}
\label{sec:LF}

We now compute the B, V, R, and I rest-frame band GLFs for 31 clusters,
using photo$-z$s to estimate the cluster membership of galaxies. 

We consider that a galaxy belongs to the cluster when its photo$-z$
is within a $\pm$ 0.2 interval centred on the cluster redshift. Once
a galaxy is identified as a potential cluster member, we set
its photo$-z$ to the cluster redshift and re-run LePhare to obtain
better estimate of redshift dependent parameters (such as absolute magnitude,
colour, k-correction, etc.). We then subtract galaxy field counts
measured in the same redshift interval using COSMOS data
  \citep{Il+09} to exclude galaxies at $\pm$0.2 from the cluster
redshift that are not cluster members.

When single band catalogues are merged to estimate galaxy photo$-z$s, 
all objects that are not detected in every band are rejected from the
catalogue. Galaxies missed in this approach are mainly very faint
galaxies, and we need to correct for this incompleteness to have the
same number of galaxies in each band than in the original single band
catalogue. Schechter functions are fitted up to the limiting magnitude
at which our galaxy catalogues are 90\% complete. We measure the
completeness independently within each image by inserting and re-detecting
stars simulated with the image point spread function (PSF). We obtain
the 90\% completeness limit in absolute magnitude by applying the
k-correction and distance modulus.

Further details of our analysis are described step by step in the rest
of this section.

\subsection{Completeness}
\label{sec:complet}

For each image, we first measure the PSF by fitting the stars with a
Gaussian light distribution using the PSFEx software
\citep{Bertin11}. With this PSF, we can model a set of Gaussian stars
of various magnitudes. For each bin of 0.5 apparent magnitude, we
simulate a hundred stars, insert them into the image, and try to
re-detect them with SExtractor. The 90\% completeness limit
corresponds to the faintest magnitude bin in which we still re-detect
at least 90 stars. This star completeness limit can be transformed to
an approximate galaxy completeness limit by subtracting 0.5 from the
magnitude \citep[e.g.][]{Adami06}.

In some cases, it is impossible to measure the PSF accurately because
there are too few stars in the field. We then take the magnitude
of the bin just brighter than the peak of the selected band magnitude
histogram to be the 90\% completeness limit. We verified
that in most cases both methods give the same completeness limit
estimate for the clusters for which it was possible to measure the PSF
in the I band. The I band completeness limits for both methods are
equal for 40\% of the clusters and always differ by less than 1
magnitude. The average galaxy 90\% completeness limit of our sample in
the I band is 23.2.

We then translate these apparent magnitude completeness limits to
absolute magnitude completeness limits by applying the k-correction
and distance modulus.  LePhare uses galaxy SED model libraries to
estimate the theoretical k-corrections, which depend on galaxy types
and redshifts. For each type, we measure the mean and the dispersion
of the k-correction over galaxy templates in a
redshift range of $\pm$0.1 around the cluster redshift. This redshift
interval is narrower than the one chosen for cluster membership to
avoid too much contamination from foreground and background
galaxies. The redshift range for cluster membership is larger as we
then subtract field counts.  We then define our corrections to be the
mean values plus 2$\sigma$ to be representative of 95\% of our galaxy
population. To keep a 90\% completeness limit for all types of
galaxies, the final k-corrections are set to the maximal values over
all types.  This step is illustrated in eq.~\ref{eq:kcorr}, where $C_X$
and $C_x$ are the completeness limit in absolute and apparent
magnitude in the x band, $DM(z)$ the distance modulus, $k_x(z)$ the
k-correction in the x band at redshift z, and $T$ the galaxy type.

\begin{equation}
\label{eq:kcorr}
C_X = C_{x} - DM(z) - \max\limits_{T}(<k_x(z)> + 2\sigma_{k_x(z)})
\end{equation}

\subsection{Computation of galaxy luminosity functions}

We use the output catalogue of LePhare with photo$-z$s, positions,
magnitudes, and absolute magnitudes for the B, V, R, I, Z, and J or Ks
band data acquired by the original telescopes, and the magnitudes computed as if
they had been observed with the VLT filters. We remove objects near
saturated stars identified by eye in all our catalogues. Some
  stars are not assigned a null photo-$z$ by LePhare. We then add an I
  band central surface brightness versus magnitude criterion to remove
  those stars. We correct for the dust extinction of the Milky Way
using the cirrus maps of \citet{Schlegel98}. We assume that this
correction is constant over each field and take the mean value of the
extinction map area corresponding to the cluster position. This
assumption is validated by the small area of our clusters compared to
the resolution of the extinction maps. A cluster rarely occupies more
than 4$\times$4 pixel area of the cirrus map.

When combining all catalogues into a single one, we delete objects
that are not detected in all 6 bands. We can estimate this loss of
galaxies by comparing the number of galaxies in the merged catalogue
to the single band catalogues. We first remove stars in single band
catalogues using a surface brightness to magnitude diagram. This step
mostly eliminates some very bright objects (bright and saturated
stars) and faint spurious detections. We then measure the ratio
of the numbers of galaxies from each of the single band
catalogues to the combined catalogue in bins of 0.5 apparent
magnitude. As we do not have the redshifts of the galaxies for which
we wish to account by applying this incompleteness correction, we apply these
ratios as a weight coefficient to all galaxies belonging to the same
apparent magnitude bin. Owing to the application of a k-correction,
galaxies in the same apparent magnitude bin do not
necessarily lie in the same absolute magnitude bin. Hence, applying
this corrective factor directly to the magnitude bins instead of
applying it to each galaxy would distort the absolute
magnitude distribution. As we perform this correction on single
band catalogues, we use the apparent magnitudes measured within the
images. All subsequent steps are done using the magnitudes
simulated by LePhare, which are as if they had been acquired with the
VLT. This allows a more reliable comparison of clusters with each other.

We select galaxy cluster members as galaxies with photo$-z$s of $\pm$
0.2 around the cluster redshift. We verify in a V-I versus I
colour-magnitude diagram that this sub-sample has a red-sequence 
that agrees with that of simulated elliptical galaxies of \citet{BC03} at
the cluster redshift.  Once this pre-selection is done, we fix galaxy
redshifts to the cluster redshift and re-run LePhare on this
sub-sample without varying photo-$z$s. This allows LePhare to determine
both the k-corrections and the absolute magnitudes of the cluster
members more accurately because they are redshift dependent properties. The k-correction
strongly depends on redshift at high redshift \citep{Chili10}. For
example, mistaking a galaxy at $z=0.4$ with a galaxy at $z=0.5$ leads
to a difference of 0.3 magnitude in the r band for an elliptical galaxy
and 0.2 for a spiral galaxy, when adopting the galaxy colors g-r=1.5 and
0.9 given in \citet{Fuku} and using the on-line 
k-correction calculator of Chilingarian (\url{http://kcor.sai.msu.ru/}).

We then perform a field galaxy background subtraction using COSMOS data
\citep{Il+09}, which are suitable for this subtraction
because they include our redshift range and have accurate photo-$z$s. We
first convert COSMOS magnitudes into our own set of filters by applying a
correction factor that depends on galaxy type and redshift.
  Magnitudes in the COSMOS catalogue are already corrected for dust
  extinction. To avoid any k-correction effect, we do the
  background subtraction in apparent magnitude. Indeed, for our clusters, we compute the
k-correction by setting all galaxies to the cluster redshift, while in
COSMOS we have access to the k-correction of galaxies at their own
photo-z. We apply the same
  photometric redshift cut applied to select our cluster members. We then count
  cluster and field galaxies in bins of 0.5 magnitude and apply a
  weight to all galaxies in each bin equal to the ratio of cluster to field
  galaxies in the bin. Field counts are first normalized to
the cluster area assuming that the COSMOS field of view is
1.73~deg$^2$ after eliminating the masked regions. This subtraction
removes line of sight galaxies that are in our cluster redshift
interval. Owing to the relatively small fields covered by our J band
data, field counts cannot be estimated from our images, hence
we use robust field counts taken from the literature. For the same reason, we are unable to investigate the
properties of clusters at very large radii. We assume that our
clusters lie in a region of radius 1~Mpc around their optical centre
(position of the BCG or in some cases barycentre of bright galaxies).
Once field counts are subtracted, we normalize GLFs by dividing by the 1~Mpc area
converted to square degrees. We choose this normalization to compare
our results to those of other authors who calculated GLFs normalized to
1~deg$^{2}$.

We study the behaviour of clusters by fitting their B, V, R, and I band
GLFs with a Schechter function (eq.~\ref{eq:schech},
\citet{Schech76}).

\begin{equation}
\label{eq:schech}
 N(M)  = 0.4\log(10)\phi^*[10^{0.4(M^*-M)}]^{\alpha+1}\exp(-10^{0.4(M^*-M)}),
\end{equation}

\noindent
where $\phi^*$ is the characteristic number of galaxies per unit
volume, $M^*$ the characteristic absolute magnitude, and $\alpha$ the
faint end slope of the GLF. We obtain these three parameters by
applying a $\chi^2$ minimization algorithm. The error bars in these 
  parameters are given by the covariance matrix (i.e. the second
  derivative matrix of the $\chi^2$ function with respect to its free
  parameters, evaluated at the best parameter values).

Since our clusters are rather distant, the numbers of points available
to fit their GLFs do not justify the inclusion of a second function (either a second
Schechter function, or a Gaussian) to fit our data, as sometimes found
in the literature. Since we are particularly interested in the faint
end slope of the GLF, a single Schechter function is therefore appropriate.

\subsection{red-sequence and blue galaxy luminosity functions}
\label{sec:rs}

To understand clearly the cluster properties, it is interesting
to study their different galaxy populations. To do so, we need to distinguish the
red-sequence (RS) from the blue galaxies. The first roughly
correspond to early type galaxies and the second to late types.

To perform this separation, we use a V$-$I versus I colour magnitude
diagram plotting only galaxies selected as cluster members based on their photo-$z$s. As it has been observed
that the RS slope does not evolve across our redshift range
\citep[e.g.][]{delucia07}, we assume a fixed slope of
$-0.0436$, as in \citet{Durret11}. For the ordinate of the RS, we first
interpolate the elliptical galaxy colour value given in \citet{Fuku}
to each cluster redshift and select a wide RS with a width of 0.6 in
magnitude. We then fit this preliminary RS with a free ordinate to get
the final RS equation on which we set the smaller width of 0.3 used in
\citet{delucia07}. We check that slightly modifying the value from
\citet{Fuku} does not significantly affect our RS selection: a shift of 0.2
to our first ordinate estimate results in only a few galaxies changing
their population type.

Once we select our two galaxy populations, we compute GLFs for
each population following the same method used for the whole
sample. Field galaxies are separated using the red-sequence
  calculated for each cluster.

  The upper absolute magnitude limit for the Schechter fit is the
    magnitude corresponding to the 90\% completeness and the lower
    limit is set to the magnitude of the cluster BCG, which is defined to be the brighest red
    sequence galaxy in the I band.
Blue galaxies brighter than the BCG are removed as they are probably
foreground galaxies incorrectly assumed to lie at the cluster
redshift.


\section{Results on Galaxy Luminosity Functions}

We present in this section our fitted Schechter
functions to the GLFs of our clusters. We first analyse our
fits to individual cluster GLFs and then study average behaviours by
stacking the GLFs of several clusters. We consider the dependence of
the Schechter parameters on redshift, mass, and cluster sub-structuring
when our GLFs are stacked. For the stacked GLFs and their dependance on environment,
we also derive GLFs for blue and red-sequence populations. To separate clusters
in terms of redshifts, masses, and substructures, we limit our analysis to the RS
  galaxy population as there are too few blue galaxies for
  the considered number of clusters. This study will be conducted in a
  future paper when we have data for more clusters in hand.



\subsection{Individual Cluster GLFs}
\label{sec:indiv}

\begin{figure*}
\begin{tabular}{cccc}
\includegraphics[width=0.17\textwidth,clip,angle=270]{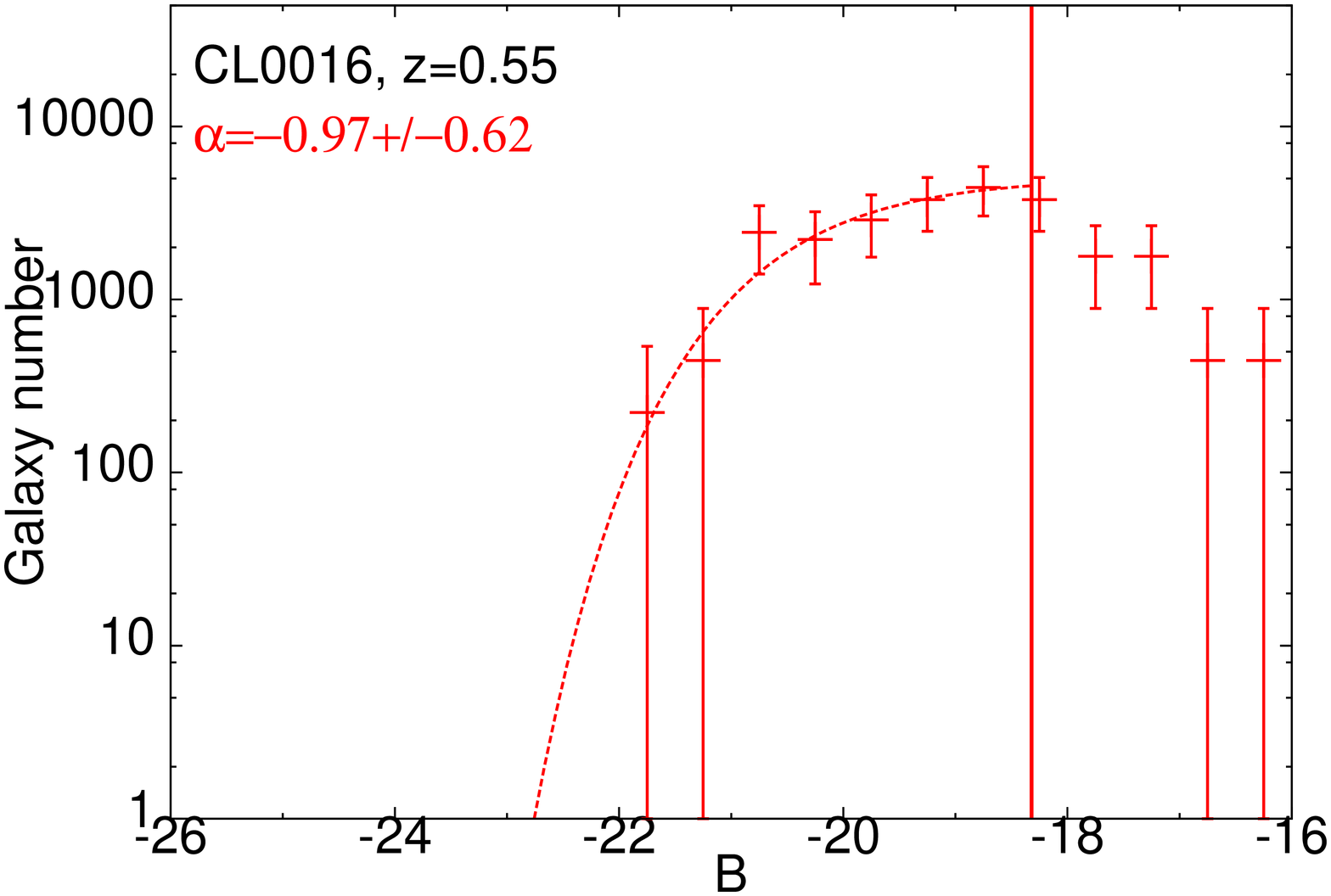}
\includegraphics[width=0.17\textwidth,clip,angle=270]{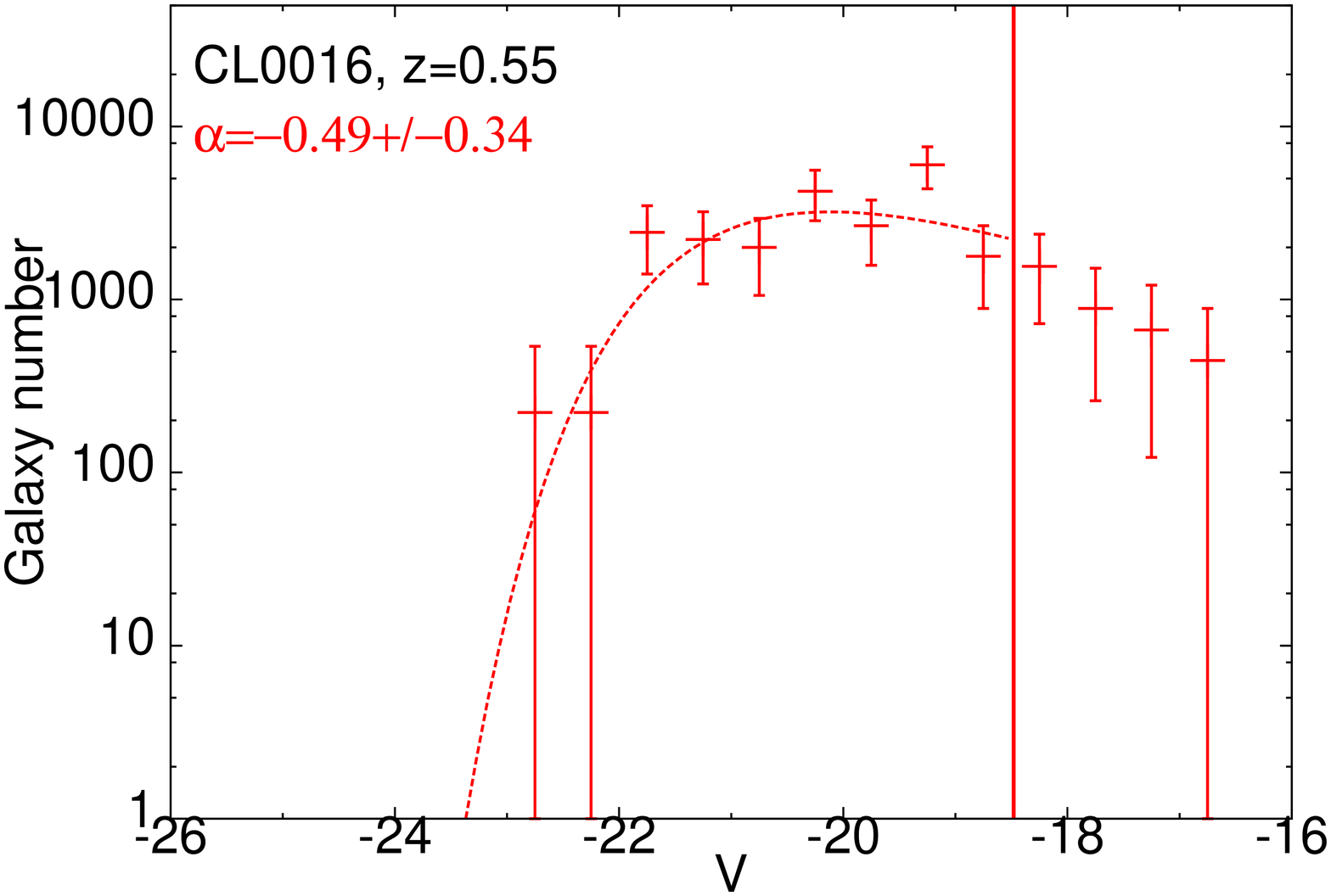}
\includegraphics[width=0.17\textwidth,clip,angle=270]{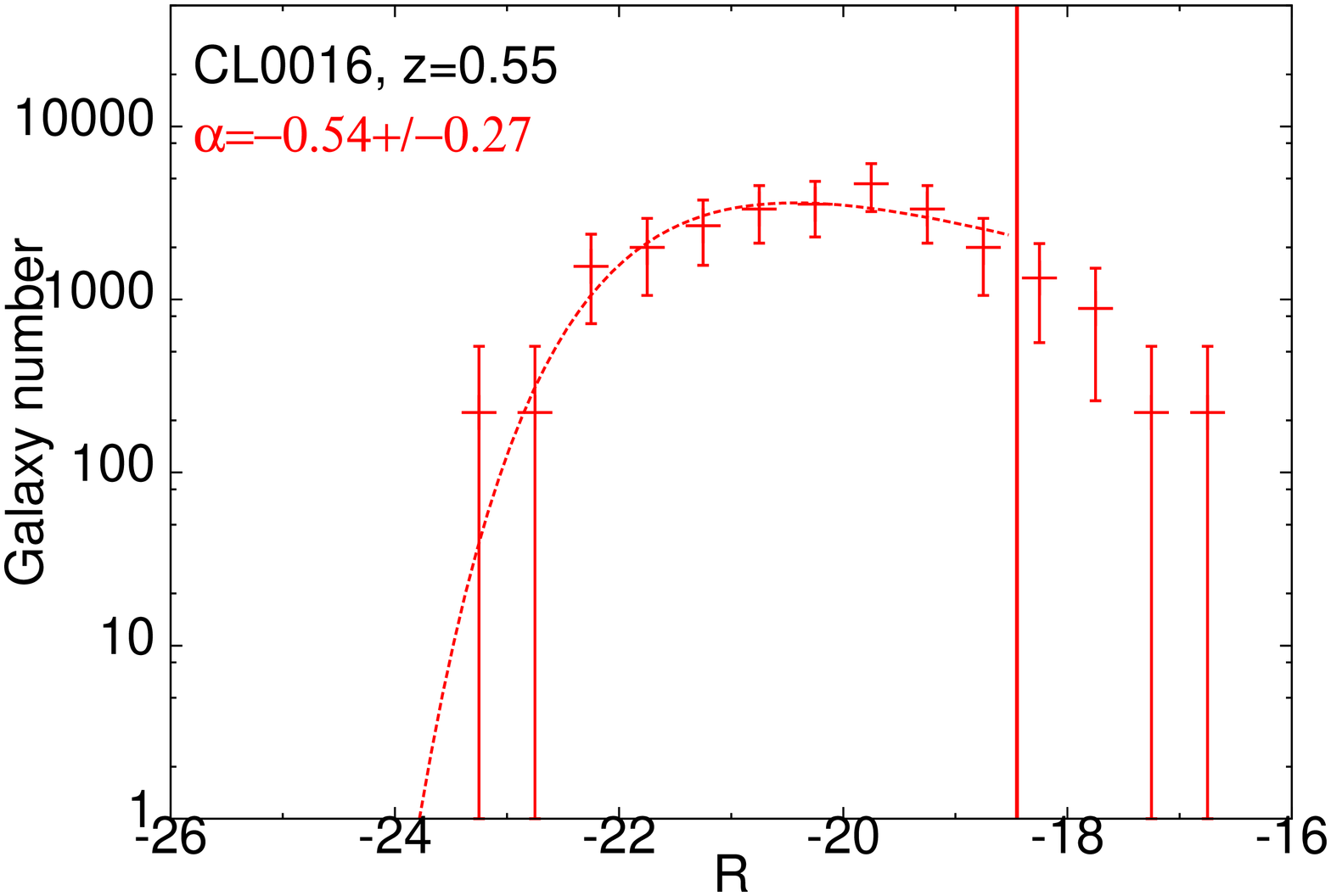}
\includegraphics[width=0.17\textwidth,clip,angle=270]{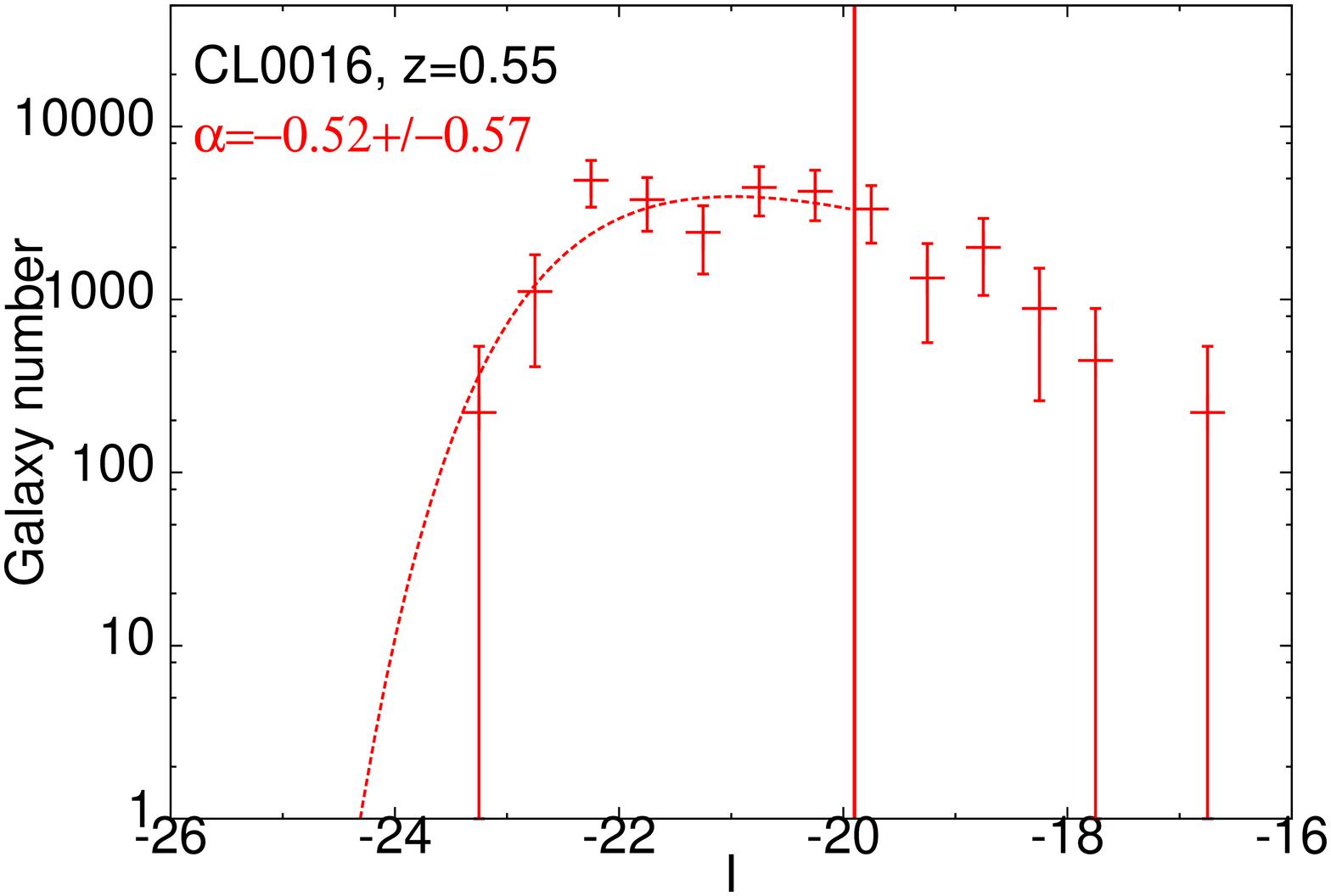} \\
\includegraphics[width=0.17\textwidth,clip,angle=270]{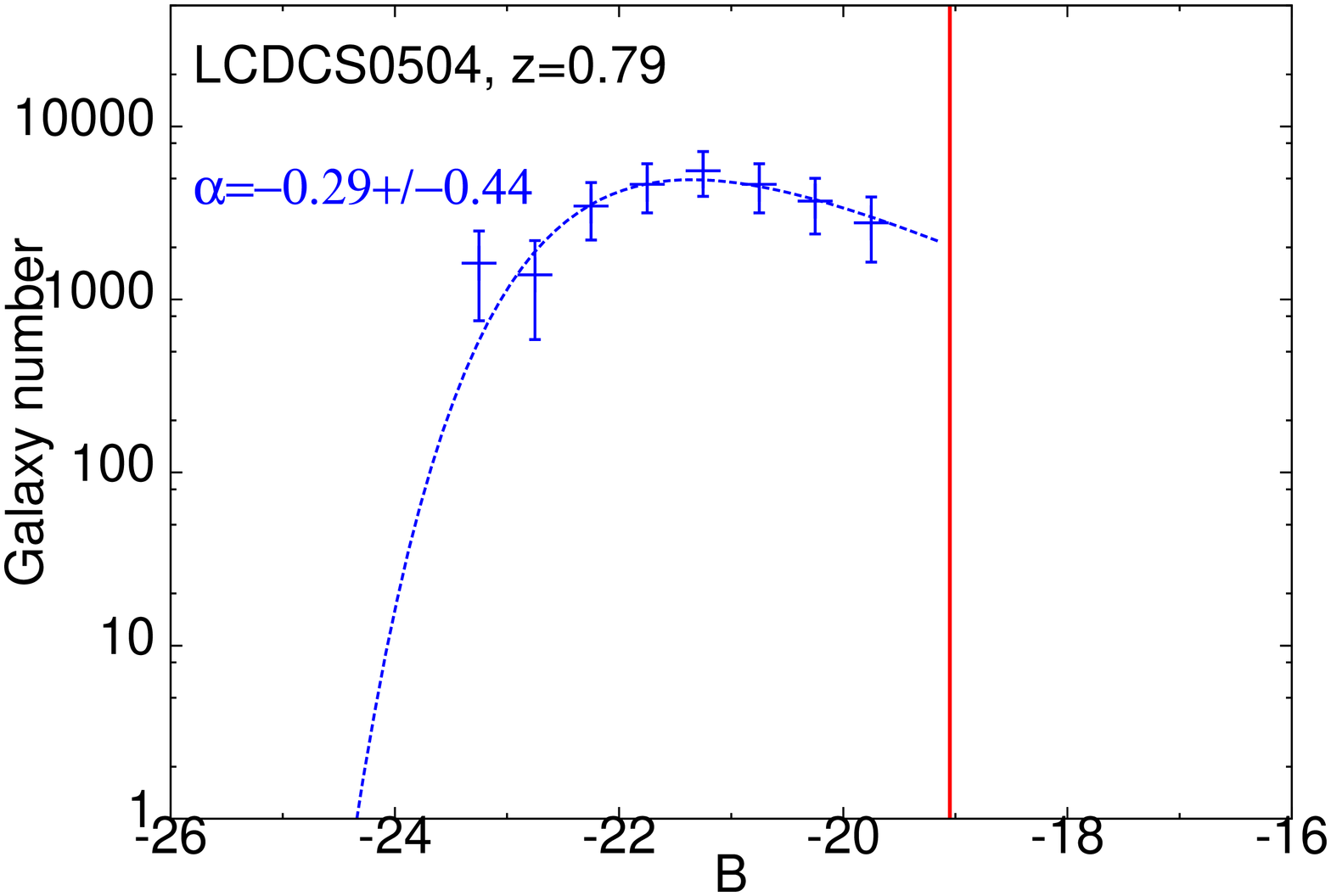}
\includegraphics[width=0.17\textwidth,clip,angle=270]{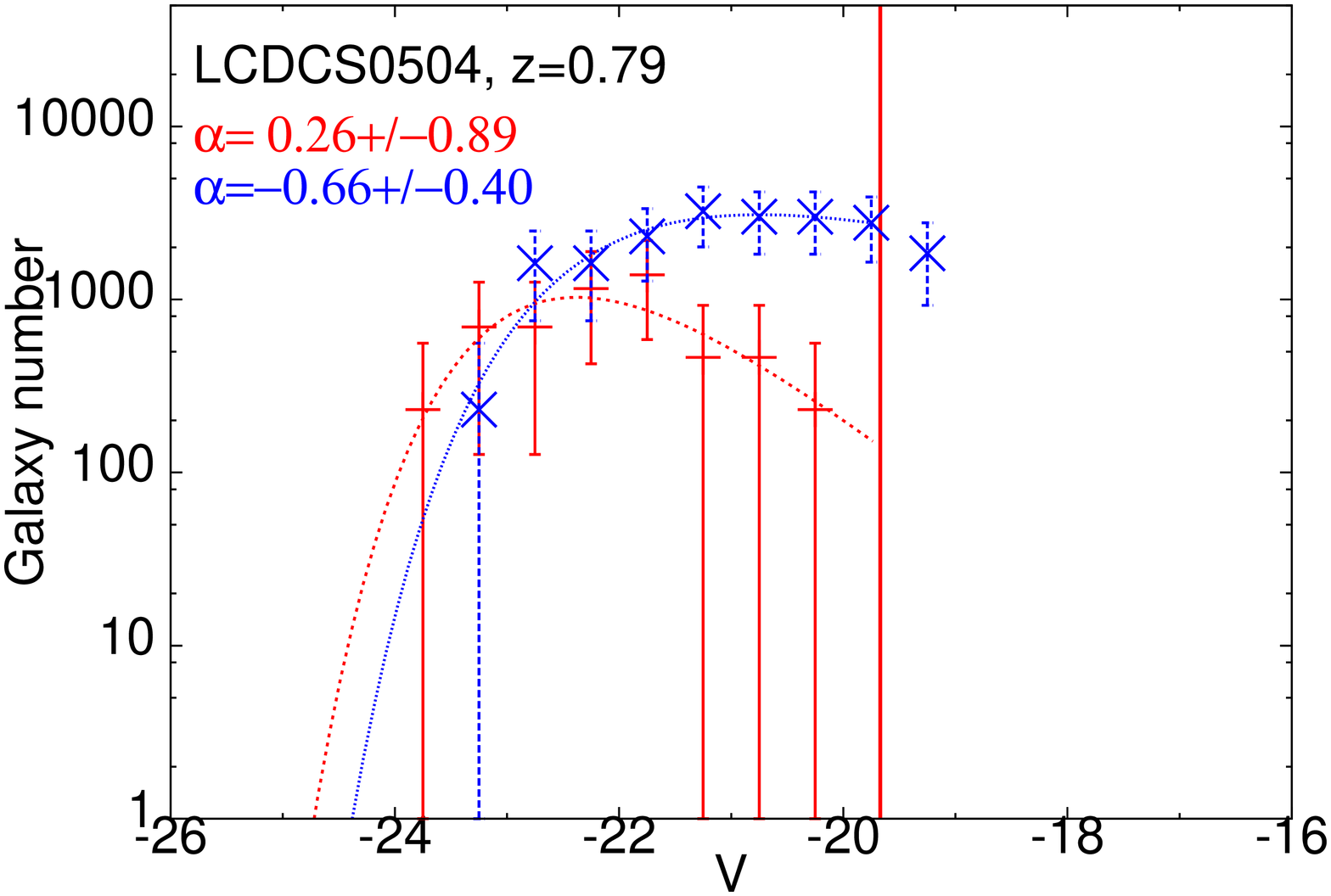}
\includegraphics[width=0.17\textwidth,clip,angle=270]{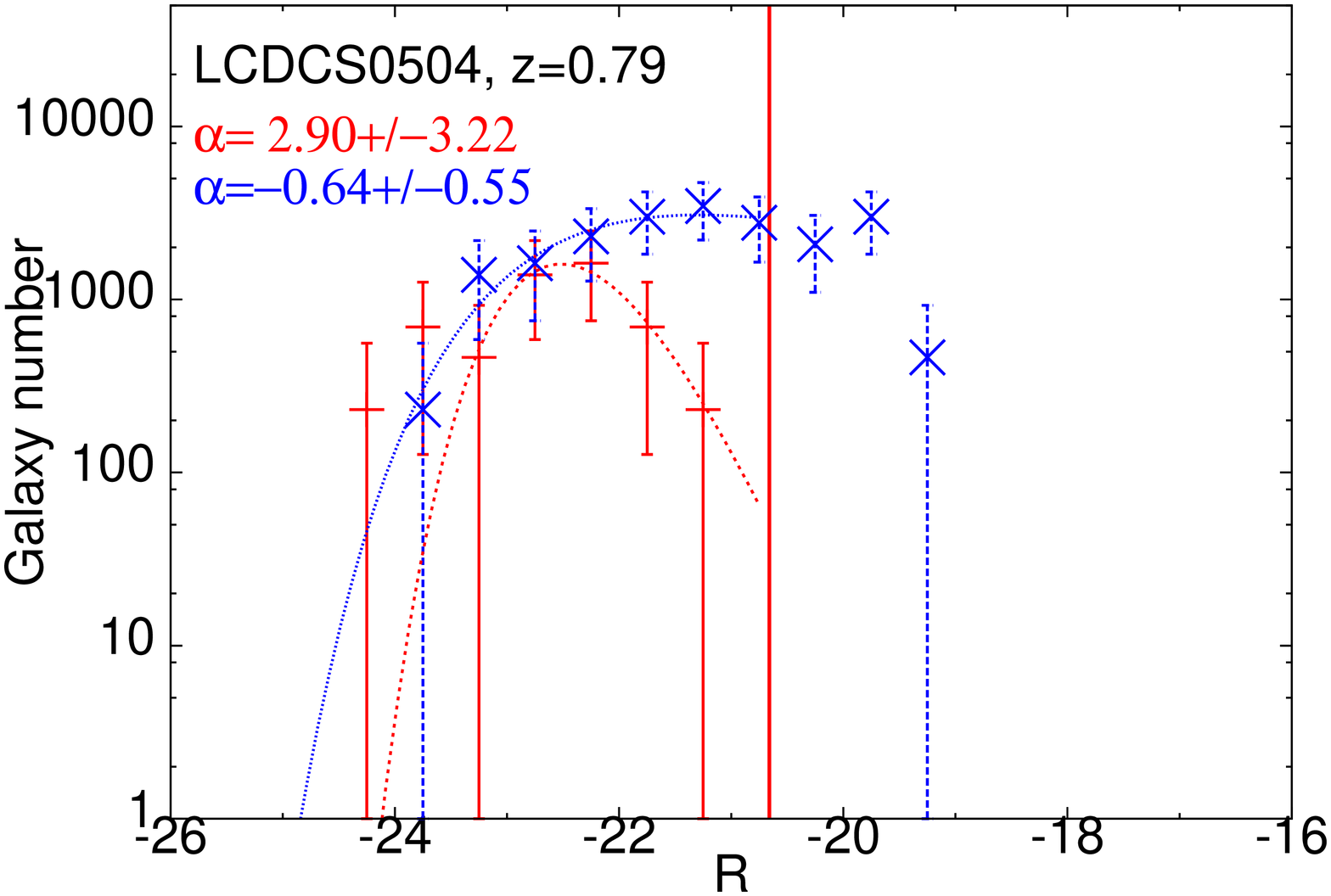}
\includegraphics[width=0.17\textwidth,clip,angle=270]{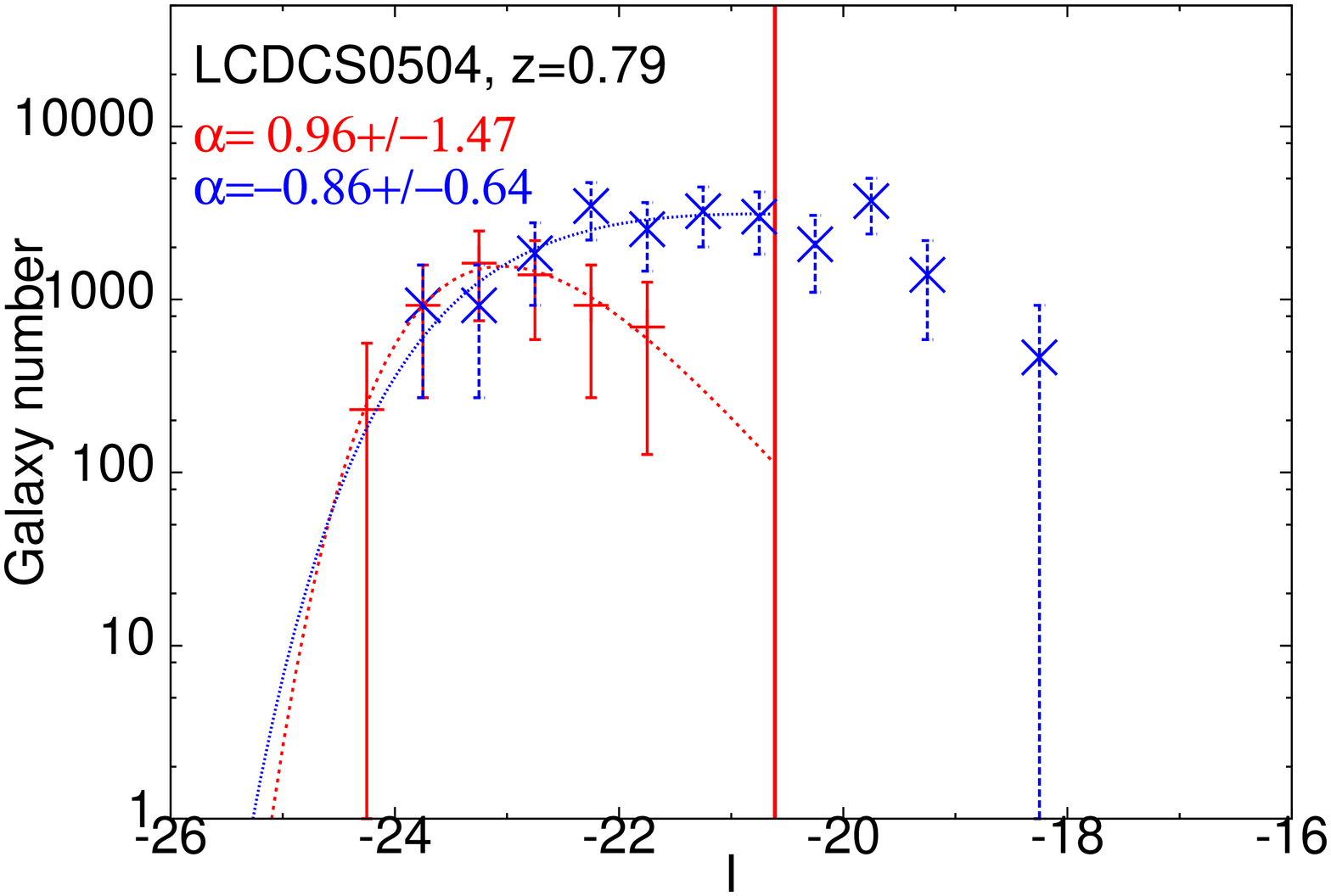}
\end{tabular}
\caption{Our GLFs in the B, V, R, and I rest-frame bands (from left to
  right) for CL0016+1609 (top) and LCDCS0504 (bottom). Red and blue
  points correspond to red-sequence and blue GLFs normalized to 1
  deg$^2$. The red vertical lines indicate the 90\% completeness
  limit. The red and blue curves show the best Schechter fits to red
  sequence and blue galaxies, and the faint end slope parameter
  ($\alpha$) is displayed in the corresponding color. Only galaxies
  brighter than the 90\% completeness limit are taken into account in
  the fits. In addition, we display GLFs only when more than 20 cluster galaxies
  are within the color RS or blue bins and after subtracting the field.}
\label{fig:glf_indiv}
\end{figure*}

We discuss individual cluster GLFs fitted with a Schechter function to
the 90\% magnitude completeness limits. These fits are done in the B,
V, R, and I restframe bands and separately for red-sequence and blue
galaxies. Two of our GLFs are shown in Fig.~\ref{fig:glf_indiv}. All
individual GLFs, and their Schechter fits when possible, are displayed
for blue and RS populations in Appendix~\ref{sec:annexe1}
(Fig.~\ref{fig:glfindiv} and Table~\ref{tab:glf_indiv}). As in
Fig.~\ref{fig:glf_indiv}, we only show galaxy counts when the selected
cluster population has more than 20 members after removal of the
background. We only display Schechter fits when they converge. In many
cases, the fits indeed do not converge, probably because the
completeness limit is too bright, even when the clusters have a
relatively high number of galaxy counts. We display these clusters
novertheless in Table~\ref{tab:glf_indiv}, as they are included in the
stacked GLFs.

Fig.~\ref{fig:glf_indiv} displays the GLFs for two clusters that span
our redshift range: CL0016+1609 at $z=0.55$ and LCDCS0504 at
$z=0.79$. CL0016+1609 represents our low-redshift clusters.  We find
rather flat RS GLFs but note the small decrease at the faint end
characterized by an $\alpha$ parameter of about $-0.5$ in the V, R,
and I restframe bands. In most of our low redshift clusters, the
observed cluster members lie primarily on the RS and too few blue
galaxies are available to produce a blue GLF. LCDCS0504 represents our
high-redshift clusters. Its RS GLFs sharply decline at their faint
end. In contrast, its blue GLFs are rather flat except in the B
band. For the high-redshift clusters, there are sometimes insufficient
RS galaxies to obtain GLFs, as here in the B band. The large
uncertainties in both counts and Schechter parameters highlight the
need for stacking to draw any clear conclusion about the GLF
behaviour.


For some clusters, the completeness limit is too bright to permit GLF
fitting, especially for high-redshift clusters. This is true for
CL\_J0152.7-1357 and LCDCS~0853 for which deeper images would be
required. Sometimes, the galaxy counts are too low to allow any fit of
the GLF, placing in doubt the high masses assumed for these
clusters. The masses we considered to select the DAFT/FADA clusters
are drawn from X-ray surveys.  The X-ray selection is often assumed to
be superior to optical selection, because the X-ray flux is
proportional to the gas density squared, while the rest-frame optical
flux is roughly proportional to the galaxy density. However, compact
X-ray sources may not have been subtracted reliably from many X-ray
observations of our clusters (especially for ROSAT data), leading to
overestimates of the cluster X-ray masses. This is true in particular
for MACS~J0647.7+7015, for which a bright source is very near the
cluster in the XMM image. Clusters with a similarly small number of
galaxy counts are F1557.19TC, MACS~J0647.7+7015, MACS~J0744.9+3927,
SEXCLAS~12, GHO~1602+4312, and GHO~2155+0334. We eliminate them from
our analysis.

  As shown in Table~\ref{tab:glf_indiv}, the Schechter parameters
  derived for individual clusters can differ, even after removing the
  problematic clusters. The large error bars are due to the use of
  Poissonian errors and the large photo-$z$ interval for cluster
  membership selection which causes more background galaxies to be
  subtracted. In the following parts we stack these clusters
  to study how the cluster properties depend on average on redshift,
  mass, substructures, and environment.

\subsection{Stacked GLFs}
\label{sec:4.2}

\begin{figure*}
\begin{tabular}{cccc}
\includegraphics[width=0.17\textwidth,clip,angle=270]{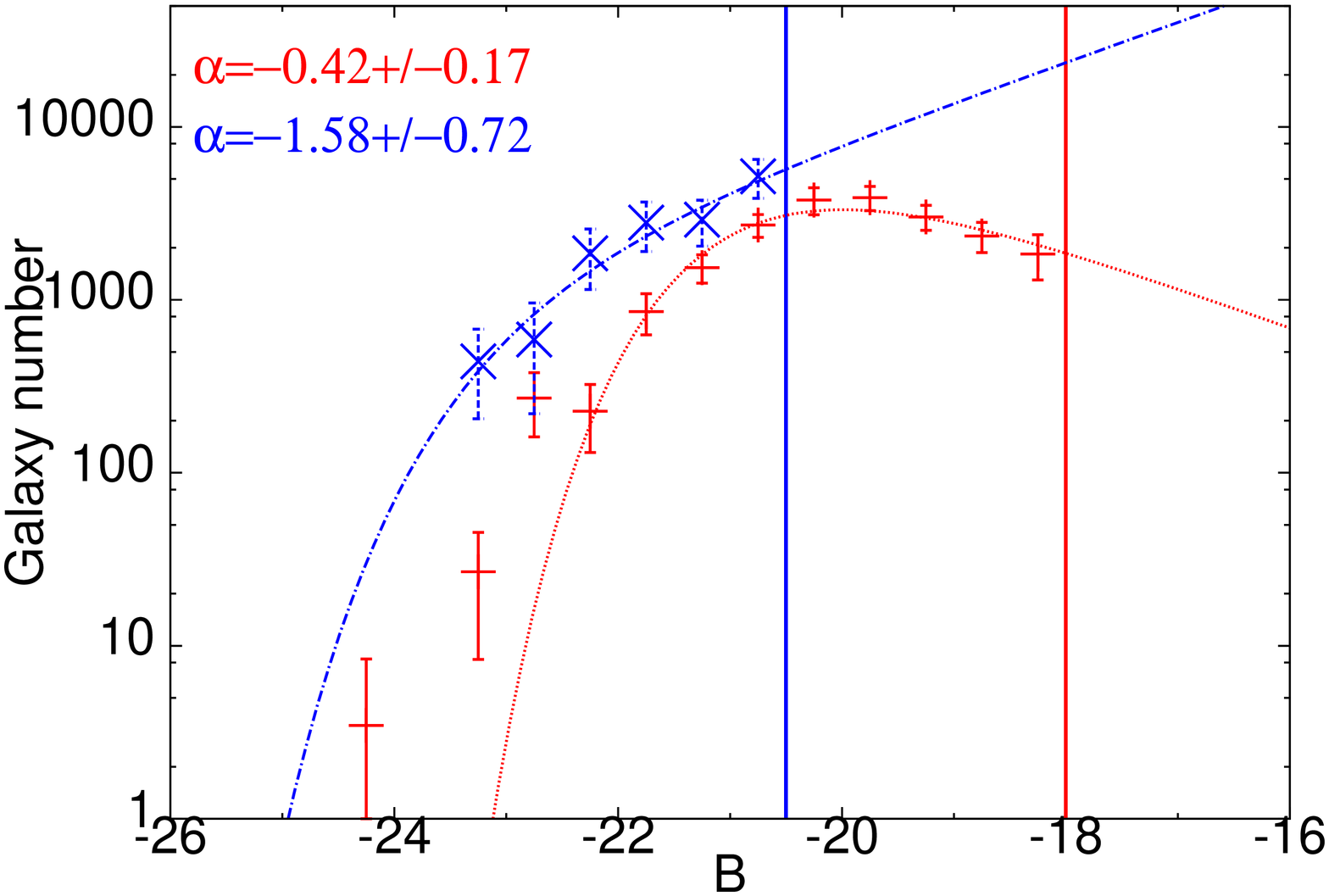}
\includegraphics[width=0.17\textwidth,clip,angle=270]{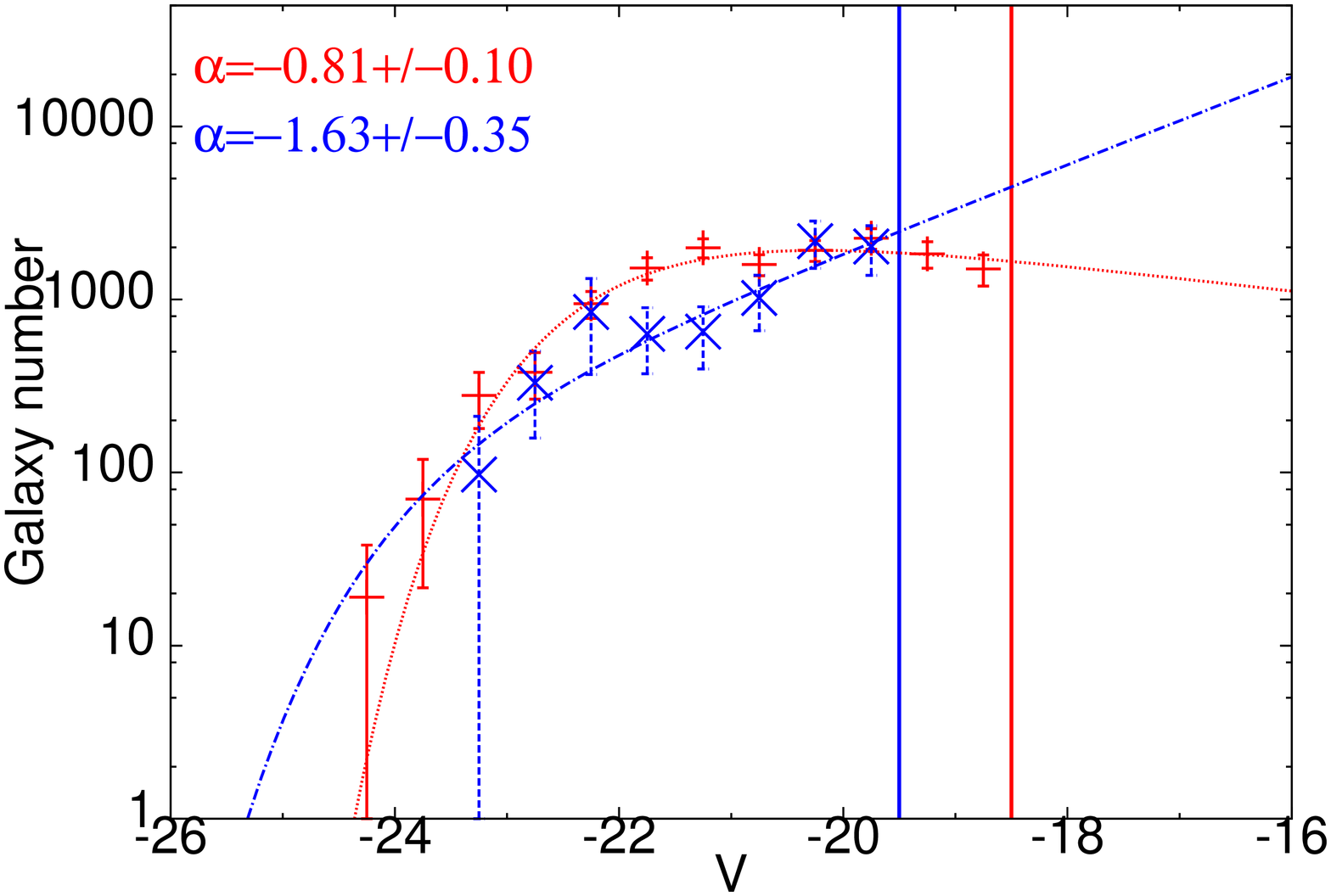}
\includegraphics[width=0.17\textwidth,clip,angle=270]{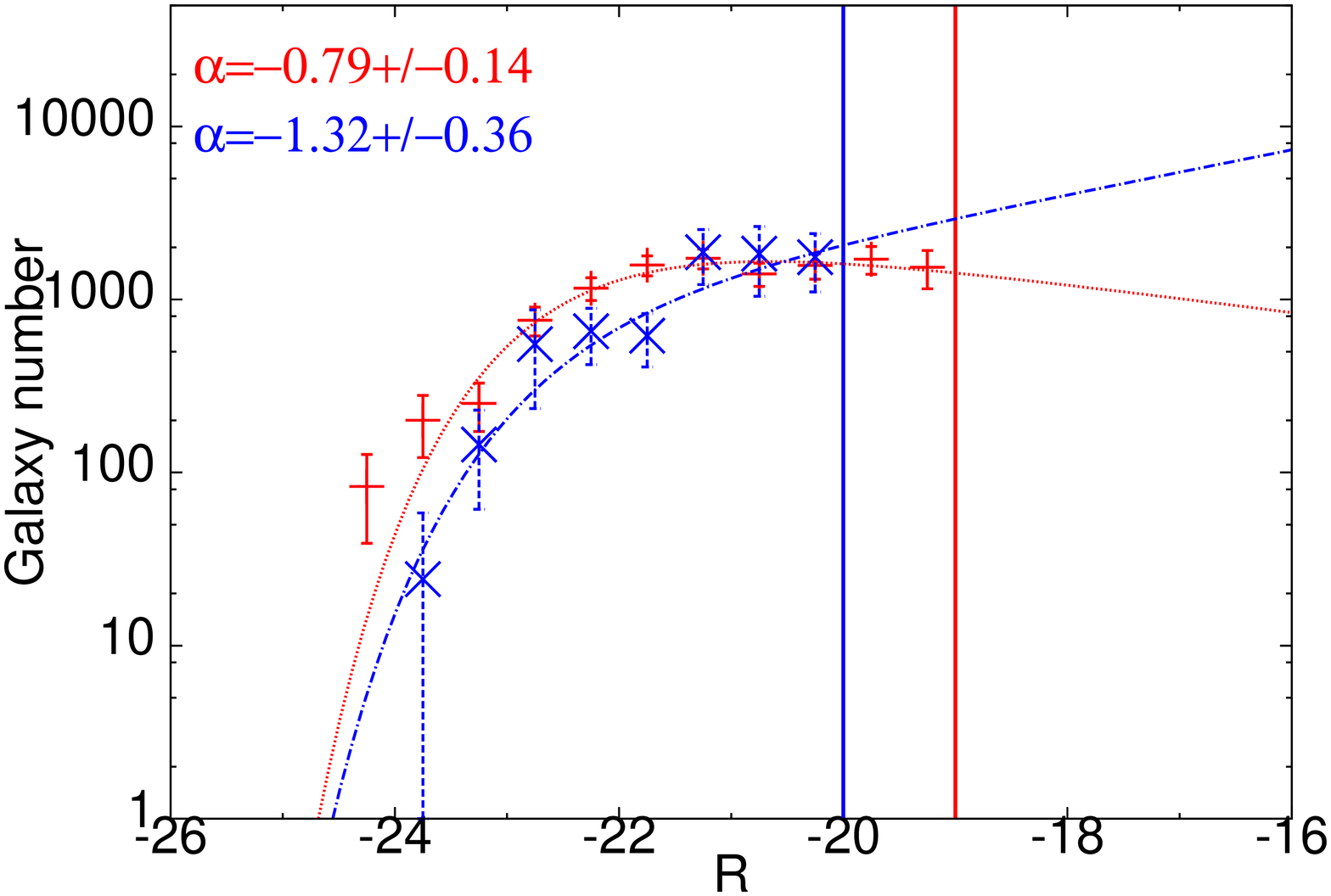}
\includegraphics[width=0.17\textwidth,clip,angle=270]{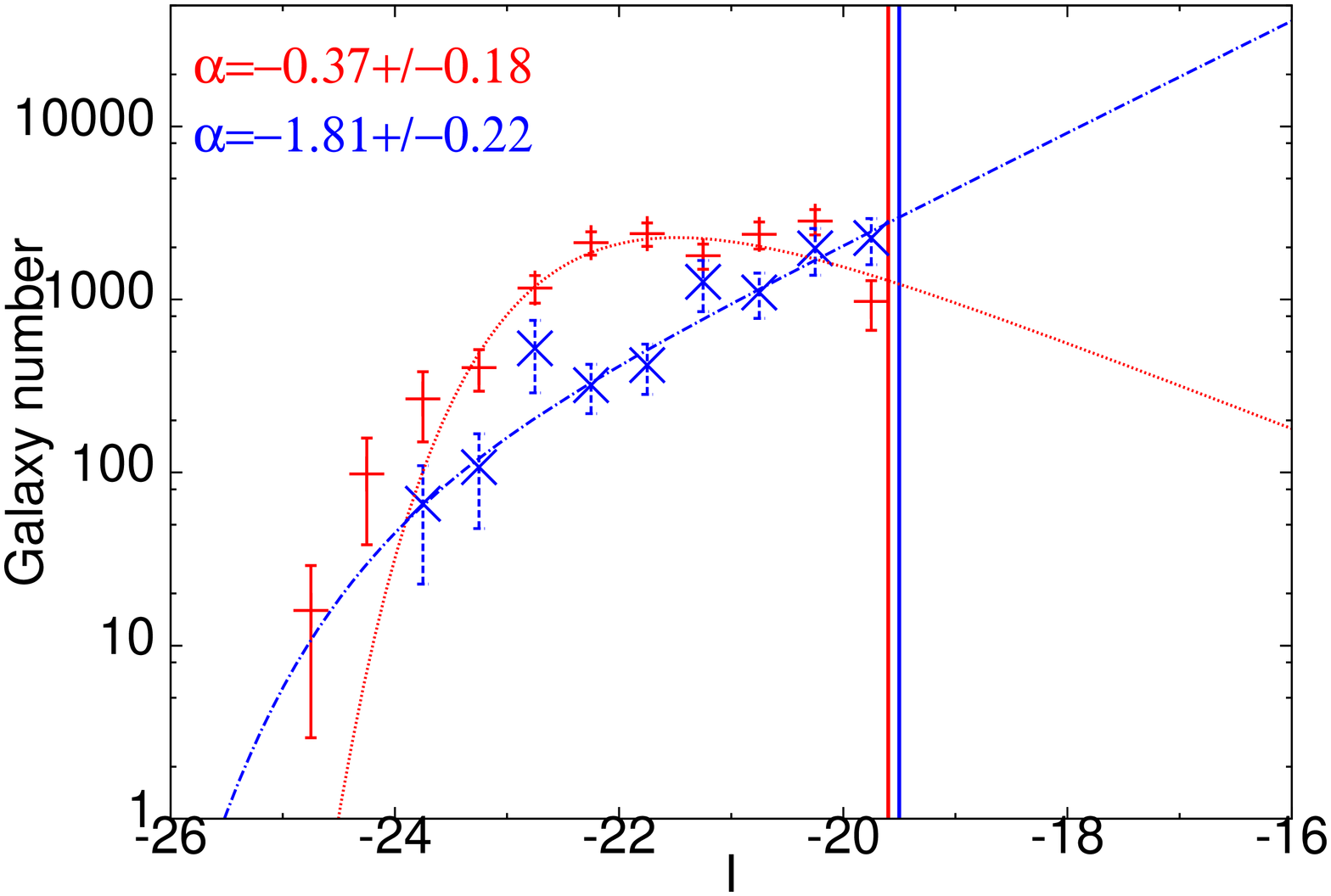} \\

\end{tabular}
\caption{Our GLFs in the B, V, R and I rest-frame bands (from left
    to right) for clusters stacked together. Red and blue points
    respectively correspond to red-sequence and blue GLFs normalized
    to 1 deg$^2$. The red and blue curves show the best Schechter fits
    to red-sequence and blue galaxies and the red and blue vertical
    lines indicate the corresponding 90\% completeness limits. The
    slope of the fit $\alpha$ is given for each population. Refer to
    Table~\ref{tab:glf_stack} for all Schechter fit parameters.}
\label{fig:glf_stack}
\end{figure*}

\begin{table*}
\centering
\caption{Parameters of the best Schechter function fits for
    stacked cluster GLFs normalized to 1 deg$^2$ for red-sequence and blue galaxies. "$N_{clus}$" is the number of clusters in the stack, "comp"
    is the 90\% completeness limit and "$<z>$" is the mean redshift of
    the stack. 
    See Fig.~\ref{fig:glf_stack} for the GLF plots.}
\begin{tabular}{l|cccccc}
\hline
\hline
&$N_{clus}$&comp&$<z>$& $\alpha$ & M*  &$\phi$* ($deg^{-2}$)\\ 
\hline
 All clusters & & & & red-sequence GLFs & & \\
\\
B & 14&-18 & 0.61 & -0.42$\pm$0.17   & -20.6$\pm$0.2  & 8837$\pm$ 1456 \\ 
V & 16&-18.5 & 0.58 & -0.81$\pm$0.10   & -22.0$\pm$0.2  & 3434$\pm$671 \\ 
R & 16&-19 & 0.58 & -0.80$\pm$0.14   & -22.4$\pm$0.2  & 3059$\pm$690   \\ 
I & 13&-19.6 & 0.54 & -0.37$\pm$0.18   & -22.0$\pm$0.2  & 6204$\pm$ 930\\ 
\hline
 All clusters & & & & blue GLFs & & \\
\\
B& 5&-20.5 & 0.70 & -1.58$\pm$0.72   & -22.9$\pm$1.7 & 1888$\pm$ 4514 \\ 
V& 6&-19.5 & 0.62 & -1.63$\pm$0.35   & -23.7$\pm$ 2.1& 238$\pm$555  \\   
R& 6&-20 & 0.62 & -1.32$\pm$0.36   & -22.4$\pm$0.5 & 1163$\pm$880  \\    
I&7 &-19.5 & 0.53 & -1.81$\pm$0.22   & -24.1$\pm$1.7 & 104$\pm$ 209 \\   
\hline
\hline
\end{tabular}
\label{tab:glf_stack}
\end{table*}

We stack our clusters using the standard Colless method
\citep{Colless89} described in \citet{Popesso06}. The idea is to
average cluster counts in each magnitude bin including all clusters that are
90\% complete in this bin. Clusters first have to be normalized to the
same area, chosen to be 1deg$^2$, and to a fixed richness.
This richness is set to the number of galaxies
detected to the completeness limit that encloses 90\% of our sample.
We do not choose our worst completeness limit because this
would result in too few galaxies for the normalization. Also, we
  only include clusters that have more than 20 galaxies above the
  background for a given galaxy population (red or blue), to avoid a
  domination of the stack by the poorest clusters.

This method allows us to use the maximum amount of information for all
our clusters. A more classical method would remove the information for
the most complete bins as we would only be able to stack clusters
reaching the same completeness limit. We could also stack different
numbers of clusters for different completeness limits. This approach
would allow to better control the evolution with completeness but it
would generate many sets of figures partially containing the same
information, thus affecting the legibility of the results.

For any method, the more complete our data, the farther we are from an
average cluster. In a standard method, this problem affects only the
stacks with the fewer clusters, while with the Colless method it
affects the faintest bins of the GLF. We investigate this bias by
stacking fixed numbers of clusters in each stack. We find that for the
same completeness limit the stacked GLFs do not change much once we
have four clusters in the stack. Hence, we require to have
at least four
clusters in each magnitude bin to take  into account in the Colless
stack, to avoid being dominated by individual cluster behaviours.



Error bars are calculated using the $\chi^2$ fit to our galaxy
counts normalized to 1~deg$^2$. Galaxy counts and their errors are
summed following eqs.~(\ref{eq:n}) and (\ref{eq:s}) below, where N(j) and
$\sigma(j)$ are the stacked galaxy counts and galaxy count errors in
magnitude bins $j$, the index $i$ indicates single cluster values, $S_i$ is the
area of cluster $i$, $N_c(j)$ the number of clusters in bin
  $j$, and $N_{0,i}$ and $\langle N_0(j) \rangle$ are the richness of the
  cluster $i$ and the mean richness of clusters in bin $j$.

\begin{equation}
\label{eq:n}
 N(j)=\frac{\langle N_0(j) \rangle}{N_c(j)}\sum_i{\frac{N_i(j)}{S_iN_{o,i}}}
\end{equation}

\begin{equation}
\label{eq:s}
 \sigma(j)=\frac{\langle N_0(j) \rangle}{N_c(j)}\sqrt{\sum_i{\left(\frac{\sigma_i(j)}{S_iN_{o,i}}\right)^2}}
\end{equation}

Individual variances are weighted by the square of the
cluster area, as for the galaxy counts, and not simply the area. This is to retain
the Poissonian distribution of the counts.  We also fit  Schechter
functions to the stacked GLFs.

Stacked GLFs are shown in Fig.~\ref{fig:glf_stack} for the RS and blue
populations of the full cluster sample in the B, V, R, and I restframe
bands. Results of their best Schechter fits are given in
Table~\ref{tab:glf_stack}, along with the 90\% completeness limit, the
numbers of clusters in the stack, and the mean redshifts of the
clusters in the stack.

  We see a common behaviour for the V, R, and I bands. The RS
  GLF is close to that at low-redshift but with a slight decline at
  the faint end. $M^*$ is almost the same for the three bands and
  $\alpha$ is slightly higher in the I band than in the other two
  bands. The blue GLFs are also very similar for these three bands, with
  steeper faint ends. In the I band, however, blue galaxy counts are smaller:
  as expected, blue galaxies are fainter in redder photometric bands. Blue and
  red GLFs cross at around V=-20, R=-20.5, and I=-20.3. These
  results repesent clusters of mean redshift about $z=0.6$. Results
  are quite different for the B band. The RS GLF also
  has a shallow decline at the faint end but the blue counts are higher
  than the red ones, implying that the 
  blue galaxies are indeed brighter in the bluer bands.

  There is also an excess of red-sequence galaxies at the bright end
  compared to the Schechter function, especially for the B and I
  bands. This kind of excess is often observed in clusters and some
  authors prefer to fit GLFs with a combination of a Schechter and a
  Gaussian \citep[e.g. ][]{Biv95}. However, this excess is puzzling
  for the B band, leading to very bright red-sequence galaxies for
  this optical band. This distribution of bright galaxies probably
  results from a complex interplay of intrinsic properties and applied
  k-correction. On examining the images, we indeed found that some
  BCGs appearing very bright in the I band are quite faint in the B
  band compared to other bright galaxies. The bright end of the
  red-sequence B-band GLF is then dominated by the k-correction
  factor, which can be as high as 3 magnitudes at these redshifts. It
  would be very useful to compare bright cluster galaxies in the B and
  I bands but this is beyond the scope of this paper. In the present
  study, we merely conclude that a Schechter function cannot
  simultaneously fit the bright and faint ends of the B band RS
  GLFs. In the rest of the paper, we concentrate mainly on the faint
  end of the GLFs, which is well constrained.

   We wish to highlight several caveats of our method. The number
  of clusters with a sufficient number of blue galaxies to be stacked
  is two to three times lower than the number for RS galaxies. Hence,
  the blue GLFs are far more poorly constrained, as can be seen from
  their larger error bars for their best fit Schechter function
  parameters. In addition, as we used only clusters with a
    sufficient number of galaxies for each population, the RS GLF is
    biased toward red-galaxy rich clusters and the blue GLF toward
    blue-galaxy rich clusters. It would be more rigorous to consider
    the same clusters in both subsamples, but this would require
    deeper images. As a sanity check, we compared our GLFs with those
    obtained by considering only the few clusters presenting both
    large red and blue populations. Results are in good agreement, but
    error bars on the latest GLFs are much larger due to the very low
    number of clusters in the stack. When applying the Colless method
  for stacking, data for different clusters are stacked in different
  magnitude bins. As our survey spans a large redshift range
    ($0.4\leq z<0.9$), each magnitude bin has a different mean
    redshift. Since the completeness limit is brighter for
  high-redshift clusters, the faint end of any stacked GLF will be
  dominated by lower redshift clusters. In the next subsection, we
  study GLFs in separate narrower redshift ranges to avoid this
  problem. We compare our stacked GLFs with field GLFs in
  Section~\ref{sec:field}.

\subsection{Evolution of GLFs with redshift}

\begin{figure*}
\begin{tabular}{cccc}
\includegraphics[width=0.17\textwidth,clip,angle=270]{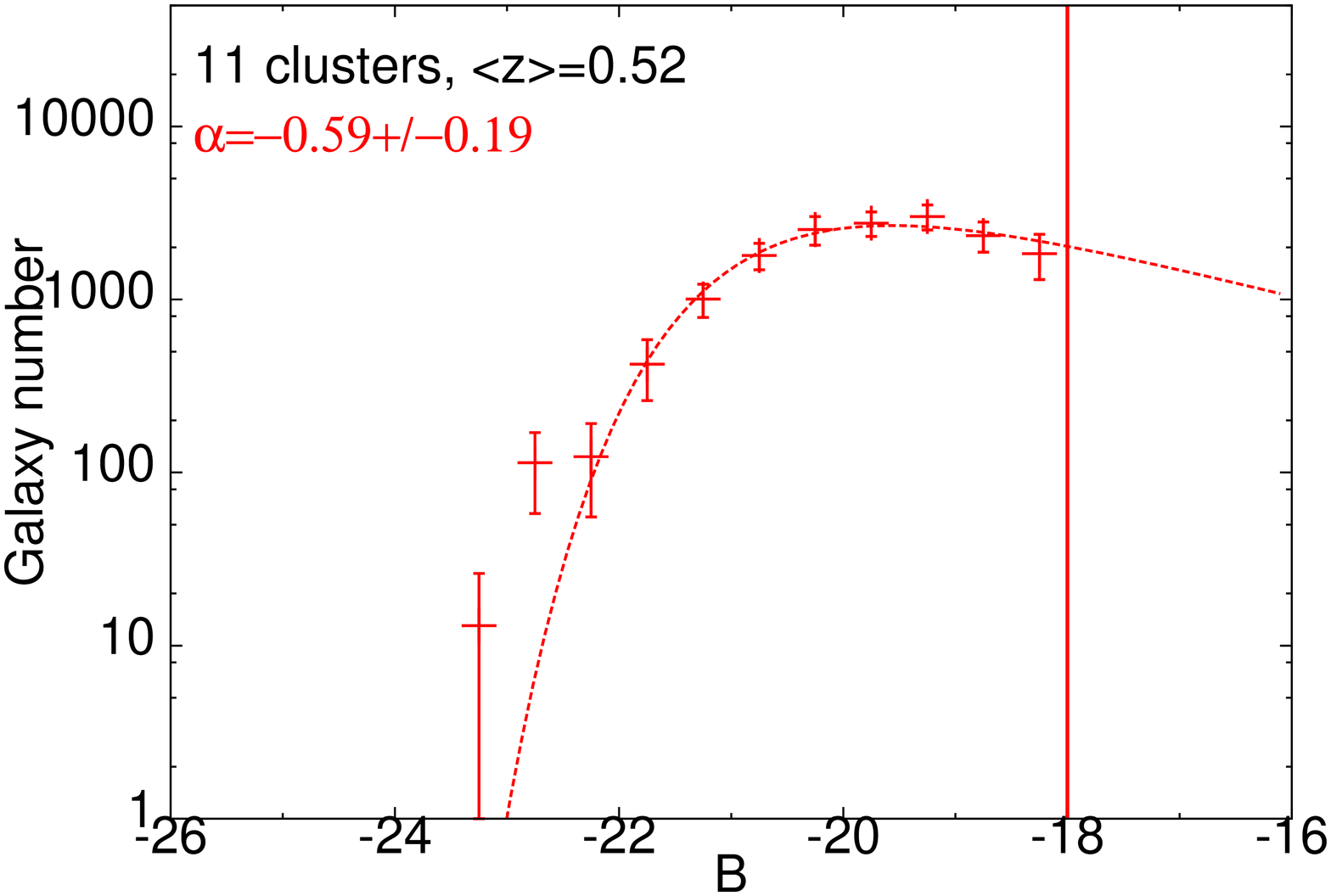}
\includegraphics[width=0.17\textwidth,clip,angle=270]{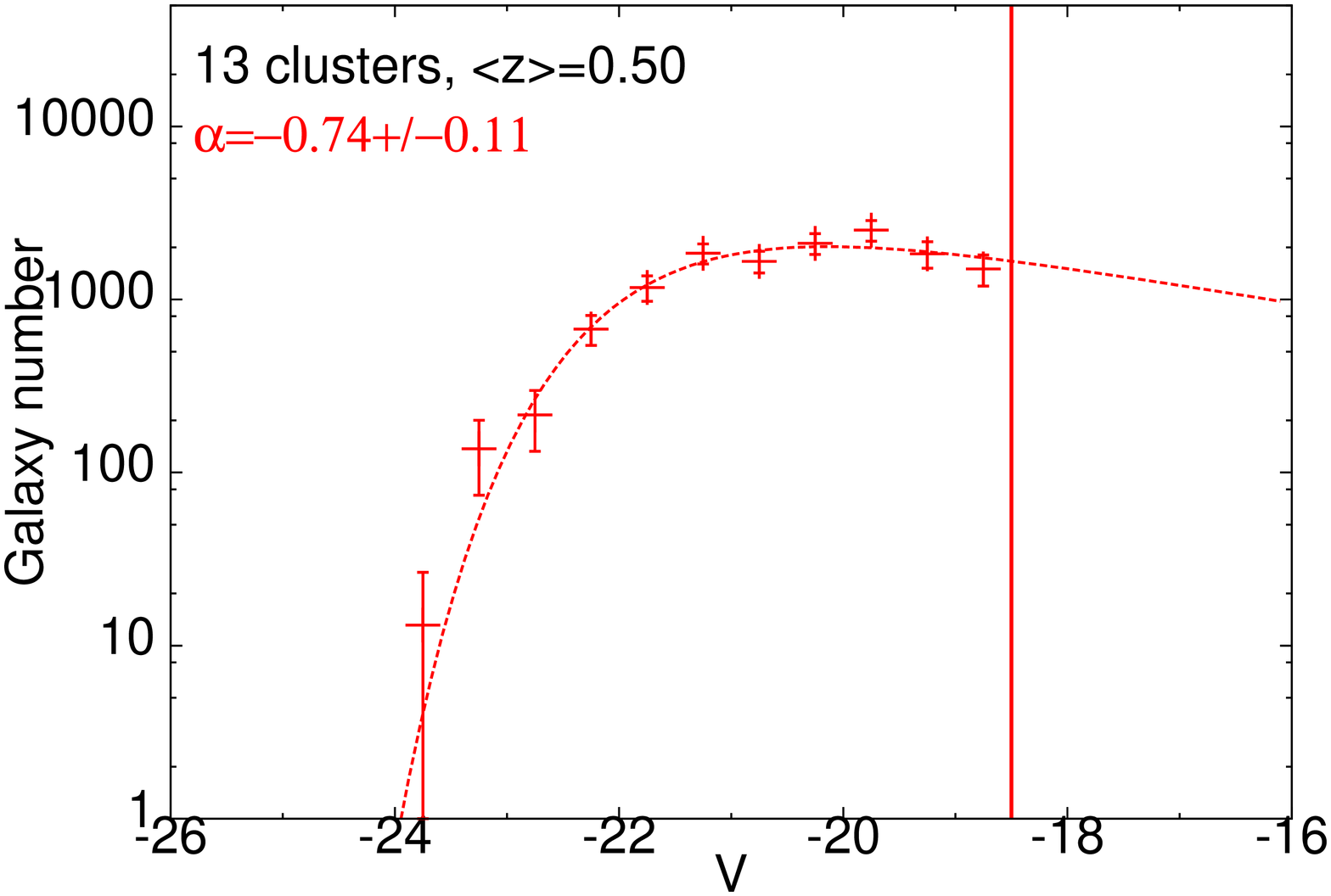}
\includegraphics[width=0.17\textwidth,clip,angle=270]{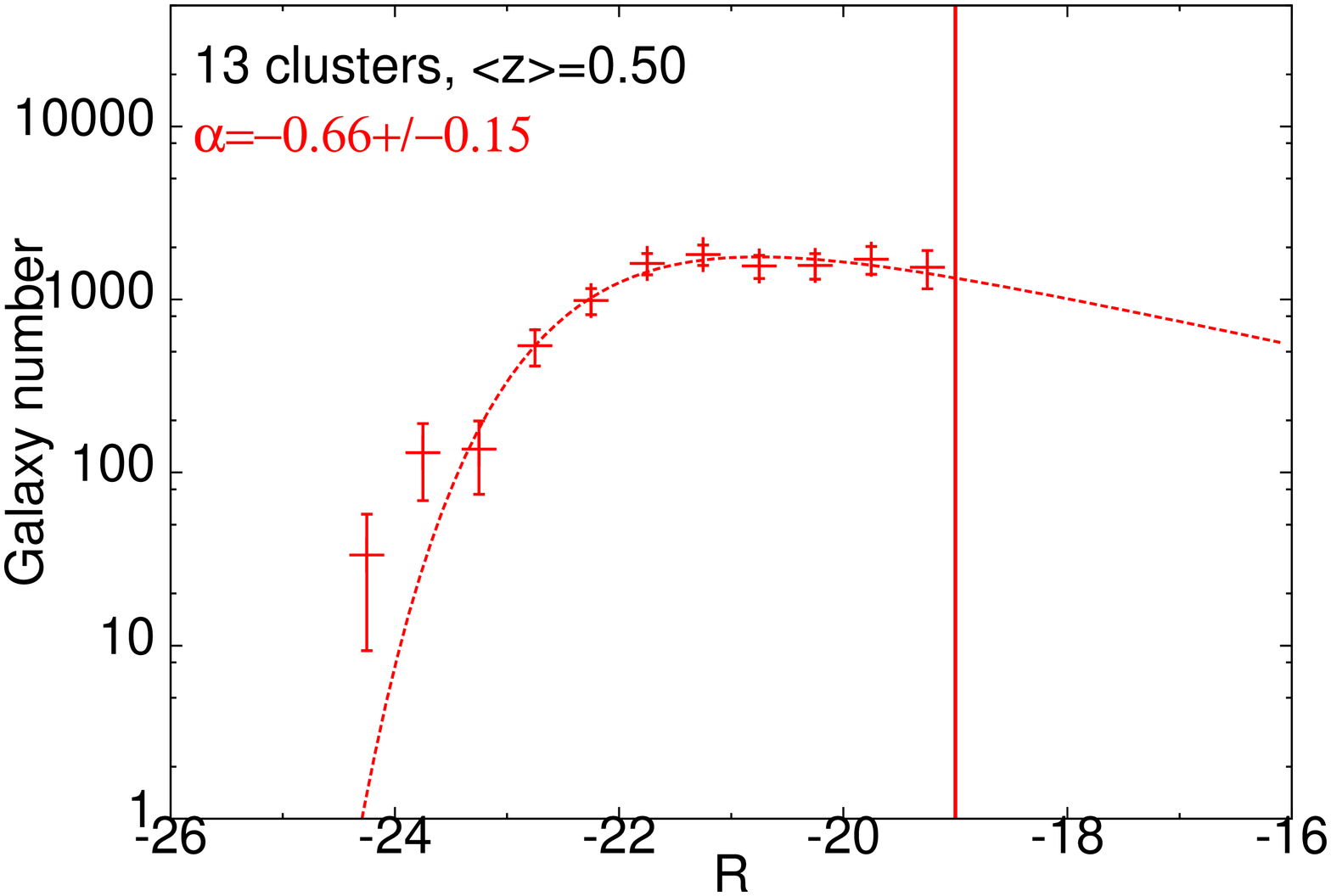}
\includegraphics[width=0.17\textwidth,clip,angle=270]{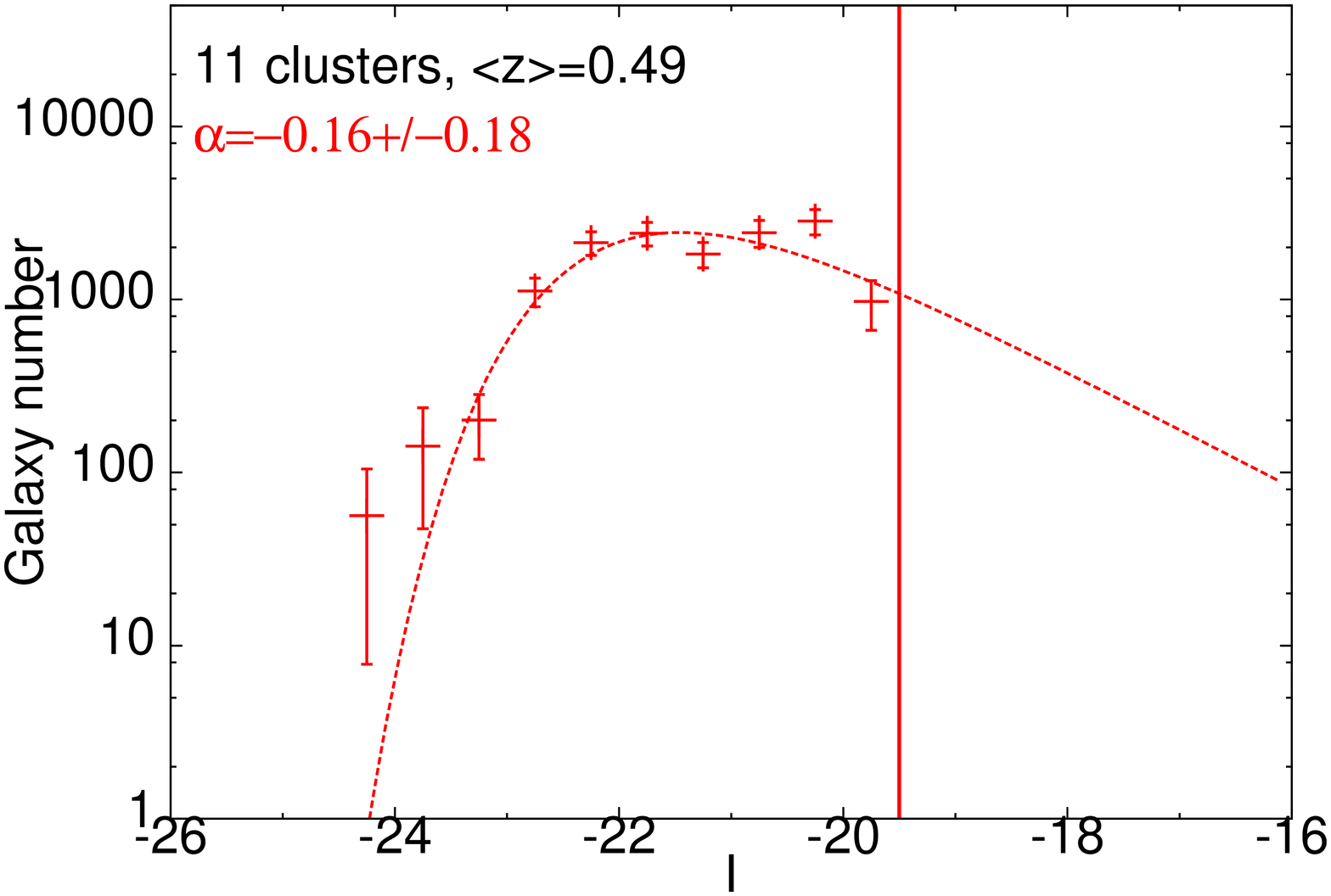} \\

\includegraphics[width=0.17\textwidth,clip,angle=270]{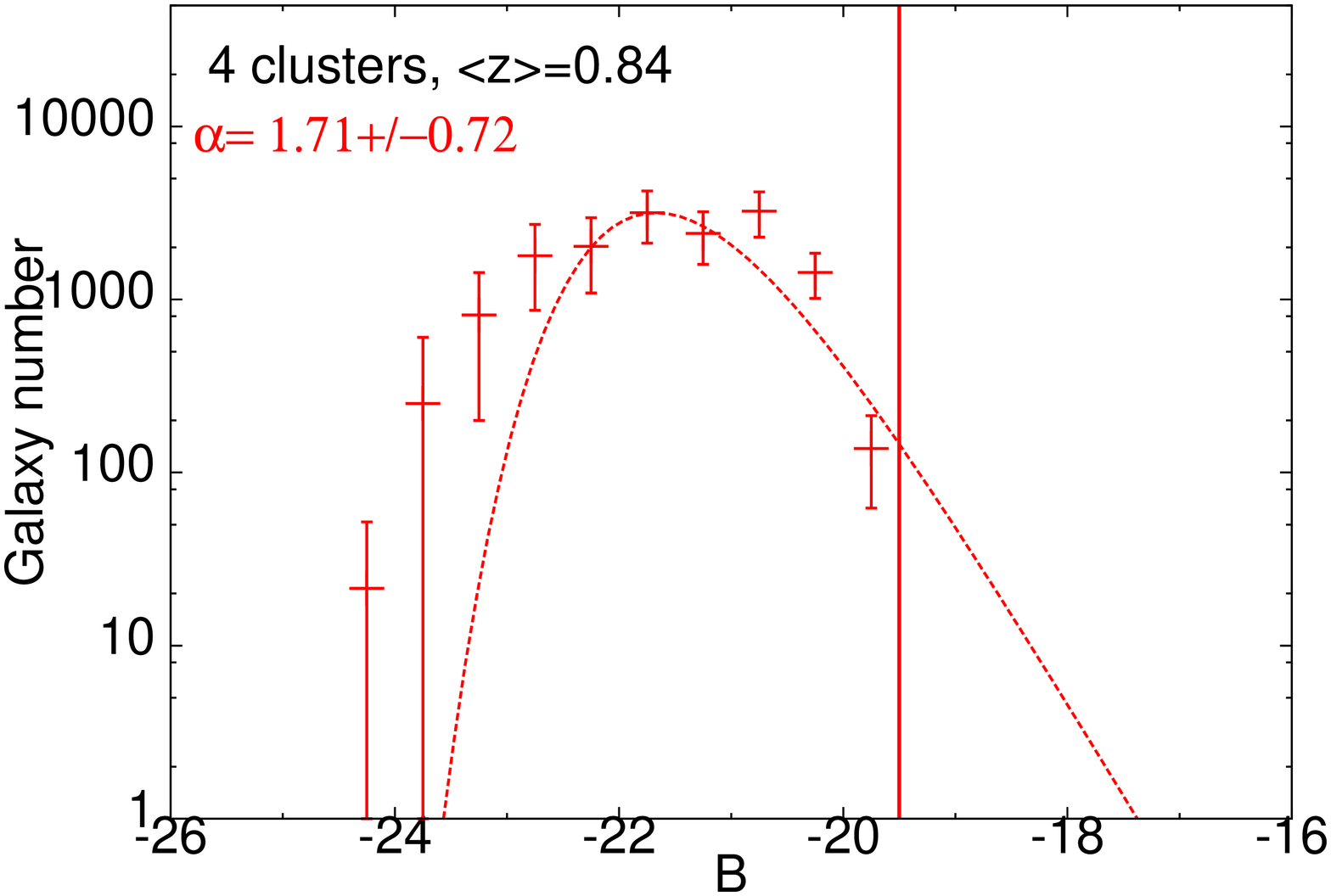}
\includegraphics[width=0.17\textwidth,clip,angle=270]{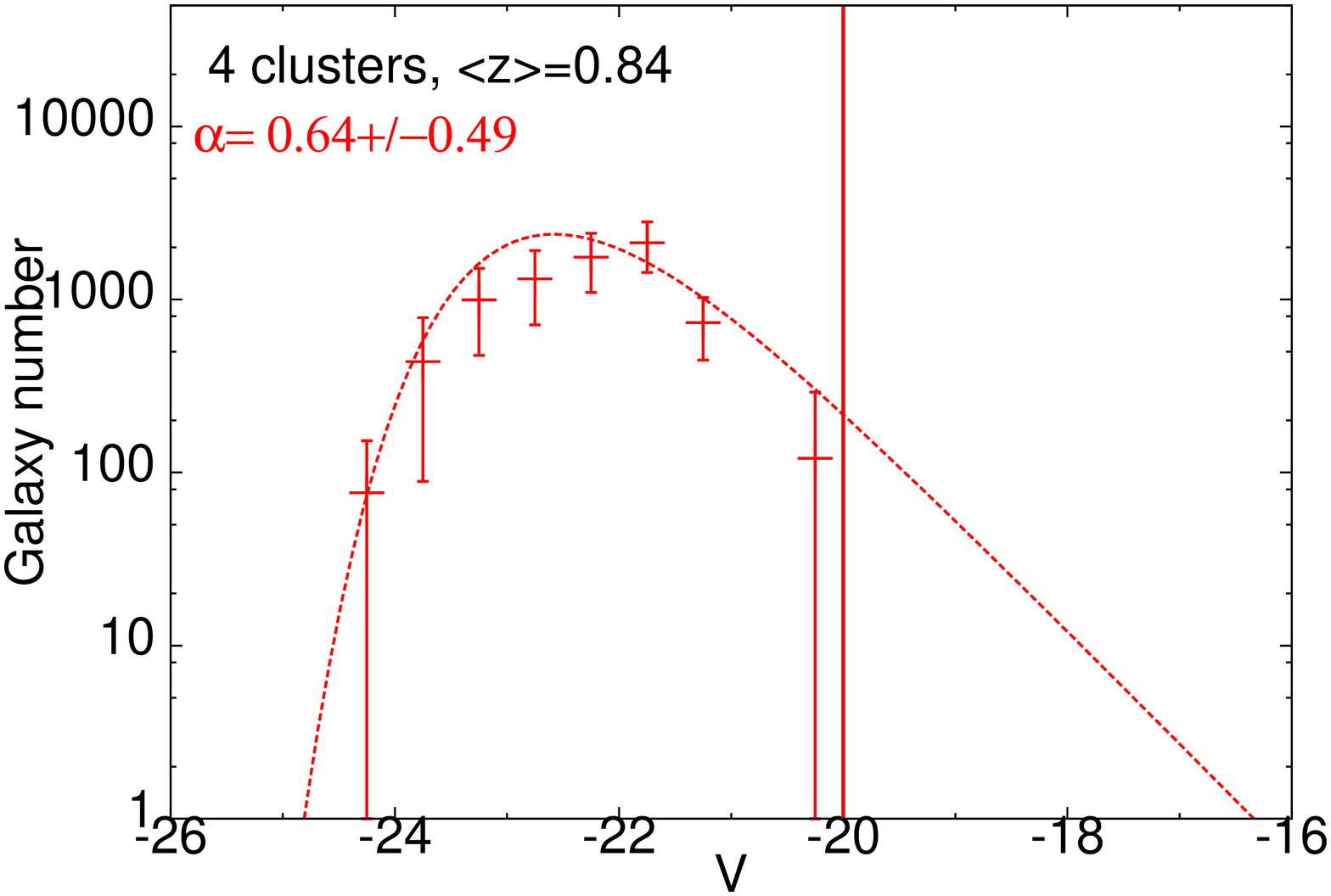}
\includegraphics[width=0.17\textwidth,clip,angle=270]{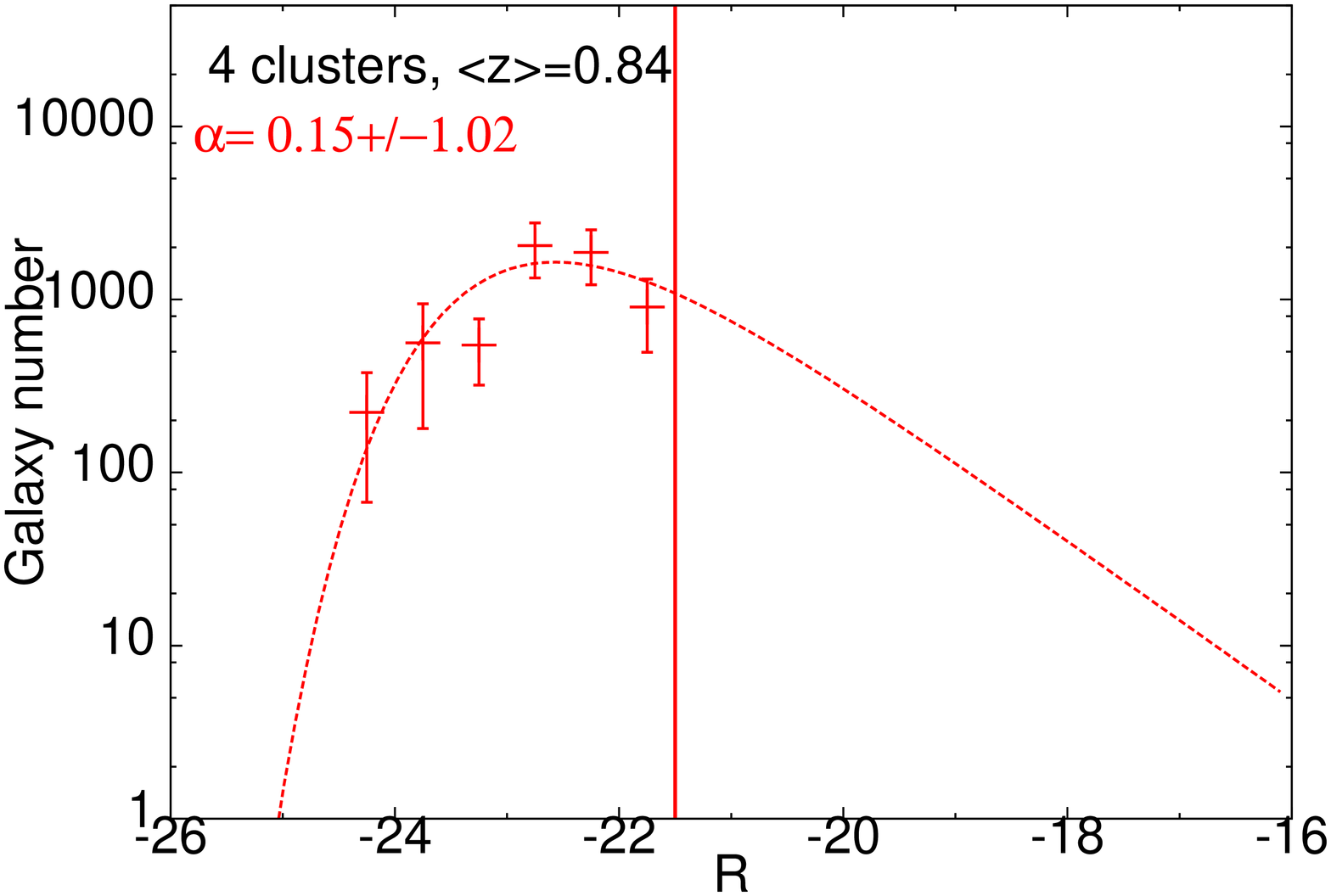}
\includegraphics[width=0.17\textwidth,clip,angle=270]{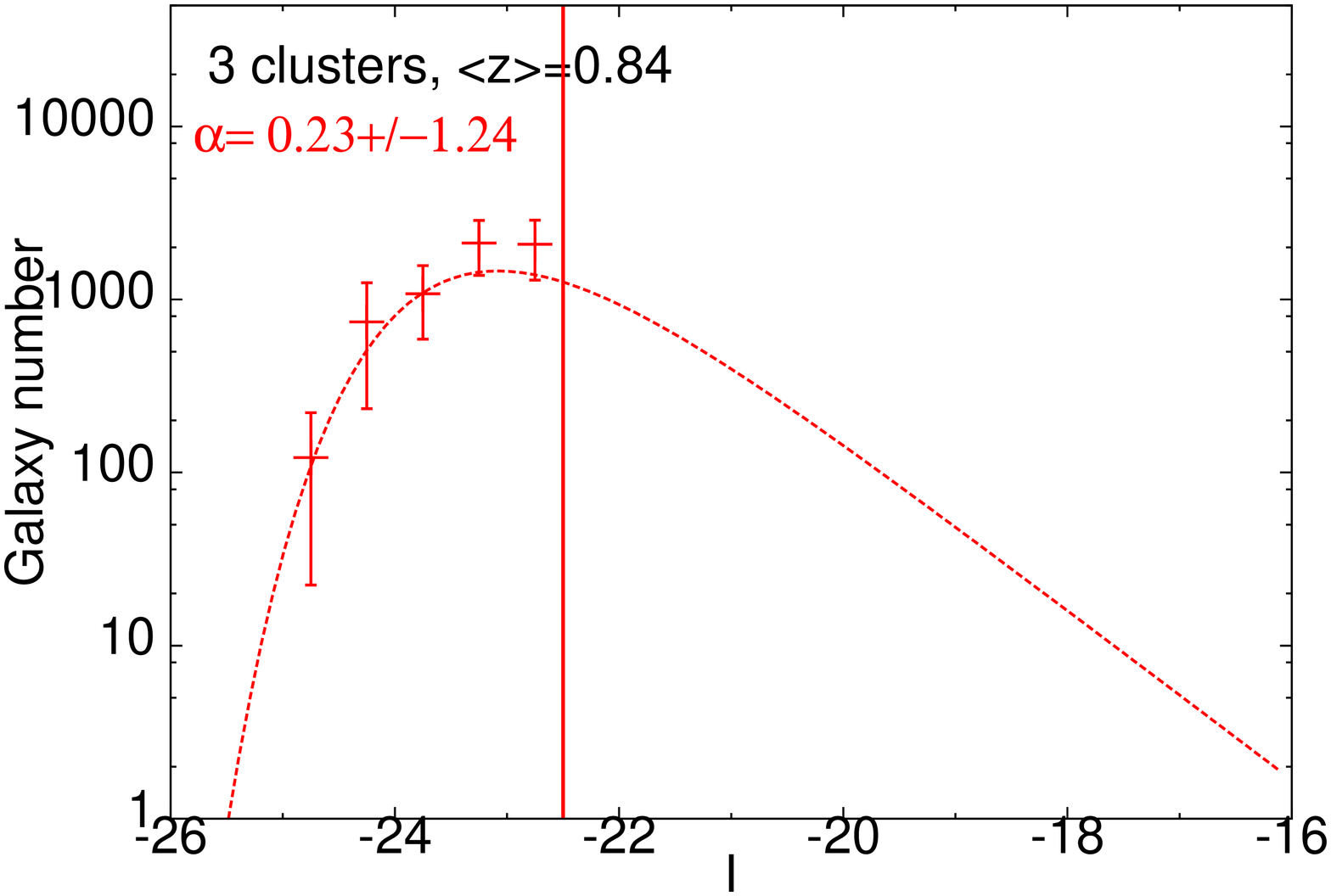} \\
\end{tabular}
\caption{Evolution of red-sequence GLFs with redshift in the B, V, R
  and I rest-frame bands (from left to right) for clusters stacked
  together.  The first line of figures is for clusters with redshifts
  $0.40\leq z<0.65$ and the last line of figures is for clusters with
  redshifts $0.65\leq z<0.90$. Red crosses are red-sequence GLFs
  normalized to 1 deg$^2$. The vertical red lines indicate the 90\%
  completeness limit. Red curves are the best Schechter fits to
  red-sequence galaxies. The slope of the fit $\alpha$ is given with
  the number of clusters and the mean redshift of the stack. Refer to
  Table~\ref{tab:glf_stackz} for all Schechter fit parameters.}
\label{fig:glf_stack_z}
\end{figure*}

\begin{table*}
\centering
\caption{Parameters of the best Schechter function fits for
    stacked cluster GLFs normalized to 1 deg$^2$ for red-sequence. The top part is for clusters with redshifts $0.40\leq z<0.65$ and the bottom is for clusters with redshifts $0.65\leq z<0.90$. "$N_{clus}$" is the number of clusters in the stack, "comp" is the 90\% completeness limit and "$<z>$" is the mean redshift of
    the stack. 
    See Fig.~\ref{fig:glf_stack_z} for the plots of the GLFs.}
\begin{tabular}{l|cccccc}
\hline
\hline
&$N_{clus}$&comp&$<z>$& $\alpha$ & M*  &$\phi$* ($deg^{-2}$)\\ 
\hline
clusters at & & & & red-sequence GLFs & & \\
$0.40\leq z<0.65 $& & & & & & \\
B & 11&-18 & 0.52 & -0.59$\pm$ 0.19    & -20.5$\pm$0.3  & 6278 $\pm$ 1366  \\
V & 13&-18.5 & 0.50 & -0.74$\pm$ 0.11  &-21.6 $\pm$0.2 & 4062$\pm$724      \\
R & 13&-19 & 0.50 & -0.66$\pm$ 0.15   & -21.9$\pm$0.2  & 3898 $\pm$734     \\
I & 11&-19.5 & 0.49 & -0.16$\pm$ 0.18   & -21.6$\pm$0.2  & 7101 $\pm$ 747  \\
\hline
 clusters at & & & & red-sequence GLFs & & \\
  $0.65\leq z<0.90 $& & & & & &  \\
B &4 & -19.5 & 0.84 & 1.71$\pm$ 0.72    & -20.6$\pm$0.5  &3476$\pm$ 2285   \\
V &4 & -20 & 0.84   & 0.64$\pm$ 0.49    & -22.0$\pm$0.4  &5936$\pm$1130    \\
R &4 & -21.5 & 0.84 & 0.15$\pm$ 1.01    & -22.4$\pm$1.0  &4795$\pm$972     \\
I &3 & -22.5 & 0.84 & 0.23$\pm$ 1.24    & -22.9$\pm$0.6  &4212$\pm$1605    \\
\hline
\hline
\end{tabular}
\label{tab:glf_stackz}
\end{table*}

To investigate the evolution of the GLF with redshift, we apply the
same analysis as previously, but separate our clusters between low
($0.4\leq z<0.65$) and high ($0.65\leq z<0.9$) redshifts. We obtain
about 13 low and 4 high-redshift clusters, which each have more than
20 red-sequence galaxies depending on the photometric
band. Unfortunatly, there are an unsufficient number of blue cluster
galaxies to fit blue GLFs for these two redshift intervals. Our
results for red-sequence galaxies are displayed in
Fig.~\ref{fig:glf_stack_z} and Table~\ref{tab:glf_stackz}.

For the low-redshift sample, the faint end of the GLFs is similar to
that of the stacked GLF for all clusters. This is evident from the
similarity of the $\alpha$ parameters for both samples
(Tables~\ref{tab:glf_stack} and \ref{tab:glf_stackz}). This means that
the faint end of our stacked GLF for all clusters is dominated by the
low-redshift clusters. This is not surprising as there are fewer high
redshift clusters than low-redshift ones. In addition, low-redshift
clusters tend to have fainter completeness limits, hence are more
likely to contribute to the faintest bins of the GLF than the
high-redshift clusters.

The red GLFs of high-redshift clusters decline far more sharply at
their faint end than the low-redshift clusters. The $\alpha$ parameter
is significantly higher even with those large error bars. On the other hand, $M^*$ is slightly brighter than at low-redshift and equal to
that for the fit of all clusters taken together, meaning that the bright end is
dominated by the high-redshift clusters. There may be
more bright galaxies in high redshift clusters but this result can
also be due to the k-correction which is higher for high-redshift
clusters and tends to distort the bright end of the GLF such that the
Schechter function is not appropriate any more. This can be clearly seen in the B band where there is a significantly larger number of red bright galaxies
than predicted by the Schechter fit. The best-fit Schechter parameters for the B band are therefore unreliable for our high-redshift sample. We also
note that there are very few clusters at high-redshift, so that
we need to target more high-redshift clusters to decrease the
error bars in both the galaxy counts and Schechter parameters. In this
particular case we took into account every magnitude bin with at least
two clusters, a number that could be increased if we observed more clusters.

These dependences of the GLF properties on redshift are interpreted in terms of
physical processes in our discussion.

We also note that overestimating the completeness
limit in apparent magnitude would also lead to a sharp decline in the
faint end of GLFs. This drop would also increase with redshift,
because the completeness limit in absolute magnitude would be brighter at
high redshift. Since we compute accurate 90\% completeness limits for
every image, we should not be affected by this effect.

\subsection{Dependence of GLFs on cluster X-ray luminosity}

\begin{figure*}
\begin{tabular}{cccc}
\includegraphics[width=0.17\textwidth,clip,angle=270]{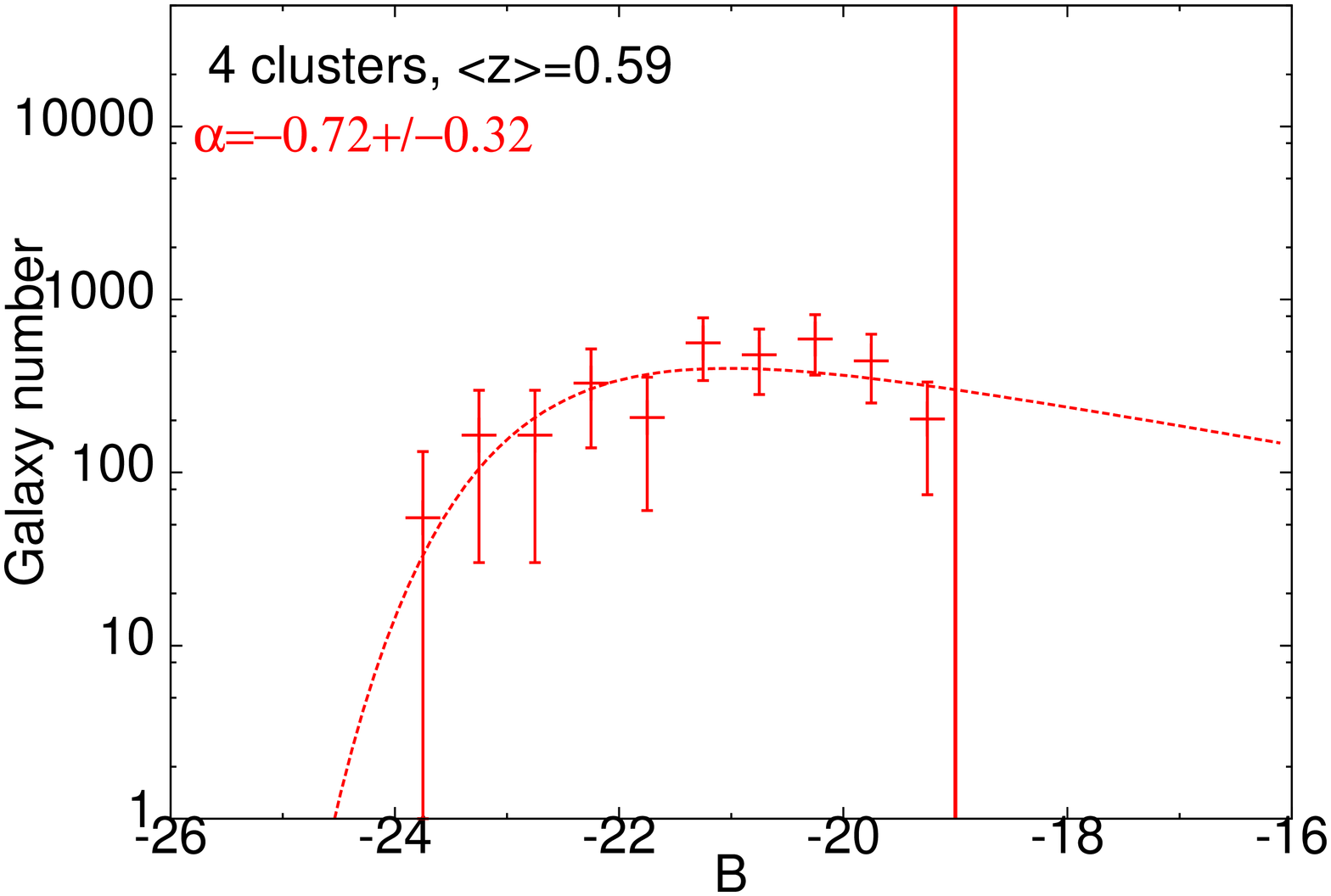}
\includegraphics[width=0.17\textwidth,clip,angle=270]{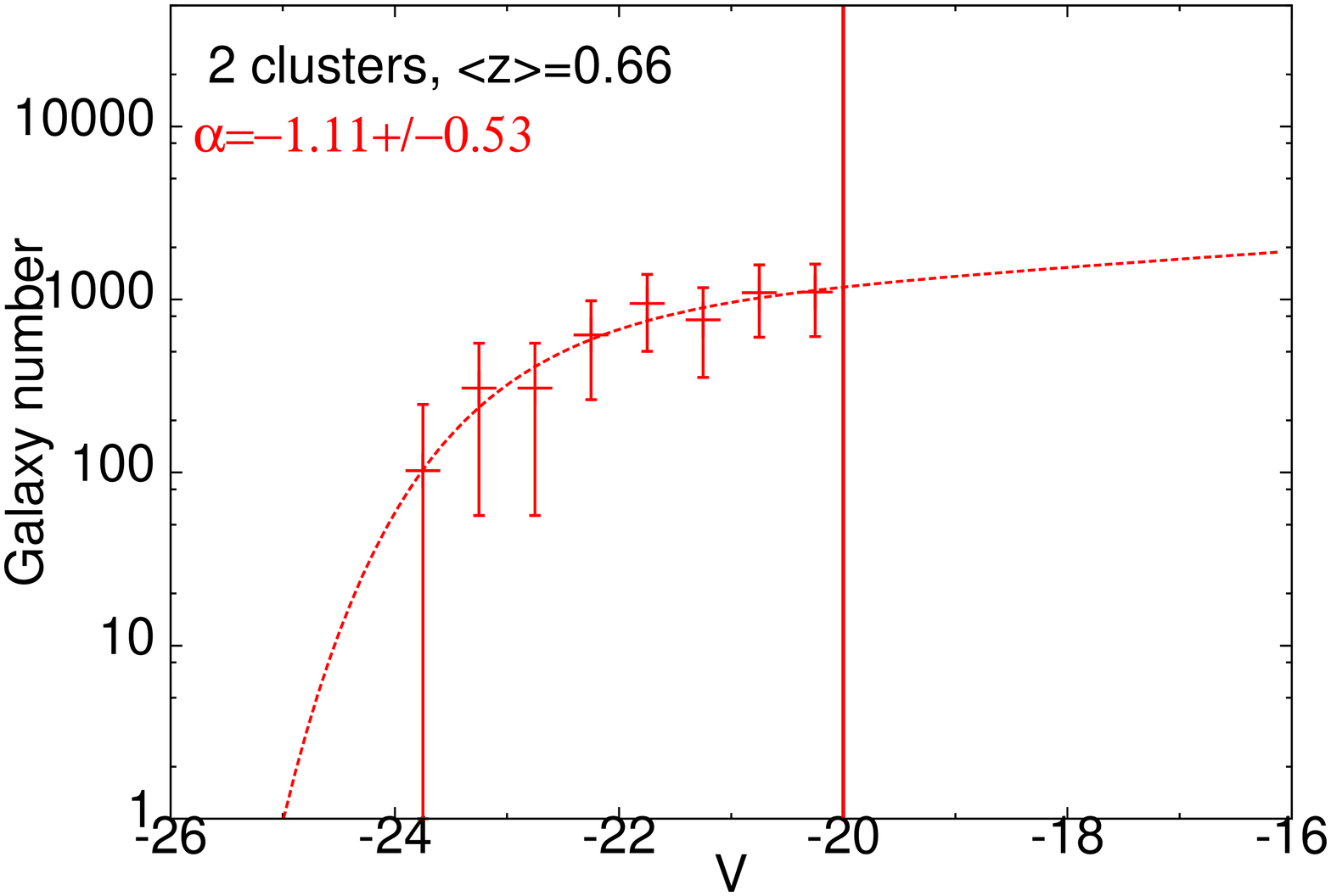}
\includegraphics[width=0.17\textwidth,clip,angle=270]{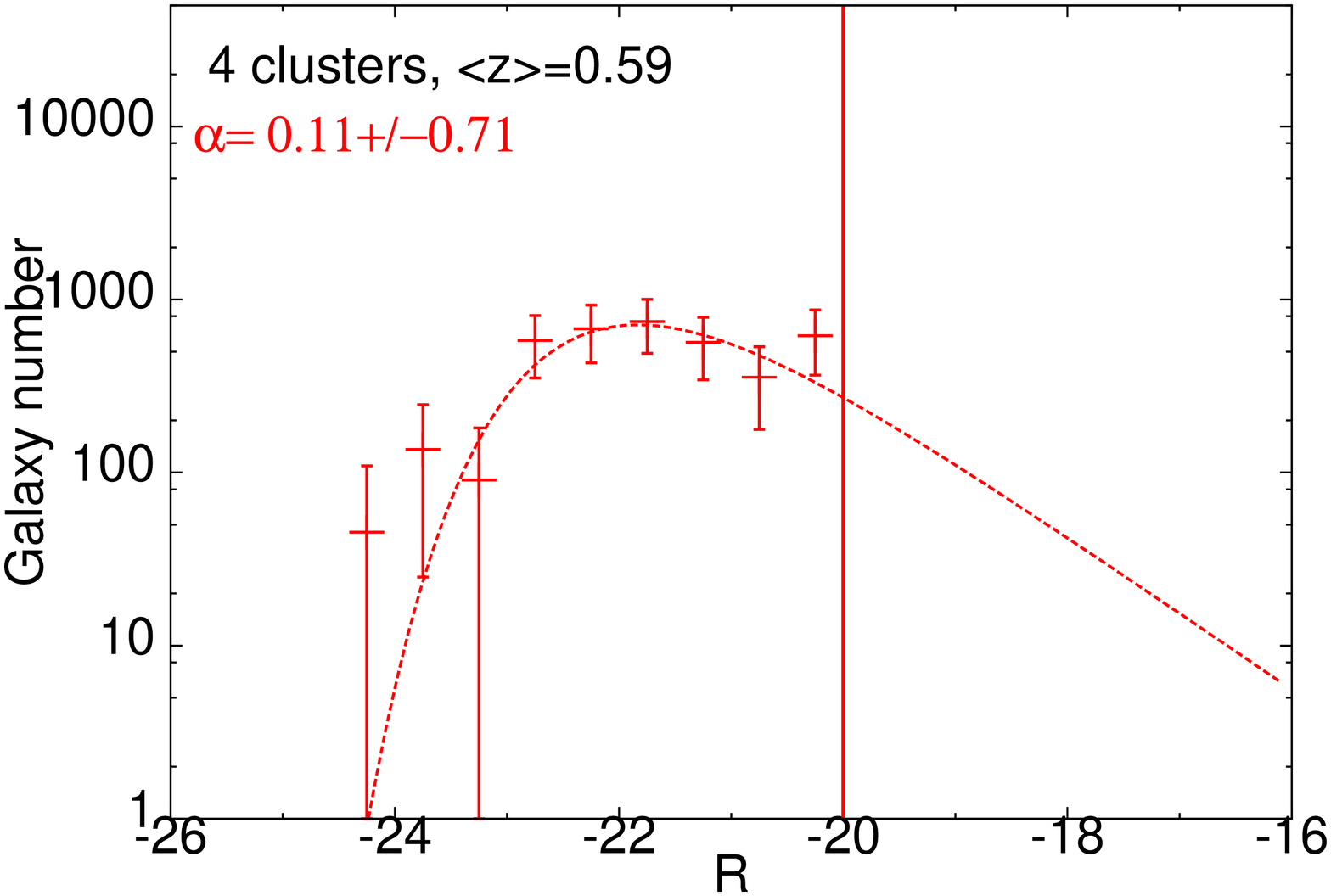}
\includegraphics[width=0.17\textwidth,clip,angle=270]{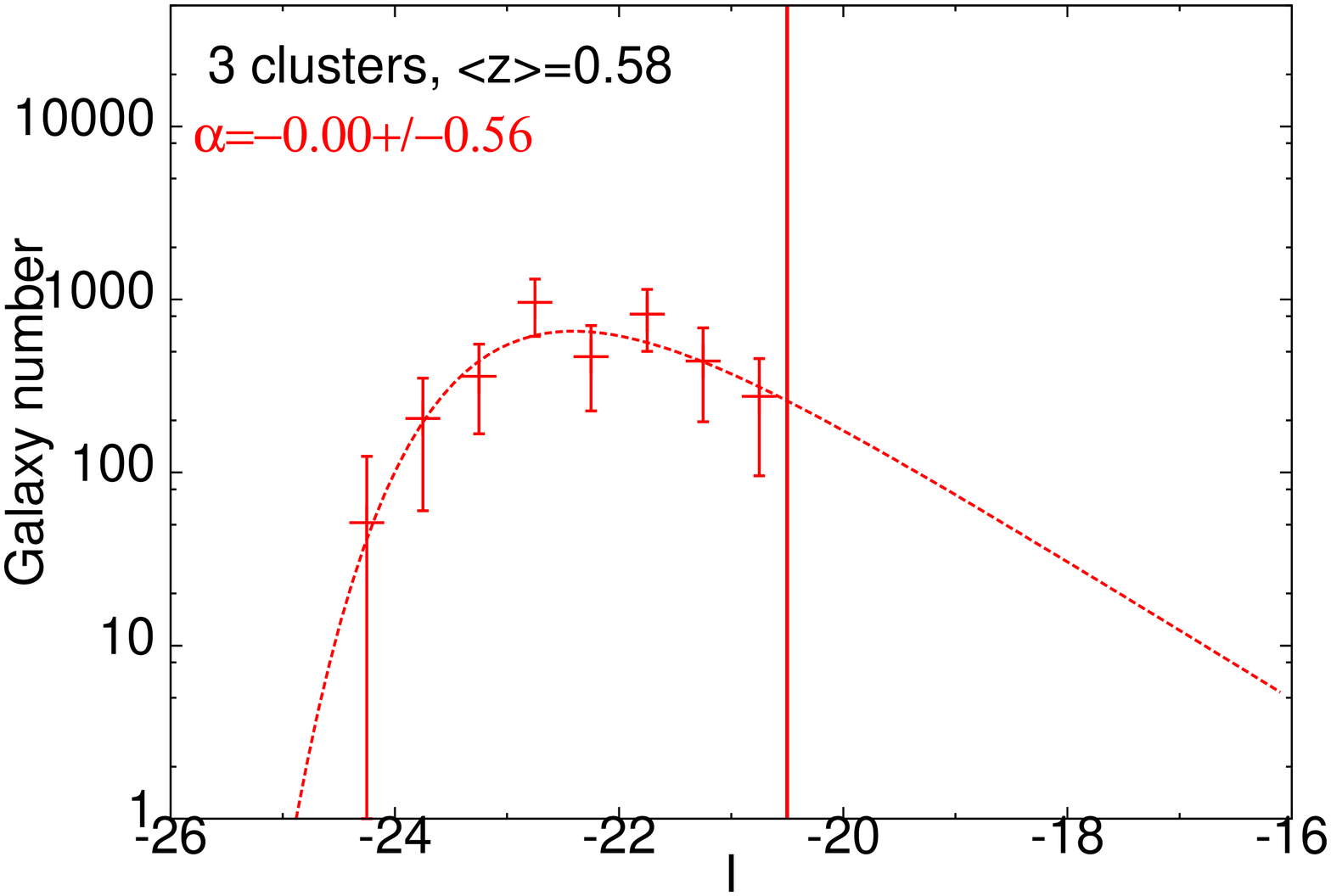} \\
\includegraphics[width=0.17\textwidth,clip,angle=270]{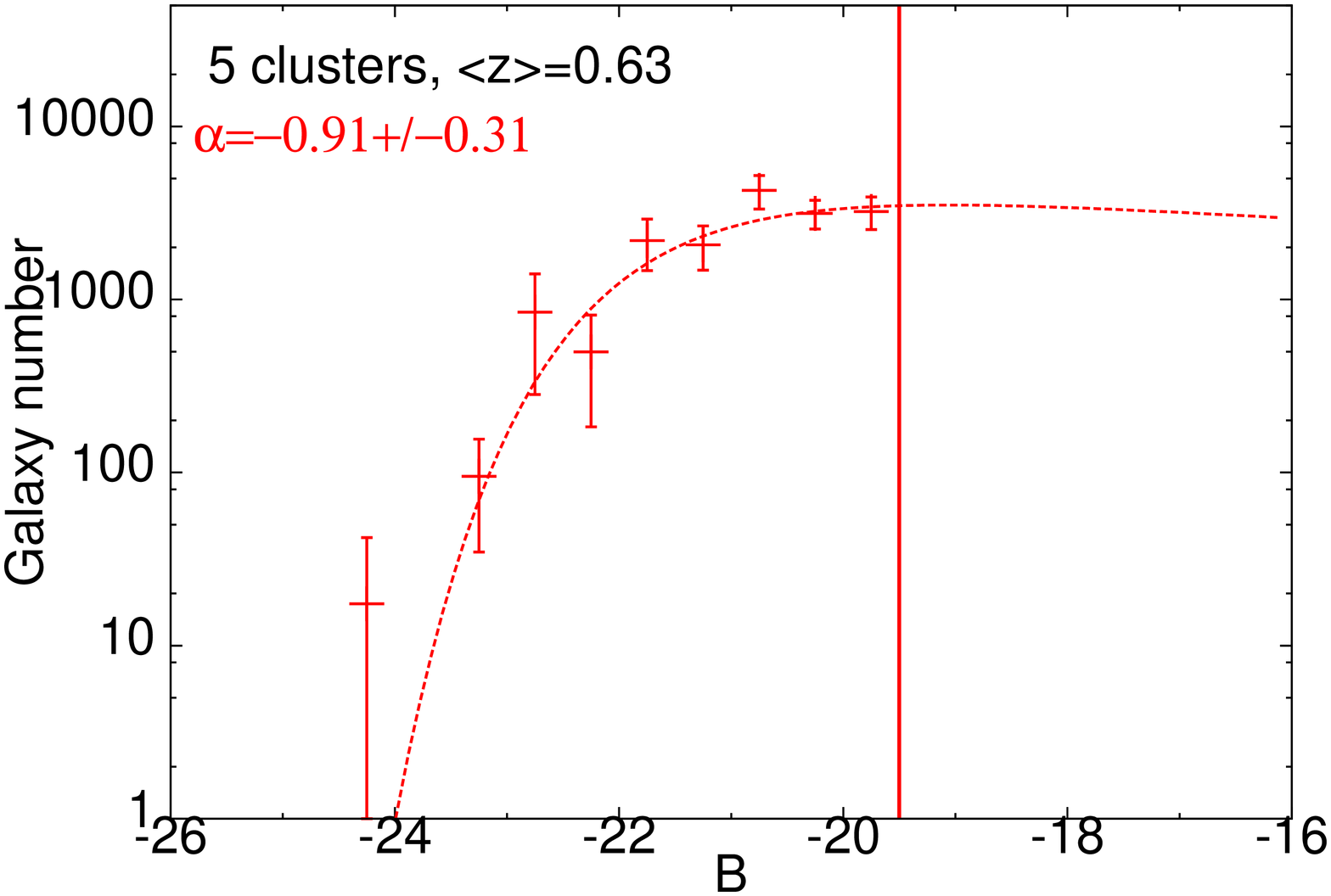}
\includegraphics[width=0.17\textwidth,clip,angle=270]{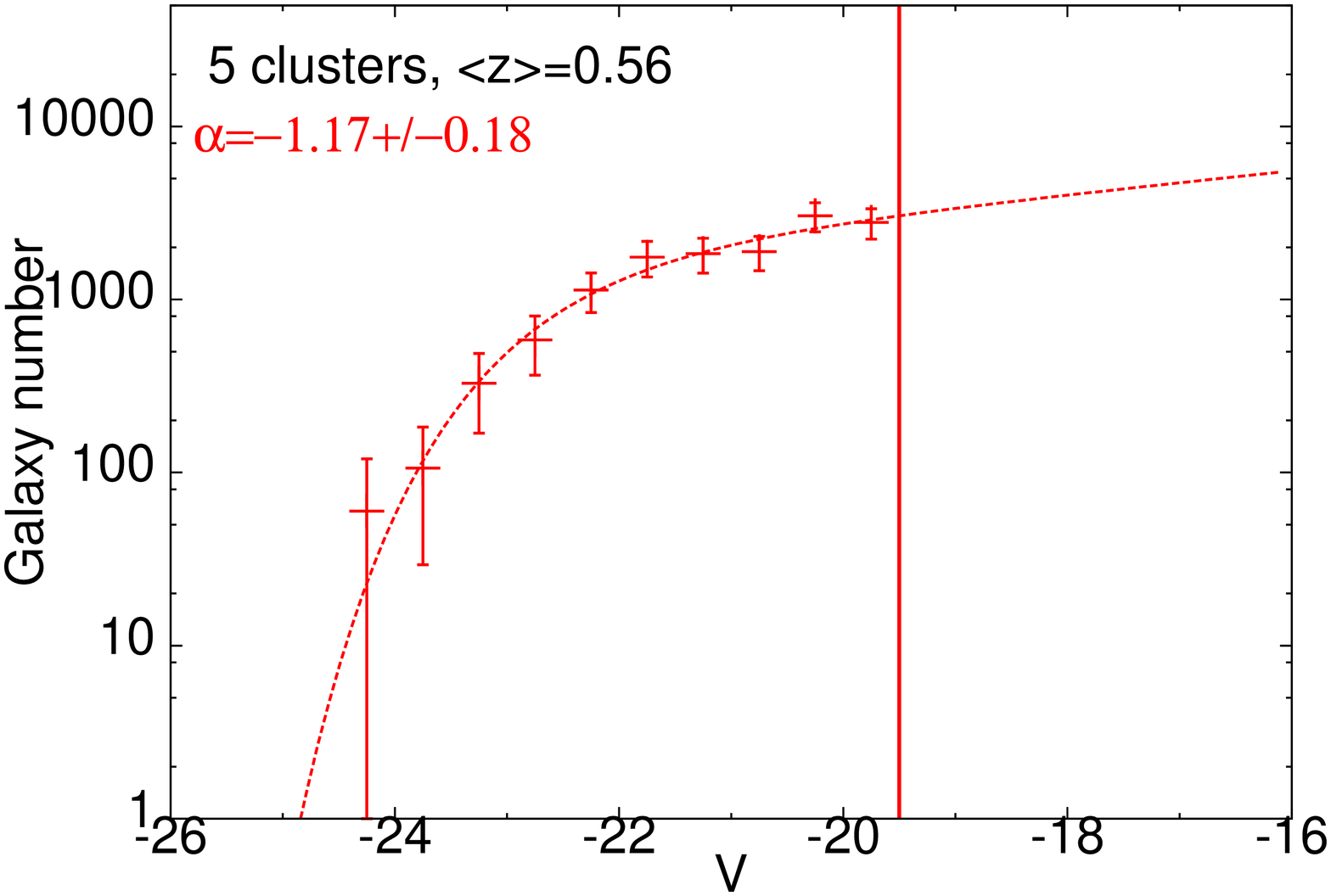}
\includegraphics[width=0.17\textwidth,clip,angle=270]{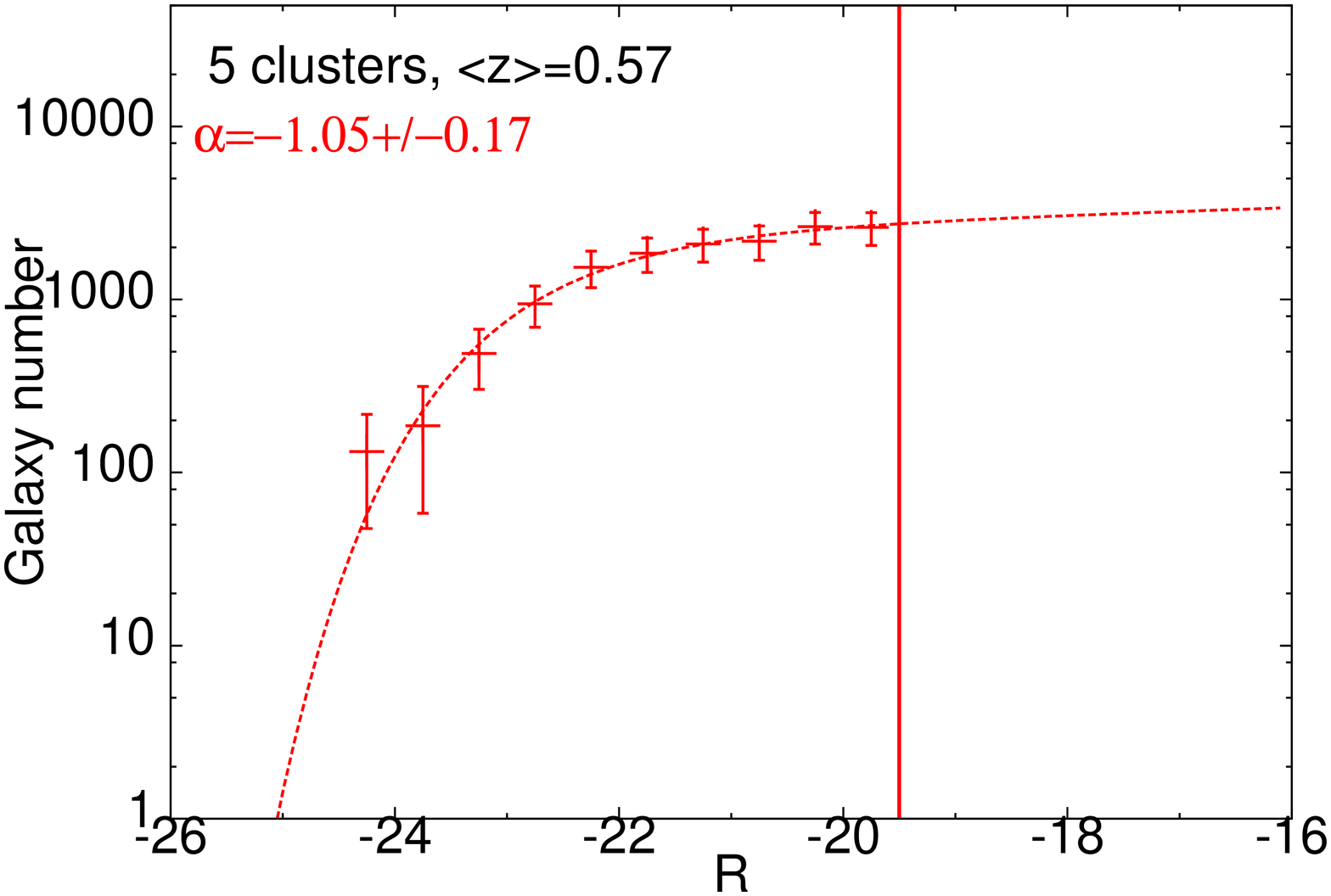}
\includegraphics[width=0.17\textwidth,clip,angle=270]{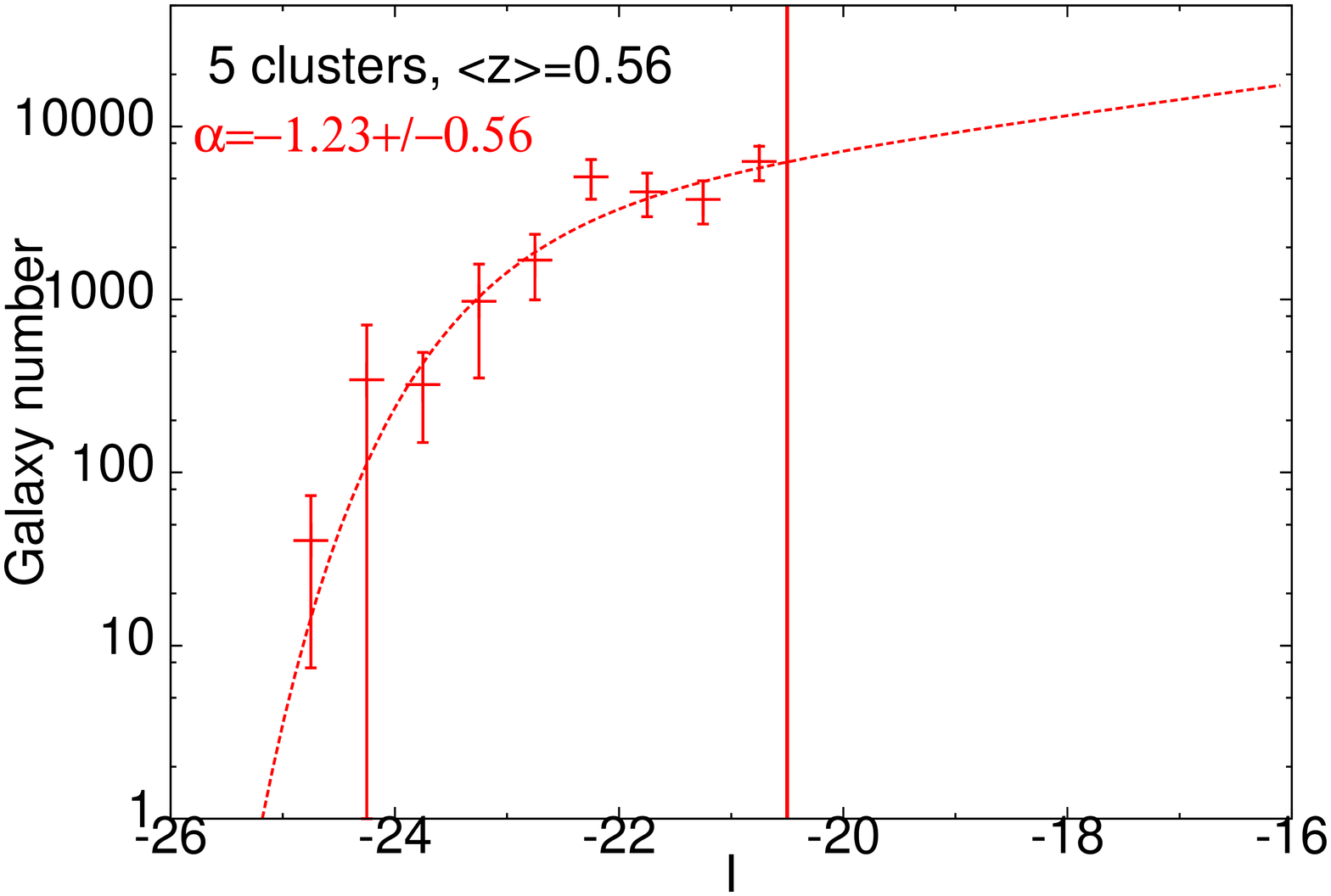} \\
\end{tabular}
\caption{Dependence of red-sequence GLFs on cluster X-ray
    luminosity in the B, V, R, and I rest-frame bands (from left to
    right) for clusters stacked together. The top line is for clusters
    with X-ray luminosities $8.10^{43}<L_X<10^{45}$~erg.s$^{-1}$ and
    the bottom line is for clusters with X-ray luminosities
    $L_X>10^{45}$~erg.s$^{-1}$.  Red crosses are red-sequence GLFs
    normalized to 1 deg$^2$. The vertical red lines indicate the 90\%
    completeness limit. Red curves are the best Schechter fits to 
    red-sequence galaxies. The slope of the fit $\alpha$ is given with the
    number of clusters and the mean redshift of the stack.}
\label{fig:glf_stack_mass}
\end{figure*}

We can similarly investigate the dependence of the
GLF on a mass proxy. To achieve this, we separate clusters according to their
X-ray luminosity, as measured in \citet{Gue13b} who analysed the
XMM-Newton data available for 42 DAFT/FADA clusters to derive their
X-ray luminosities and temperatures, and search for substructures. We
have about 5 clusters with a luminosity greater than
$10^{45}$~erg.s$^{-1}$ and 4 with a lower luminosity depending on the
considered optical band. These numbers are for RS galaxies ;
we do not have enough clusters with more than 20 blue galaxy members to compile
blue GLFs in this case. We do not have accurate X-ray luminosities for
the remaining clusters.

We find a steeper faint end for high-mass clusters than low-mass ones in every photometric band (Fig.~\ref{fig:glf_stack_mass}). This could
mean that the drop at the faint end of RS GLFs is
essentially due to low mass clusters. In addition, the number of member galaxies is
much larger for high X-ray luminosity clusters, which seems logical.
However, we note that the number of clusters, especially for low-mass
clusters, is small. In this particular case we recall that we consider every
magnitude bin with at least two clusters.  More data are needed to
produce larger samples that cover a wide range of mass but similar in
redshifts, and also to study the variations in the blue and red GLFs
with mass. When our DAFT/FADA sample of about 90 clusters is complete,
we should be able to draw conclusions about the GLF dependence on mass
and redshift, provided that we have the same proportion of clusters
with good completeness as in the present sub-sample. We also
note that our clusters all have quite high X-ray luminosities. Our
results therefore only concern clusters with X-ray luminosities ${\rm
  L_X}>8\times10^{43}$~erg.s$^{-1}$.

\subsection{Dependence of GLFs on substructures}

\begin{figure*}
\begin{tabular}{cccc}
\includegraphics[width=0.17\textwidth,clip,angle=270]{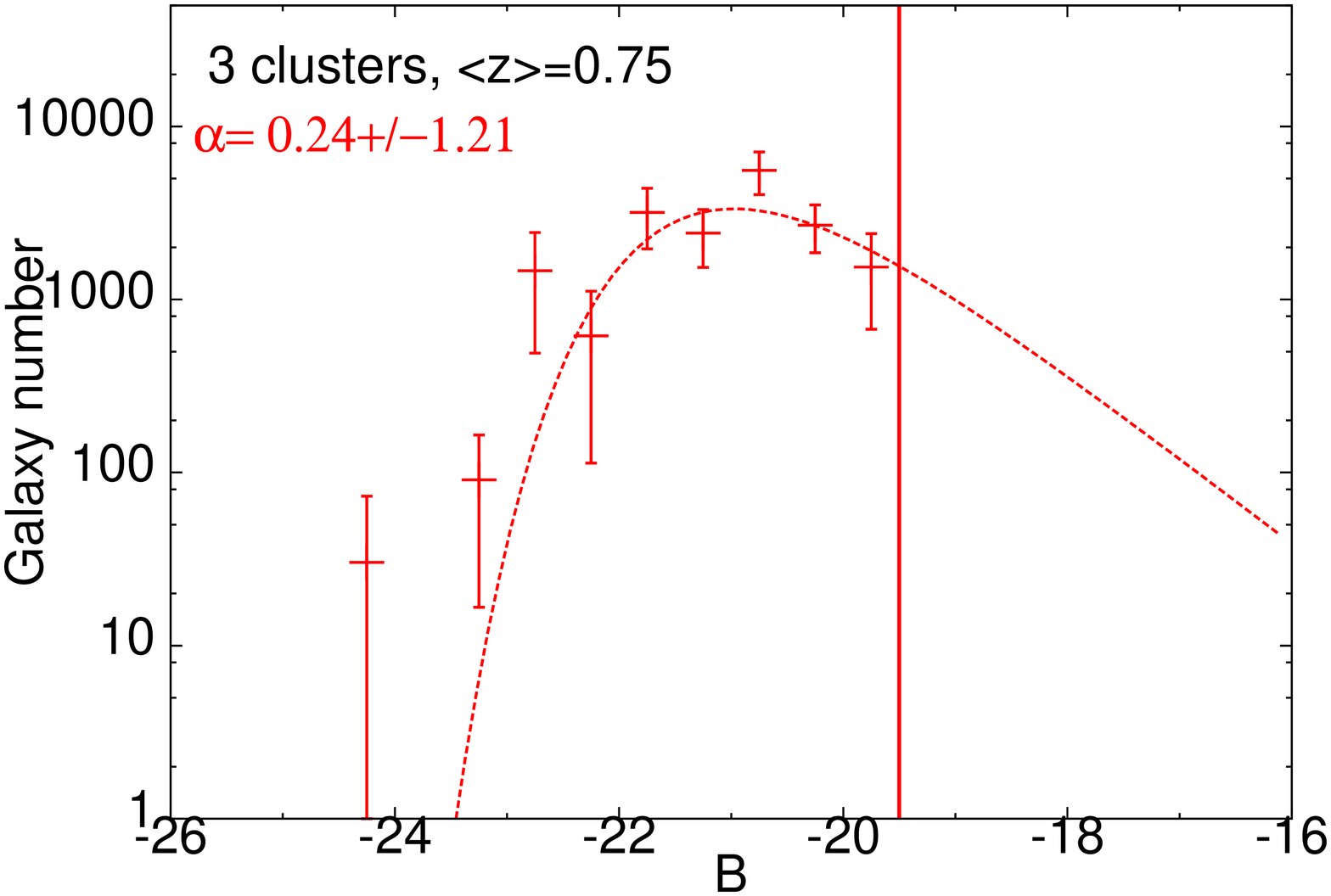}
\includegraphics[width=0.17\textwidth,clip,angle=270]{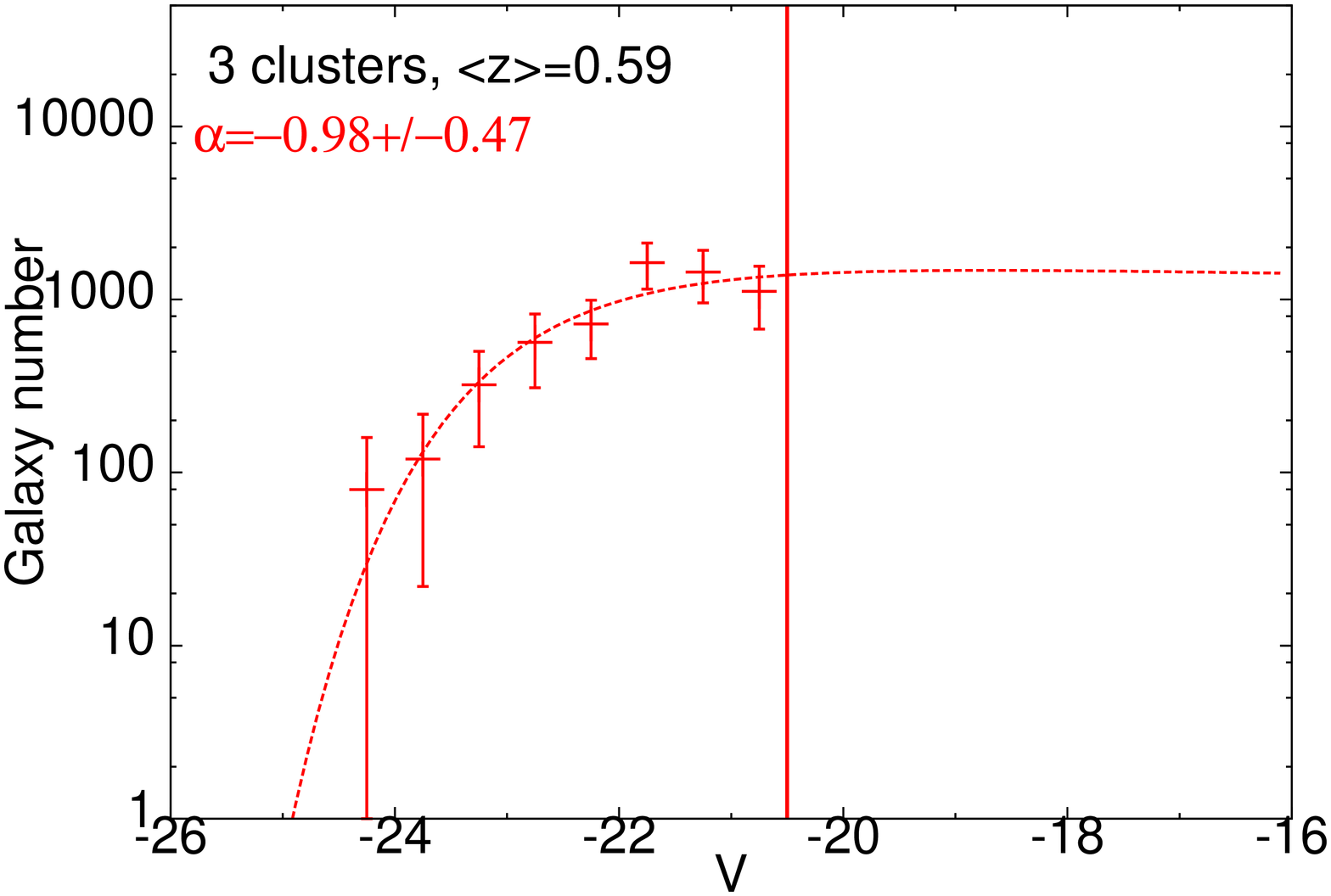}
\includegraphics[width=0.17\textwidth,clip,angle=270]{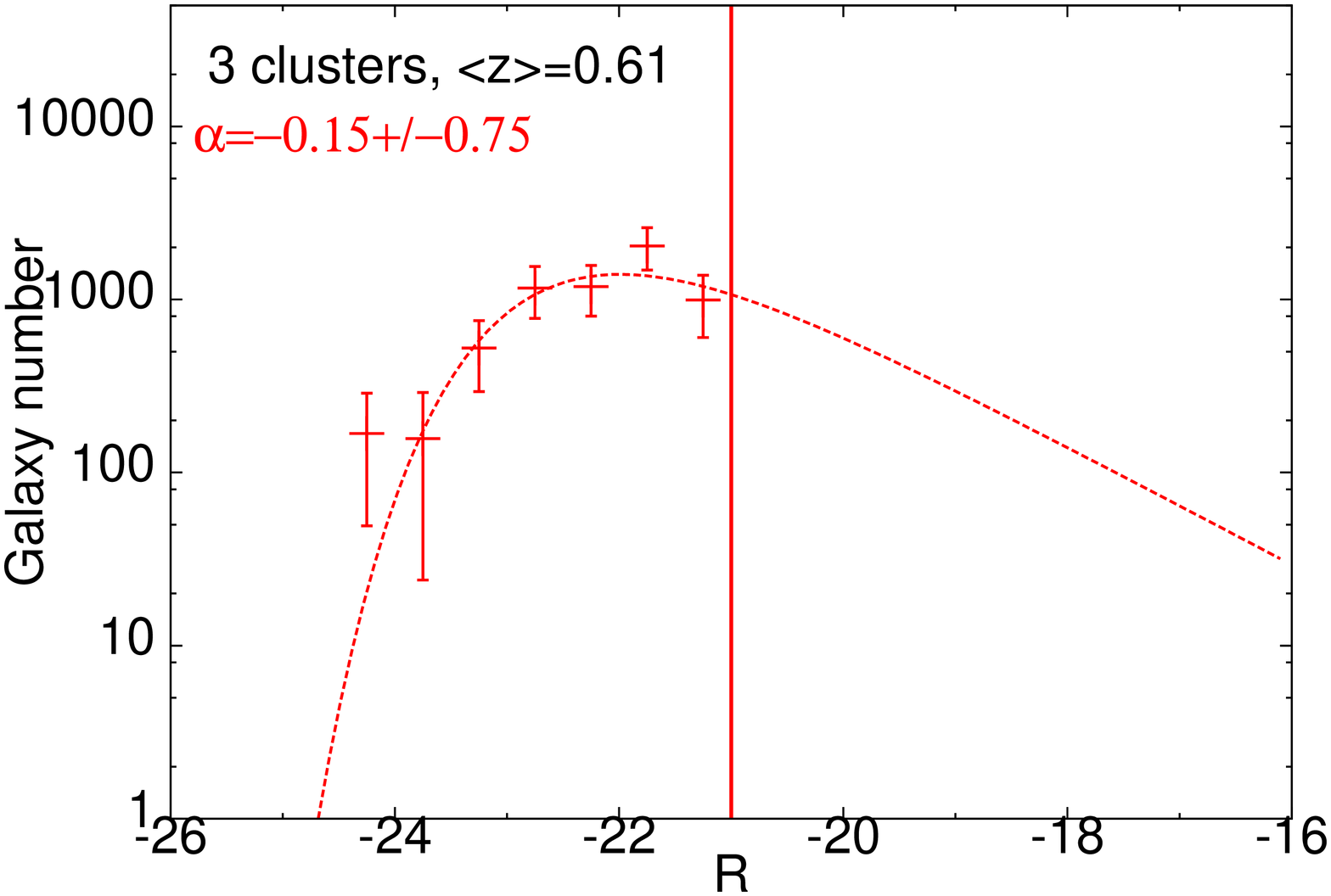}
\includegraphics[width=0.17\textwidth,clip,angle=270]{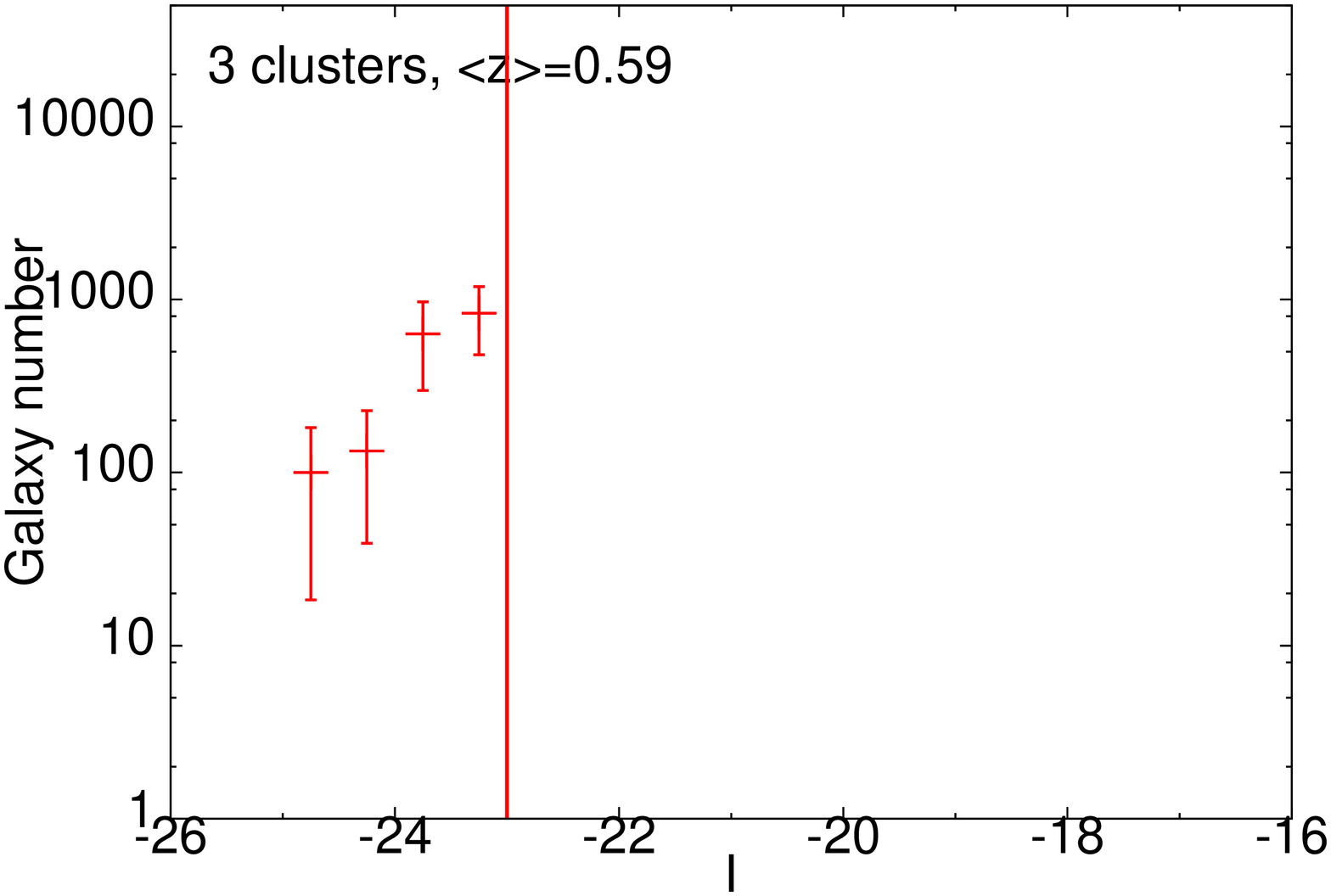}
\end{tabular}
\caption{Dependence of red-sequence GLFs on cluster
  substructures in the B, V, R, and I rest-frame bands (from left to
  right) for clusters stacked together. Only clusters with detected
  substructures are considered here \citep[cf.][]{Gue13b}. Red crosses
  are red-sequence GLFs normalized to 1 deg$^2$. The vertical red
  lines indicate the 90\% completeness limit. Red curves are the best
  Schechter fits to red sequence galaxies. The slope of the fit
  $\alpha$ is given with the number of clusters and the mean redshift
  of the stack.}
\label{fig:glf_stack_ss}
\end{figure*}

We also search for differences between clusters with and
without substructures. We consider clusters with
substructures detected both with optical spectroscopy and X-ray data
by \citet{Gue13b}. We have 3 clusters with substructures and 2 that
are relaxed and sufficiently rich in red galaxies. For the remaining
clusters, we have been unable to robustly confirm either the presence
or absence of substructures and therefore discard these clusters in this
subsection.

It is difficult to draw any conclusions about relaxed clusters as
they are too few in number here and there completeness limit is too bright. Hence, we
only study stacked clusters with substructures
(Fig.~\ref{fig:glf_stack_ss}). In this particular case
  we allow some bins to contain as few as two clusters to be able to
  draw the red-sequence GLFs. There is no clear difference in either the slope
  or $M^*$ of the Schechter function from those parameters for stacks containing all
  clusters, given the large error bars caused by the low number of clusters. In
  the I band, the very bright completeness limit does not allow us to
  study the faint part of the GLF.

For stacks of clusters with substructures, we also have higher counts for B-band data than for other bands. This is consistent
with a burst of star formation being produced as the clusters merge. These faint blue
galaxies might also be the debris of any merging processes. However, we need data with a fainter
completeness limit to investigate whether these debris dwarf galaxies exist. 

Given the error bars in our Schechter parameters, our interpretations
of the analysis of our sub-structured clusters are not statistically
significant. We could reduce our error bars by either reducing the
number of background galaxies, i.e. adding more clusters to the
stacks, or computing more accurate photometric redshifts for field and
cluster galaxies.

\subsection{Cluster cores and outskirts}

\begin{figure*}
\begin{tabular}{cccc}
\includegraphics[width=0.17\textwidth,clip,angle=270]{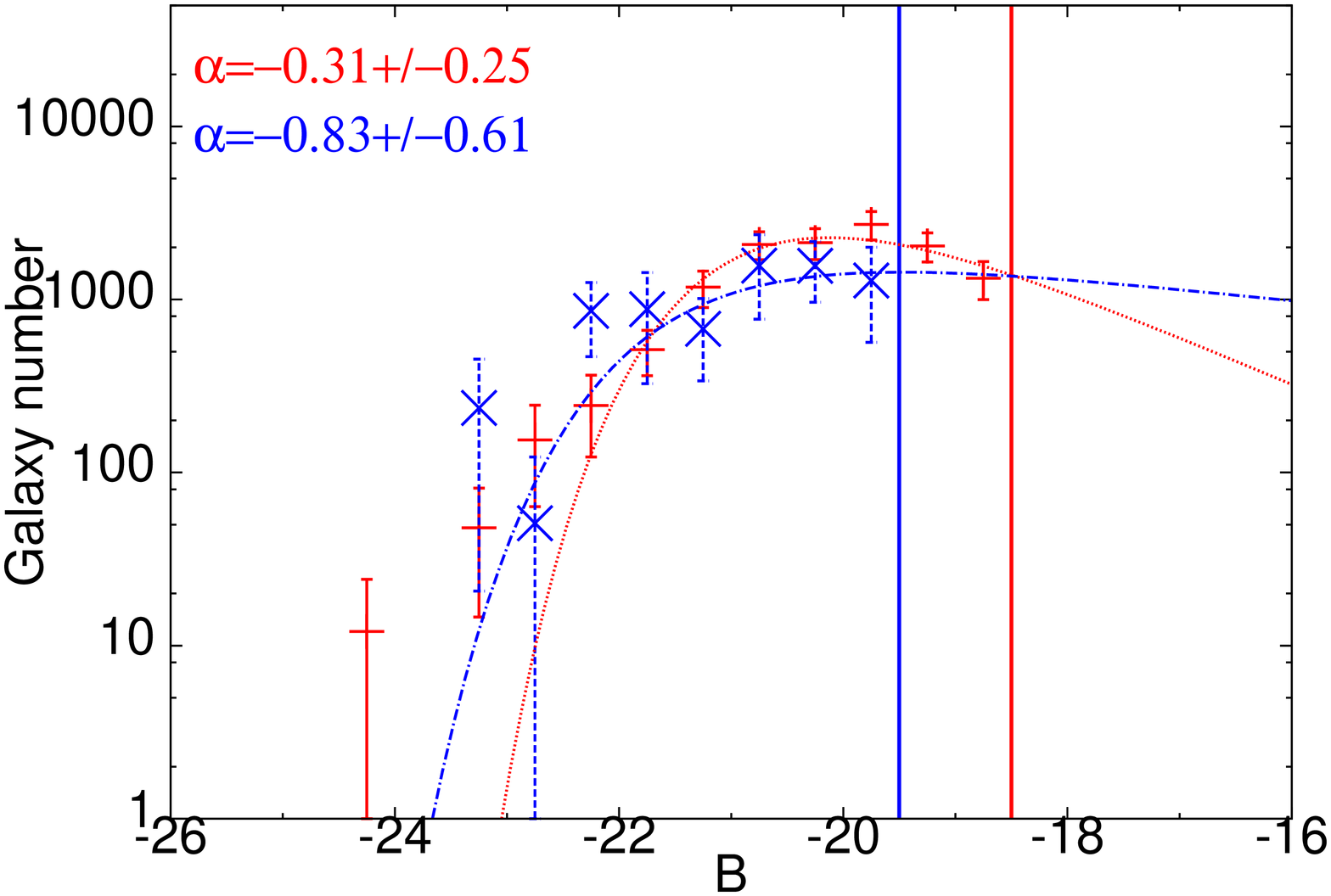}
\includegraphics[width=0.17\textwidth,clip,angle=270]{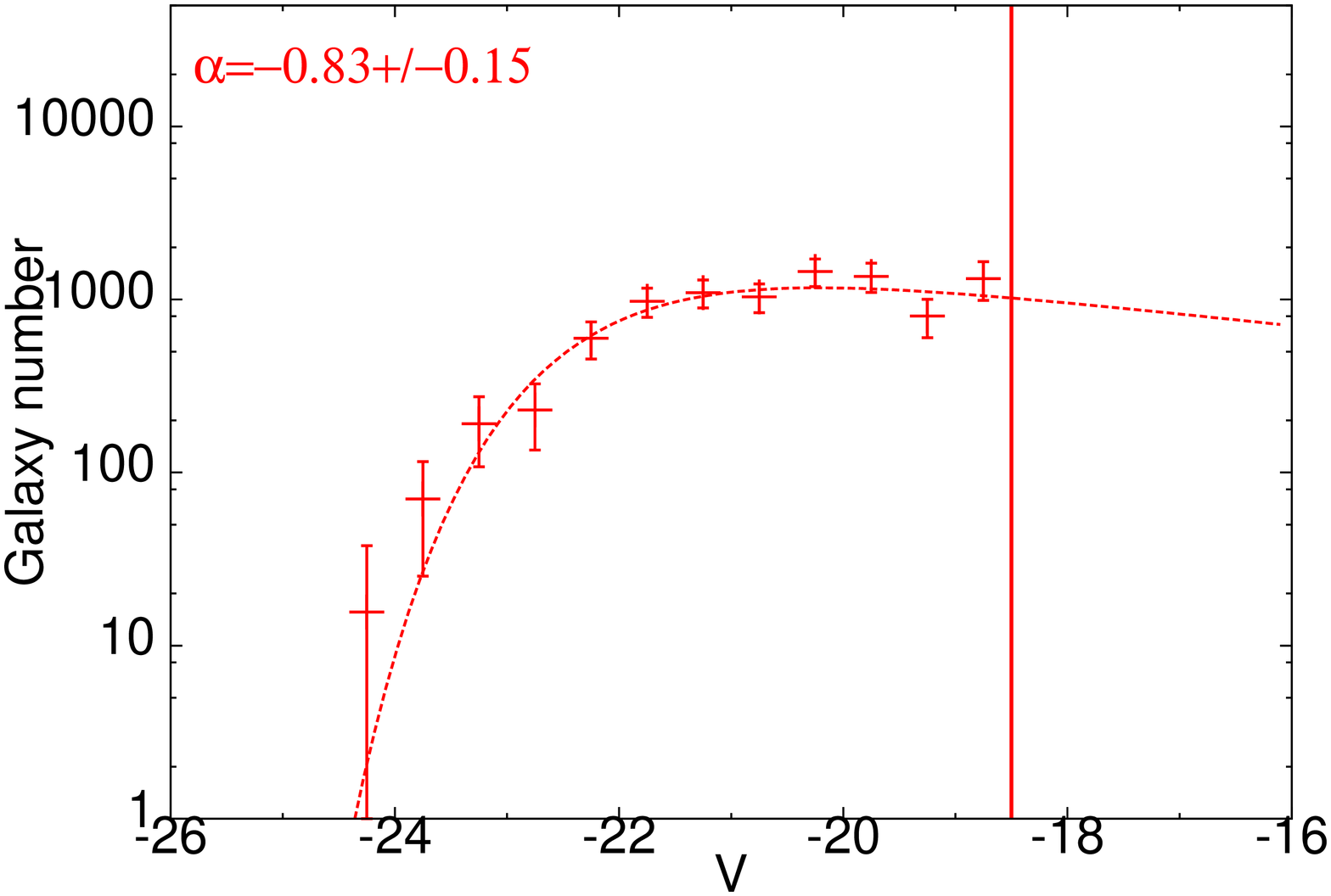}
\includegraphics[width=0.17\textwidth,clip,angle=270]{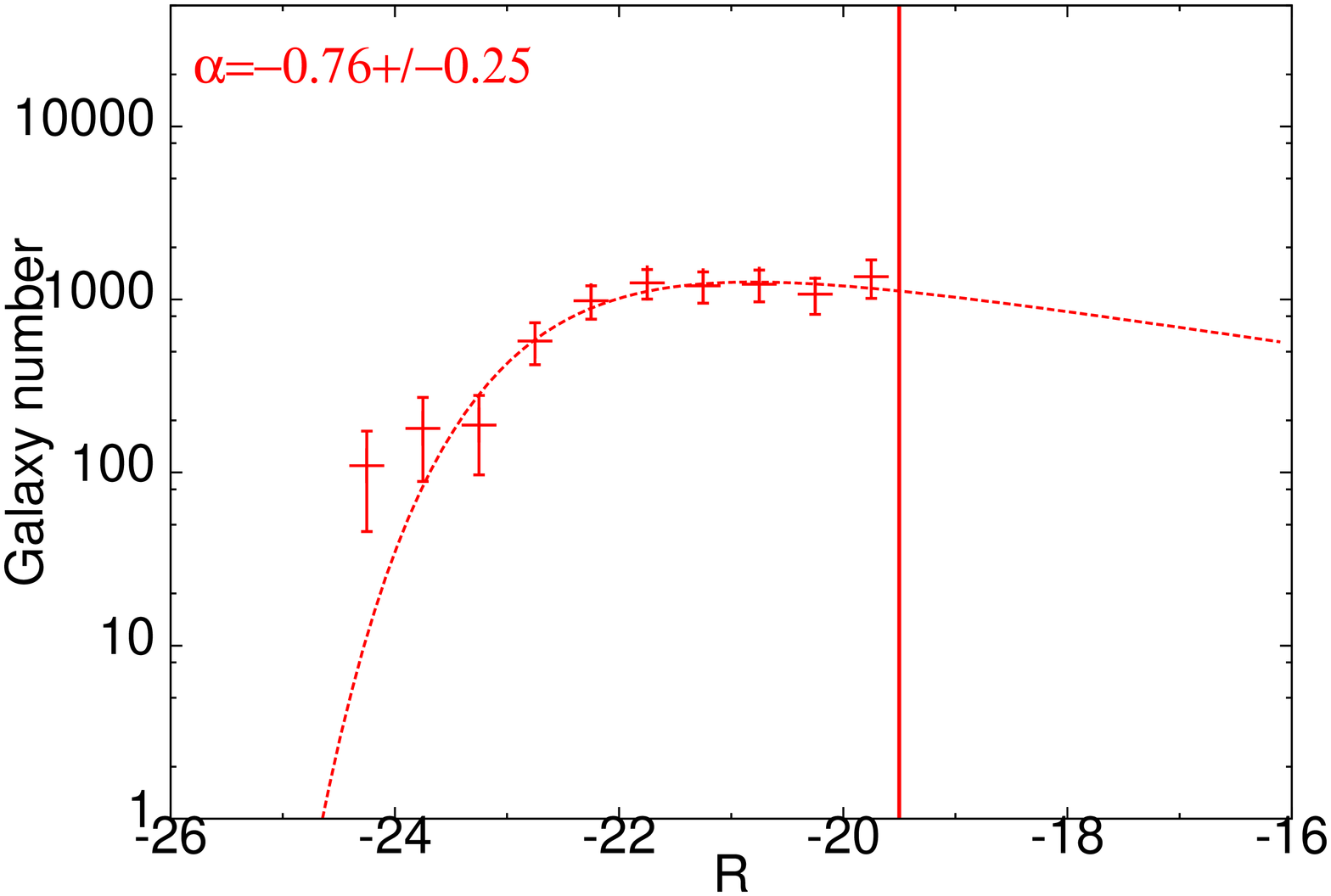}
\includegraphics[width=0.17\textwidth,clip,angle=270]{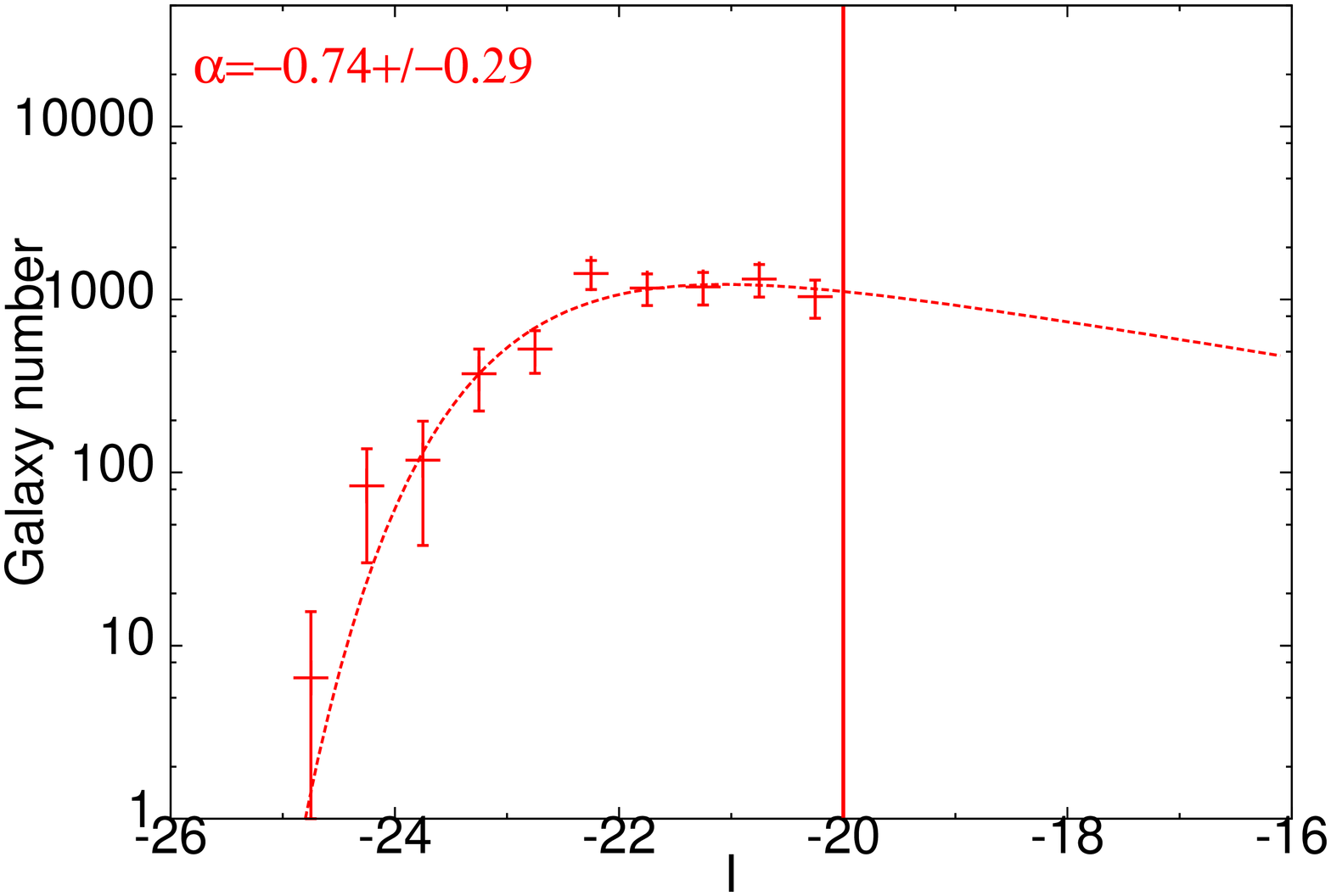} \\
\includegraphics[width=0.17\textwidth,clip,angle=270]{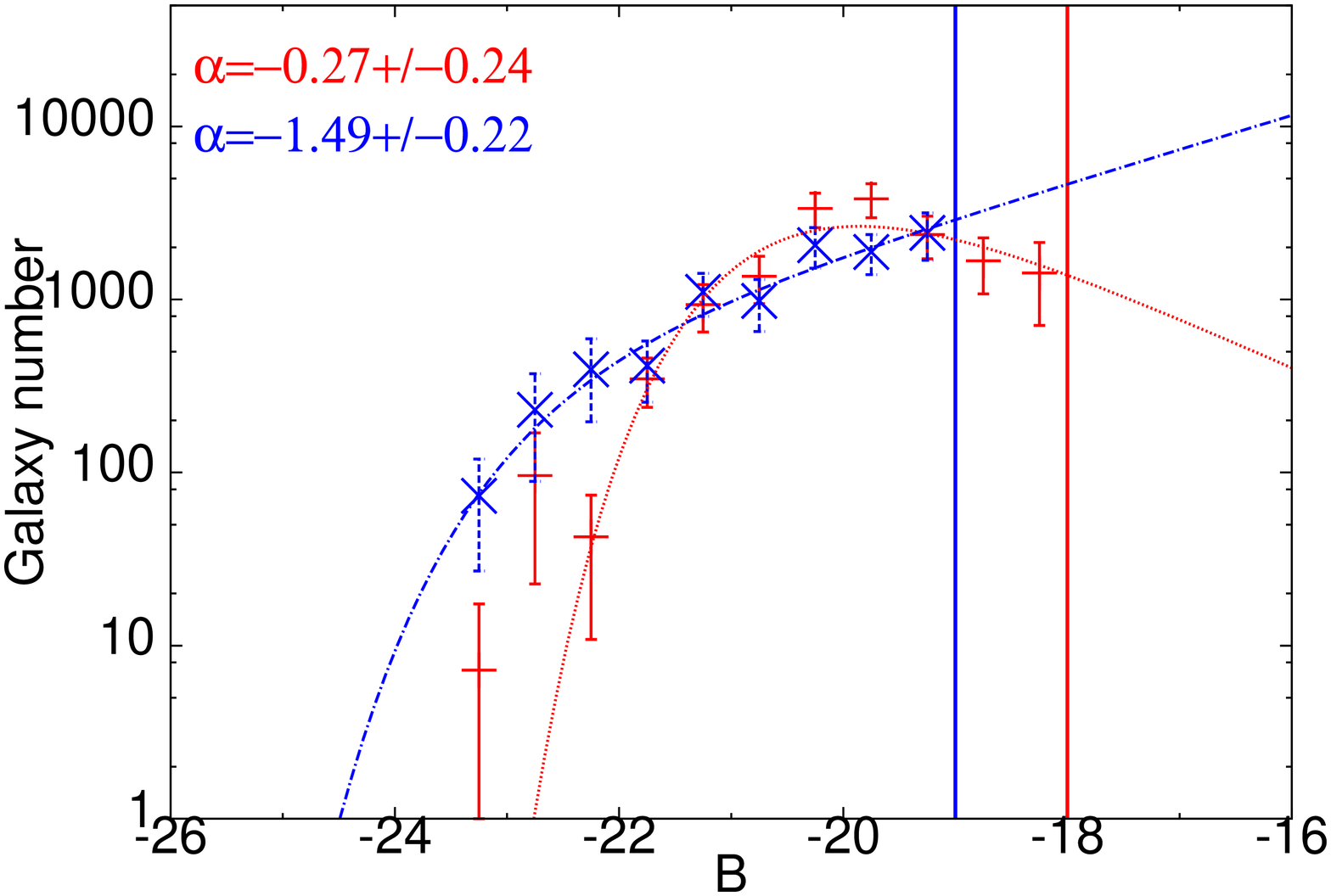}
\includegraphics[width=0.17\textwidth,clip,angle=270]{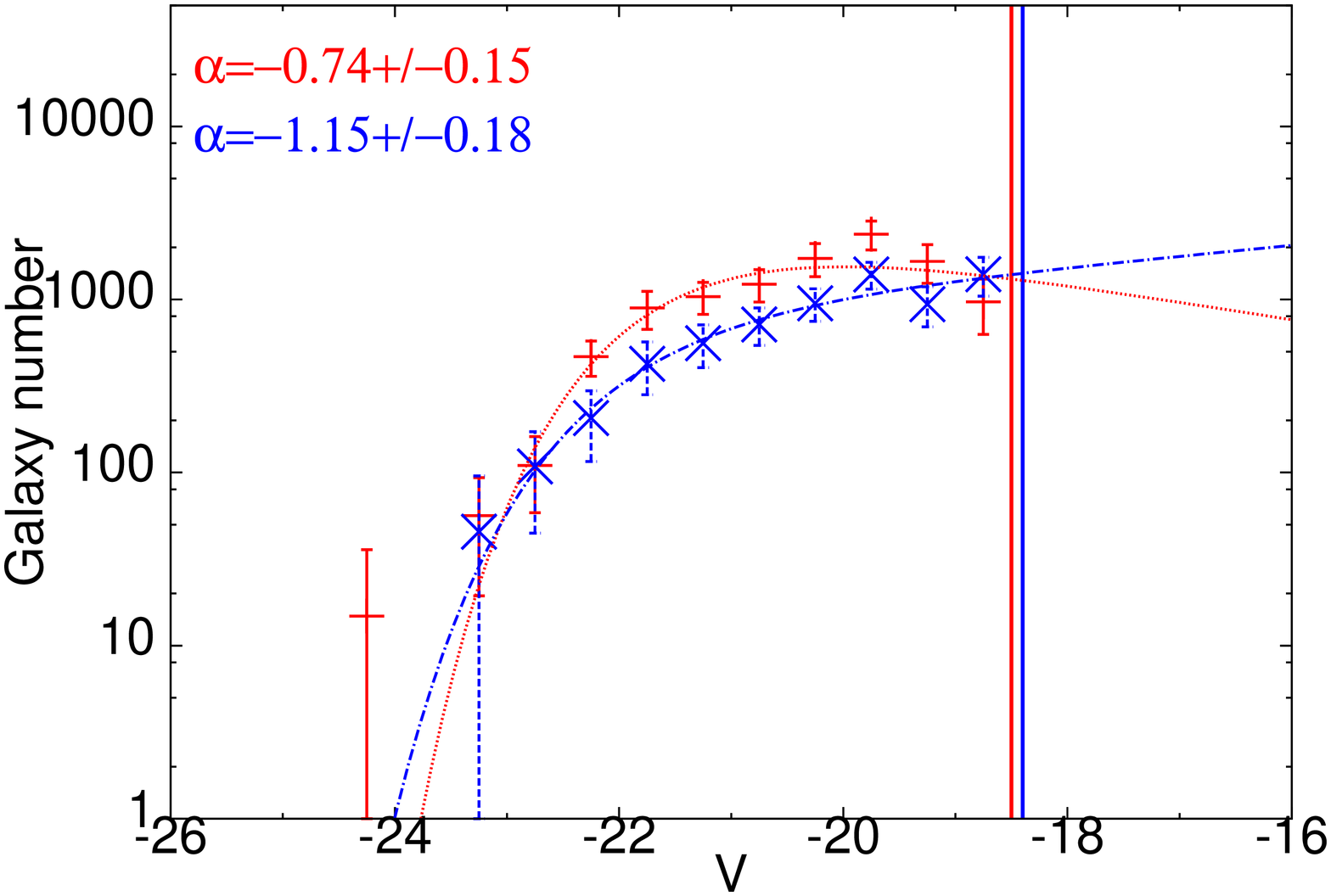}
\includegraphics[width=0.17\textwidth,clip,angle=270]{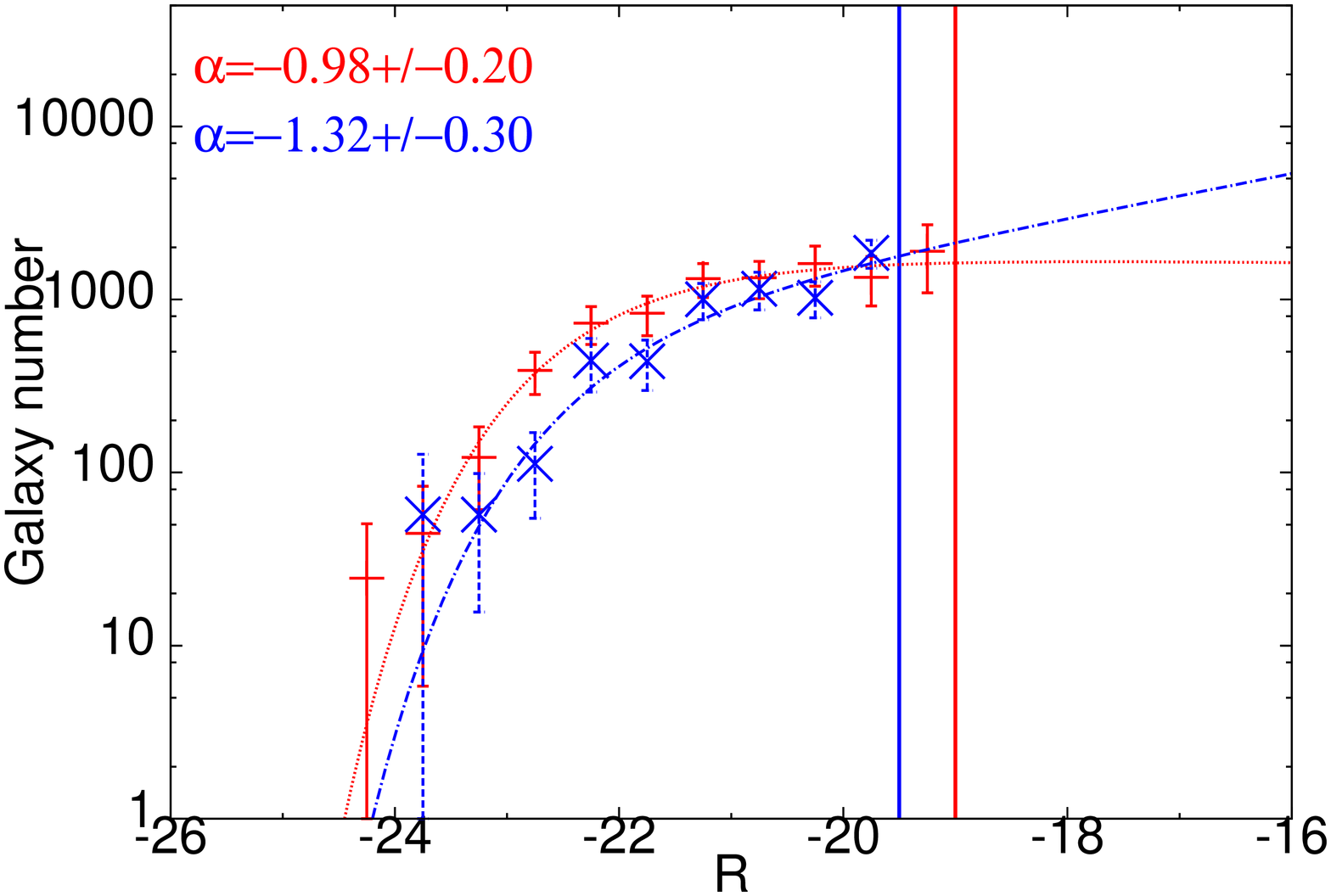}
\includegraphics[width=0.17\textwidth,clip,angle=270]{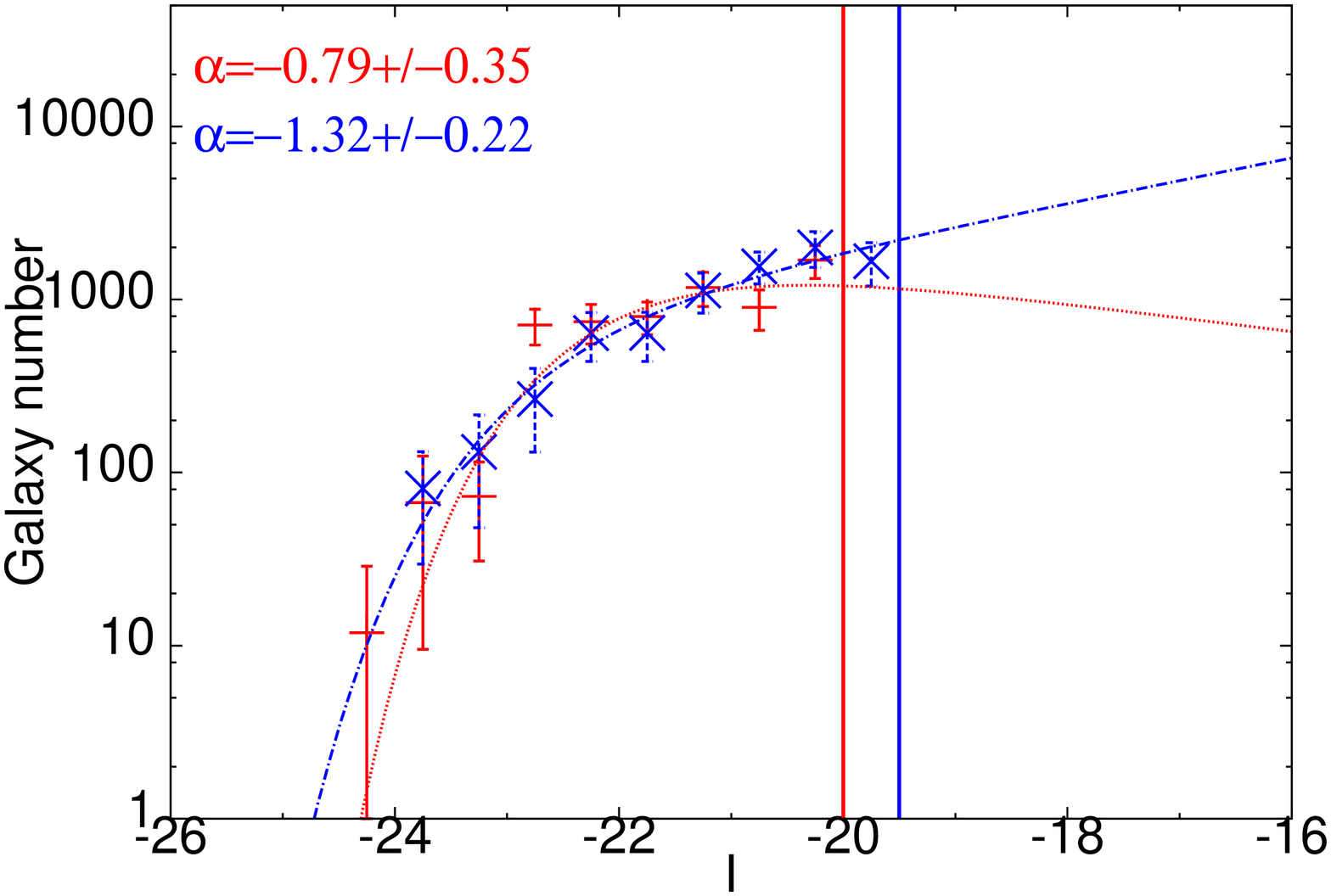} \\

\end{tabular}
\caption{Galaxy luminosity functions in the B, V, R, and I rest-frame
  bands (from left to right) for stacked clusters in cores and
  outskirts. The first line of figures is for cluster cores
  ($r\leq500$~kpc) and the second line of figures for cluster
  outskirts ($500<r\leq1000$~kpc). Red and blue points, respectively,
  correspond to red-sequence and blue GLFs normalized to 1 deg$^2$.The
  red and blue curves show the best Schechter fits to red-sequence and
  blue galaxies and the red and blue vertical lines indicate the
  corresponding 90\% completeness limits. The slope of the fit
  $\alpha$ is given for each population.}
\label{fig:paramocomp2}
\end{figure*}

  In some cases, stellar formation can be triggered by in-fall in
  the outskirts of clusters \citep[e.g.][]{Biv11}. To investigate whether this is true for clusters in general, we compute GLFs for the core
  ($r\leq500$~kpc) and outskirts ($500<r\leq1000$~kpc) of clusters. We
  present the stacked GLFs for blue and red-sequence galaxies in
  different environments in Fig.~\ref{fig:paramocomp2}. This figure is
  to be compared with Fig.~\ref{fig:glf_stack} which displays cluster
  GLFs for the same cluster galaxies but in both the core and outskirt regions.

  We only consider clusters which are richer than 20 galaxies once
  background galaxies have been subtracted, in the particular cluster area and for the selected
  colour population. For red-sequence GLFs, we see no difference between the
  faint ends for each cluster region. The $\alpha$ parameters are the same within the error bars for the cluster cores, outskirts, and both regions combined. However, the brighest galaxies tend to lie in cluster cores, so the excess seen at the bright end of GLFs diminishes in the cluster outskirts.

  We find that there are more blue galaxies in the outskirts than in
  the cores, so blue GLFs in cluster core can only be plotted for the
  B band for our data. In the B band, the blue core stacked GLF is
  much closer to the red-sequence GLF than when taking galaxies from
  all the regions together. In the outskirts, blue and red-sequence
  galaxies seem to equally contribute to the cluster population at any
  magnitude to our completeness limits for V, R, and I bands. However,
  the faint end of the GLF is steeper for blue galaxies, implying that
  at fainter magnitude, blue galaxies are more numerous than red ones
  in cluster outskirts. In the B band, we detect far more bright blue
  galaxies than red-sequence galaxies, which indicates that the bright
  end of the blue B-band GLF seen in Fig.~\ref{fig:glf_stack} is dominated
  by the outskirts of clusters.

  To conclude, we find an excess of blue galaxies in
  the outskirts compared to the core of clusters but in the cluster outskirts the GLFs of blue
  and red galaxies are very similar. This can be
  interpreted as an infall of blue galaxies on cluster outskirts from
  the field populations or by a burst of stellar formation.

\section{Discussion}
\label{sec:discu}

\subsection{The faint end of the GLF}
\label{sec:faintend}

  The GLF faint end depends on both colour and redshift. We have
  investigated this evolution by stacking cluster counts for blue and
  RS galaxies separated within color-magnitude diagrams, and at either low-redshift ($0.40\leq z<0.65$) or high
  redshift ($0.65\leq z<0.90$). We now interpret Fig.~\ref{fig:glf_stack} and Table~\ref{tab:glf_stack} in terms of the
  colour evolution, and Fig.~\ref{fig:glf_stack_z} and
  Table~\ref{tab:glf_stackz} in terms of redshift evolution.

  Taking our full redshift cluster sample, we find steep blue GLFs
  with $\alpha_{blue} \sim -1.6$ for all bands, owing to the large error
  bars caused by the small amount of clusters with a
  sufficient number of blue galaxies. This is more or less consistent
  with similar analyses for clusters at lower redshifts. For red-sequence galaxies, we see
  a small drop at the faint end with a slope $\alpha_{red} \sim -0.4$
  for B and I bands and $\alpha_{red} \sim -0.8$ for V and R, while
  lower-redshift clusters usually present a flat faint end for the red-sequence
  population. In our redshift range, red galaxies dominate the blue
  population for magnitudes brighter than $V=-20$, $R=-20.5$, and
  $I=-20.3$. Above these magnitudes, blue faint counts become higher
  than red ones. In the B band, the blue galaxies dominate over red
  ones at all magnitudes, possibly because blue galaxies are
  brighter in the blue band. We note however that we draw this conclusion for only
  five clusters with sufficient blue galaxies in the B band.

  If we now separate our clusters between high and low redshift, we
  find that the red-sequence faint end drop is more important at high
  redshift. At low redshift, the slope is comparable to that for all clusters combined. This is because we perform our stacking using the
  Colless method, in which the faint end of the GLF is
  dominated by the low-redshift clusters in the stack. At higher
  redshifts, we find slopes of between $\alpha_{red} \sim 0.1$ and
  $\alpha_{red} \sim 0.7$. We have only a few clusters at high
  redshift, though the error bars in the slope are of the order of from 0.5
  to 1. Data for more high-redshift clusters are needed to fully investigate
  this behaviour. We also have insufficient clusters with enough blue
  galaxies to produce blue GLFs at these redshifts.

  When interpreting these results in terms of galaxy evolution, we can conclude that blue star forming galaxies are quenched in dense cluster environments to
  enrich the red-sequence population between high redshifts and
  today. However, that the properties of blue galaxies in clusters are similar at
  $z \sim 0$ and our redshift range ($0.4\leq z<0.9$) implies that clusters
  continue to accrete galaxies from the field across a wide range of redshift.




  This deficit of faint red galaxies at $z\sim0.7$ has already been
  observed by many authors \citep[e.g.][]{Tanaka+05, delucia07, Rud09,
    Stott+07, Gilbank+08, Vulcani+11}. In particular, our strong
  change in the RS faint end slope is in good agreement with that found for EDisCs clusters (see Fig. 5 of \citet{Rud09} and our
  Fig.~\ref{fig:glf_stack_z}). \citet{Kodama04} also observed a
  compensation for the decrease in the faint red galaxy counts by those of bluer
  galaxies. Furthermore, this evolution agrees with the
  empirical model of \citet{Peng+10} which predicts a difference
  between the blue and red faint end slopes of the order of unity owing
  to mass quenching being proportional to the star formation rate
  (SFR) of galaxies in our redshift range.

  On the other hand, some authors find no evolution in the RS GLFs
  with redshift \citep[e.g.][]{Andreon06, deprop07,
    deprop13}. \citet{deprop13} wrote that surface brightness
  selection effects could account for claims of evolution at the faint
  end. Observations with various surface brightness limits are
  required to confirm this hypothesis.



We can only compare our results with GLFs that have been fitted by a
single Schechter function. It is sometimes useful to fit GLFs with both a
Gaussian and a Schechter function \citep{Biv95} or with two Schechter
functions \citep{Popesso06}. The first case allows to better account
for the excess of very bright galaxies observed in certain clusters,
while the second fits well the upturn of very faint counts that can
exist for dwarf galaxies at fainter magnitudes than the usual GLF flat
faint end. In the present study, we chose to use a single Schechter
fit, as our data are insufficiently complete to investigate the upturn of
very faint galaxies found in \citet{Popesso06}. More
sophisticated fitting with a higher number of degrees of freedom for
the fit would require a larger number of data points and a fainter
completeness limit. We cannot compare our results with those of the following
authors because their approaches differ from ours: \\
- \citet{Mancone12} studied GLFs only in apparent magnitude. Hence, the
k-correction is not taken into account and it is difficult to know
exactly which population is studied and the precise completeness limit,
particularly since they consider high-redshift clusters ($1<z<1.5$); \\
- \citet{Muzzin08} fixed the slope of the faint end $\alpha = -0.8$
and fit the two other parameters ($\phi^*$ and $M^*$).

\subsection{Dependence on environment}
\label{sec:field}



Another important debate concerns the interaction between clusters and
their environment. To properly address this problem, it is necessary
to investigate it on three different scales. First, we compare
cluster GLFs to field GLFs, then study the dependence of GLFs on various cluster
properties before finally studying the variations inside clusters.

\begin{figure*}
\begin{tabular}{cccc}
\includegraphics[width=0.17\textwidth,clip,angle=270]{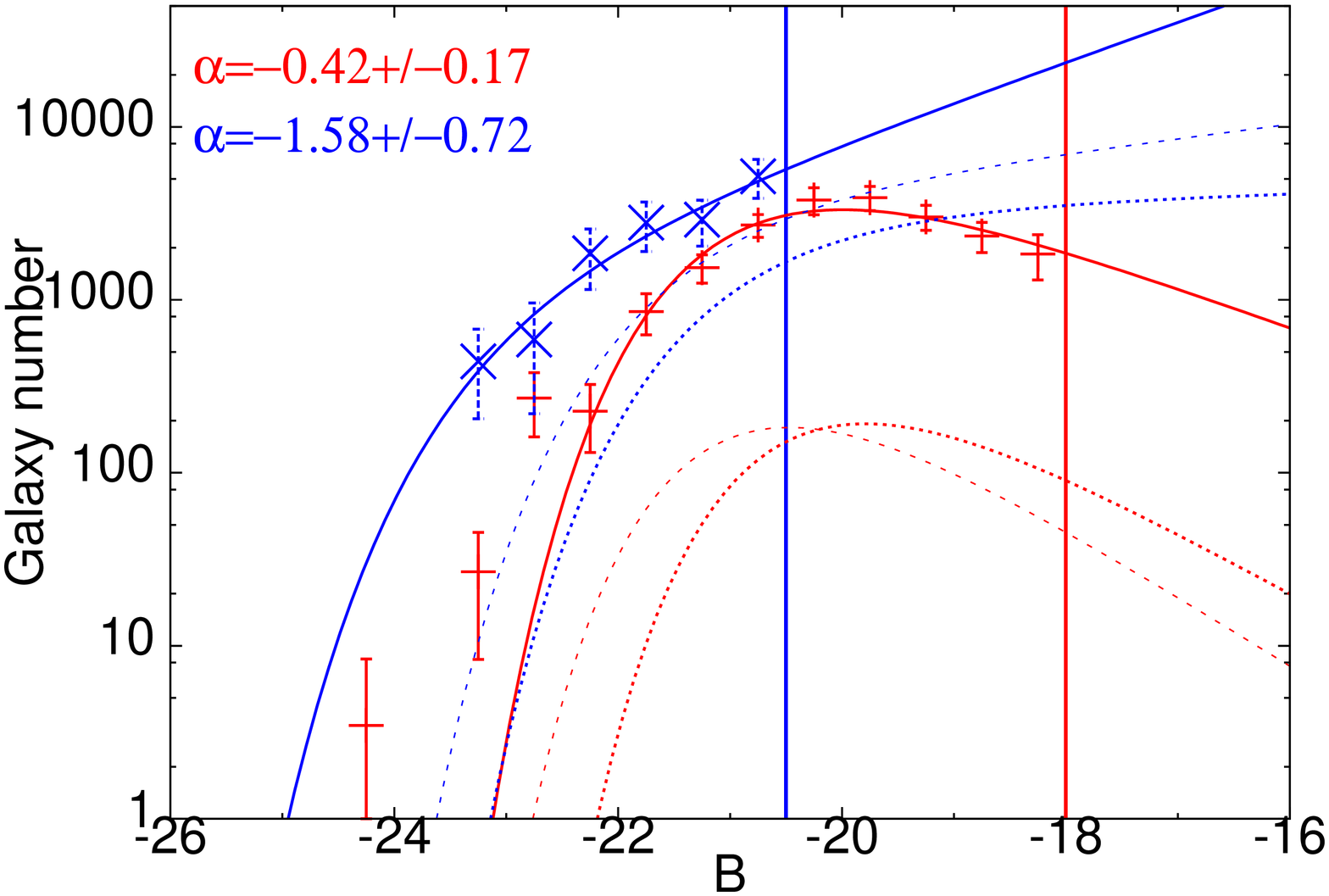}
\includegraphics[width=0.17\textwidth,clip,angle=270]{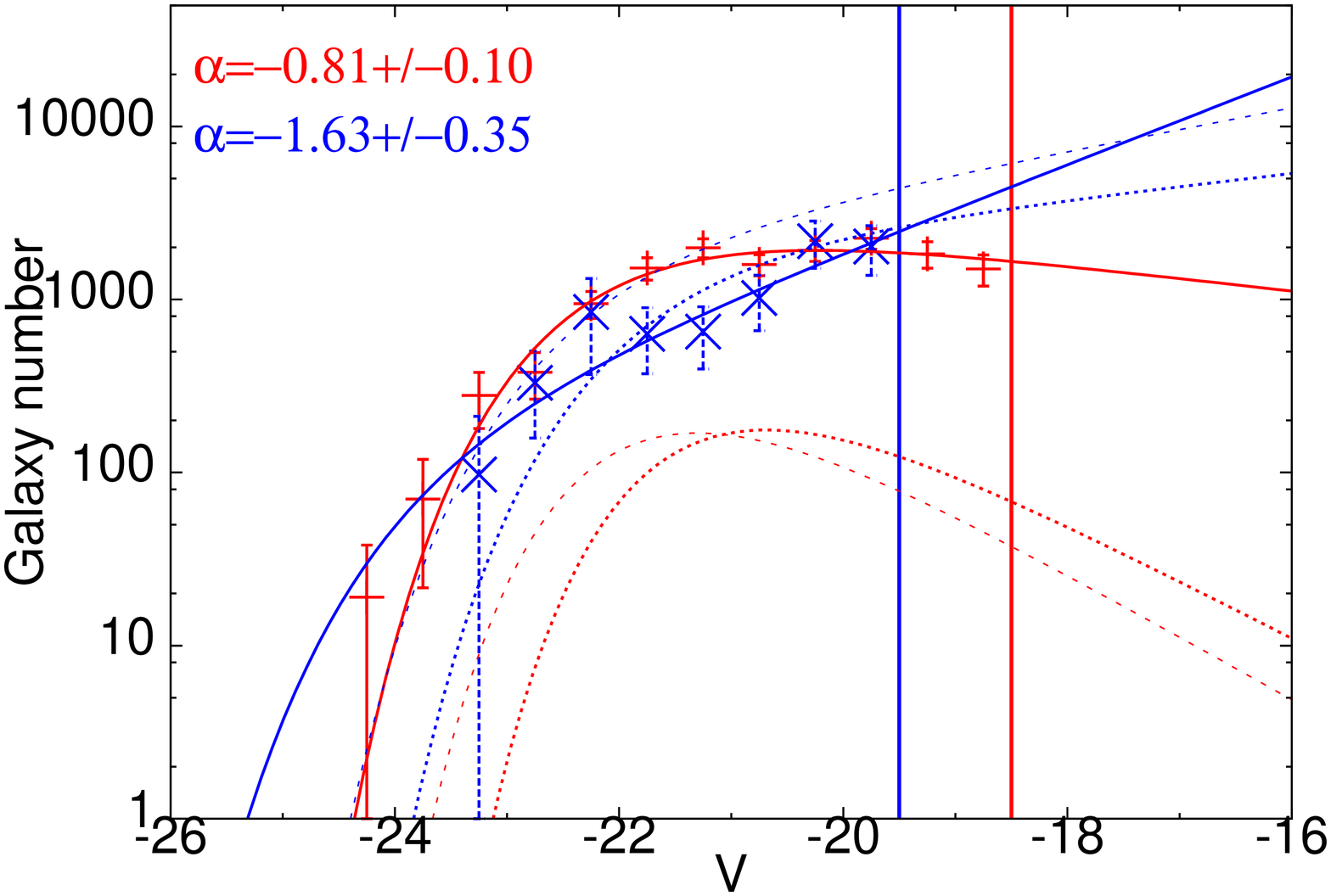}
\includegraphics[width=0.17\textwidth,clip,angle=270]{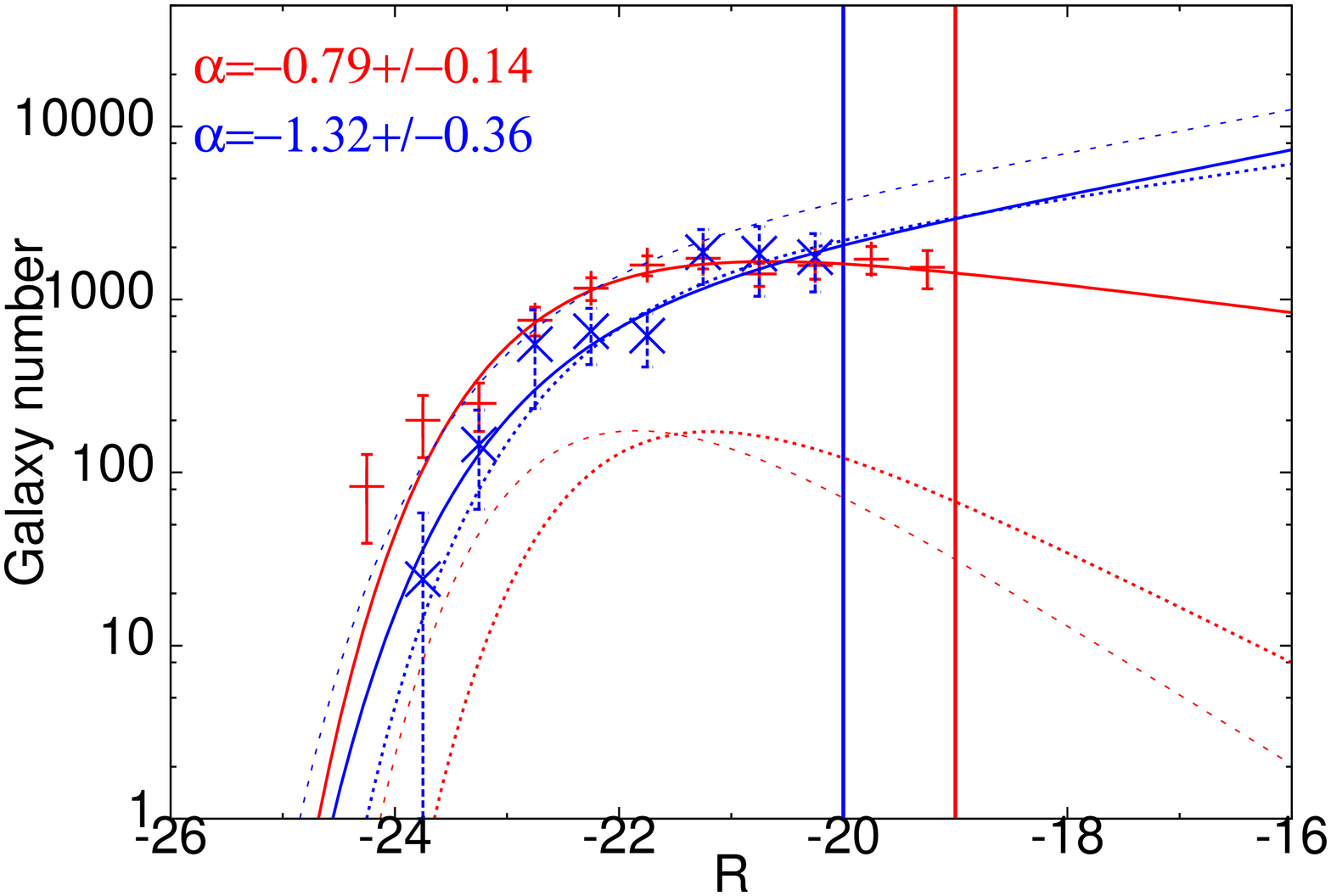}
\includegraphics[width=0.17\textwidth,clip,angle=270]{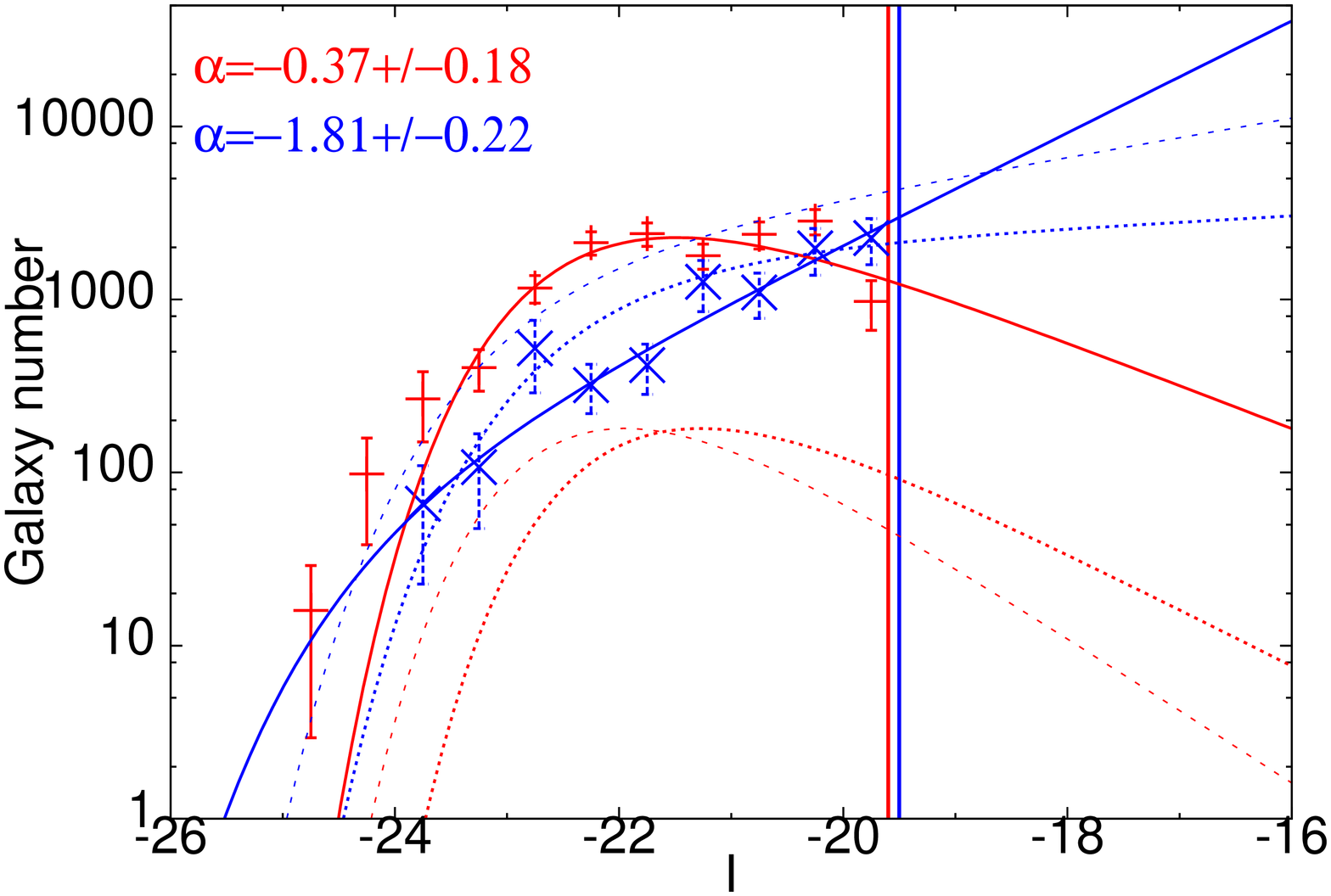} \\
\end{tabular}
\caption{Cluster and field GLFs in the B, V, R, and I rest-frame
    bands (from left to right). Red and blue points respectively
    correspond to red-sequence and blue stacked cluster GLFs
    normalized to 1 deg$^2$. The red and blue plain curves show the
    best Schechter fits to red-sequence and blue galaxies and the red
    and blue vertical lines indicate the corresponding 90\%
    completeness limits. The slope of the fit $\alpha$ is given for
    each population. Refer to Table~\ref{tab:glf_stack} for cluster best
    Schechter fit parameters.  The thin dotted and dashed curves
    correspond to the COSMOS field GLFs centered at redshifts $z=0.5$
    and $z=0.7$ normalized to 1 deg$^2$. The separation between red
    and blue field galaxies is done the same way than for clusters at
    the corresponding redshift (see text for details).}
\label{fig:glf_stack_field}
\end{figure*}

  We first compare our cluster GLFs calculated in section
  \ref{sec:4.2} with field GLFs derived from COSMOS data \citep{Il+09}. We
  compute two field GLFs for redshifts of 0.5 and 0.7 with a width of
  $\pm0.2$ around these redshifts to be consistant with the way we
  made our cluster GLFs. We separate blue from red field galaxies by
  applying a color magnitude relation similar to the one used for our
  clusters. The ordinate of this red-sequence is equal to the colour
  of elliptical galaxies at the central redshift taken from
  \citet{Fuku} and the width of the RS is $\pm0.3$ in color. This
  allows us to compare cluster and field GLFs computed in the same way
  with the same separation between red and blue galaxies. Results are shown in
  Fig.~\ref{fig:glf_stack_field}.

  The GLFs of blue galaxies are similar for the field and clusters in
  the V, R, and I bands, while we find more blue galaxies in our
  clusters than the field for the B band but with a similar shape. The
  shape of the red-sequence GLFs are also almost identical. However,
  there are about ten times more red galaxies in clusters than the
  field. Another difference is the GLF of cluster RS galaxies has a
  sharper drop at the faint end at high redshift, while the field red
  GLFs remain unchanged across our redshift range. This apparent lack
  of evolution in early type field galaxies was assessed in
  \citet{Zucca+06}, who proposed that it highlights an efficient
  transformation of blue to red galaxies in higher density
  environments. Inside clusters, interactions between galaxies are
  more likely to happen, boosting this evolution, while in the field
  galaxy interactions are less frequent and the red population
  increases at a far lower rate.

  We wrote above that the average
  cluster GLF of RS galaxies depends on mass. GLFs of more
  massive clusters ressemble more the GLFs of nearby clusters with a flat faint
  end. This implies that these high-redshift massive clusters are
  more evolved than their companions at the same redshift
  either because they formed earlier or in a denser environment.

  We find no remarkable difference between the properties of general GLF and those of the GLF of only substructured clusters. However, we have only three clusters
  that can be studied in this way, leading to large error bars. We know
  that the merging of galaxy clusters can strongly affect the slope of
  the GLF, as illustrated by studies of cluster pairs or violently merging
  clusters \citep[e.g.][]{Durret10, Durret11, Durret14}. Hence we
  would expect that substructured clusters present a variation of
  their faint end slope compared to others. More data at different
  stages of the merging process are needed before we can draw stronger conclusions.


  Finally, we find differences in GLF behaviours between cluster cores
  and outskirts. We find more bright galaxies in the core compared to
  the outskirts, in agreement with CDM models that predict the most
  massive galaxies to lie at the cluster cores. We also find more blue
  galaxies in the outskirts than in the core. This larger number of
  blue galaxies in the outskirts could be explained by infalls from
  the field. However, the red GLF faint end remains the same in any
  part of the cluster. Some authors found steeper faint end slopes in
  cluster outskirts \citep[e.g.][]{Adami08,Boue08}. In particular,
  strong variations in $\alpha$ have been observed in the highly
  structured Coma cluster, which can be probed with high completeness
  and quality due to its proximity \citep{Adami07a, Adami07b}.


\subsection{Evolution of cluster galaxy types with redshift}
\label{sec:types}

 With colour-selected populations, we analyse the variations
  of the galaxy types within clusters.  We consider blue and red galaxies
  selected in a colour-magnitude diagram and for which field galaxies
  have been subtracted.  We also remove galaxies that are outside
  disks of 1Mpc radius centered on cluster optical centers.  We
  compute the percentages of each type for every cluster and then
  average them over clusters by stacks of four clusters. Error bars
  correspond to the dispersion in values over all clusters within a
  stack. This allows us to study the evolution of cluster galaxy types
  with redshift from $z=0.4$ to $z=0.9$. We limit our sample to
  galaxies brighter than I=$-21$ and only consider clusters that are at least
  90\% complete at this magnitude.

\begin{figure}
\centering
 \includegraphics[width=0.3\textwidth,clip,angle=270]{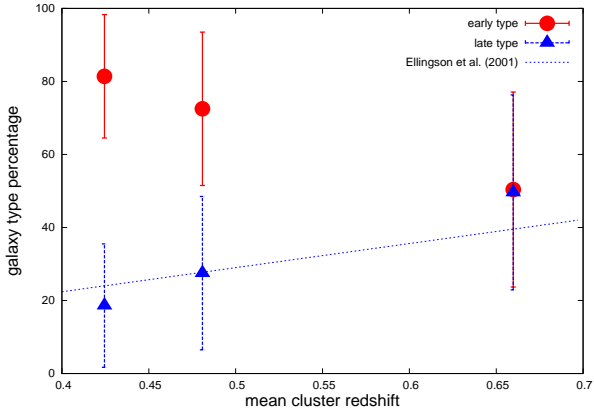}
 \caption{Evolution of cluster galaxy type percentages with
   redshift. Each point represents the mean value of a percentage over
   four clusters. Red dots are red (early type) galaxies and blue
   triangles correspond to blue (late type) galaxies. The blue dashed
   line corresponds to the blue fraction from \citet{Ellingson+01}.}
\label{fig:typespec}
\end{figure}

  Looking at Fig.~\ref{fig:typespec}, we note a clear decrease in the
  fraction of early-type galaxies from low to high
  redshift, while the fraction of late-type galaxies increases with
  redshift. This scenario agrees with galaxy-evolution scenarios where spiral galaxies evolve into
  ellipticals. Furthermore, it is consistent with the evolution of
  early-type and late-type GLF faint ends that we discussed in section
  ~\ref{sec:faintend}.

Our early-type fraction decreases from $f_{\rm red}=0.81\pm0.17$ at
$z=0.42$ to $f_{\rm red}=0.50\pm0.27$ at $z=0.66$. We compare these
results to spectral classifications from \citet{Ellingson+01}, and
also to the morphology-density relation found by \citet{Smith+05} and
\citet{Postman+05}. In the first case, we overplot on
Fig.~\ref{fig:typespec} their best fit to the blue fraction of
clusters between $0.18<z<0.55$. The blue galaxies taken into account
in this study are spectroscopic cluster members and are separated in a
colour magnitude diagram. They are also brigther than $R=-20$, such
that it is very close to the low redshift sample of galaxies we
use. We find a very good agreement between our blue fraction and the
one from \citet{Ellingson+01} at all redshifts. We do not attempt to separate
our sample into high-density (cluster cores) and medium-density
regions, so our results should compare with the
$\Sigma=100$~Mpc$^{-2}$ and $\Sigma=1000$~Mpc$^{-2}$ curves in Fig.~3
of \citet{Smith+05} and Fig.~13 of \citet{Postman+05}. We find an
overall good agreement. At intermediate redshift ($z=0.4$), our early-type fraction lies within the fractions of medium and high density of
the cited authors ($f_{E+S0}=0.6$ and $f_{E+S0}=0.85$). At higher
redshifts ($z=0.7$), our values are also consistent with the interval
$0.55<f_{E+S0}<0.8$ found by the previous authors, but only when
taking our error bars into account.  However, one must note that
previous authors used a classification based on galaxy shapes while
our work relies on the galaxy colours. The trend of a decreasing early
type fraction with increasing redshift is clearly seen whichever
method is used. A comparison of both spectral and morphological
methods for the same galaxy sample would help understanding the
different biases of each method. Apart from possible biases,
the morphological and spectral evolutions might also be different.

\subsection{A scenario for the evolution of clusters}

\citet{Peng+10} empirically showed that the red-sequence is fed by two
different types of quenching that happen at different redshifts. At
high-redshift ($z>2$), environmental effects dominate and the 
red-sequence grows through the quenching of blue star forming galaxies
that fall into the dark matter halo of the forming group
(``environment quenching''). At lower redshift ($0<z<1$), galaxies are
quenched proportionally to their star forming rate and progressively
enrich the red-sequence (``mass quenching'').

Cosmological models of cluster mass assembly predict the most intense
mass growth of clusters at redshifts earlier than $z\sim0.8$ to $1$
(e.g. \citet{Adami+13} from the Millennium simulation). In this
redshift interval, clusters grow through accretion of major groups,
and this already provides a pre-processed galaxy population formed by
the cited ``environment quenching''. This explains the fact that
clusters at $z>1$ already exhibit a red-sequence \citep{Gobat+11, Fassbender+11}.

At lower redshifts ($z<0.8$), group accretion only concerns more
modest groups in terms of relative mass and \citet{Gue13b} have shown
in our survey that this accretion only involves less than 10\% to 15\%
of the cluster mass. In the $0<z<0.8$ redshift interval, the galaxies
accreted by clusters are therefore not only coming from red
populations preprocessed in groups but also from regular blue field
galaxy populations. In the meantime, blue galaxies evolve into red
ones following the ``mass quenching'' defined in \citet{Peng+10}.

We can show that this evolutionary scenario for clusters is assessed
by the main results of our paper as follows:

  1) Red-sequence cluster galaxies show a drop at their faint end
  which is more significant at higher redshift.

  2) Blue cluster GLFs are steeper than those of RS galaxies and are similar across
  our redshift range and at lower redshift.

  3) There is a large excess of red galaxies in clusters compared to
  the field while the blue galaxies behave more or less in the same
  way. The red GLFs of clusters continue to evolve across our redshift
  range, while for the field there is little evolution.

  4) There is a strong decrease of the early type fraction in clusters
  with increasing redshift.

  5) There might be infalls of blue galaxies from the field to the
  cluster outskirts. This could explain why we find so few blue
  galaxies in cluster cores compared to the outskirts.

  6) When considering our more massive clusters, we find a red-sequence GLF
  that is consistent with those observed at $z=0$ with a flat faint end.

  The result 1) shows that clusters are formed at redshifts higher than
  $z=0.9$. A possible explanation of the redshift dependent drop at
  the faint end of the red GLFs (point 1) would reside in the blue to
  red colour evolution in cluster galaxies populating the faint part
  of the GLF (point 4). This agrees with the mass
  quenching expected at these redshifts from \citet{Peng+10}. The
  evolution of the red cluster GLFs with redshift compared to the
  field GLFs (point 3) suggests that red galaxy formation is more
  efficient in high-density environments. At the same time, a non-negligible infall of faint galaxies from the field (point 5) could
  explain how the blue GLF faint end remains the same from $z\sim0.9$
  to 0 (point 2). Our discovery that very massive clusters have
  the same red GLF faint end as clusters in the nearby Universe (point 6) indicates
  that cluster evolution can be faster in denser environments or that
  some clusters formed earlier than others.







\section{Conclusion}
\label{sec:ccl}

We have computed GLFs in the B, V, R, and I rest-frame bands for 31
clusters of the DAFT/FADA survey using photo-$z$s, the largest
medium-to-high redshift ($0.4 \leq z<0.9$) cluster sample to date.

To overcome the problem of lower photometric redshift
precision in clusters mainly due to a lack of red enough spectral
templates, we have artificially allowed the inclusion of extinction in
early-type galaxies.  This process does not affect drastically
photometric redshifts outside clusters but increases their quality
by $\sim50\%$ inside clusters, allowing us to reach the same
precision inside and outside the cluster.

Another result of this paper is that GLFs are strongly correlated to
the completeness of the data. This should be kept in mind when
comparing GLFs from different studies.

  We have shown that GLFs have similar properties for the B, V, R,
  and I rest-frame bands with small differences for the B-band blue
  GLFs. We found a sharp decline in the red faint end that increases with
  redshift: $\alpha_{red} \sim -0.5$ at $0.40\leq z<0.65$ and $\alpha_{red}
  > 0.1$ at $0.65\leq z<0.90$. High mass clusters appear to have a flat
  faint-end which may indicate that galaxy evolution is more rapid in denser
  environments or different formation epochs for clusters of different masses. Blue GLFs
  are steeper with $\alpha_{blue} \sim -1.6$ and do not seem to evolve
  with redshift.

Our study of galaxy types with redshift shows an evolution of late-type galaxies to early types from high z until today that could
account for the drop found at the red faint end.

  We also found an excess of red galaxies in clusters compared to
the field, while blue galaxies have more or less identical GLFs.

Our results imply that clusters have formed at high-redshift ($z>0.9$)
and that blue cluster galaxies are efficiently quenched into red ones
between $z\sim0.9$ and today. During this time interval, galaxy clusters continue
to accrete faint galaxies from the field environment.

Finally, we note an inversion of the red to blue population dominance
at magnitudes $V=-20$, $R=-20.5$, and $I=-20.3$ at redshift $0.40\leq z<0.90$. We plan to compute
stellar mass functions (SMFs) in a future paper, to see whether the
blue and red populations have comparable behaviours in mass and
luminosity. This would allow us to compare our results with simulations of
galaxy cluster formation.

\begin{acknowledgements}

  We thank Greg Rudnick for useful discussions. We also thank Eric
  Jullo, Marceau Limousin, Dennis Zaritsky for comments on earlier
  versions of this paper. We are grateful to the referee for
  interesting comments. F.D. acknowledges long-term financial support
  from CNES. I.M. acknowledges financial support from the Spanish
  grant AYA2010-15169 and from the Junta de Andalucia through TIC-114
  and the Excellence Project P08-TIC-03531.

\end{acknowledgements}

\begin{appendix}

\section{Individuel cluster GLFs}
\label{sec:annexe1}

\begin{figure*}[ht!!]
\begin{tabular}{cccc}
\includegraphics[width=0.17\textwidth,clip,angle=270]{new_plots/INDIV/CL0016_B.ps}
\includegraphics[width=0.17\textwidth,clip,angle=270]{new_plots/INDIV/CL0016_V.ps}
\includegraphics[width=0.17\textwidth,clip,angle=270]{new_plots/INDIV/CL0016_R.ps}
\includegraphics[width=0.17\textwidth,clip,angle=270]{new_plots/INDIV/CL0016_I.ps} \\
\includegraphics[width=0.17\textwidth,clip,angle=270]{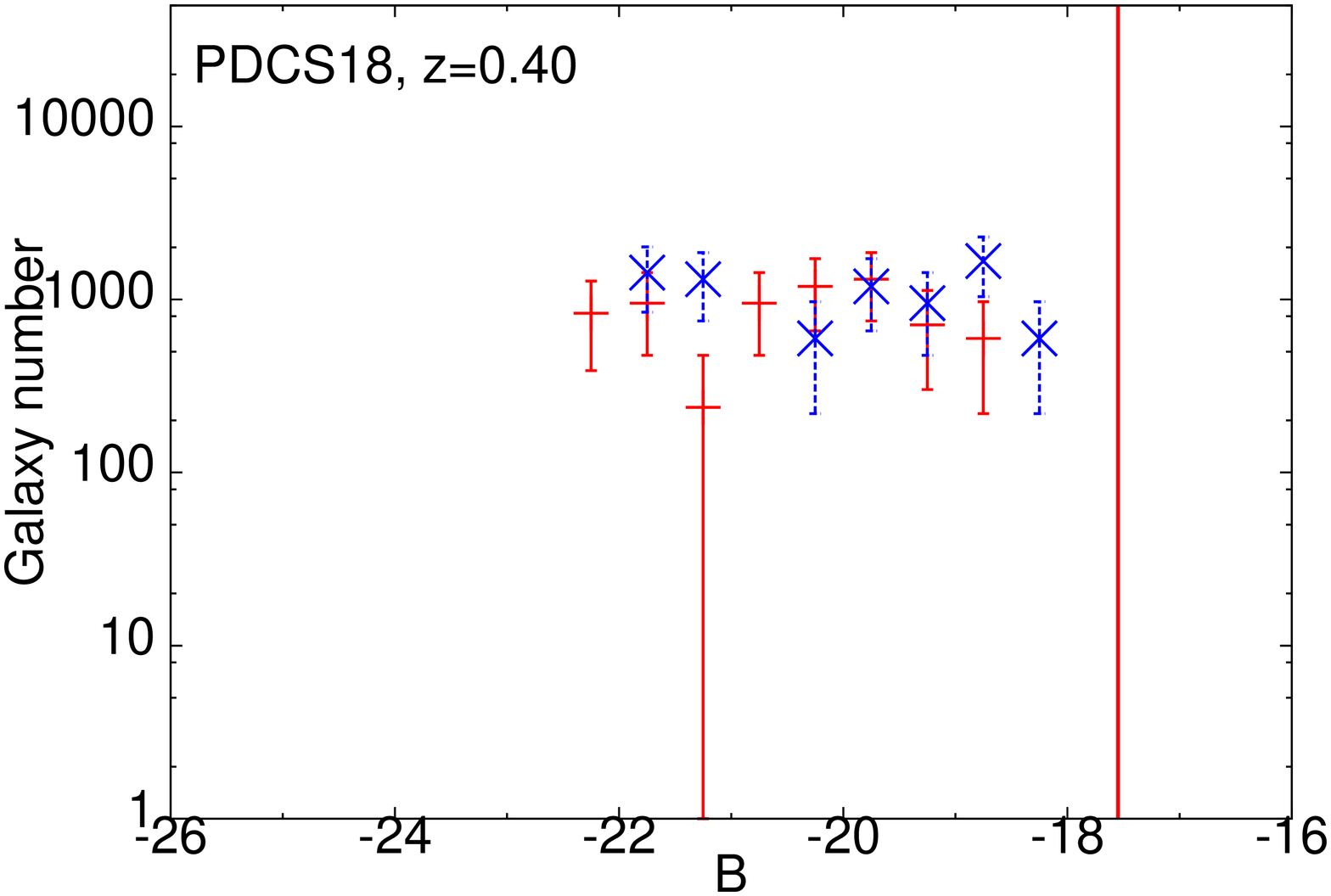}
\includegraphics[width=0.17\textwidth,clip,angle=270]{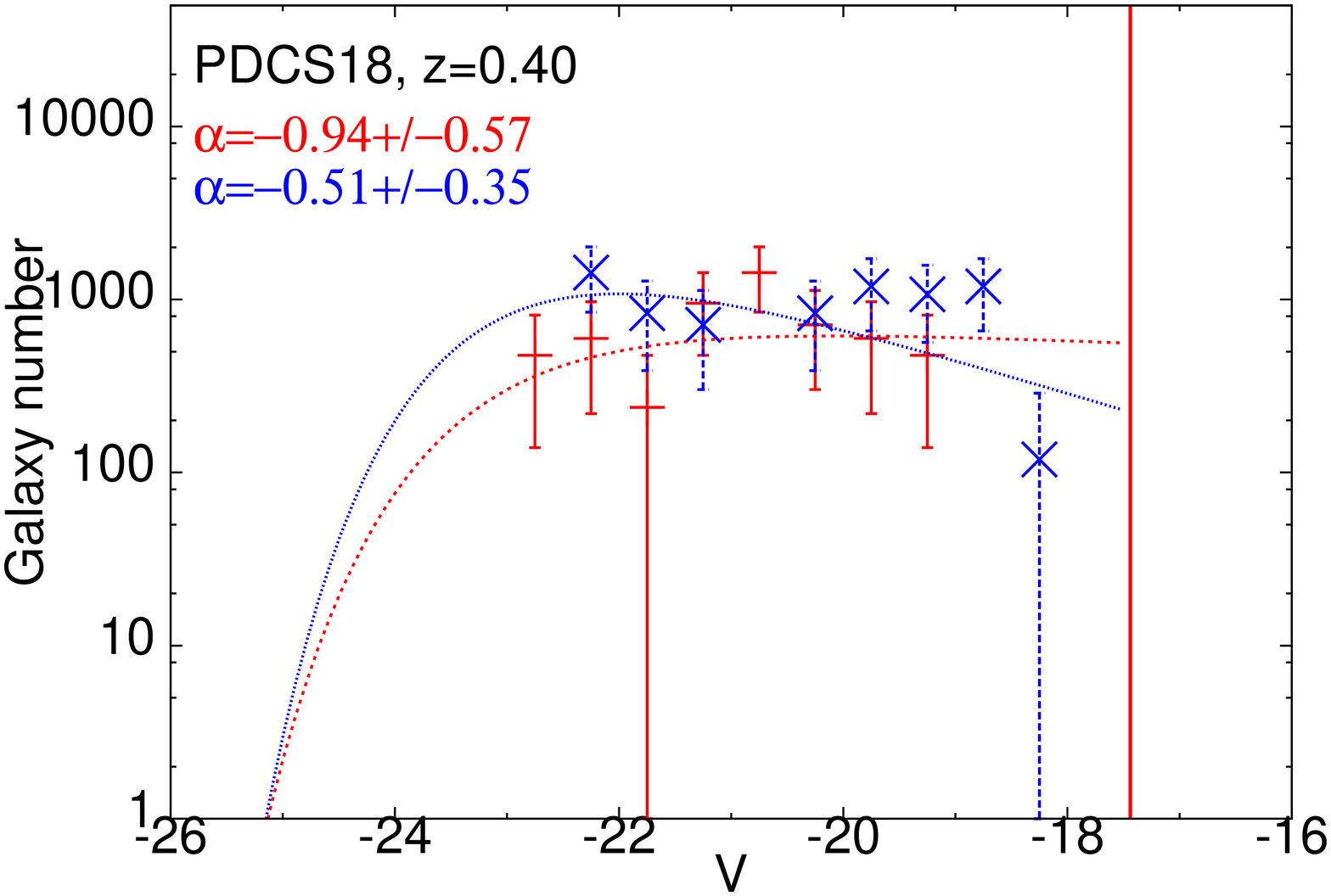}
\includegraphics[width=0.17\textwidth,clip,angle=270]{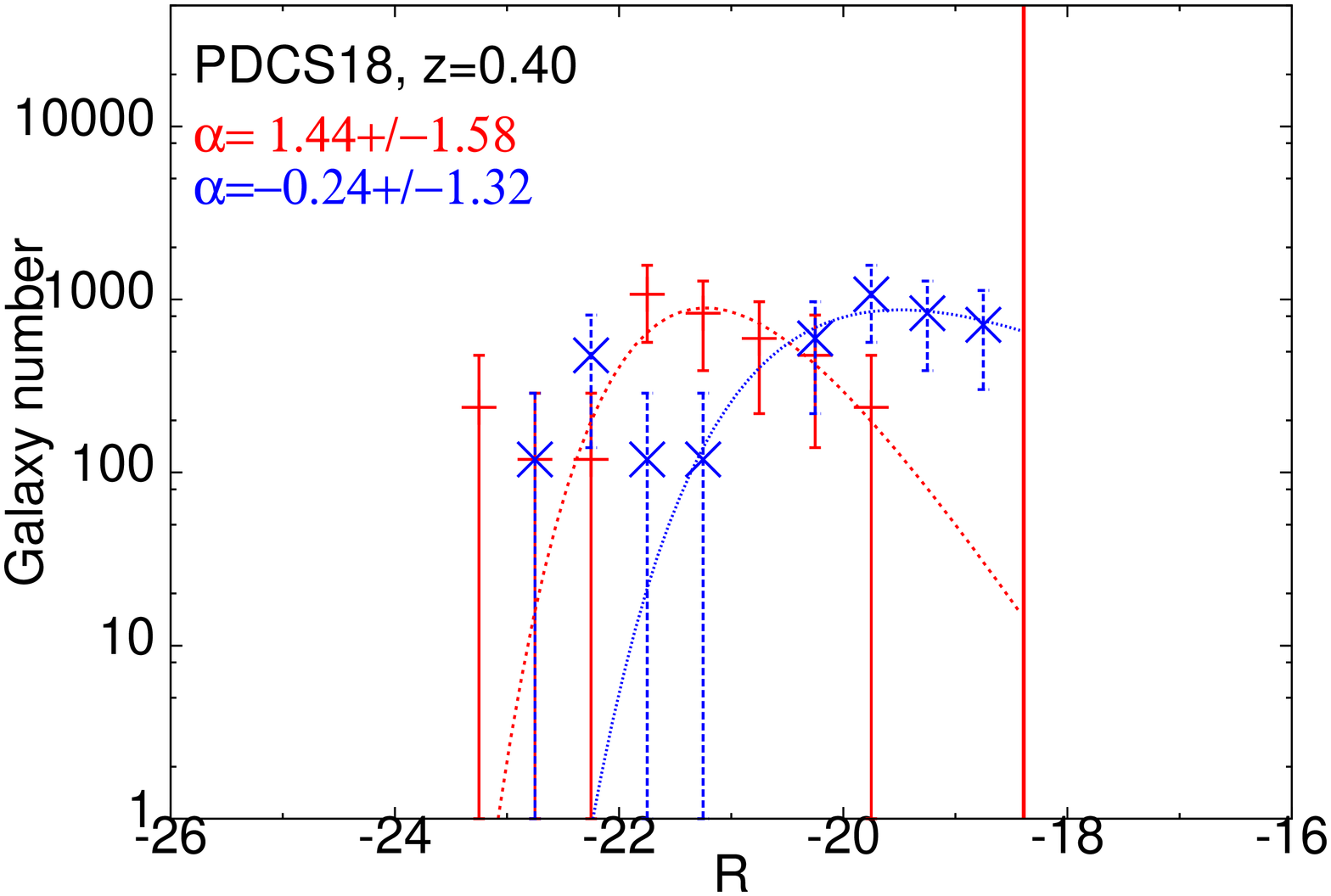}
\includegraphics[width=0.17\textwidth,clip,angle=270]{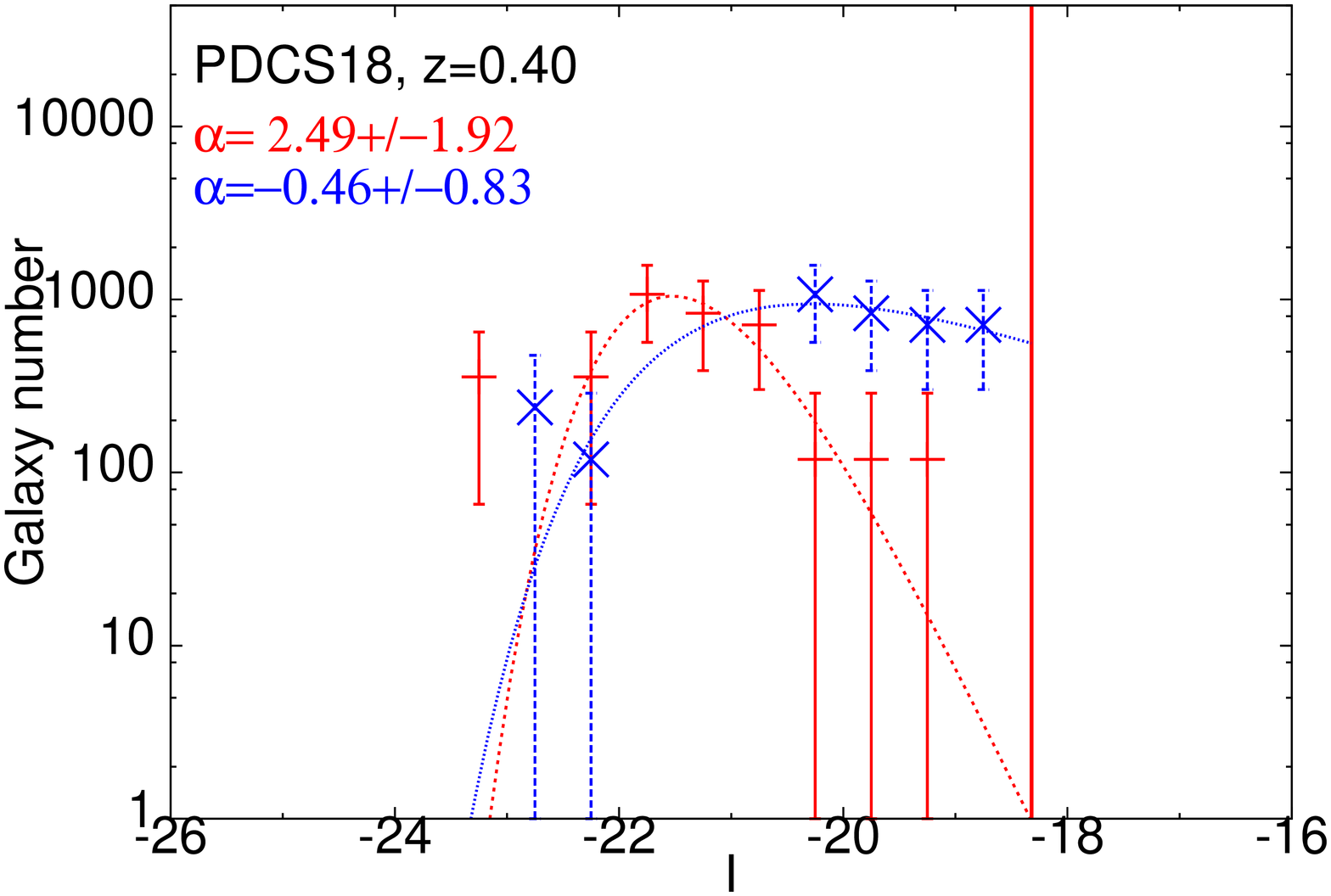} \\
\includegraphics[width=0.17\textwidth,clip,angle=270]{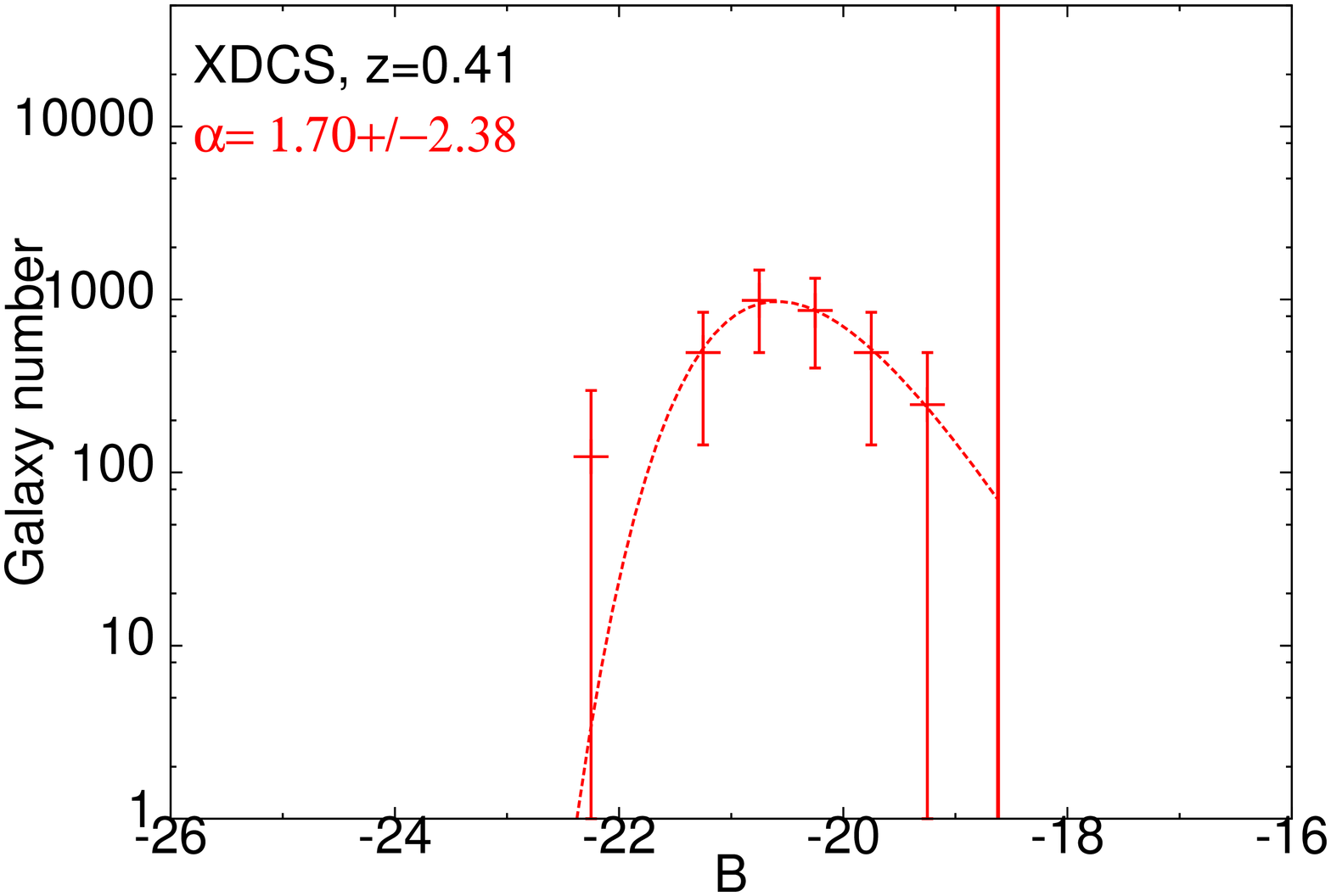}
\includegraphics[width=0.17\textwidth,clip,angle=270]{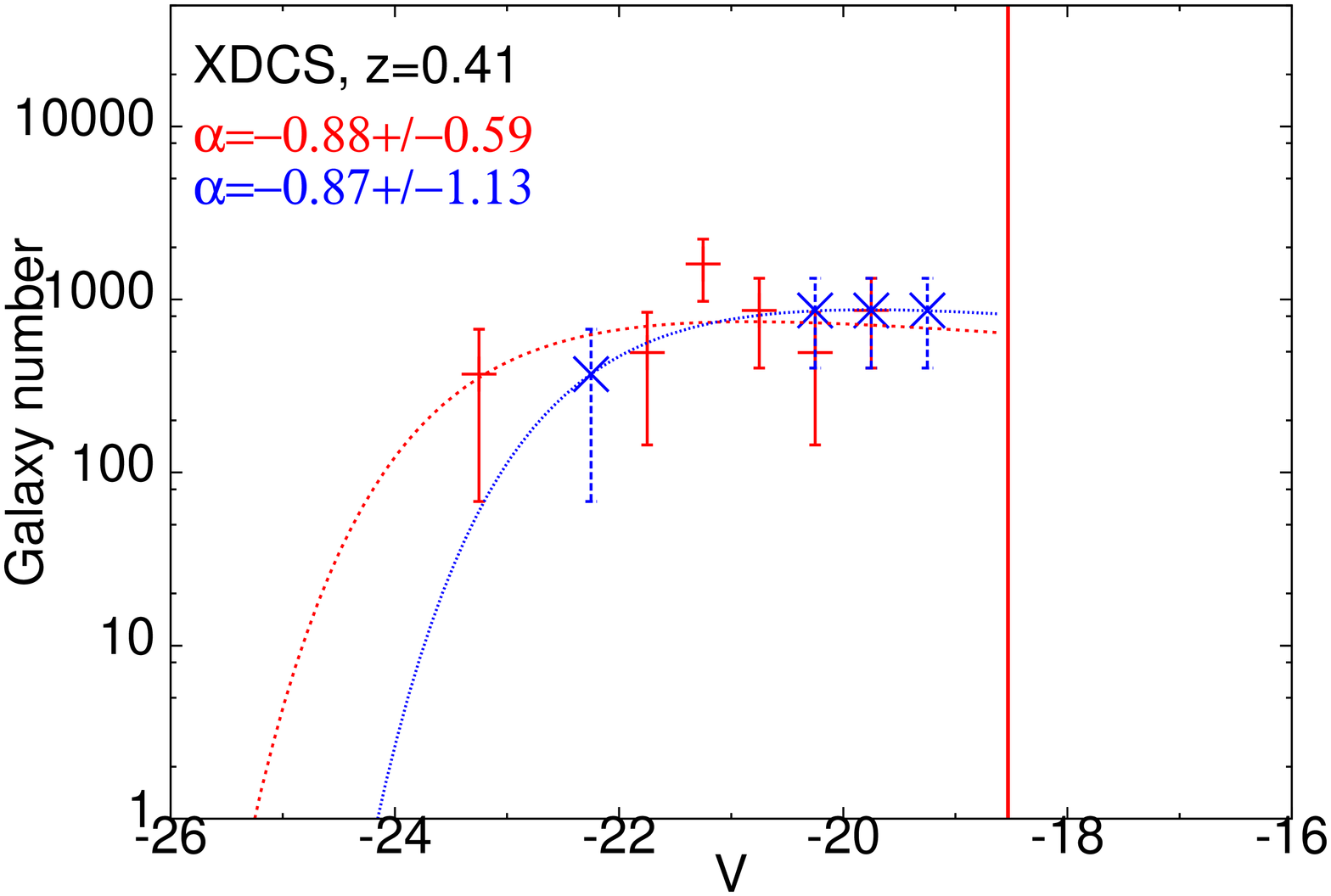}
\includegraphics[width=0.17\textwidth,clip,angle=270]{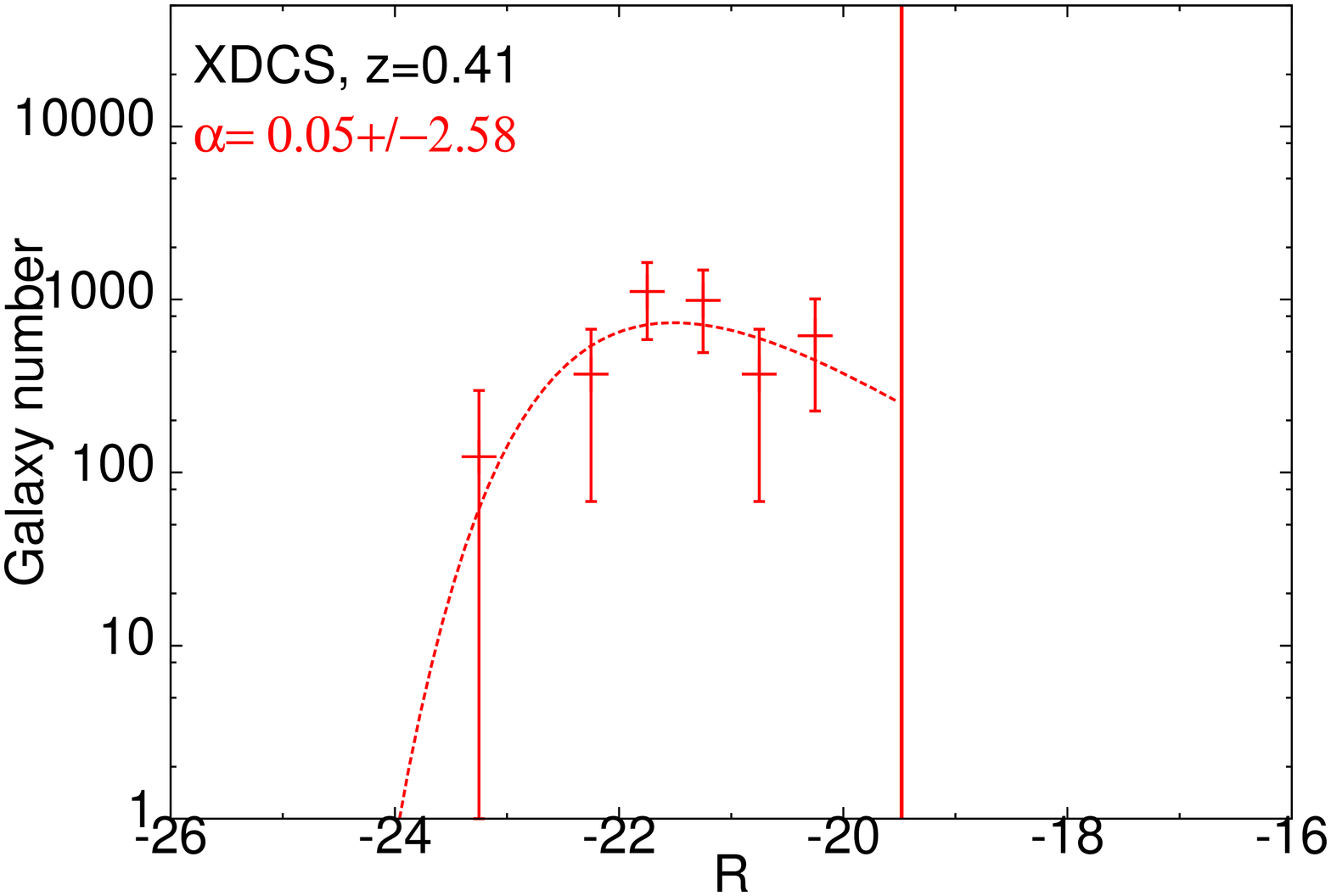}
\includegraphics[width=0.17\textwidth,clip,angle=270]{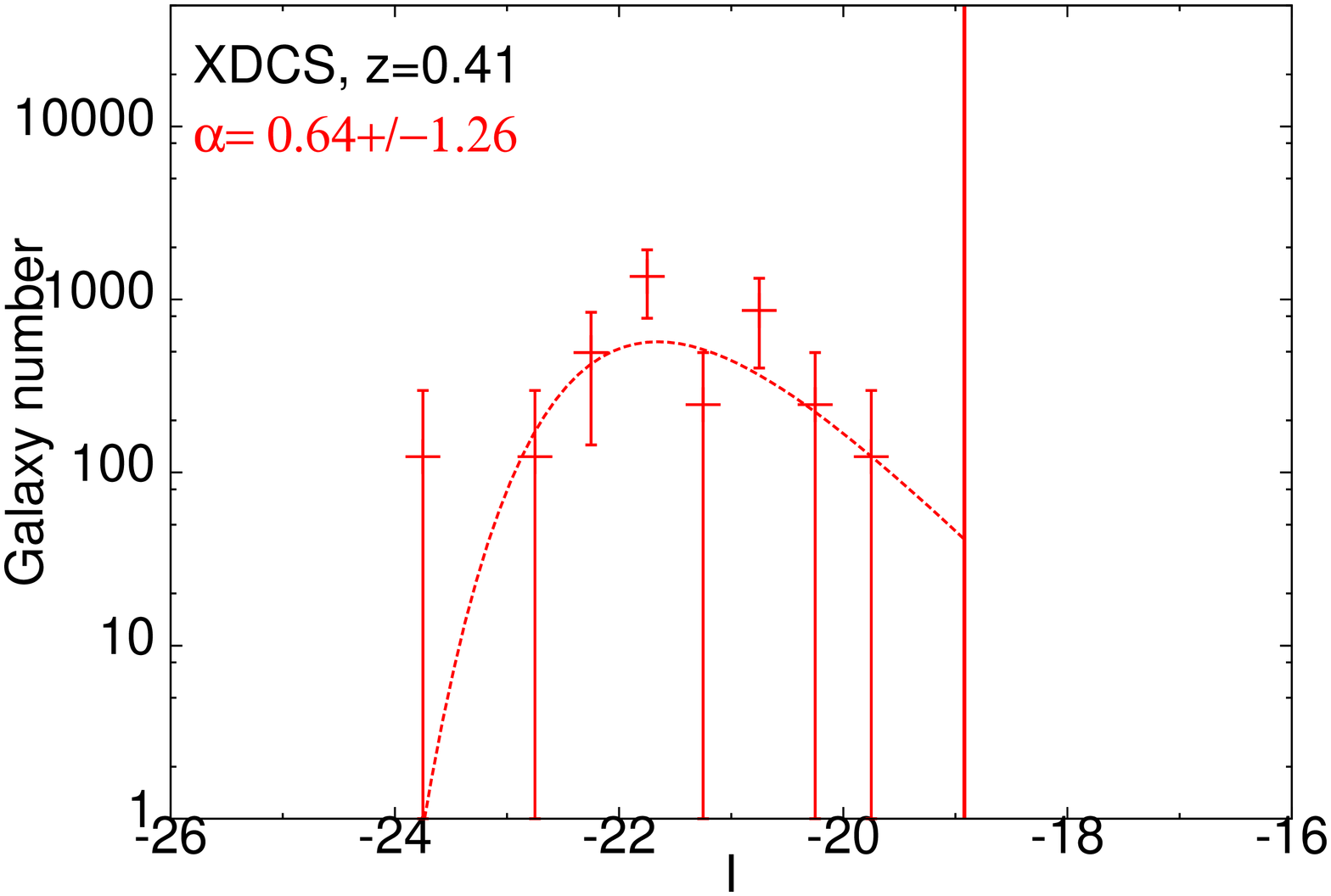} \\
\includegraphics[width=0.17\textwidth,clip,angle=270]{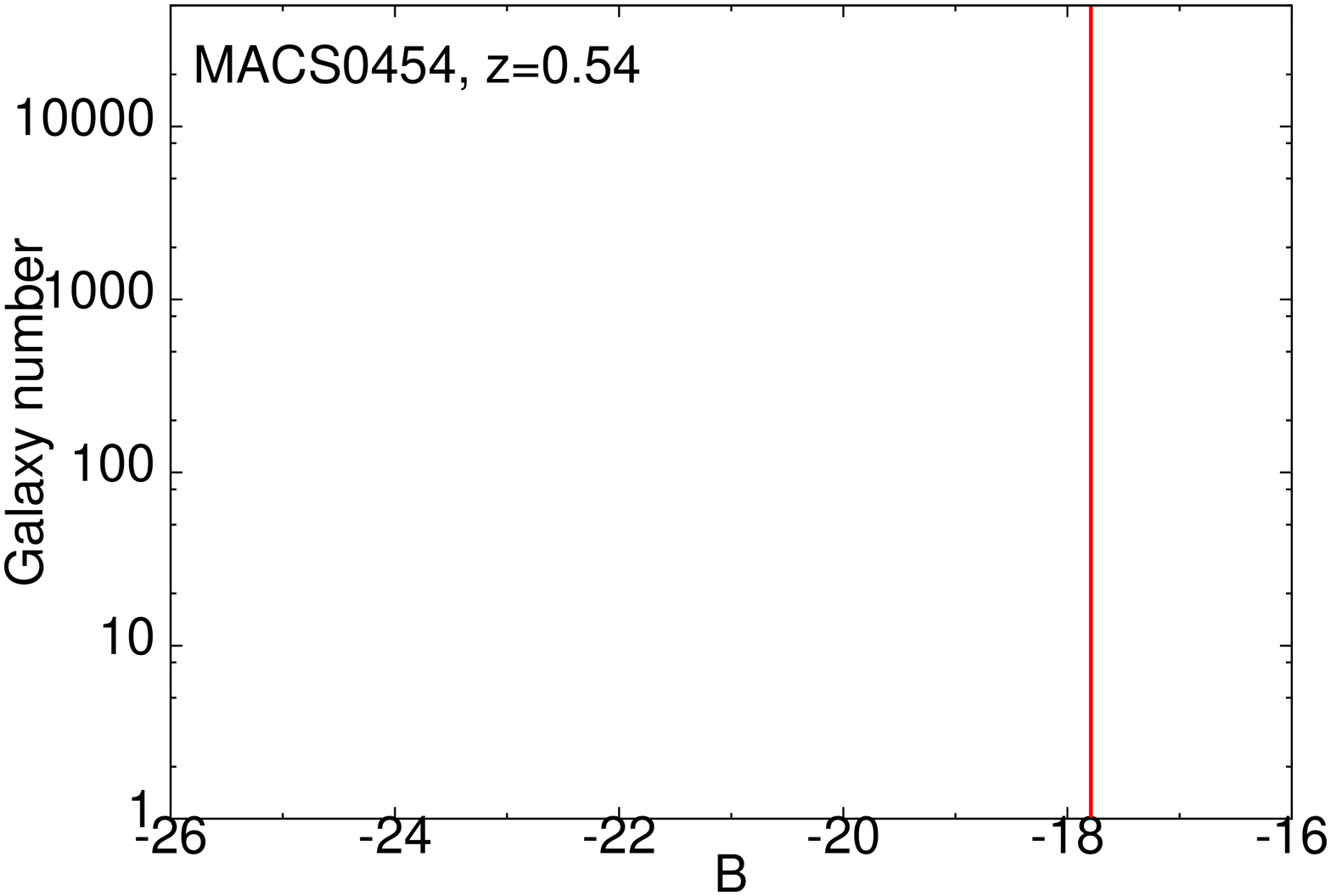}
\includegraphics[width=0.17\textwidth,clip,angle=270]{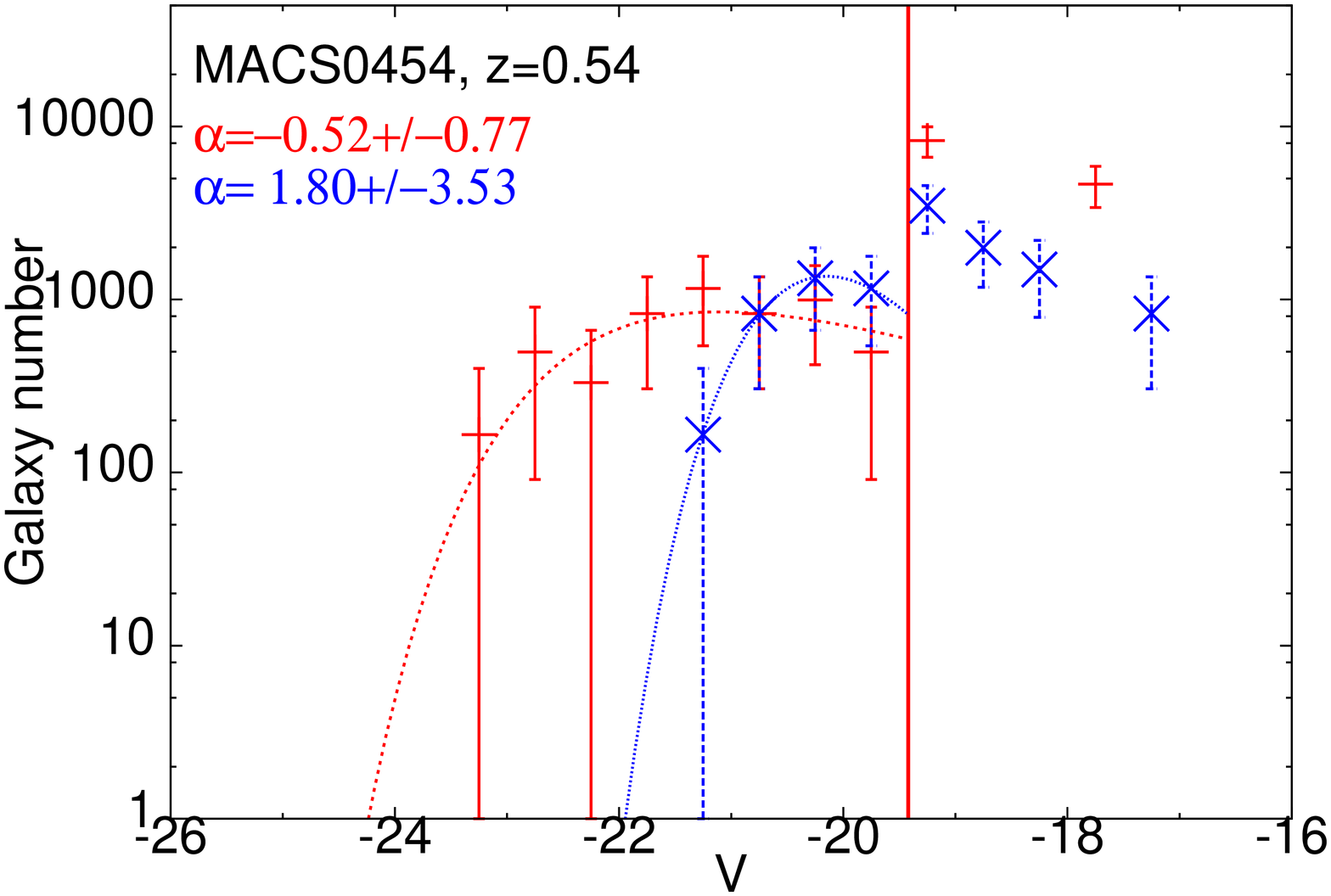}
\includegraphics[width=0.17\textwidth,clip,angle=270]{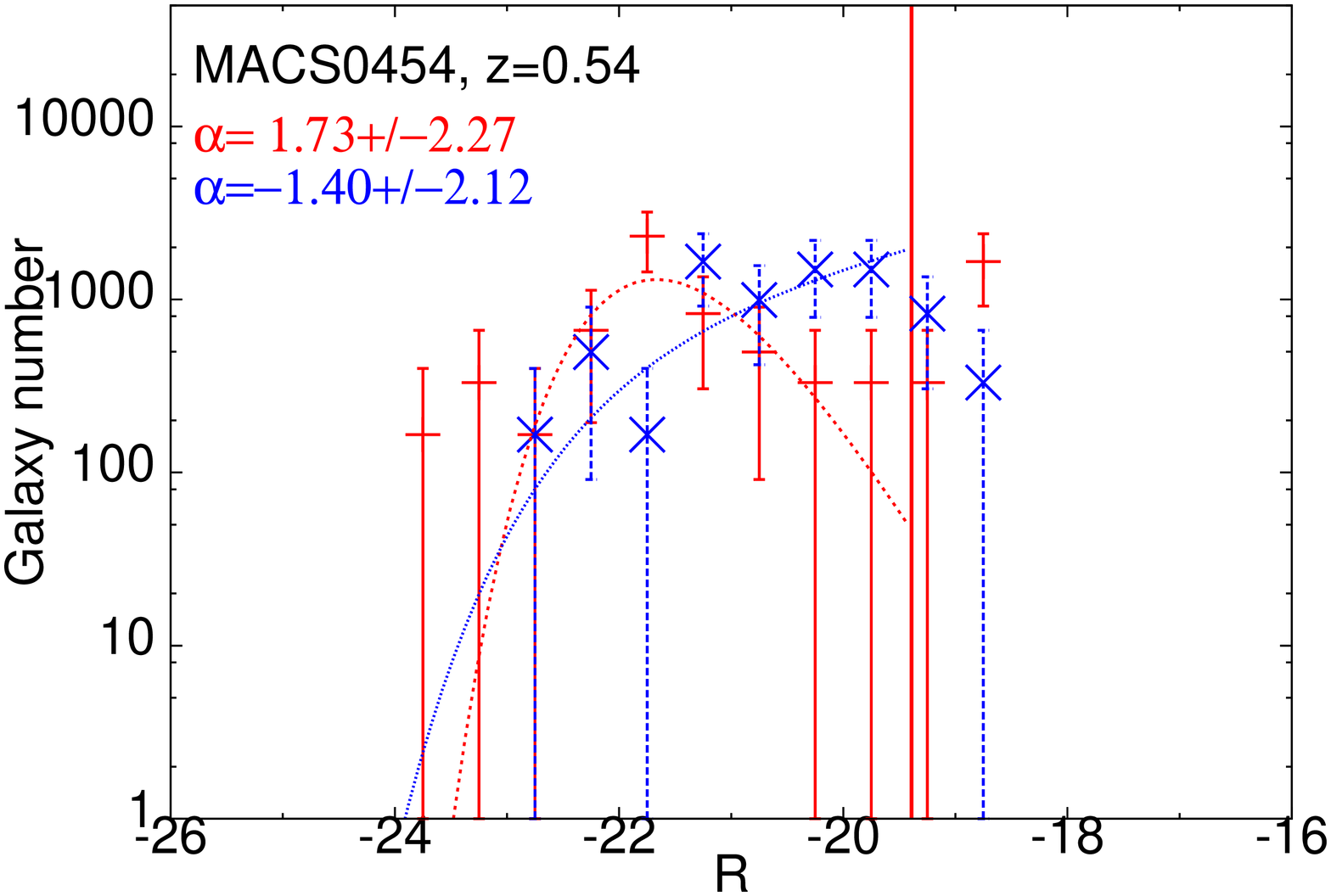}
\includegraphics[width=0.17\textwidth,clip,angle=270]{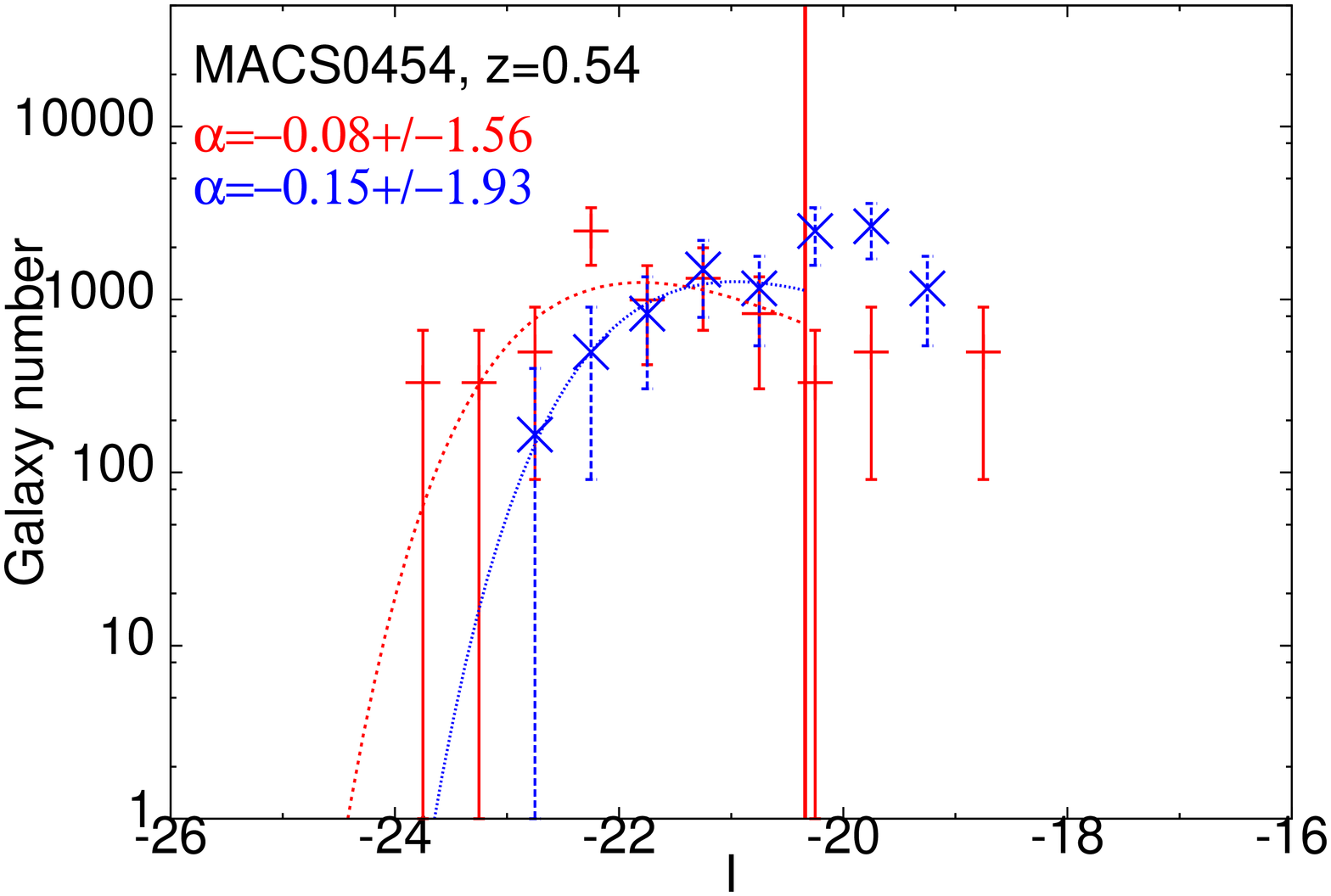} \\
\includegraphics[width=0.17\textwidth,clip,angle=270]{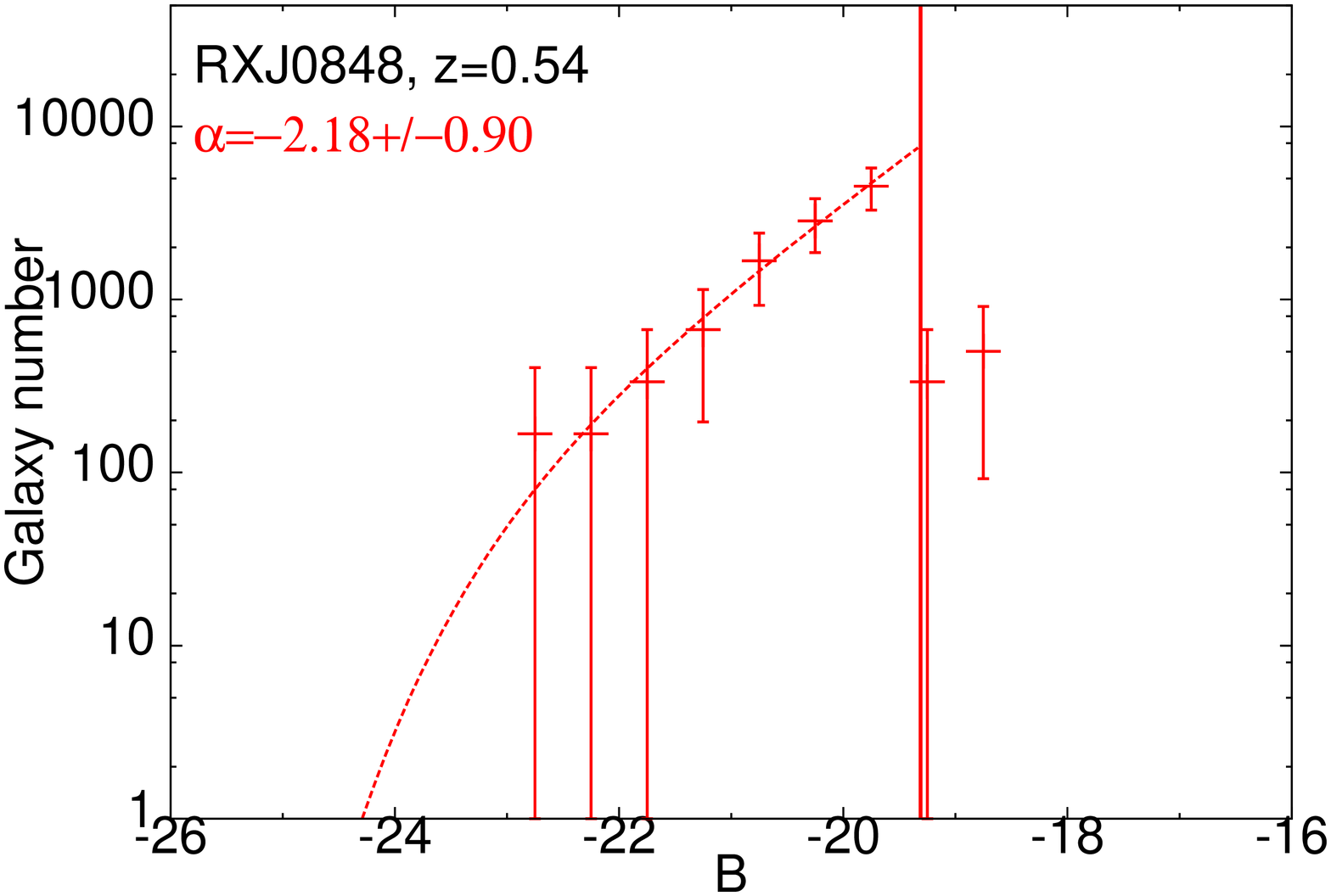}
\includegraphics[width=0.17\textwidth,clip,angle=270]{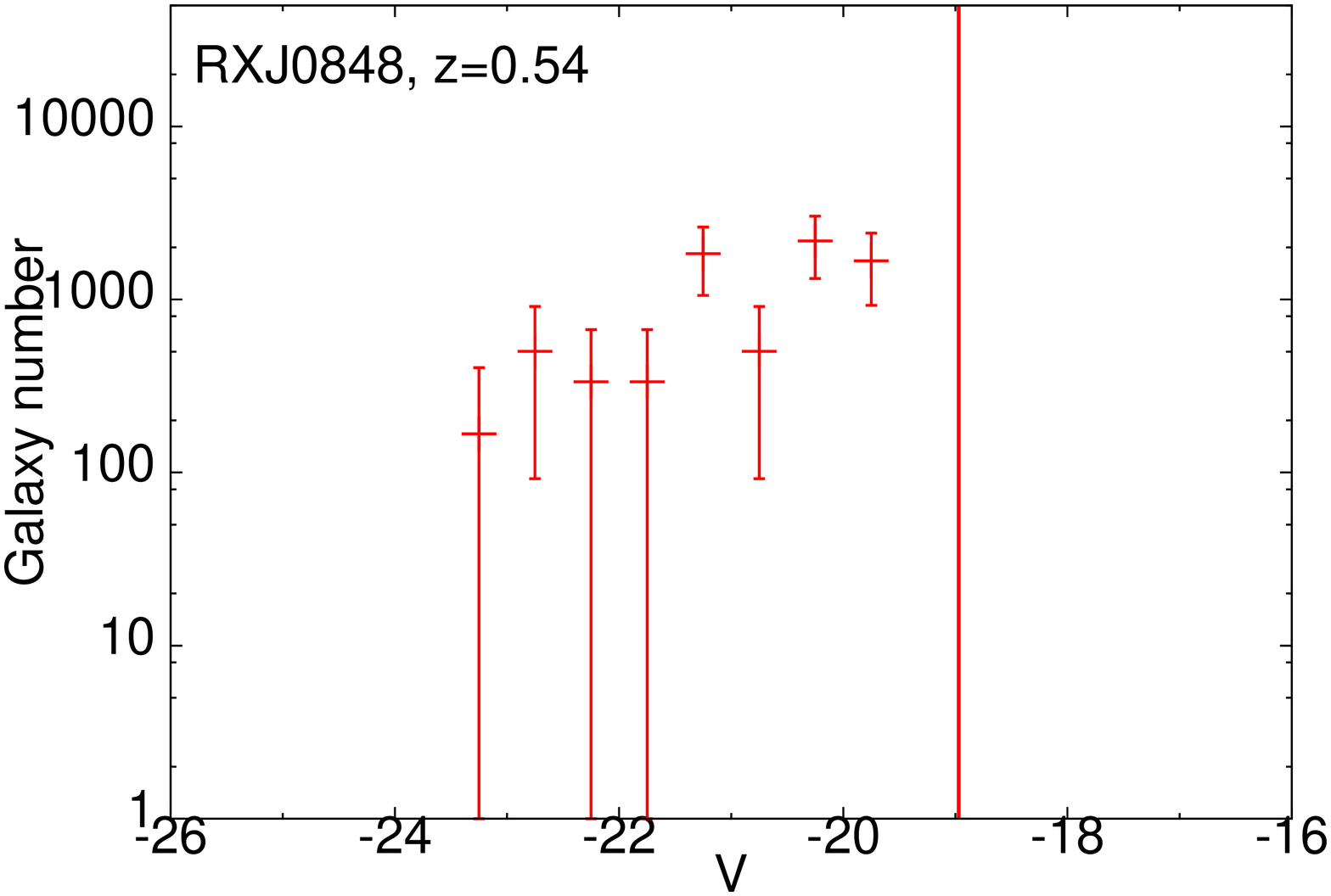}
\includegraphics[width=0.17\textwidth,clip,angle=270]{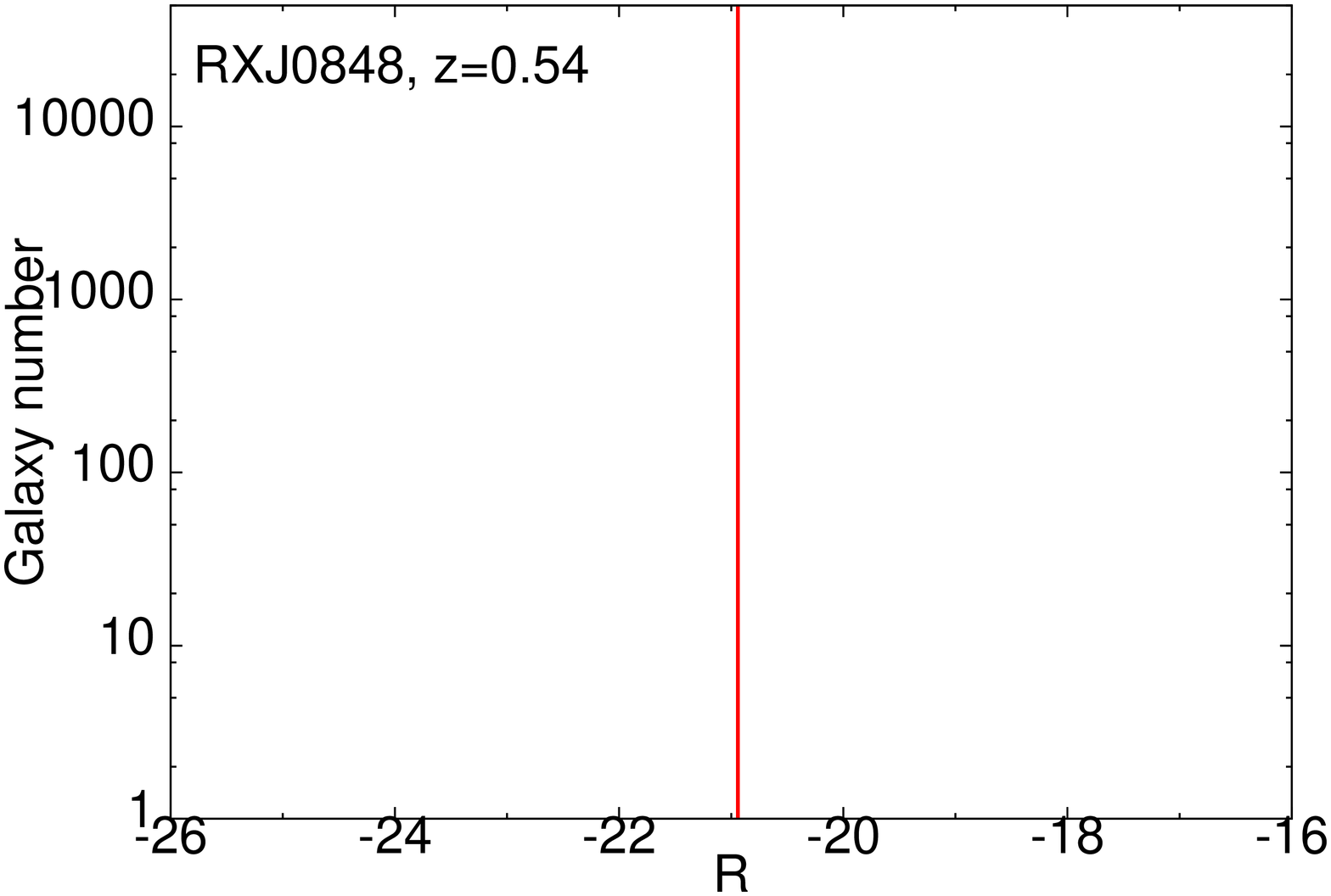}
\includegraphics[width=0.17\textwidth,clip,angle=270]{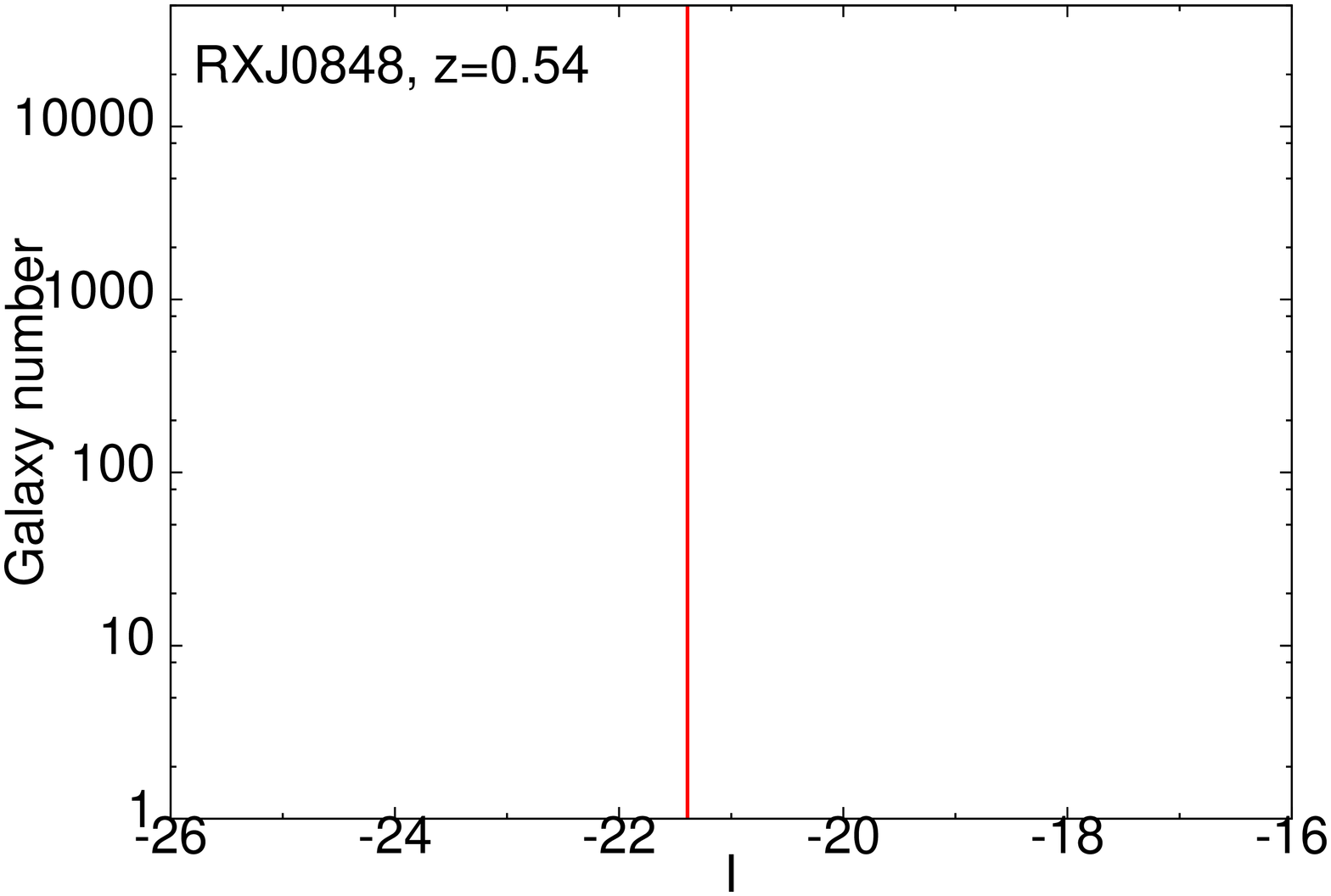} \\
\includegraphics[width=0.17\textwidth,clip,angle=270]{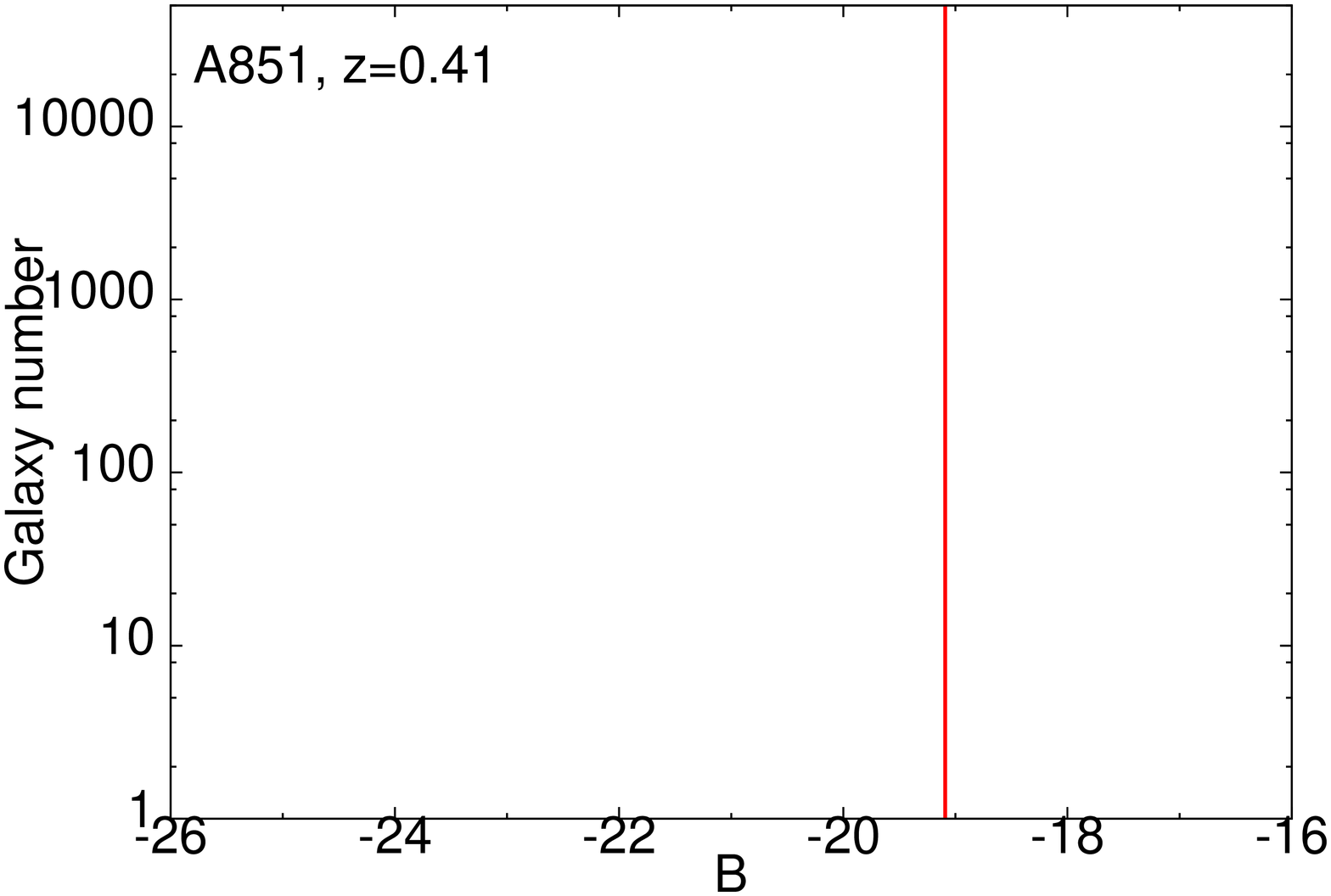}
\includegraphics[width=0.17\textwidth,clip,angle=270]{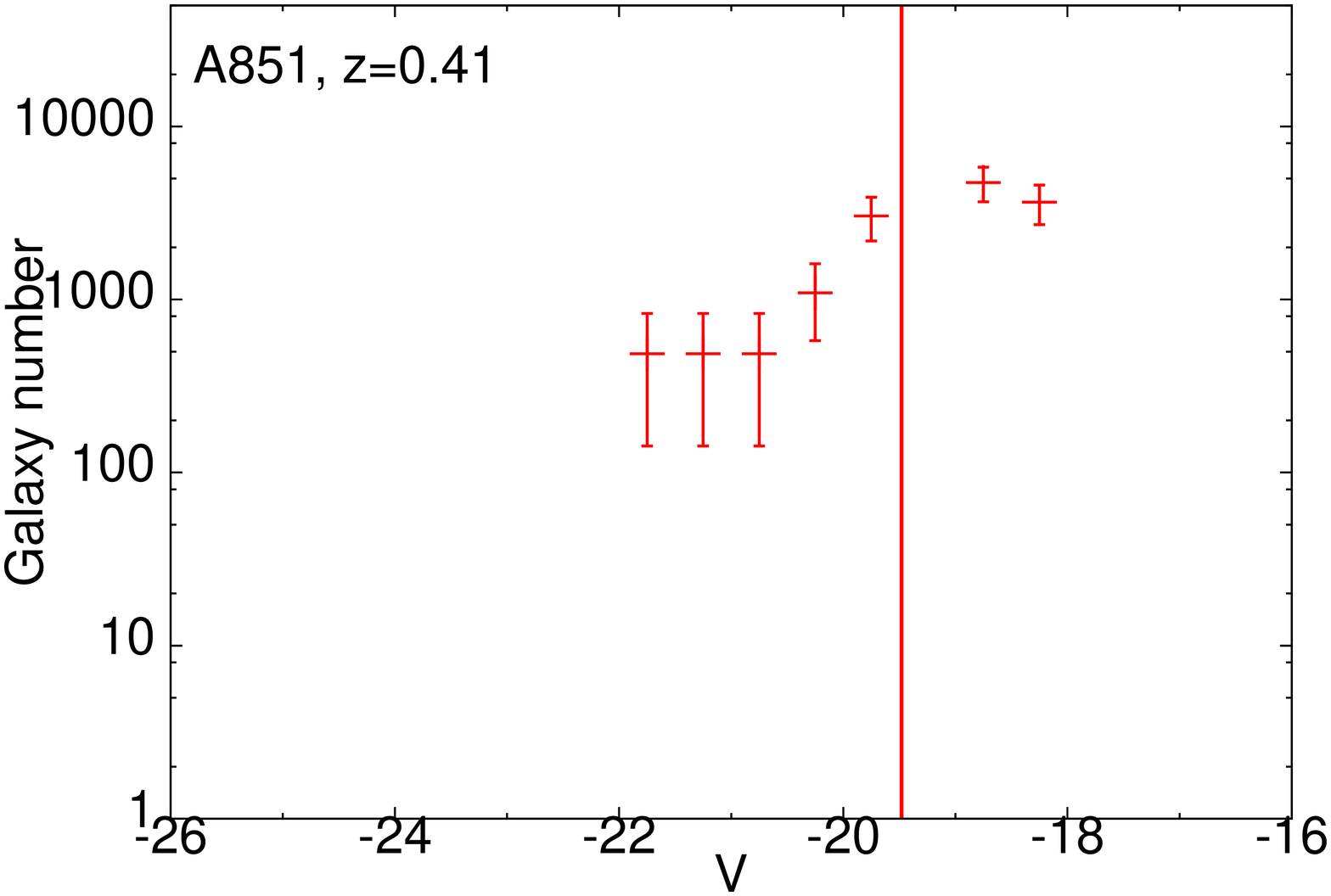}
\includegraphics[width=0.17\textwidth,clip,angle=270]{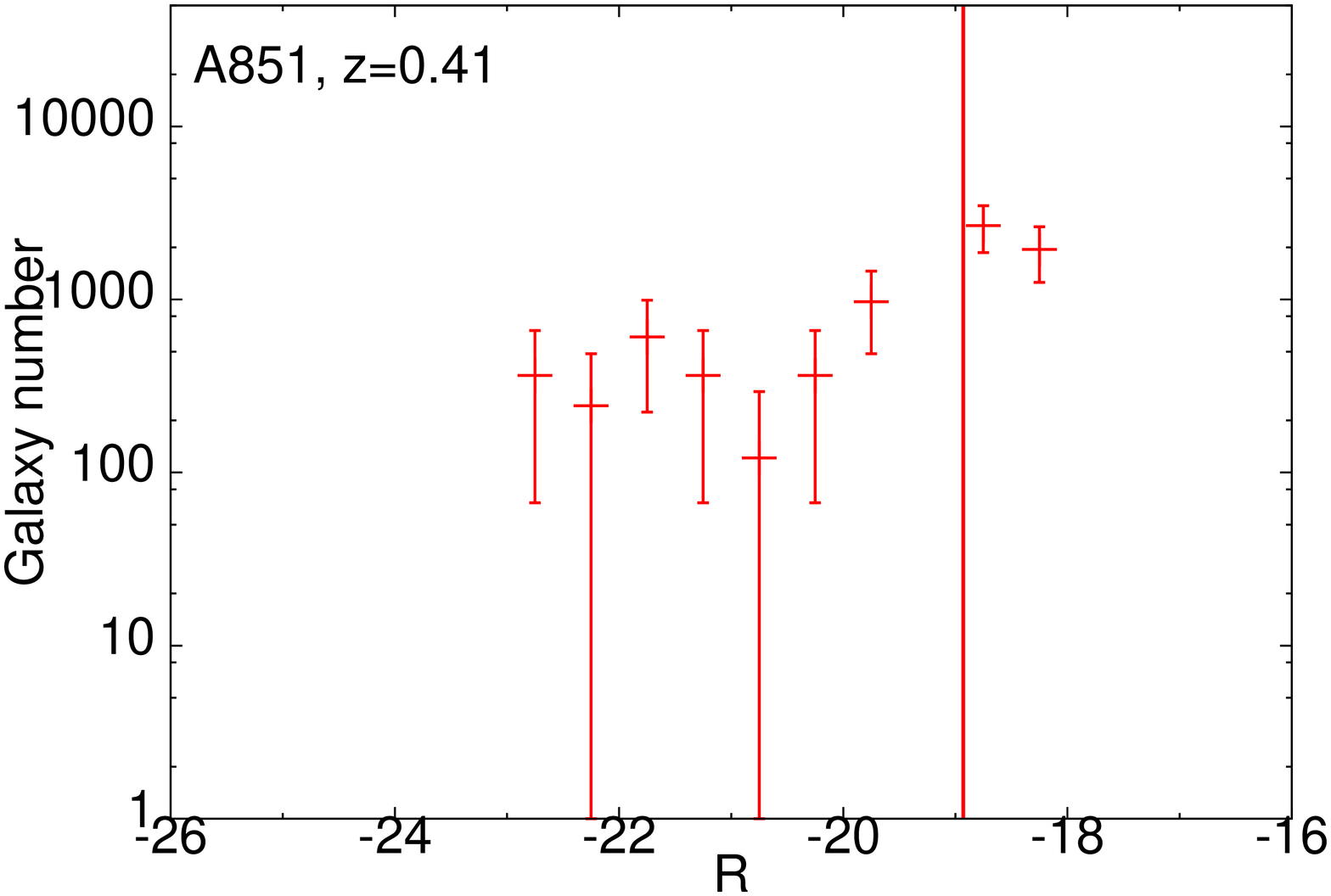}
\includegraphics[width=0.17\textwidth,clip,angle=270]{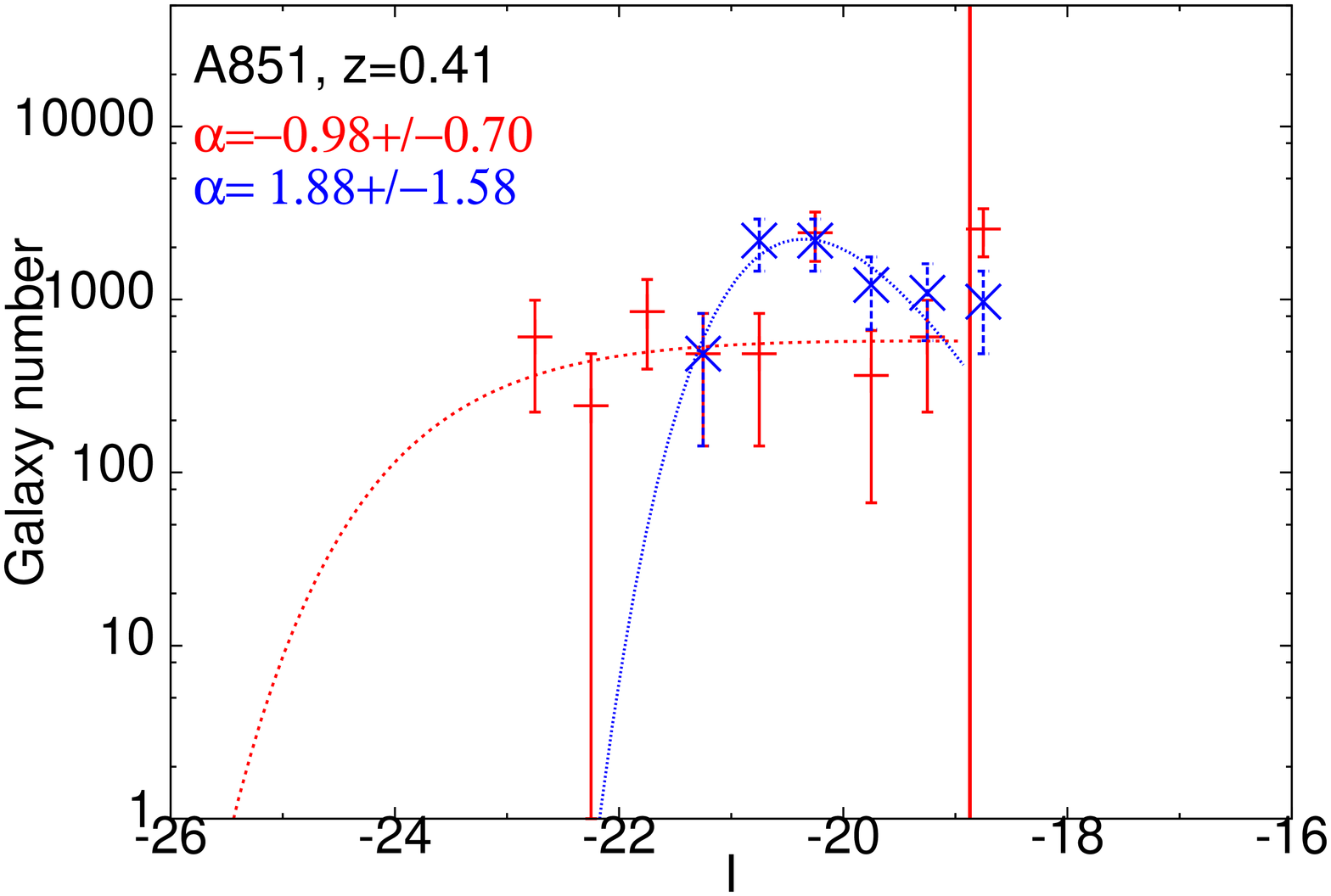} \\
\includegraphics[width=0.17\textwidth,clip,angle=270]{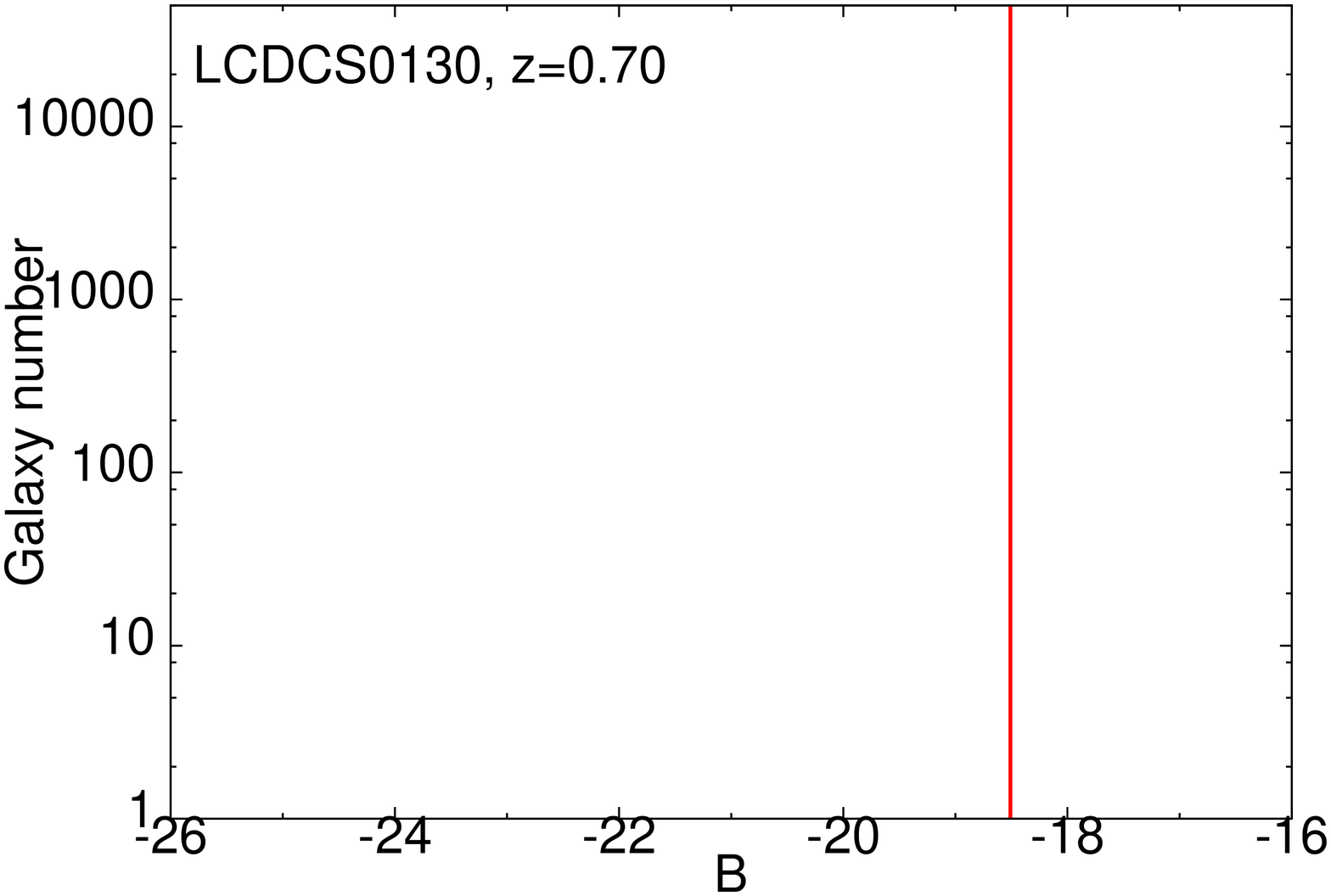}
\includegraphics[width=0.17\textwidth,clip,angle=270]{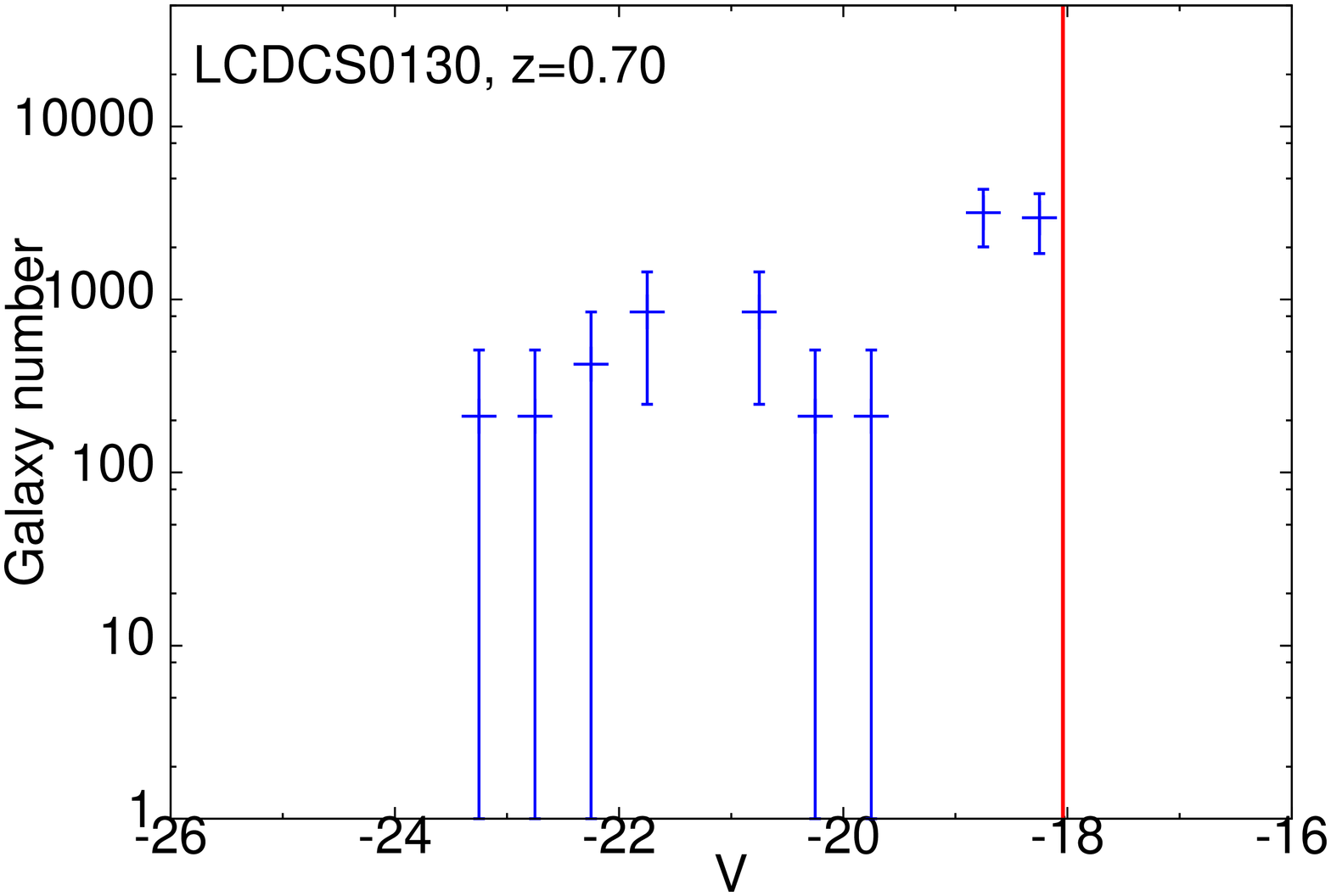}
\includegraphics[width=0.17\textwidth,clip,angle=270]{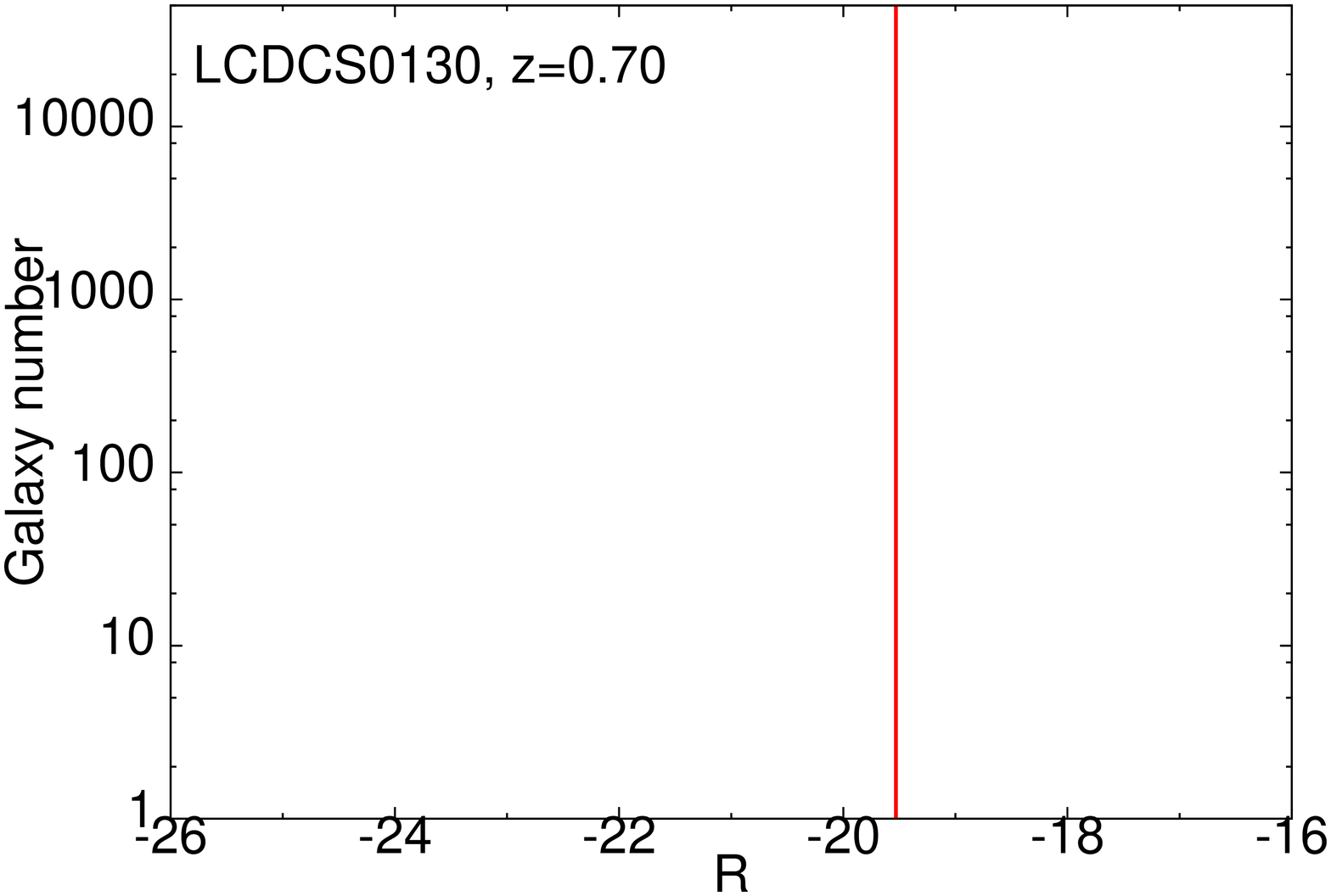}
\includegraphics[width=0.17\textwidth,clip,angle=270]{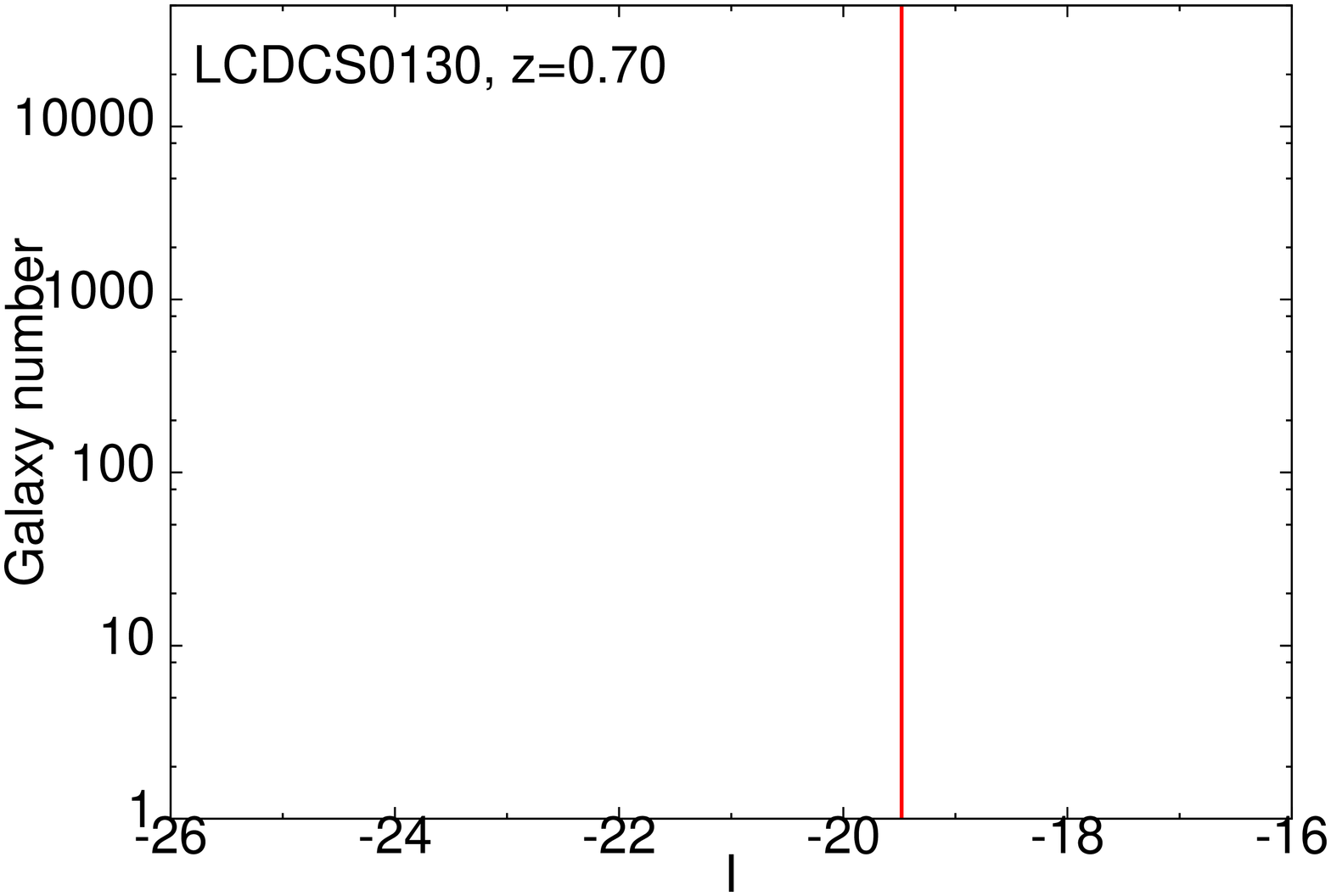} \\

\end{tabular}
\caption{GLFs in the B, V, R and I rest-frame bands (from left to
  right) for individual clusters ordered by right ascension. Red and blue
  points correspond to red-sequence and blue GLFs normalized to 1
  deg$^2$. The red vertical lines indicate the 90\% completeness
  limit. The red and blue curves show the best Schechter fits to red
  sequence and blue galaxies, and the faint end slope parameter
  ($\alpha$) is displayed in the corresponding color. Only galaxies
  brighter than the 90\% completeness limit are taken into account in
  the fits. Also, we only show GLFs richer than 20 galaxies
  after the colour separation and after subtracting the field.}
\label{fig:glfindiv} 
\end{figure*}

\begin{figure*}[h!!]
\setcounter{figure}{0}
\begin{tabular}{cccc}

\includegraphics[width=0.17\textwidth,clip,angle=270]{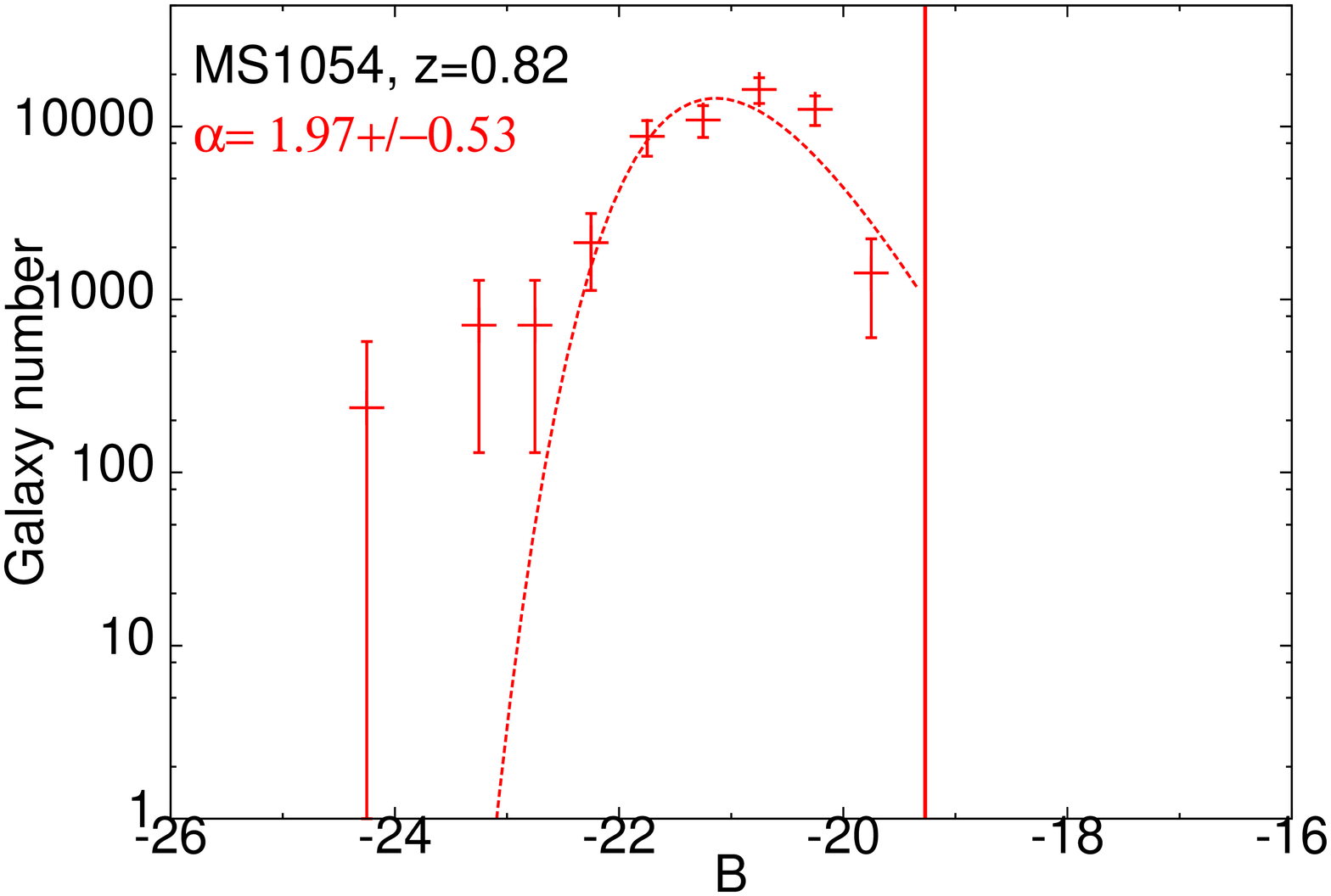}
\includegraphics[width=0.17\textwidth,clip,angle=270]{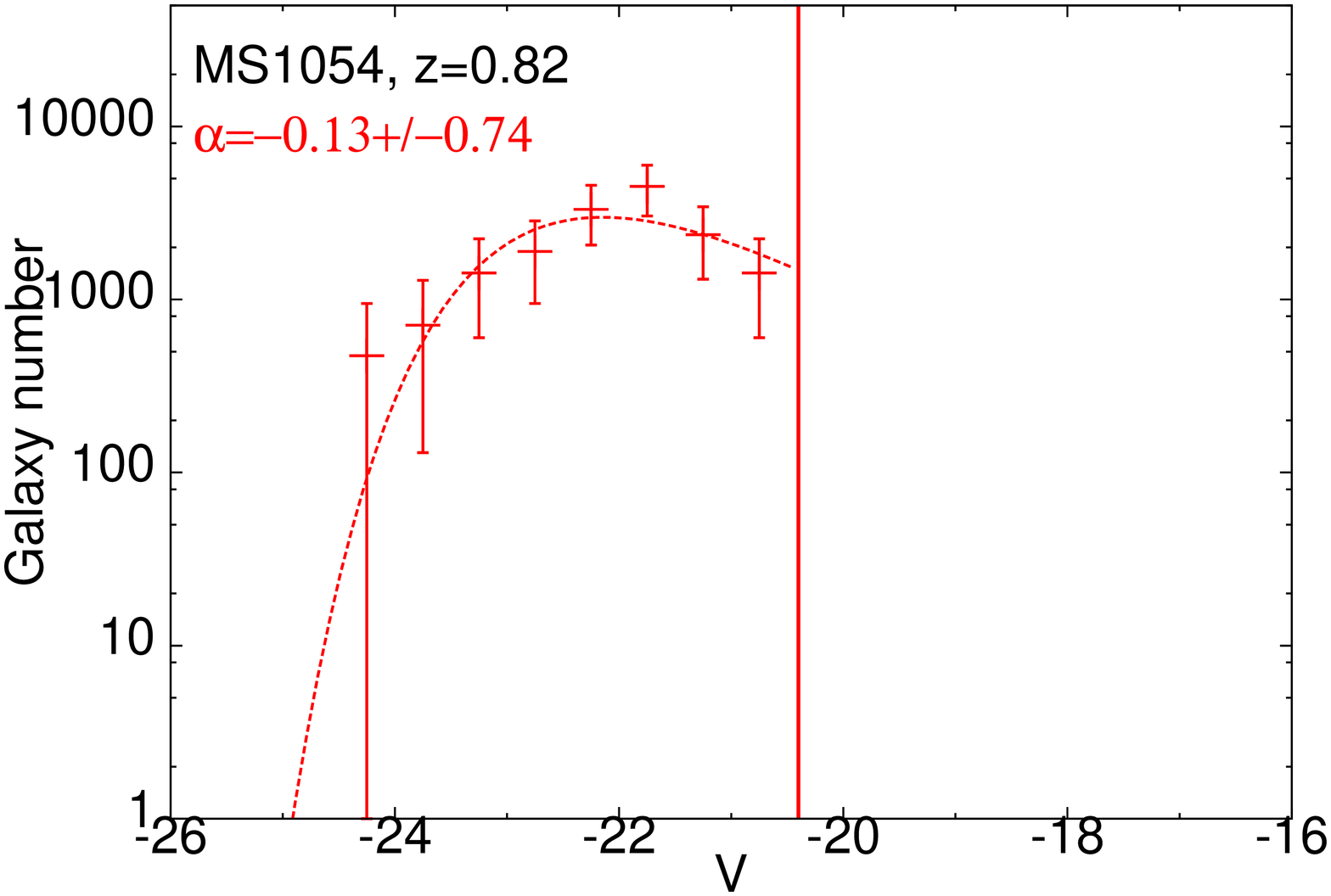}
\includegraphics[width=0.17\textwidth,clip,angle=270]{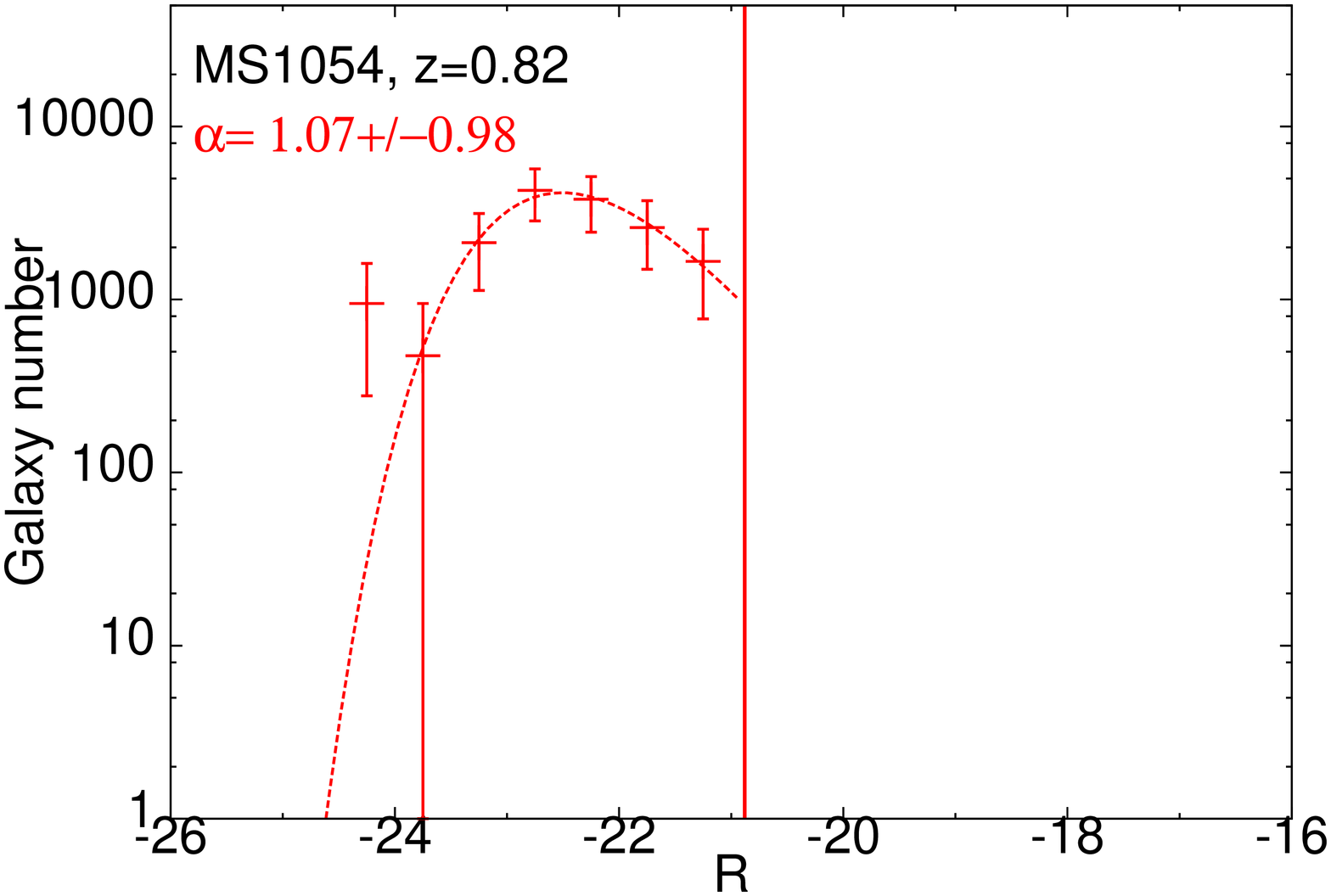}
\includegraphics[width=0.17\textwidth,clip,angle=270]{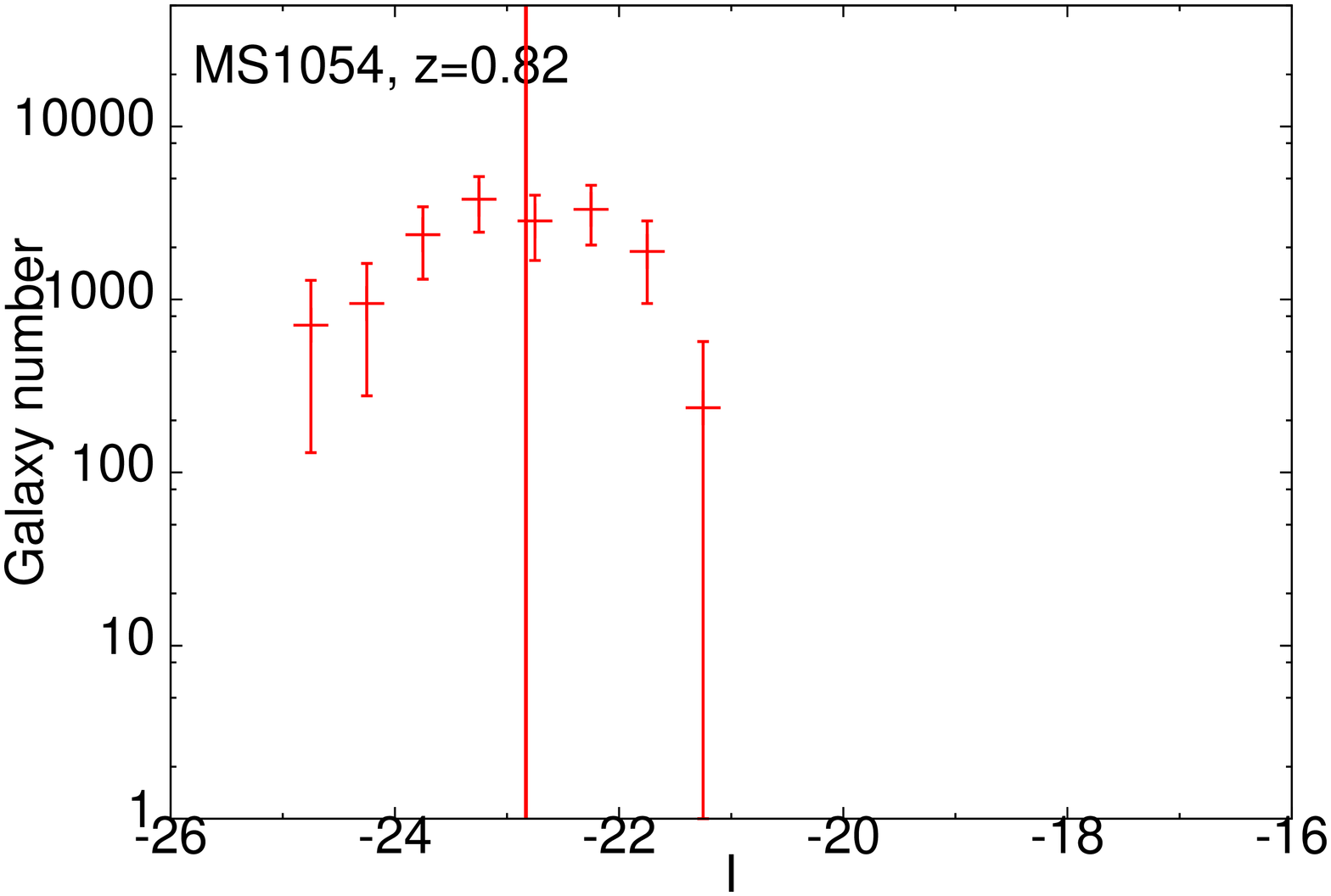} \\
\includegraphics[width=0.17\textwidth,clip,angle=270]{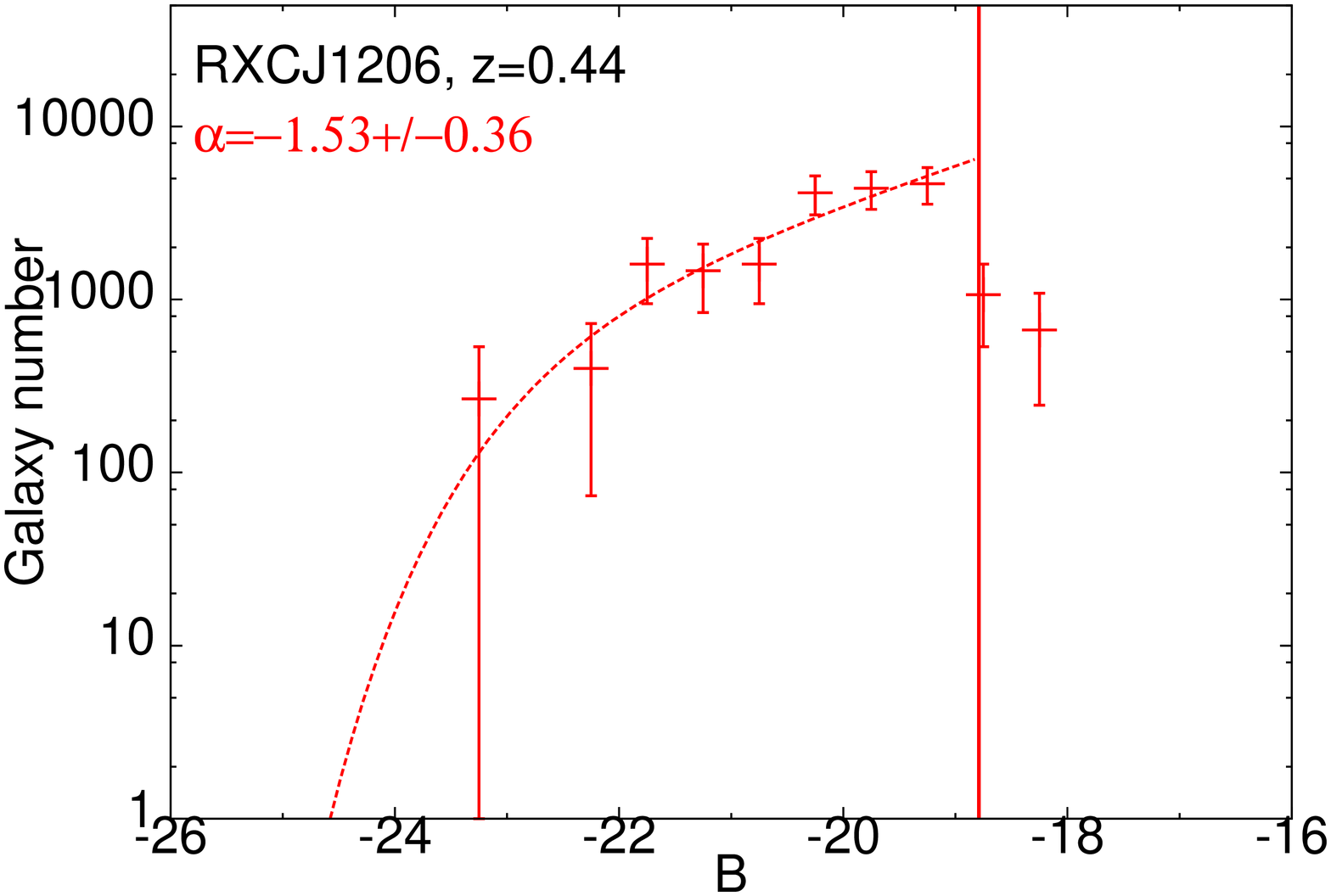}
\includegraphics[width=0.17\textwidth,clip,angle=270]{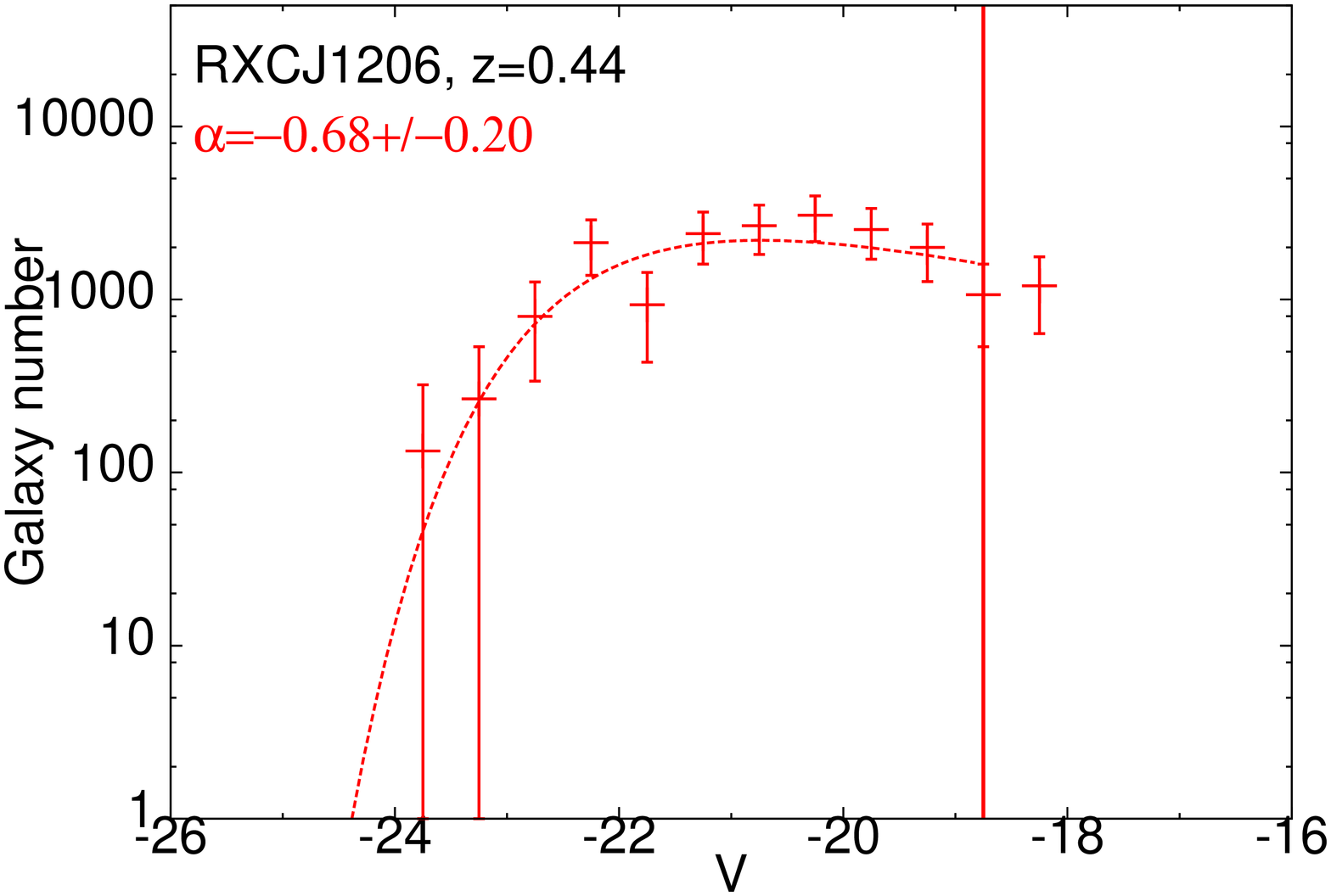}
\includegraphics[width=0.17\textwidth,clip,angle=270]{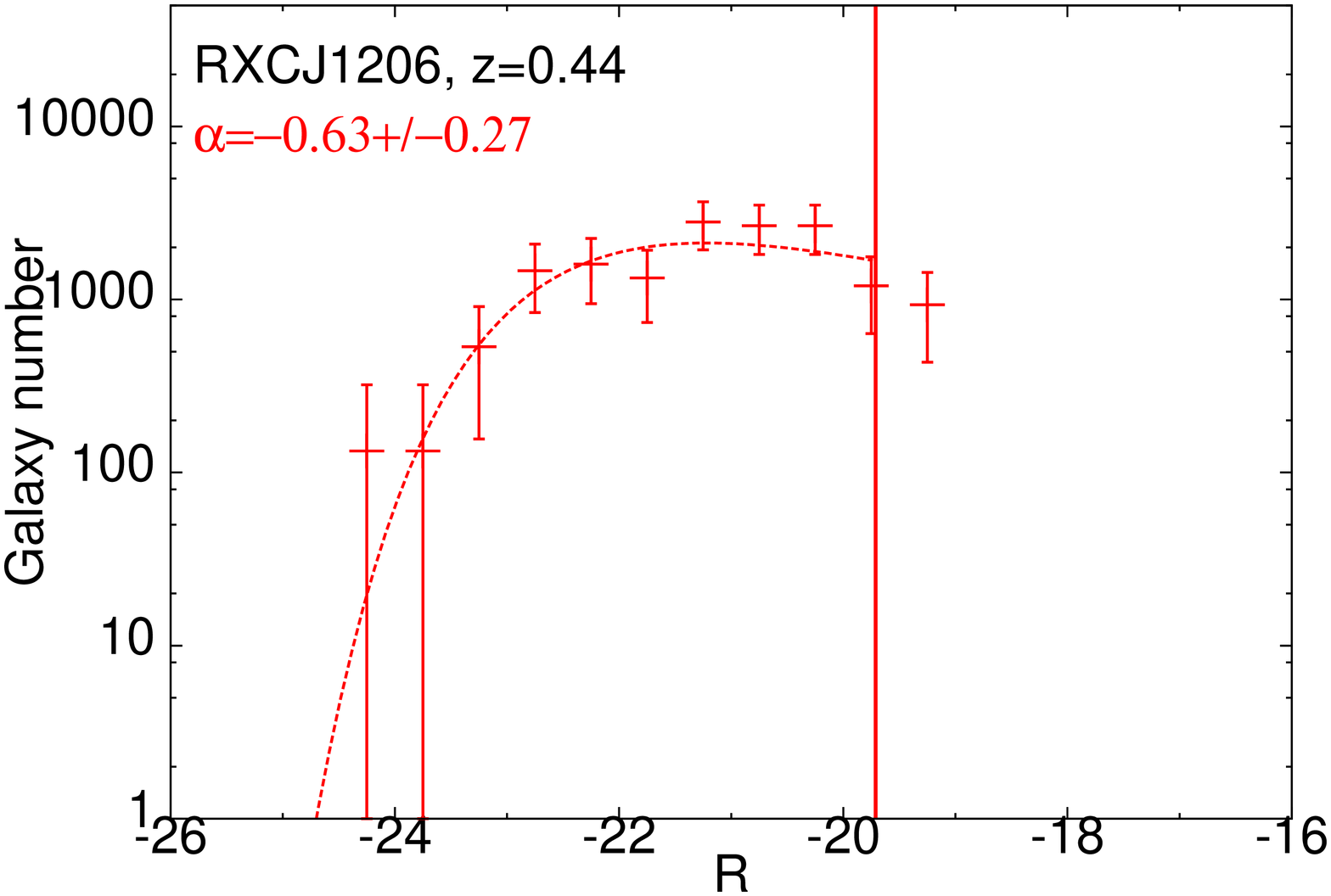}
\includegraphics[width=0.17\textwidth,clip,angle=270]{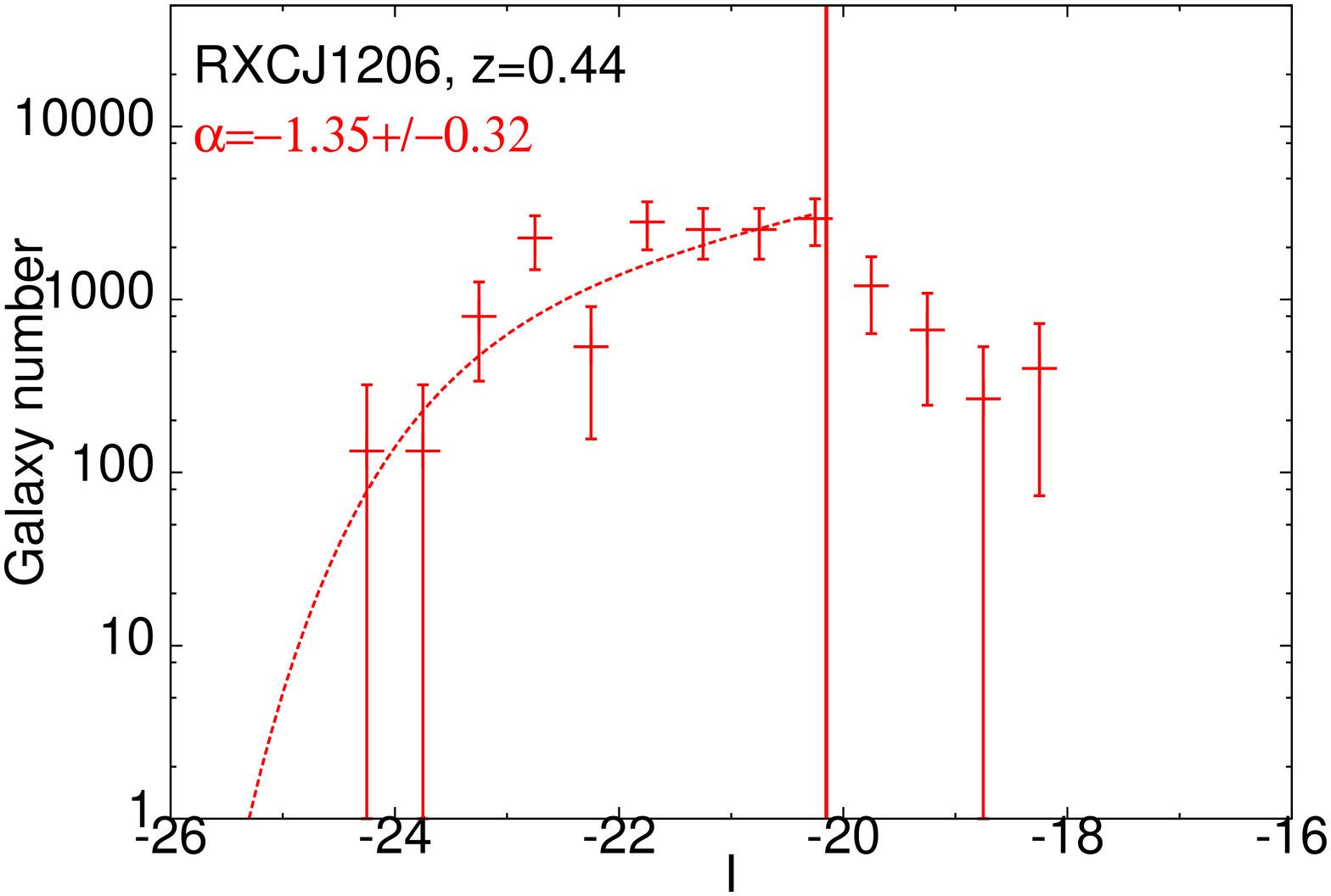} \\
\includegraphics[width=0.17\textwidth,clip,angle=270]{new_plots/INDIV/LCDCS0504_B.ps}
\includegraphics[width=0.17\textwidth,clip,angle=270]{new_plots/INDIV/LCDCS0504_V.ps}
\includegraphics[width=0.17\textwidth,clip,angle=270]{new_plots/INDIV/LCDCS0504_R.ps}
\includegraphics[width=0.17\textwidth,clip,angle=270]{new_plots/INDIV/LCDCS0504_I.ps} \\
\includegraphics[width=0.17\textwidth,clip,angle=270]{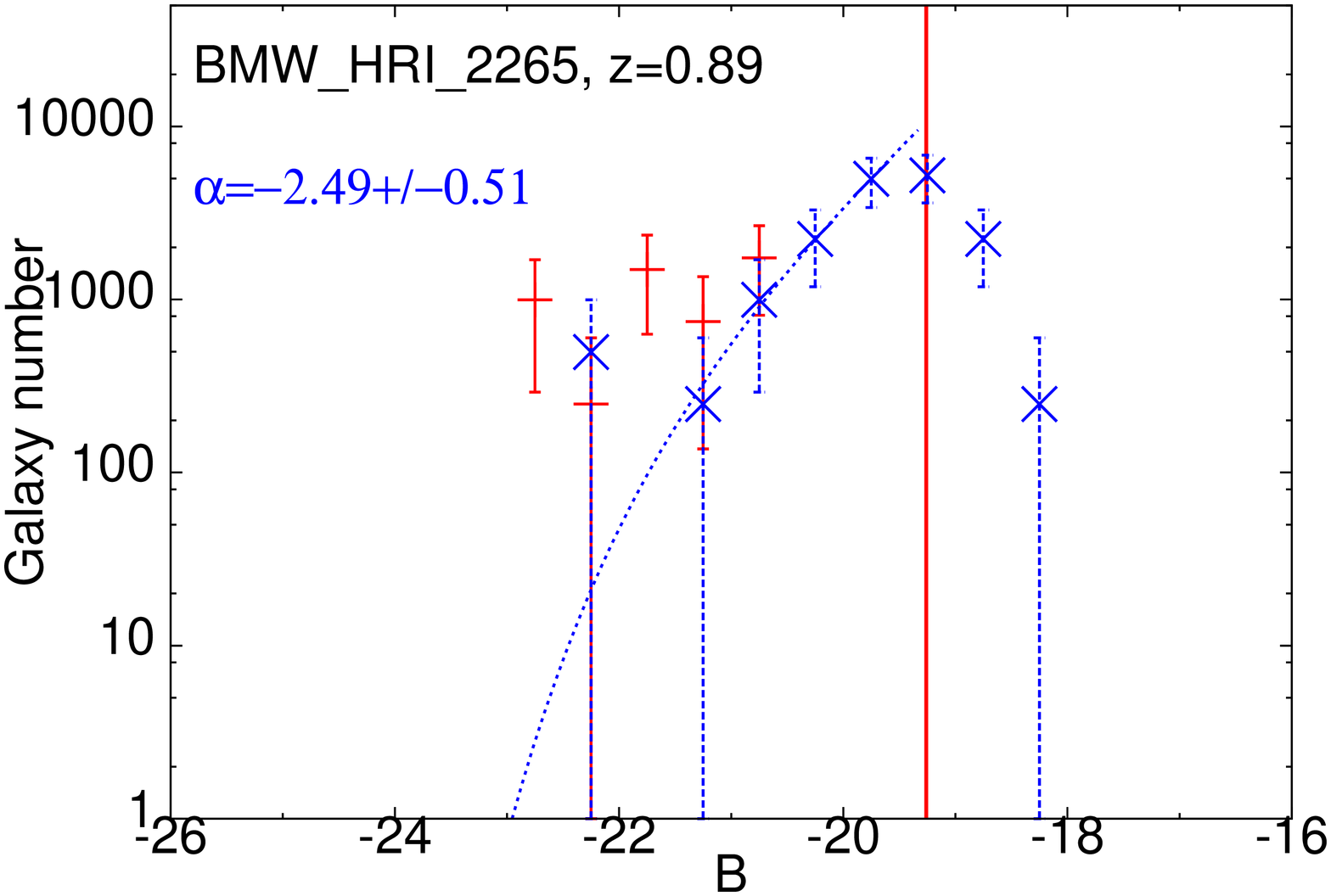}
\includegraphics[width=0.17\textwidth,clip,angle=270]{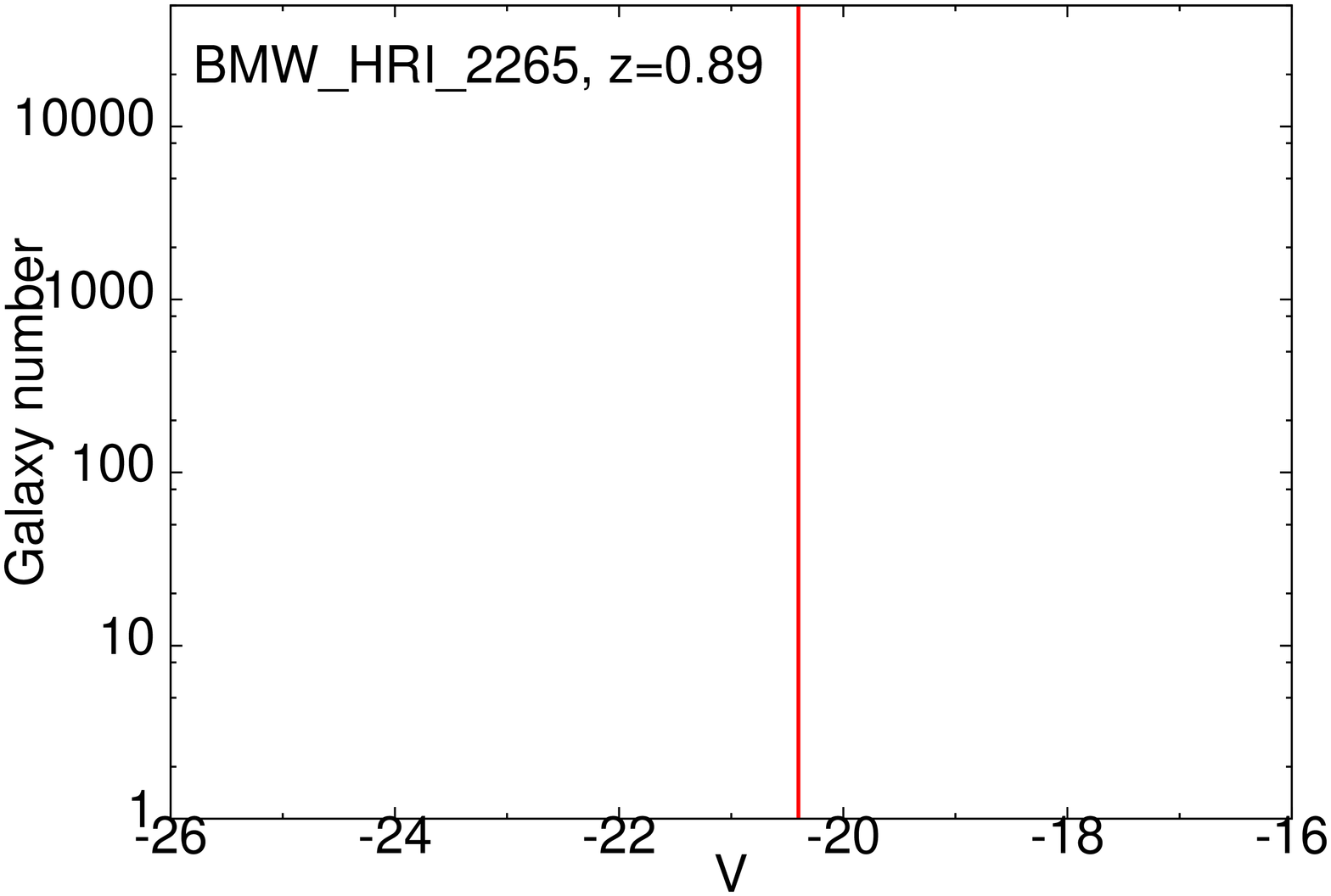}
\includegraphics[width=0.17\textwidth,clip,angle=270]{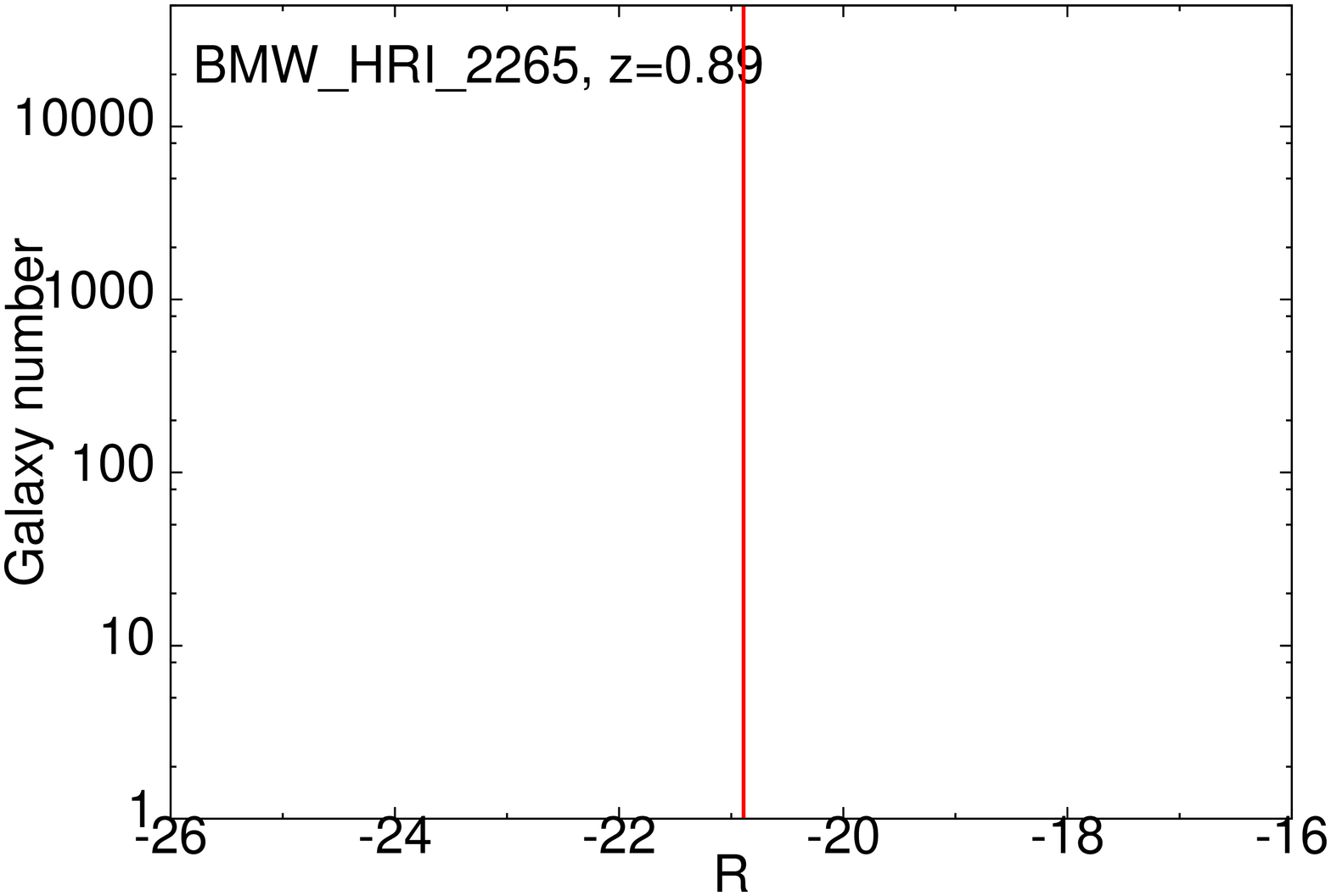}
\includegraphics[width=0.17\textwidth,clip,angle=270]{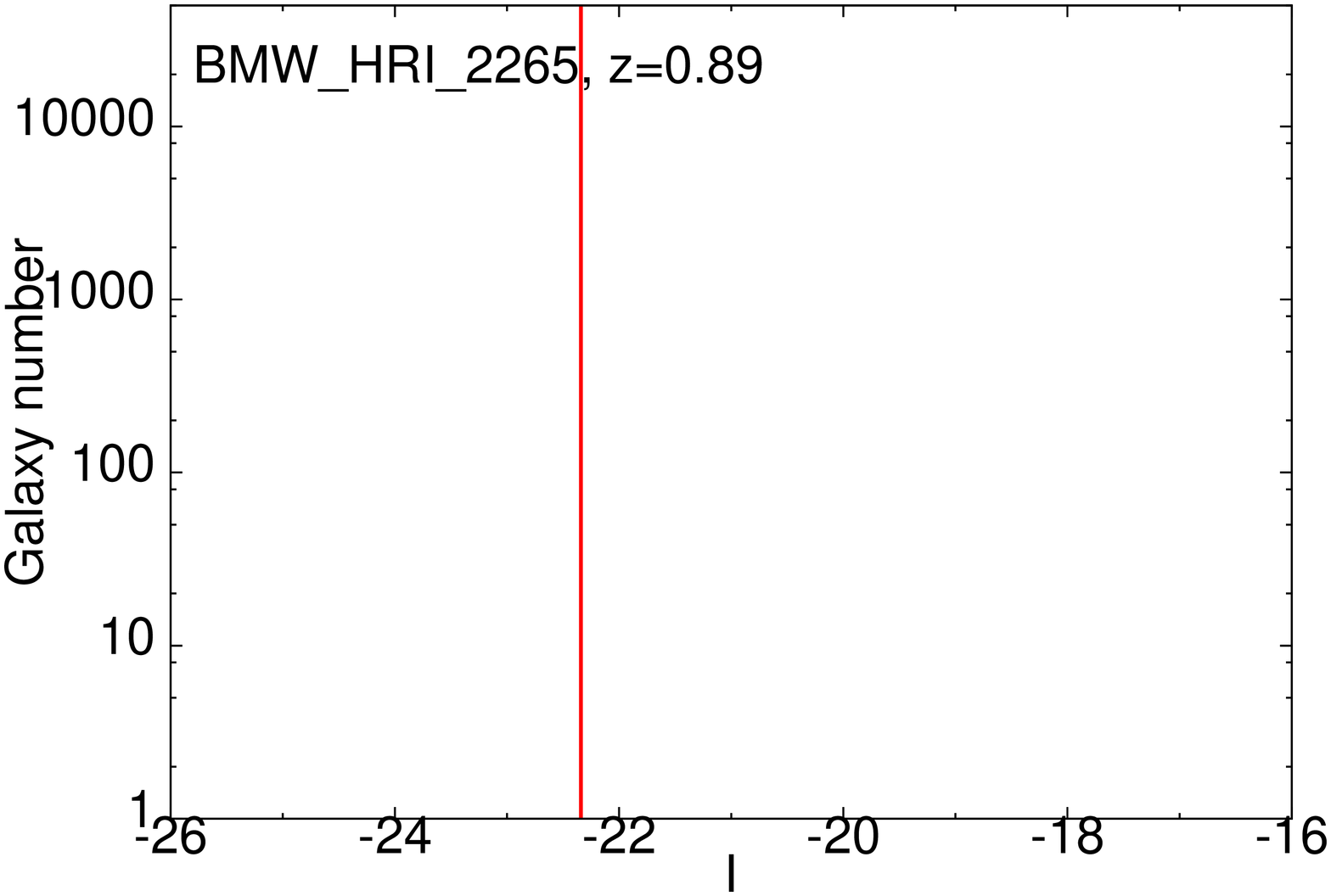} \\
\includegraphics[width=0.17\textwidth,clip,angle=270]{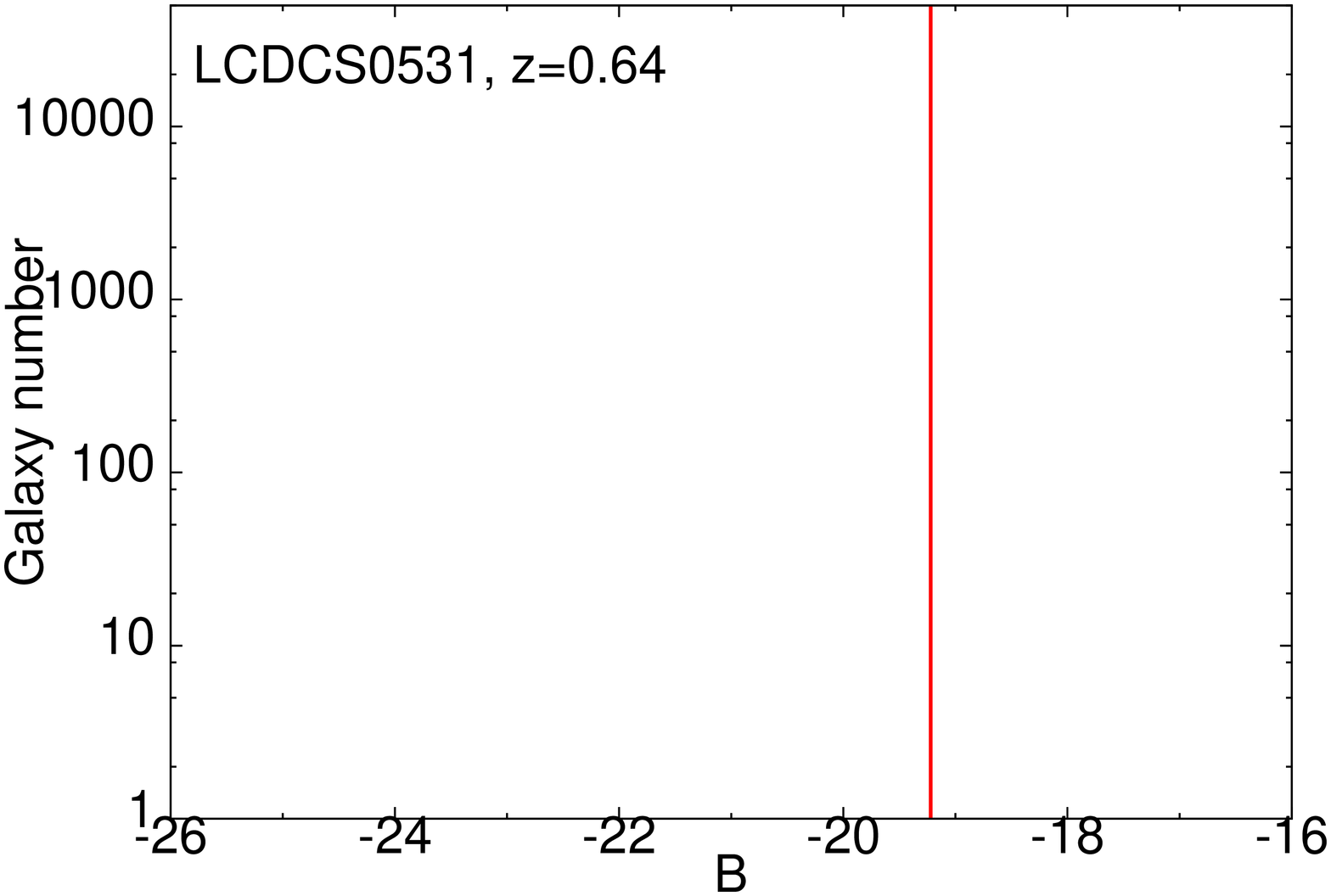}
\includegraphics[width=0.17\textwidth,clip,angle=270]{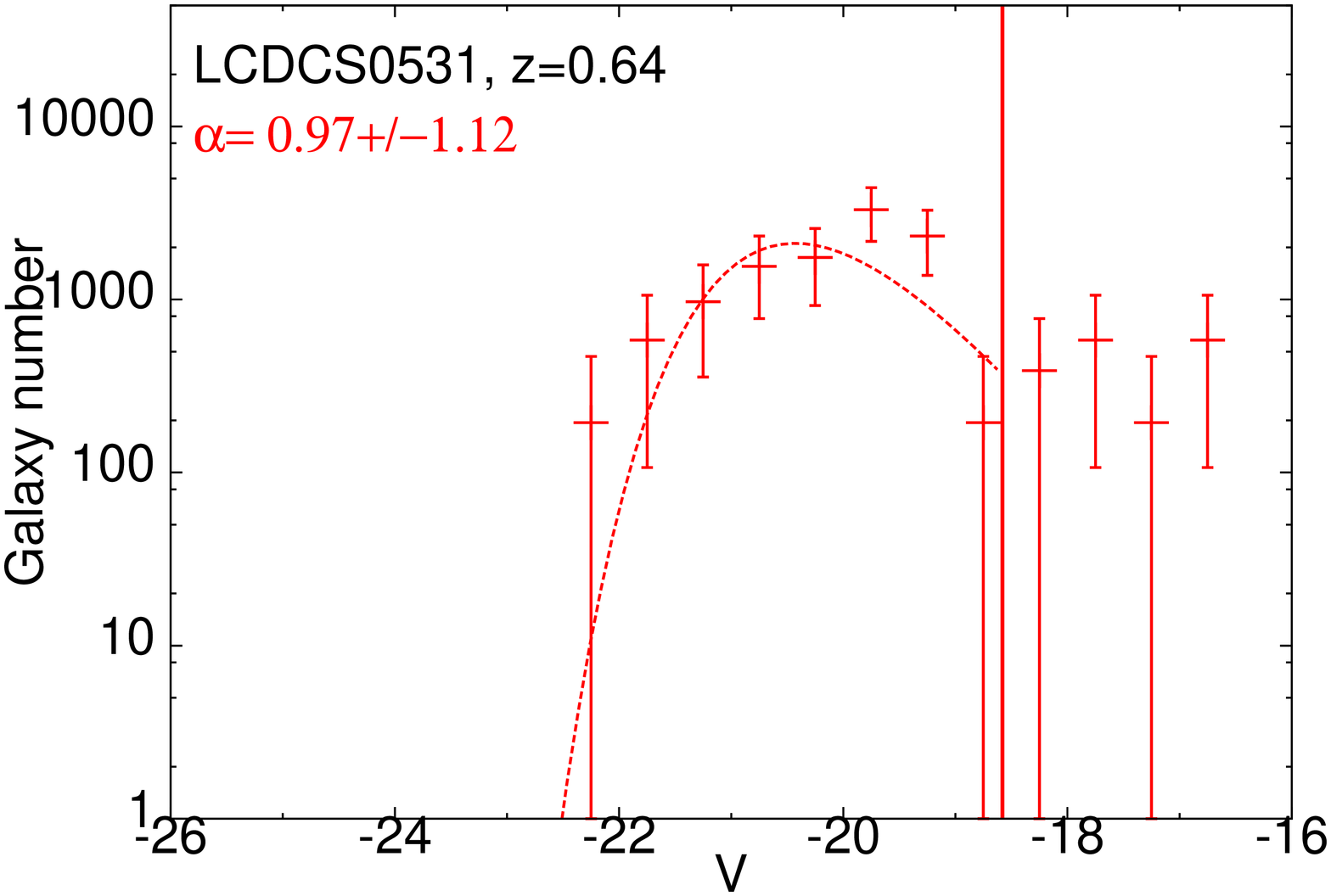}
\includegraphics[width=0.17\textwidth,clip,angle=270]{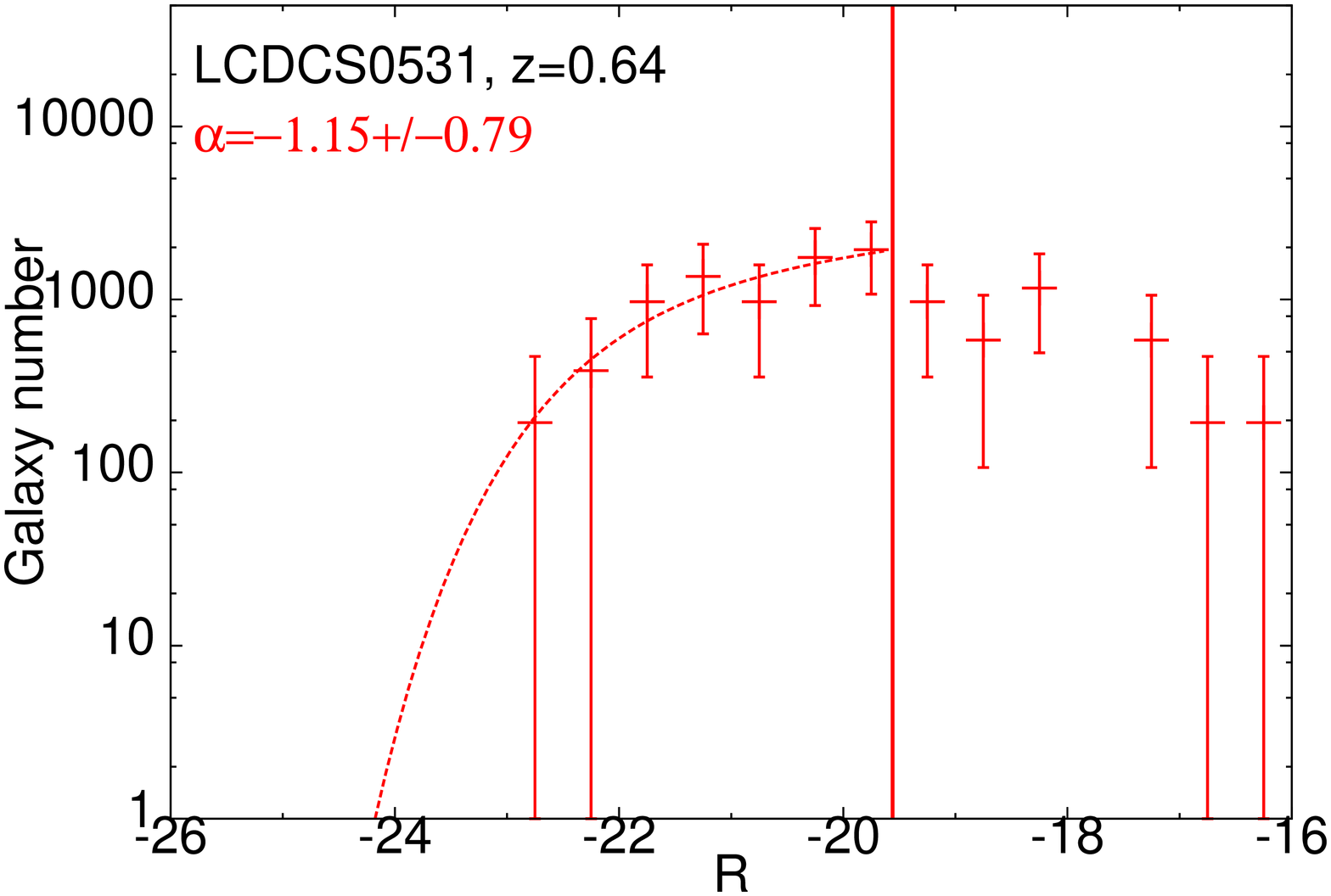}
\includegraphics[width=0.17\textwidth,clip,angle=270]{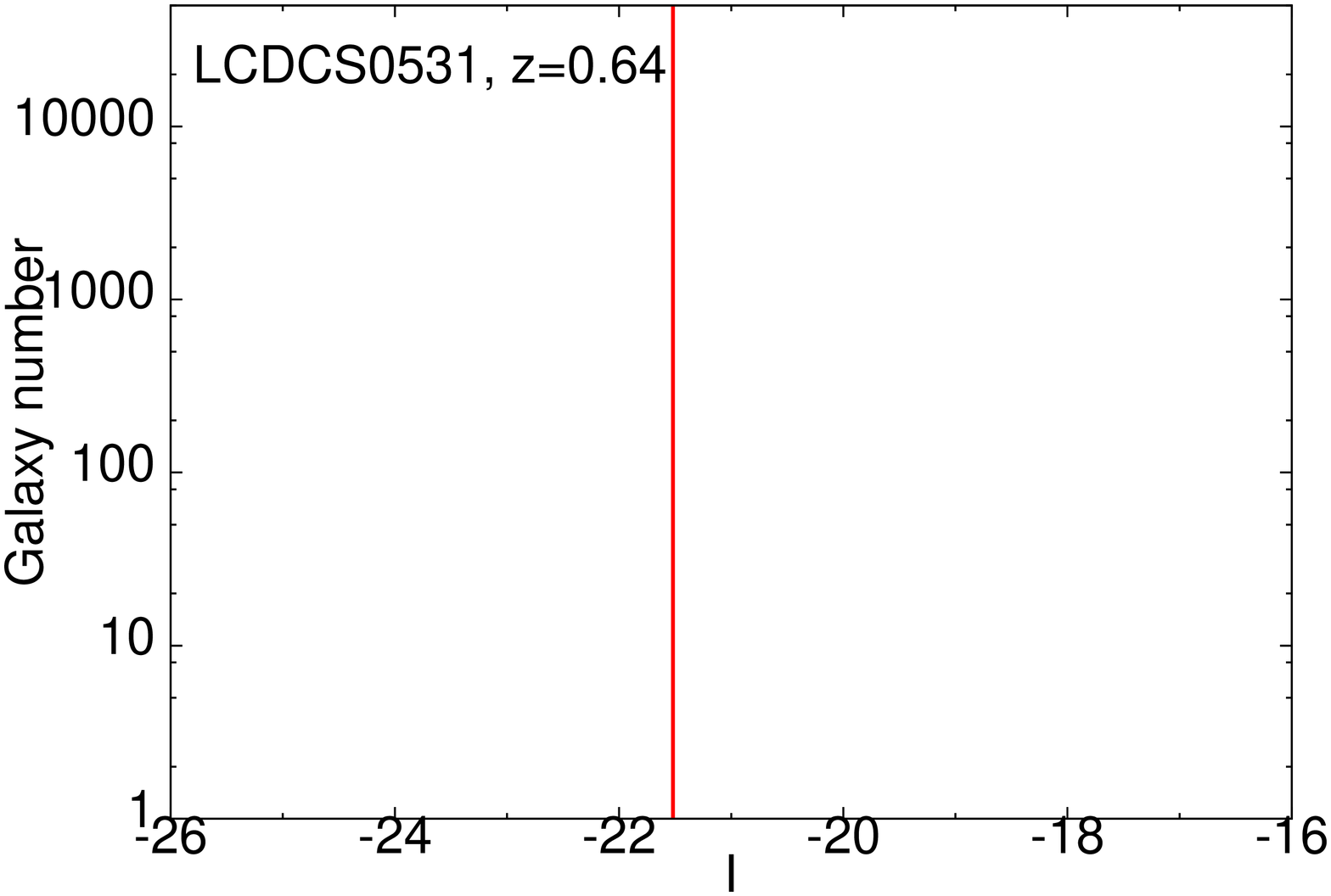} \\
\includegraphics[width=0.17\textwidth,clip,angle=270]{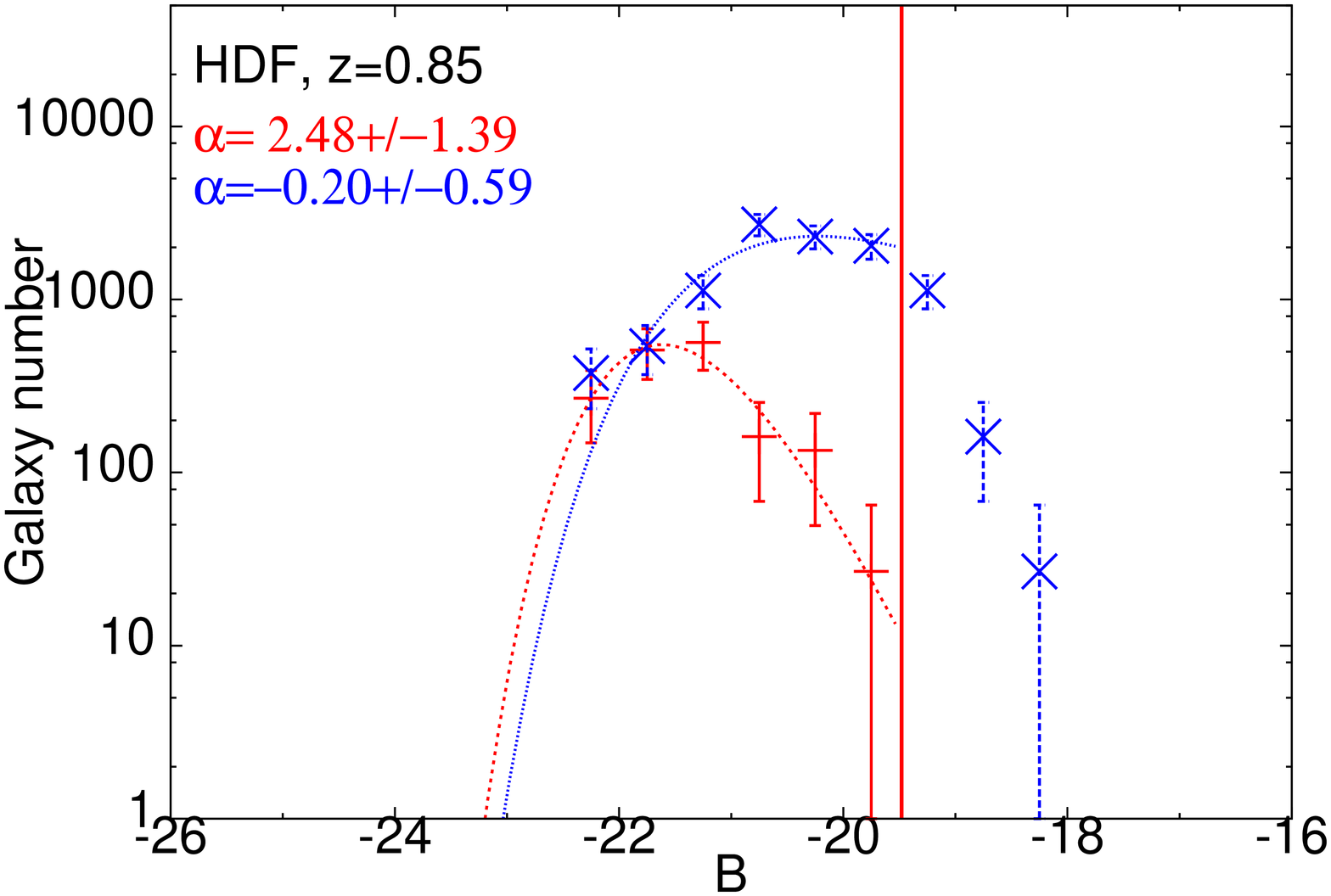}
\includegraphics[width=0.17\textwidth,clip,angle=270]{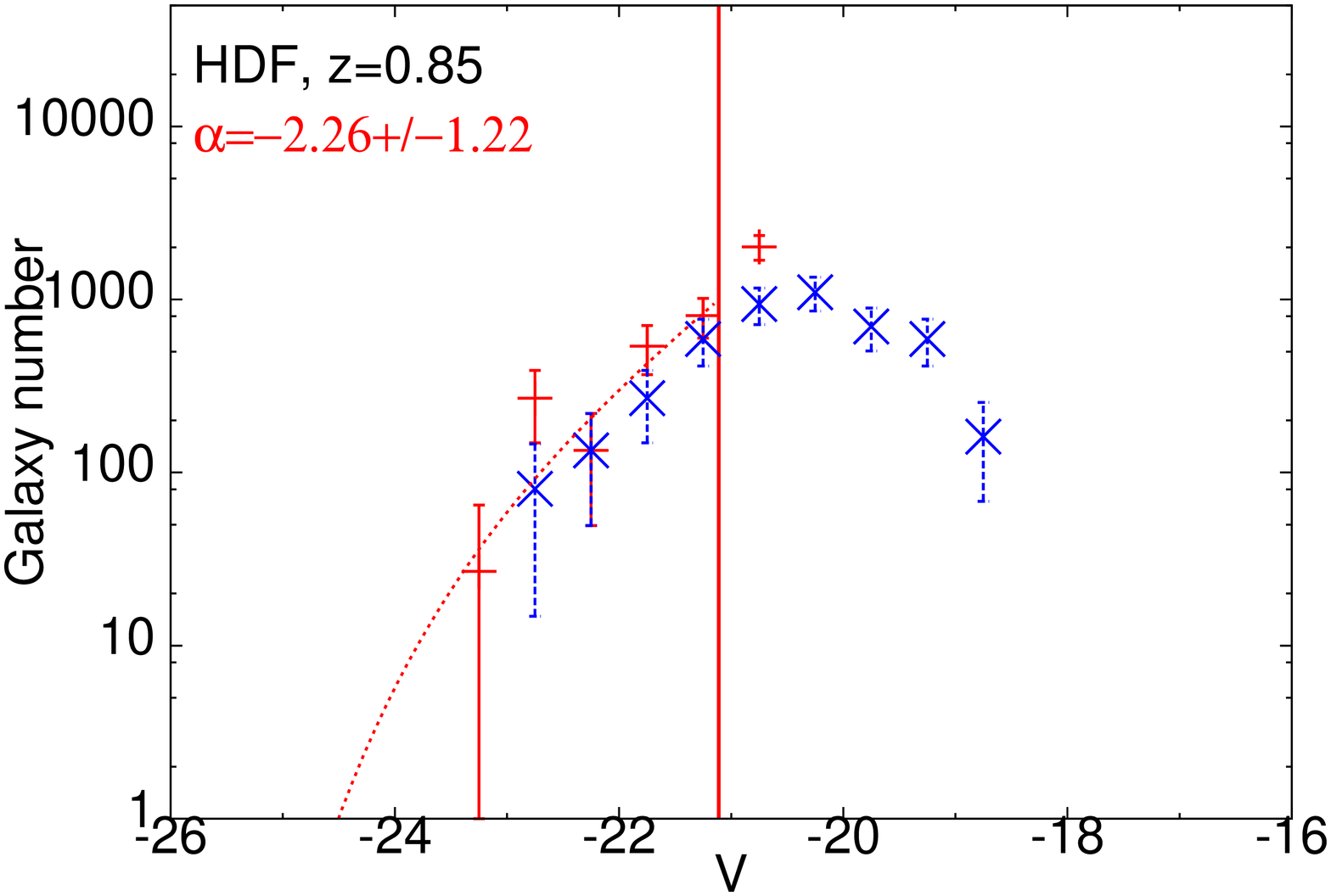}
\includegraphics[width=0.17\textwidth,clip,angle=270]{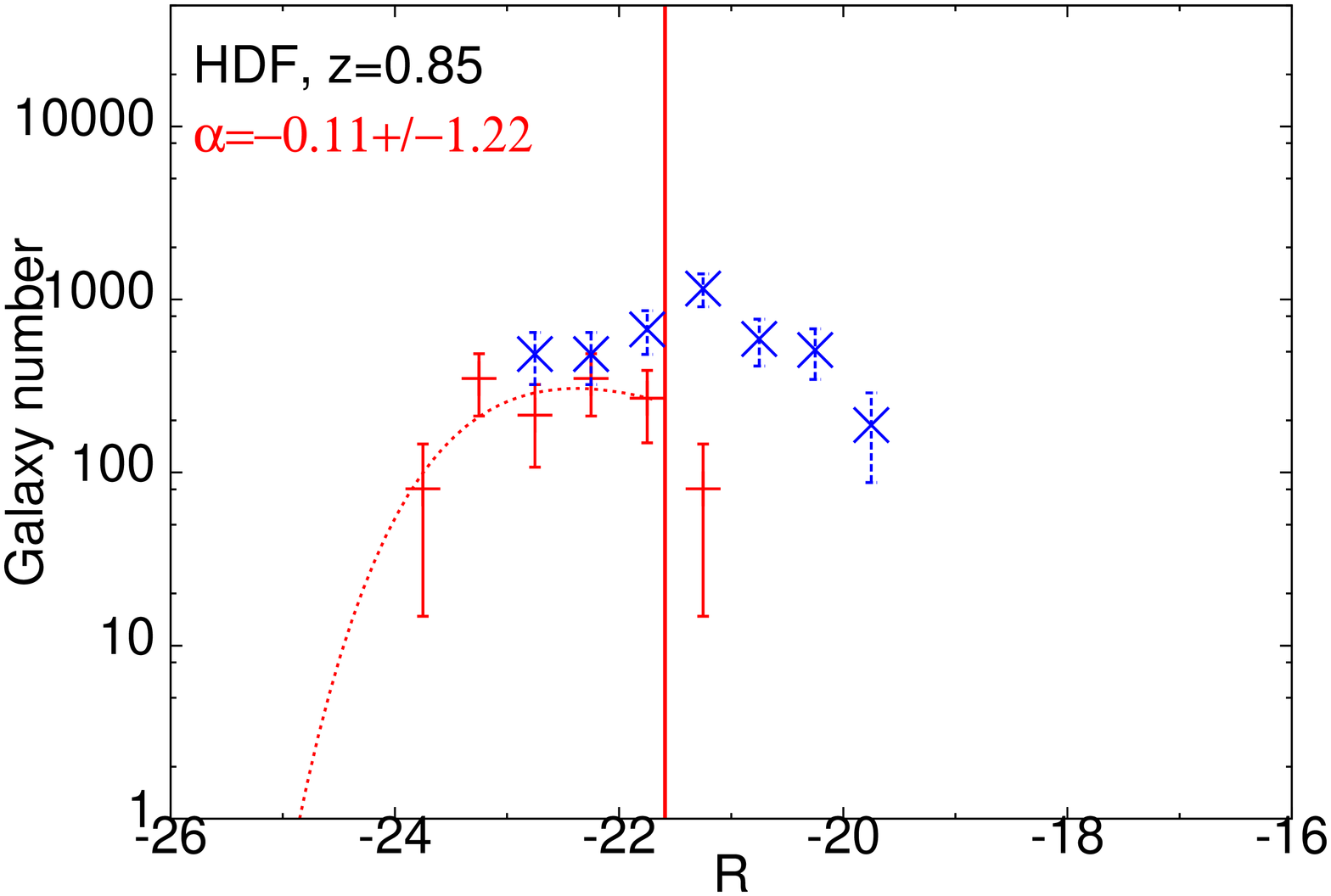}
\includegraphics[width=0.17\textwidth,clip,angle=270]{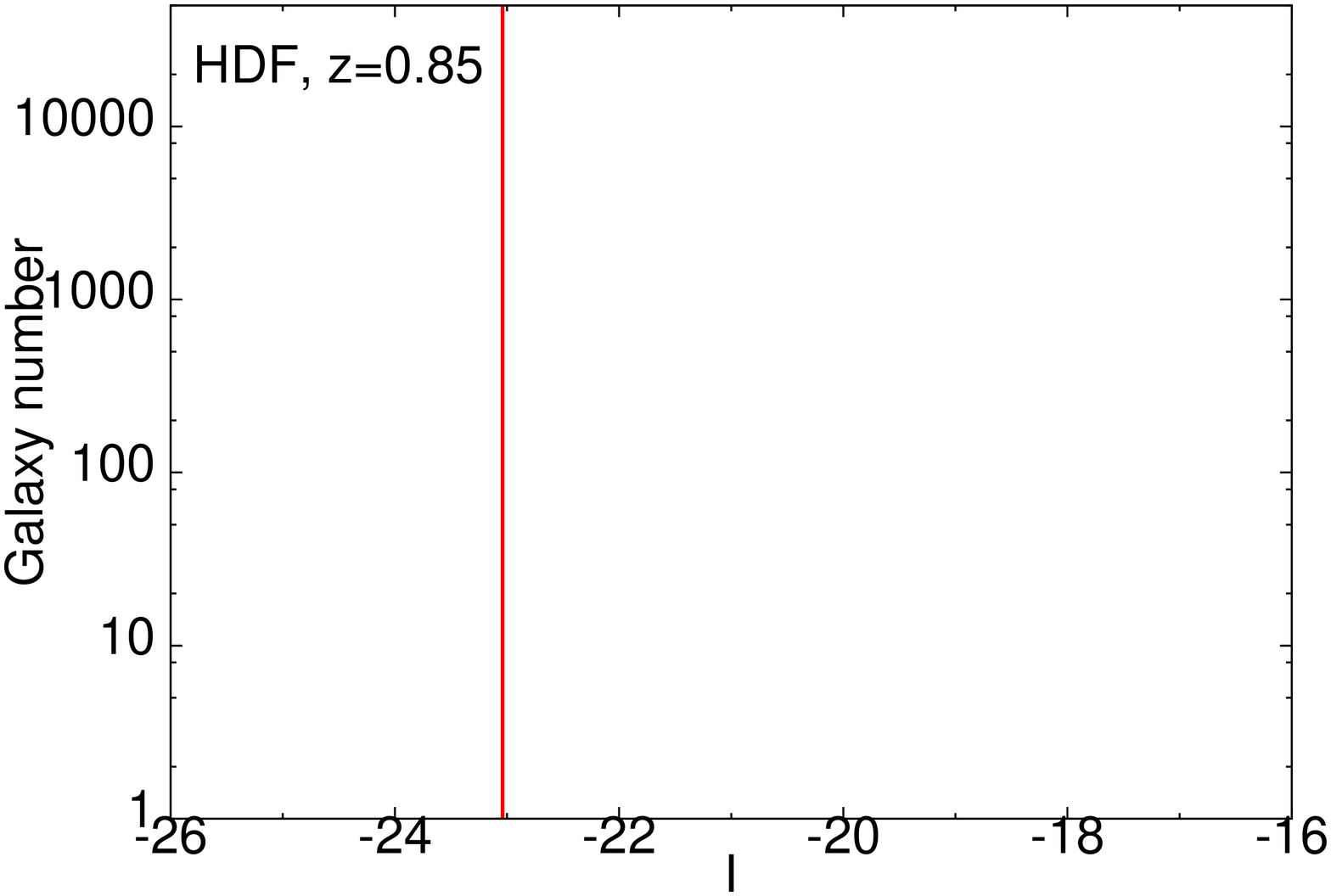}\\
\includegraphics[width=0.17\textwidth,clip,angle=270]{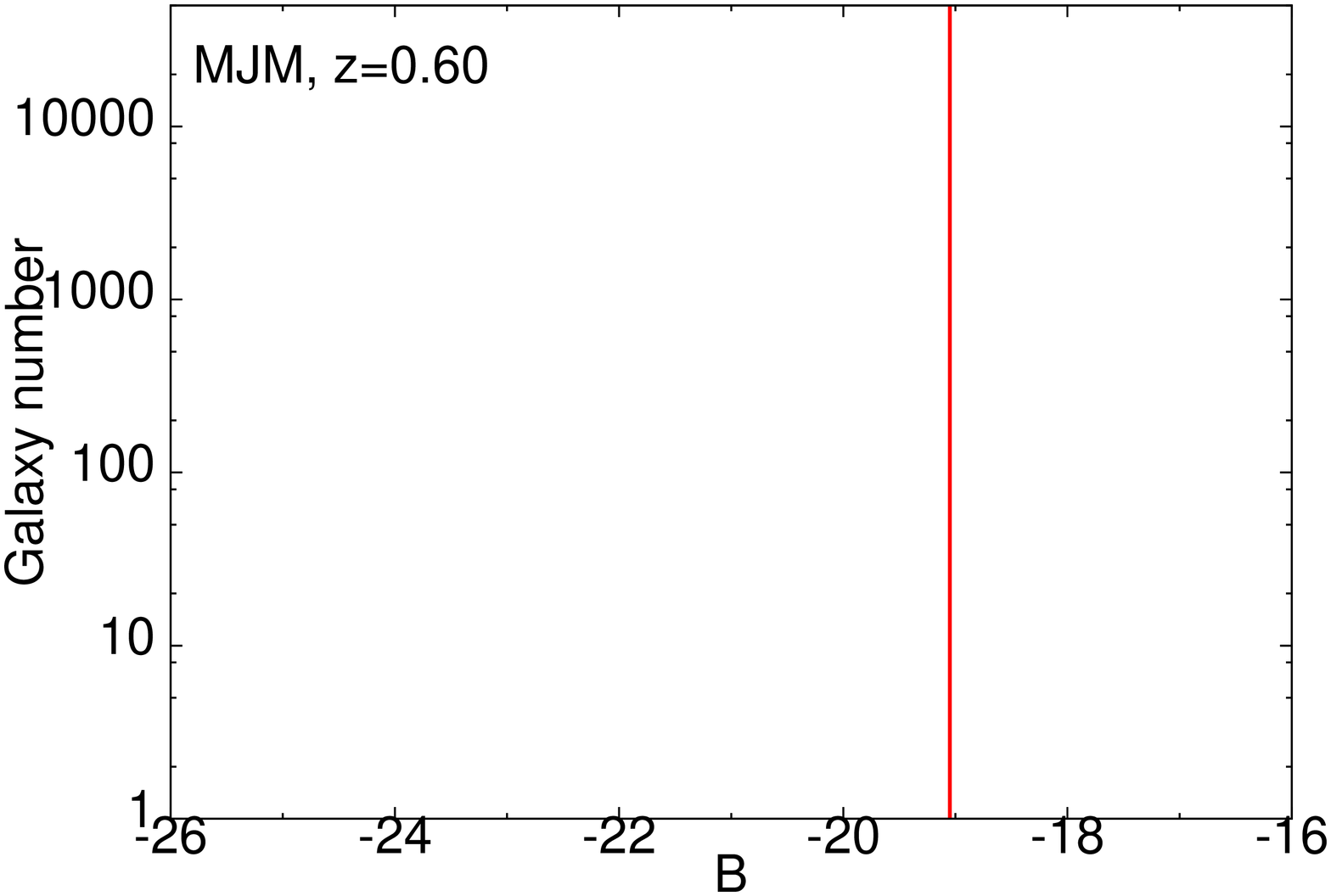}
\includegraphics[width=0.17\textwidth,clip,angle=270]{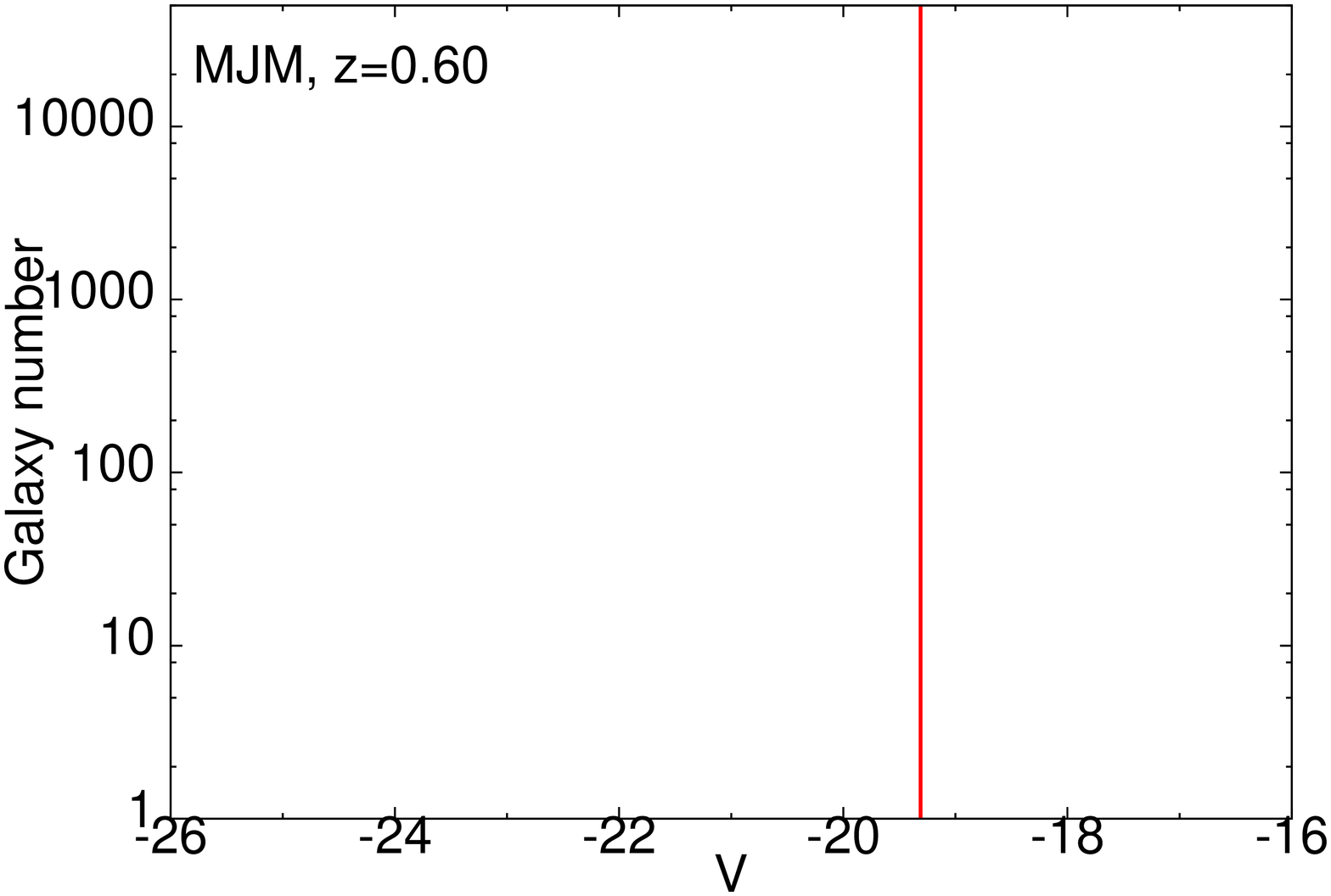}
\includegraphics[width=0.17\textwidth,clip,angle=270]{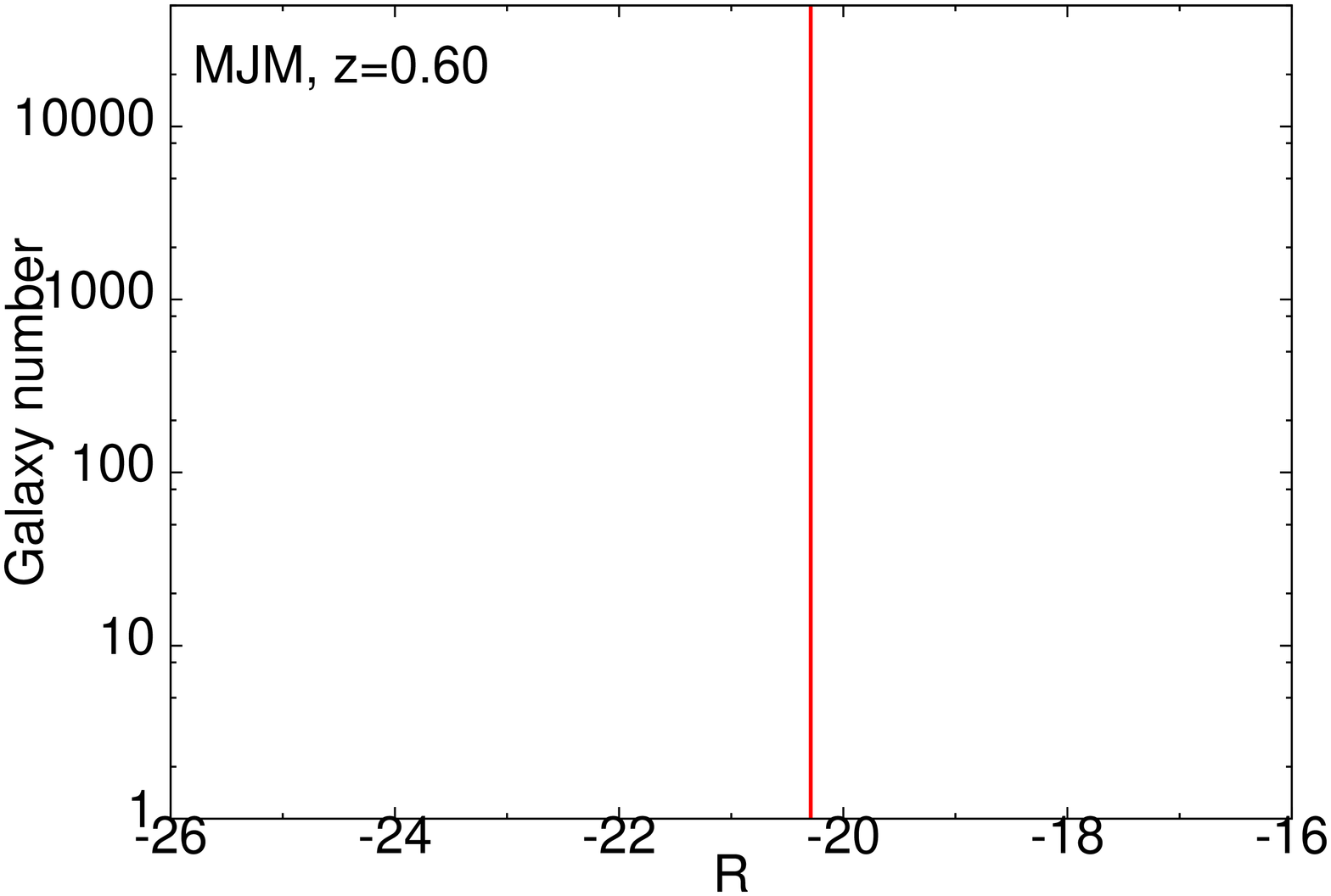}
\includegraphics[width=0.17\textwidth,clip,angle=270]{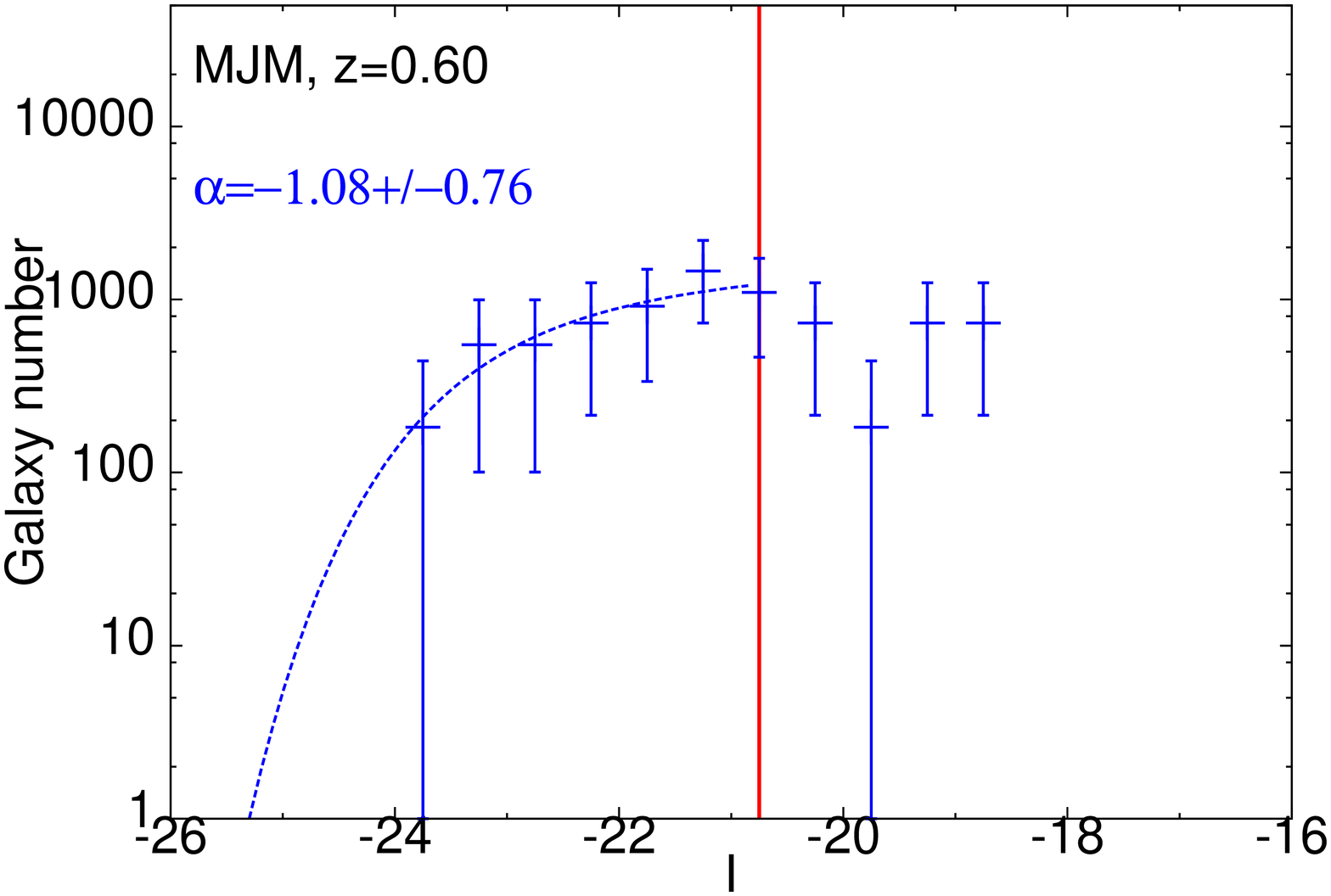} \\

\end{tabular}
\caption{Continued.}
\end{figure*}

\begin{figure*}[h!!]
\setcounter{figure}{0}
\begin{tabular}{cccc}

\includegraphics[width=0.17\textwidth,clip,angle=270]{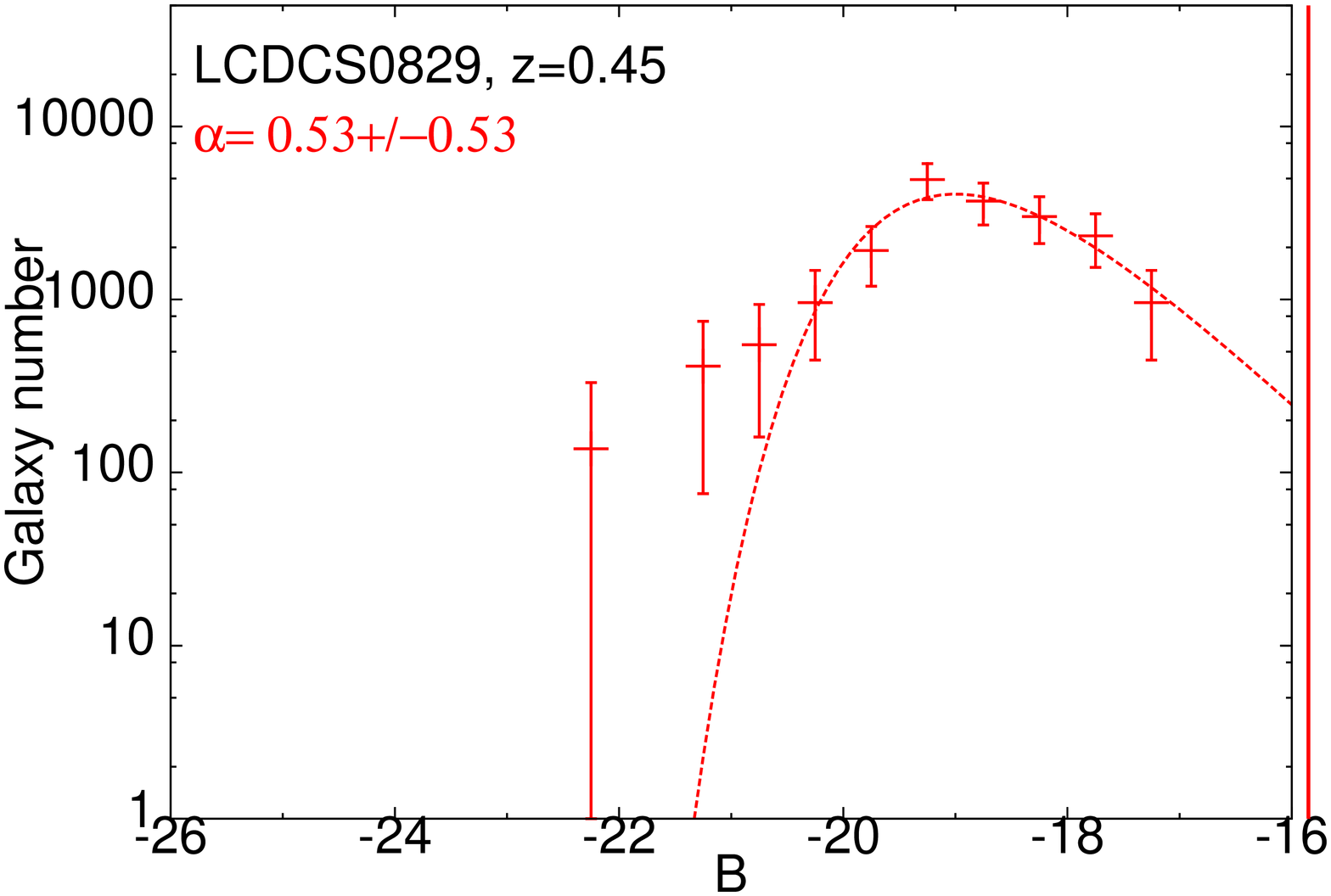}
\includegraphics[width=0.17\textwidth,clip,angle=270]{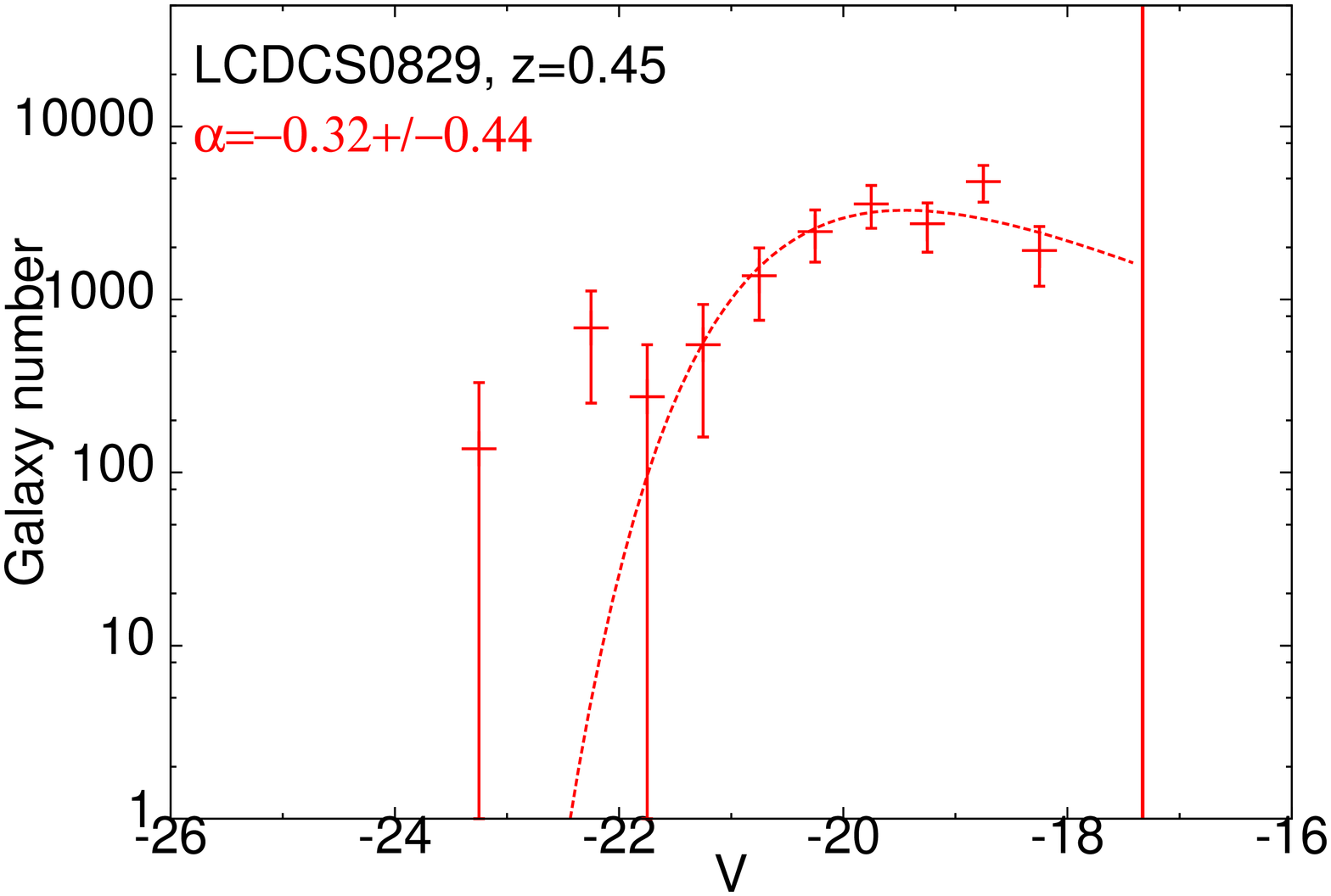}
\includegraphics[width=0.17\textwidth,clip,angle=270]{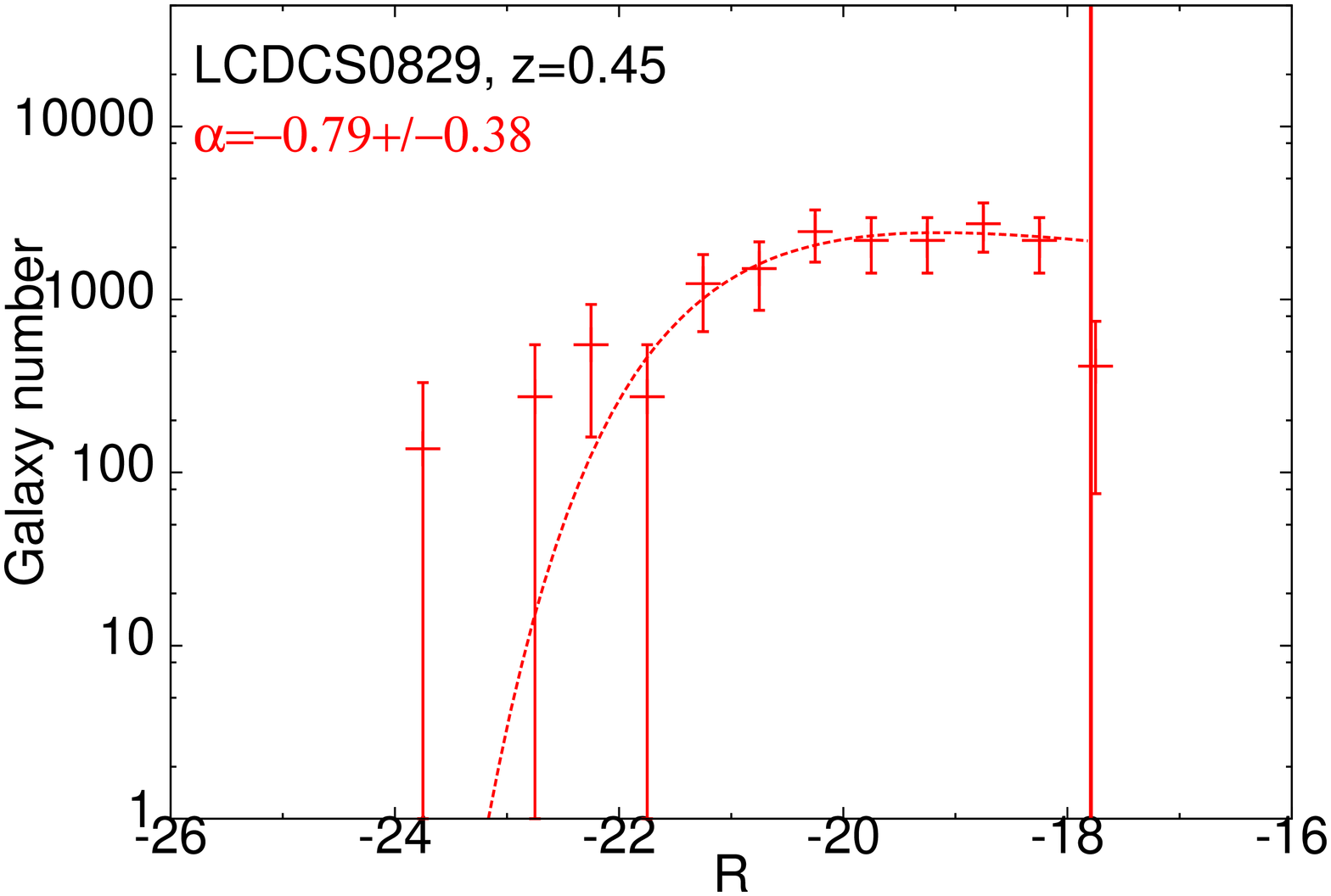}
\includegraphics[width=0.17\textwidth,clip,angle=270]{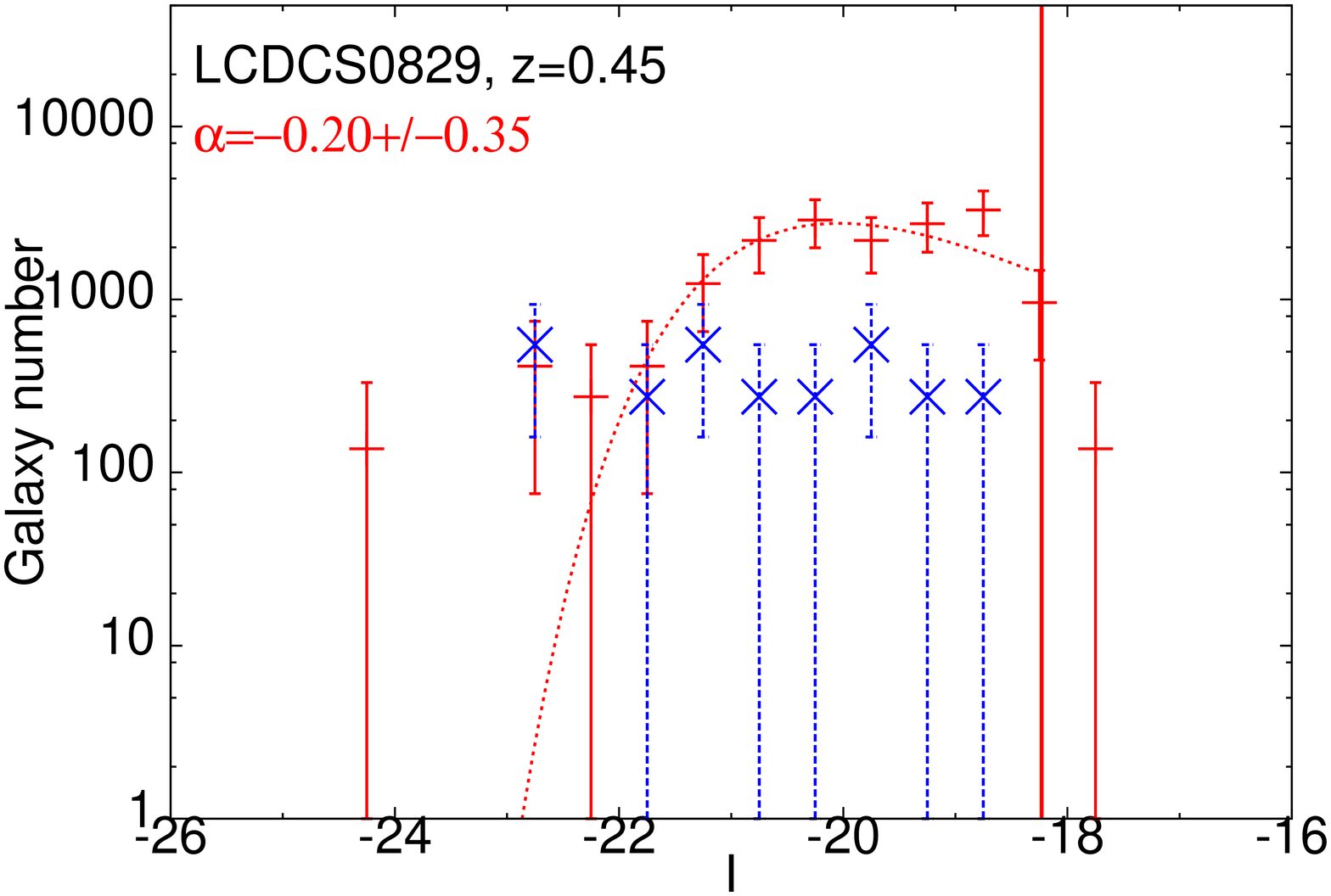} \\
\includegraphics[width=0.17\textwidth,clip,angle=270]{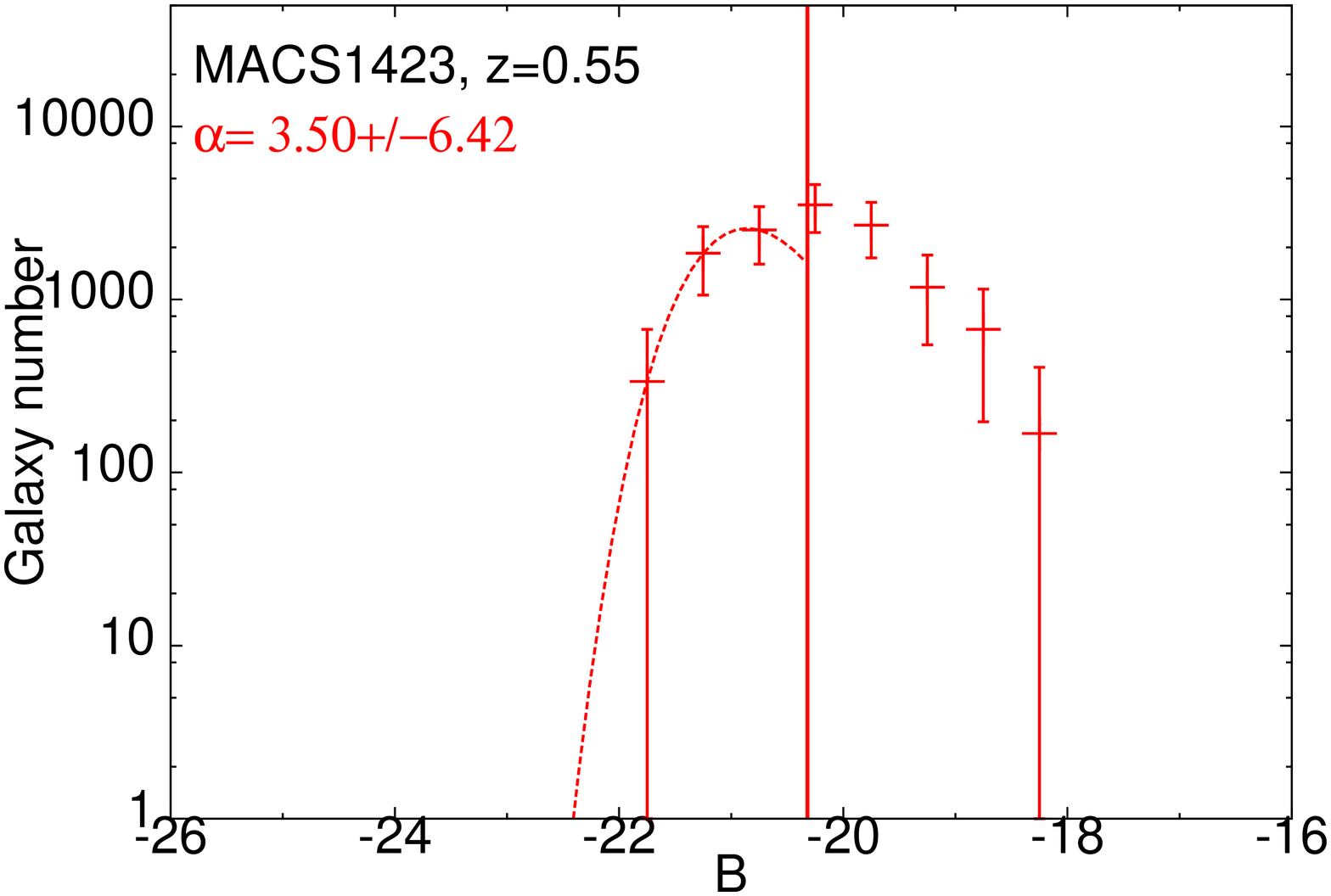}
\includegraphics[width=0.17\textwidth,clip,angle=270]{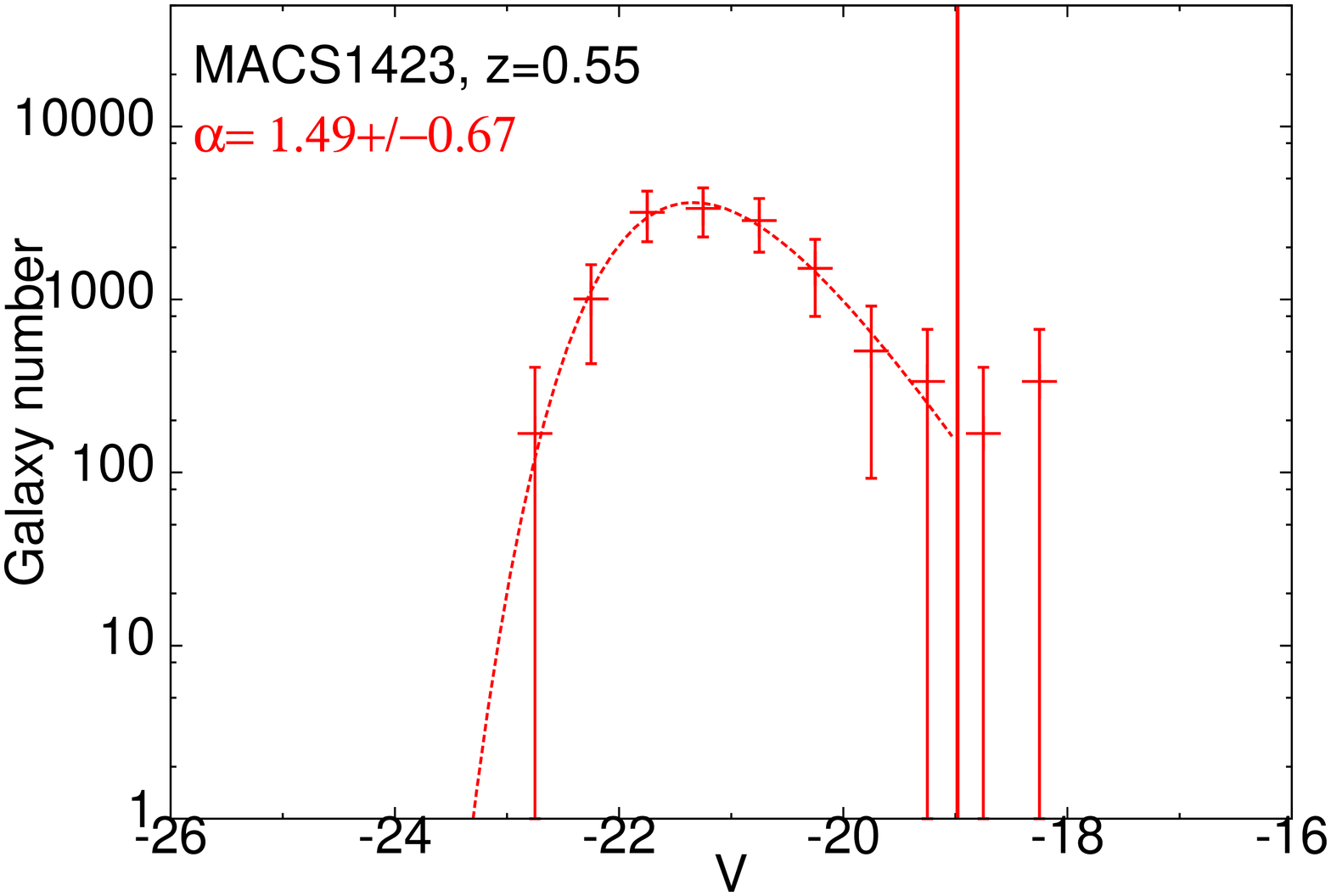}
\includegraphics[width=0.17\textwidth,clip,angle=270]{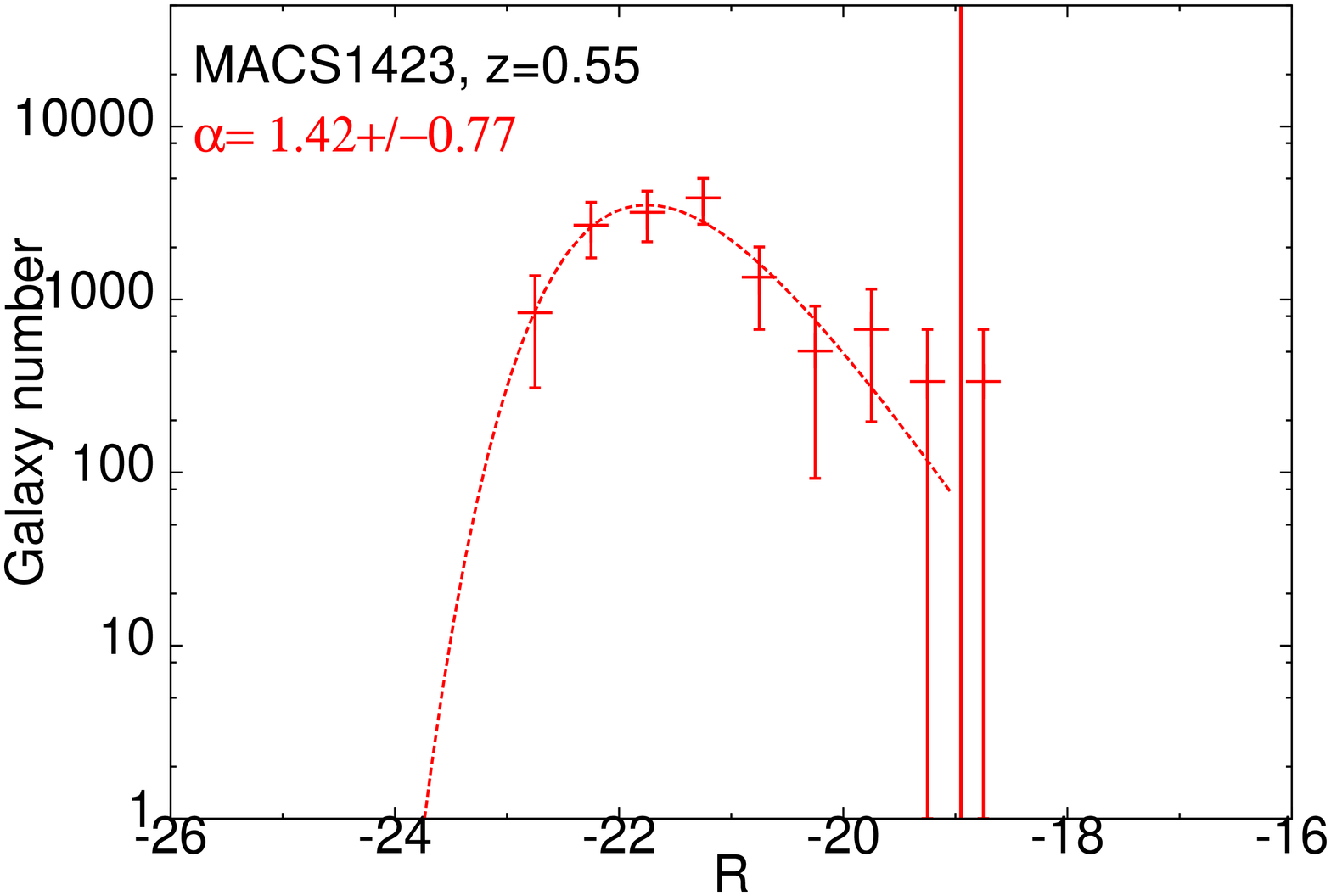}
\includegraphics[width=0.17\textwidth,clip,angle=270]{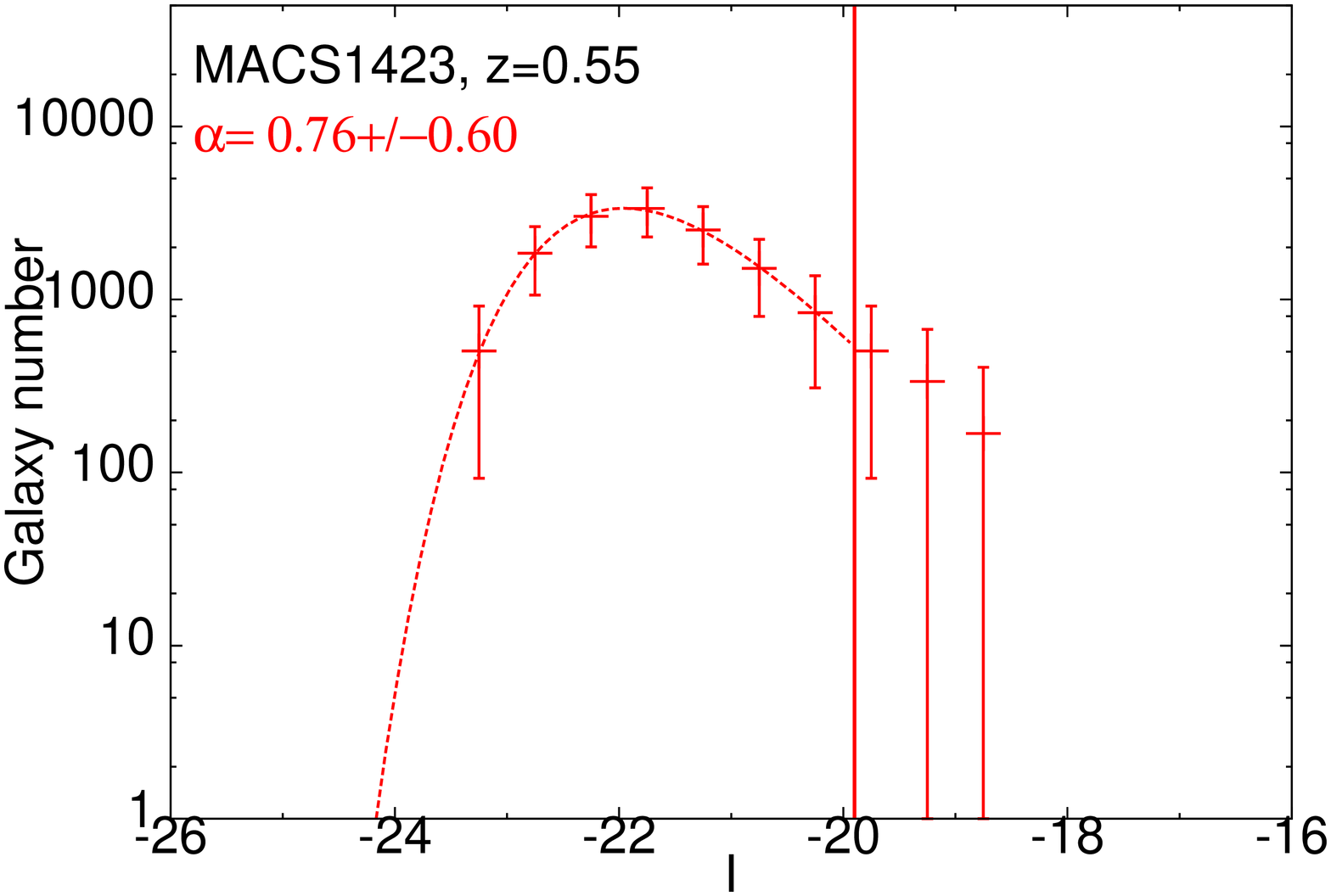} \\
\includegraphics[width=0.17\textwidth,clip,angle=270]{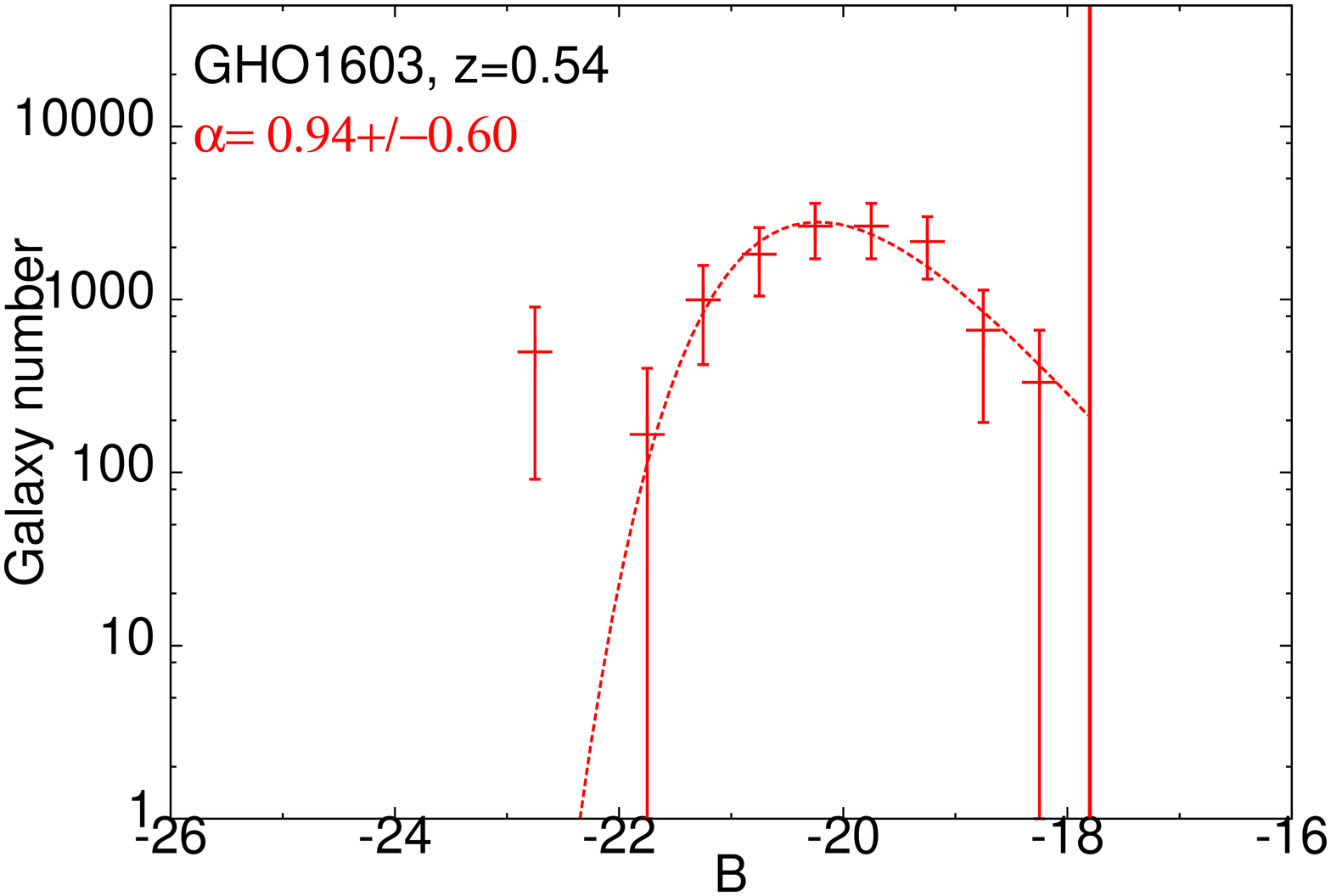}
\includegraphics[width=0.17\textwidth,clip,angle=270]{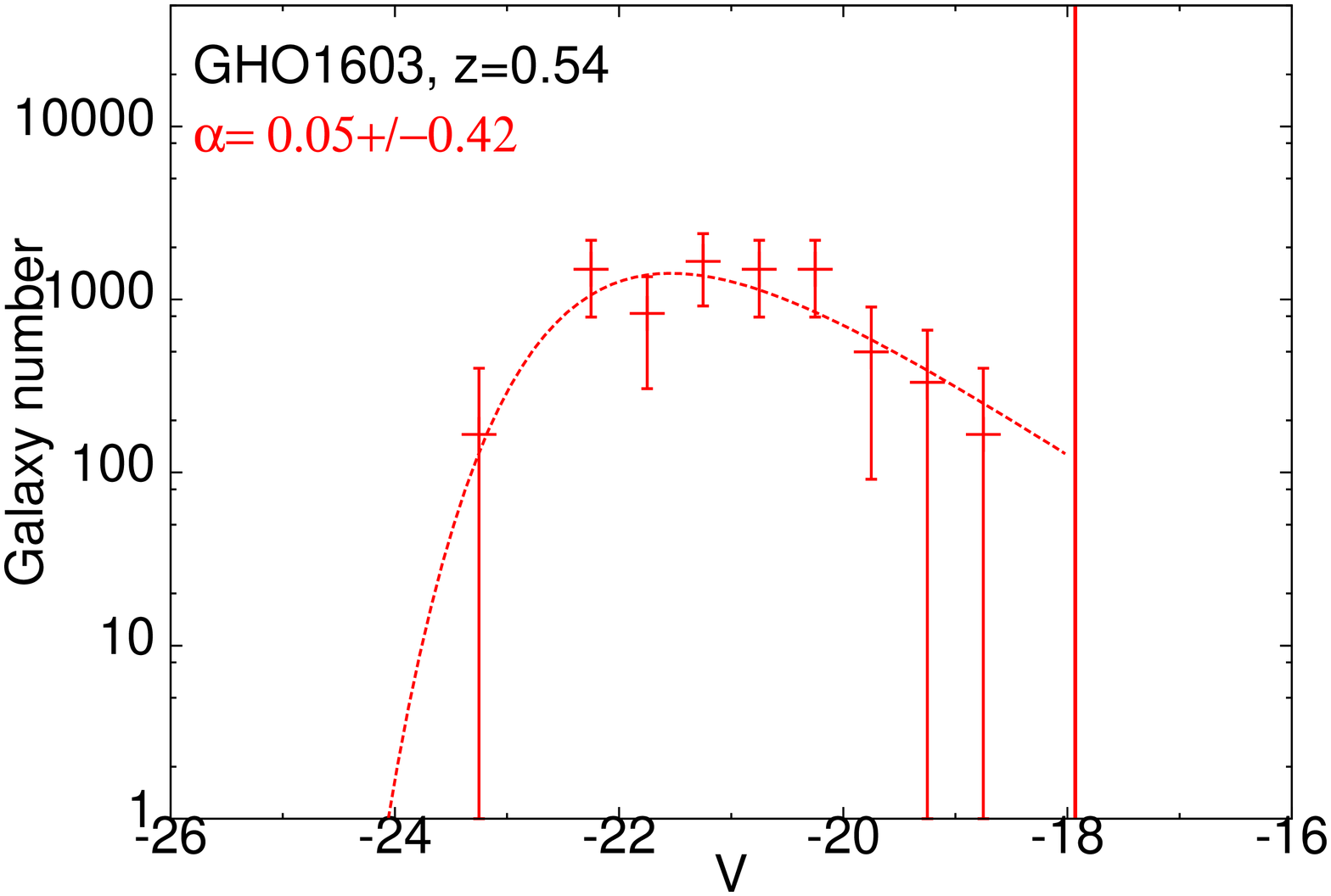}
\includegraphics[width=0.17\textwidth,clip,angle=270]{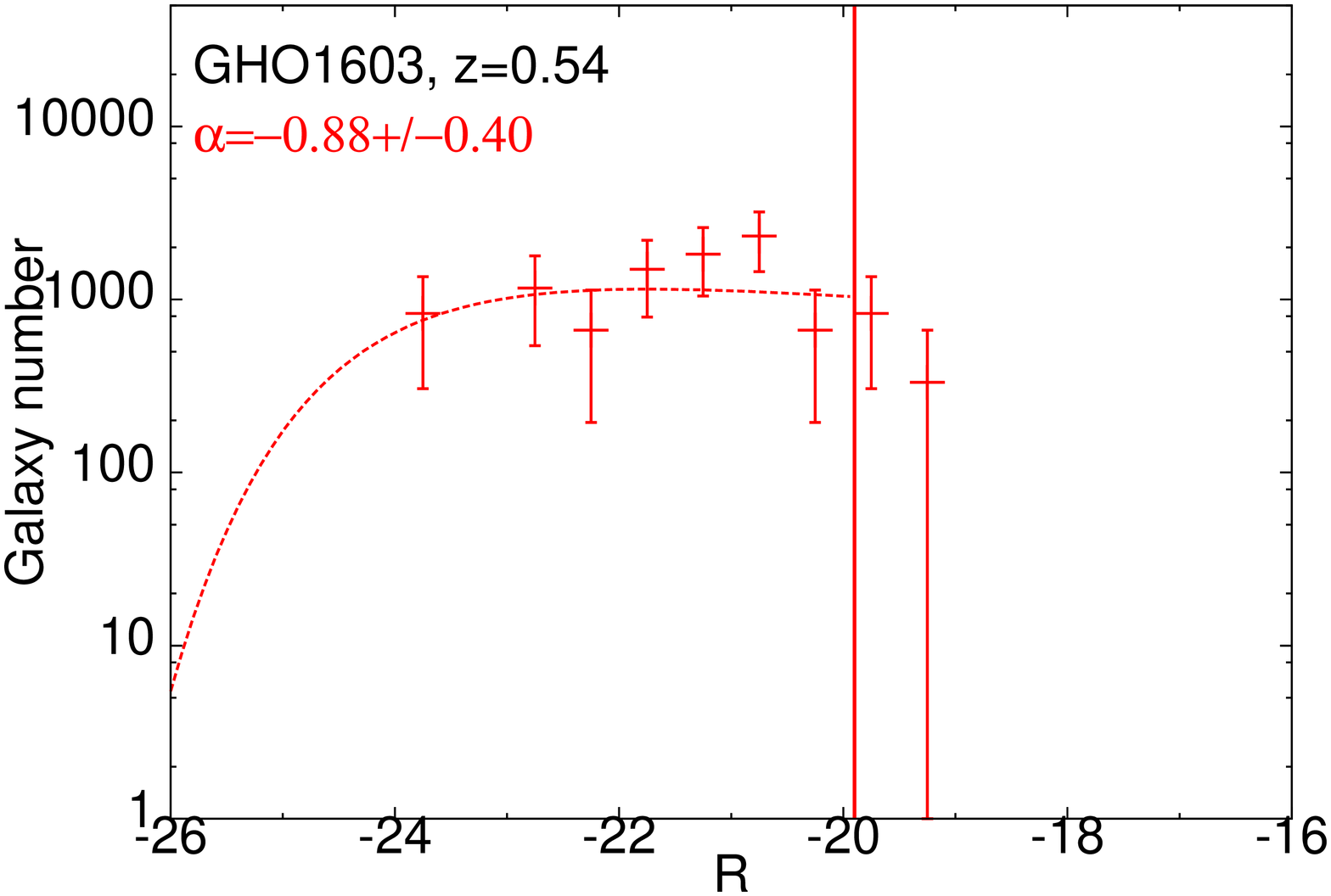}
\includegraphics[width=0.17\textwidth,clip,angle=270]{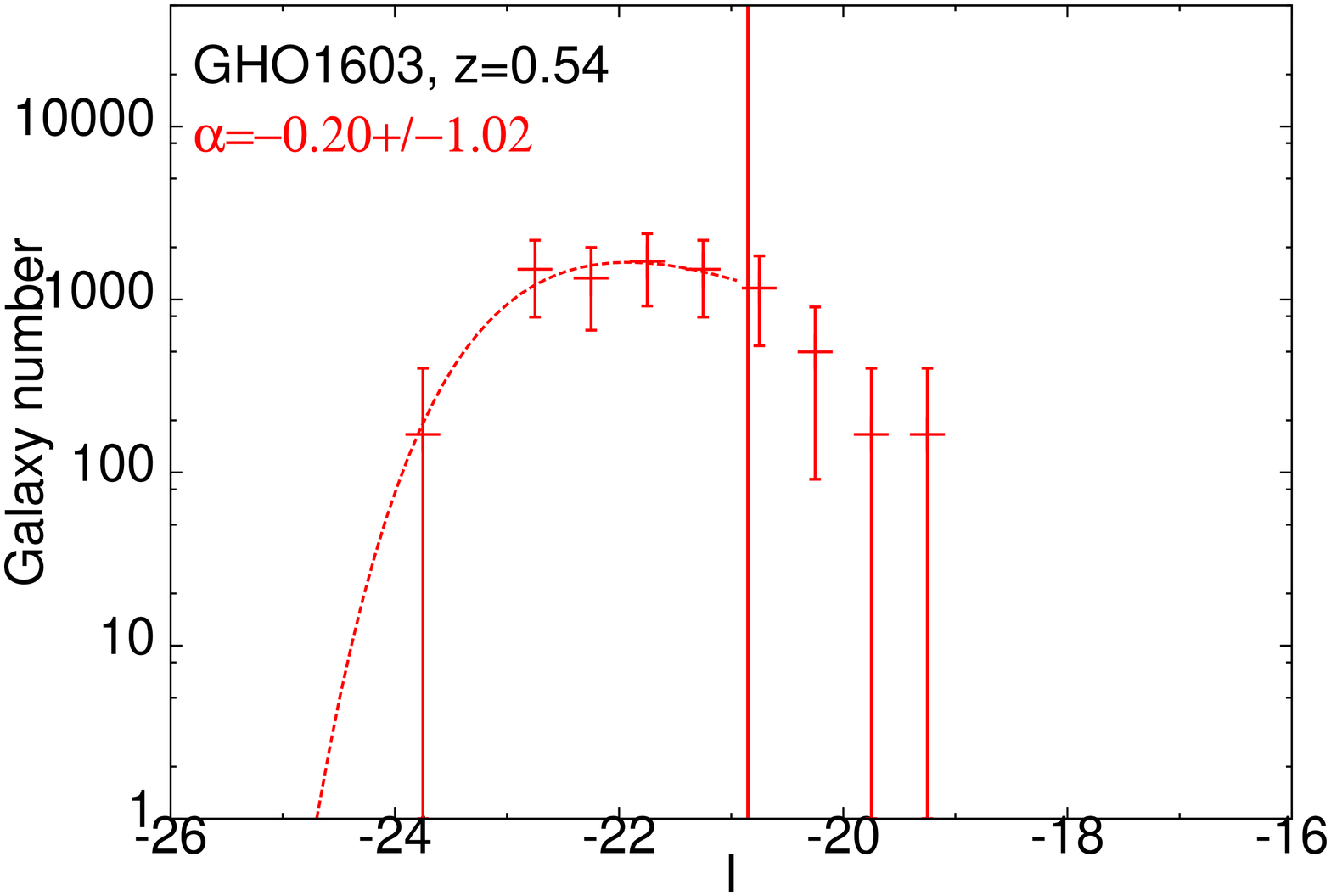} \\
\includegraphics[width=0.17\textwidth,clip,angle=270]{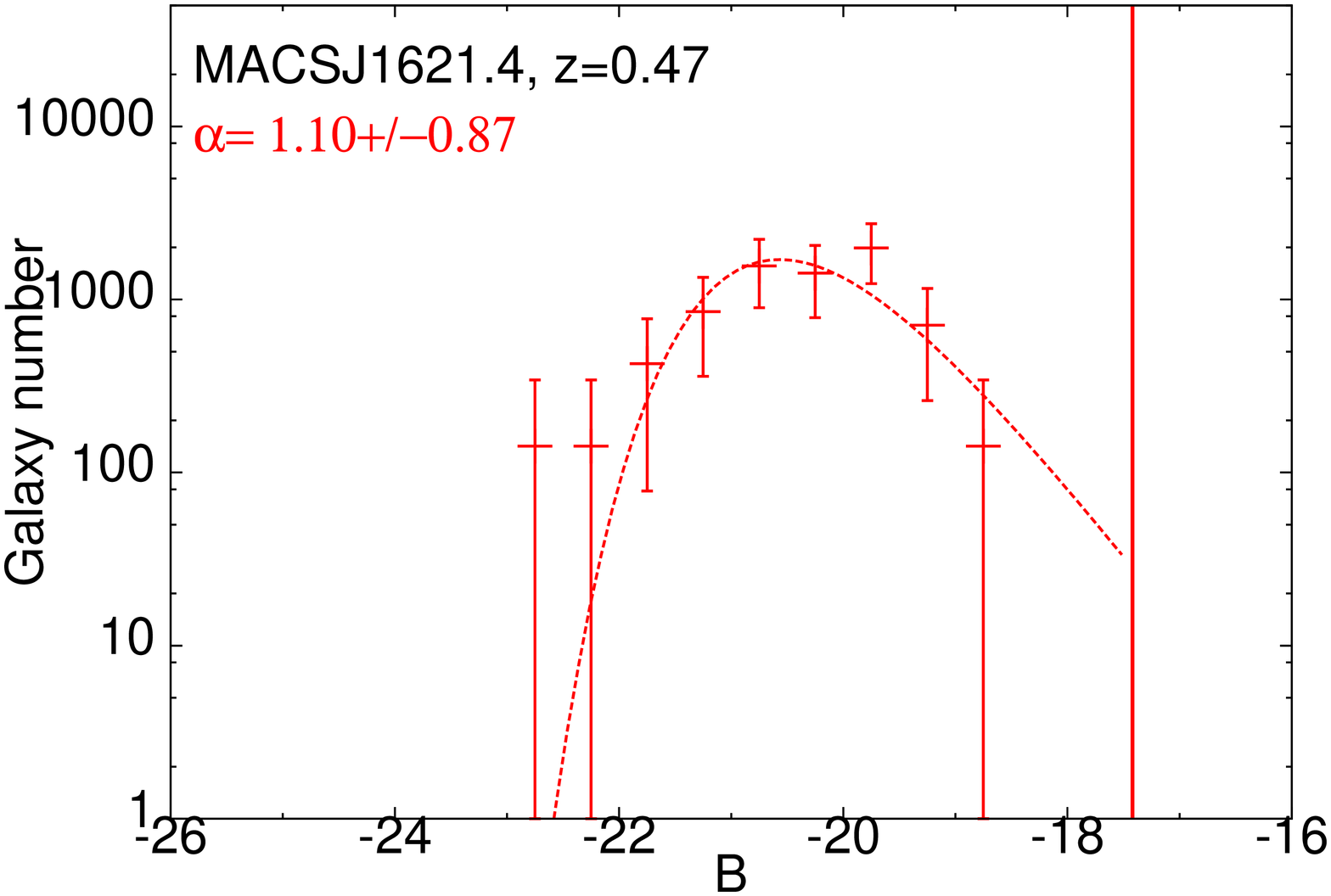}
\includegraphics[width=0.17\textwidth,clip,angle=270]{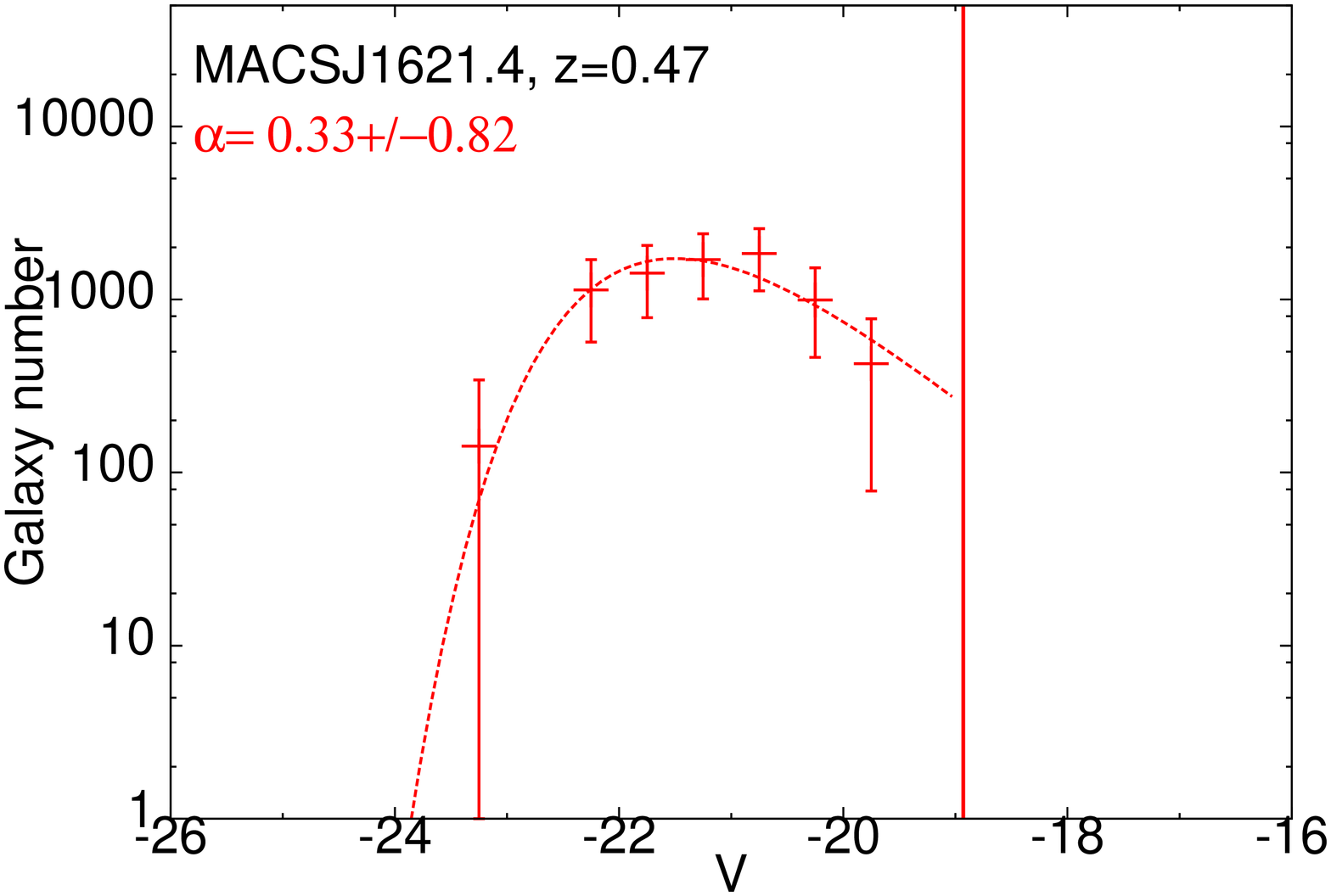}
\includegraphics[width=0.17\textwidth,clip,angle=270]{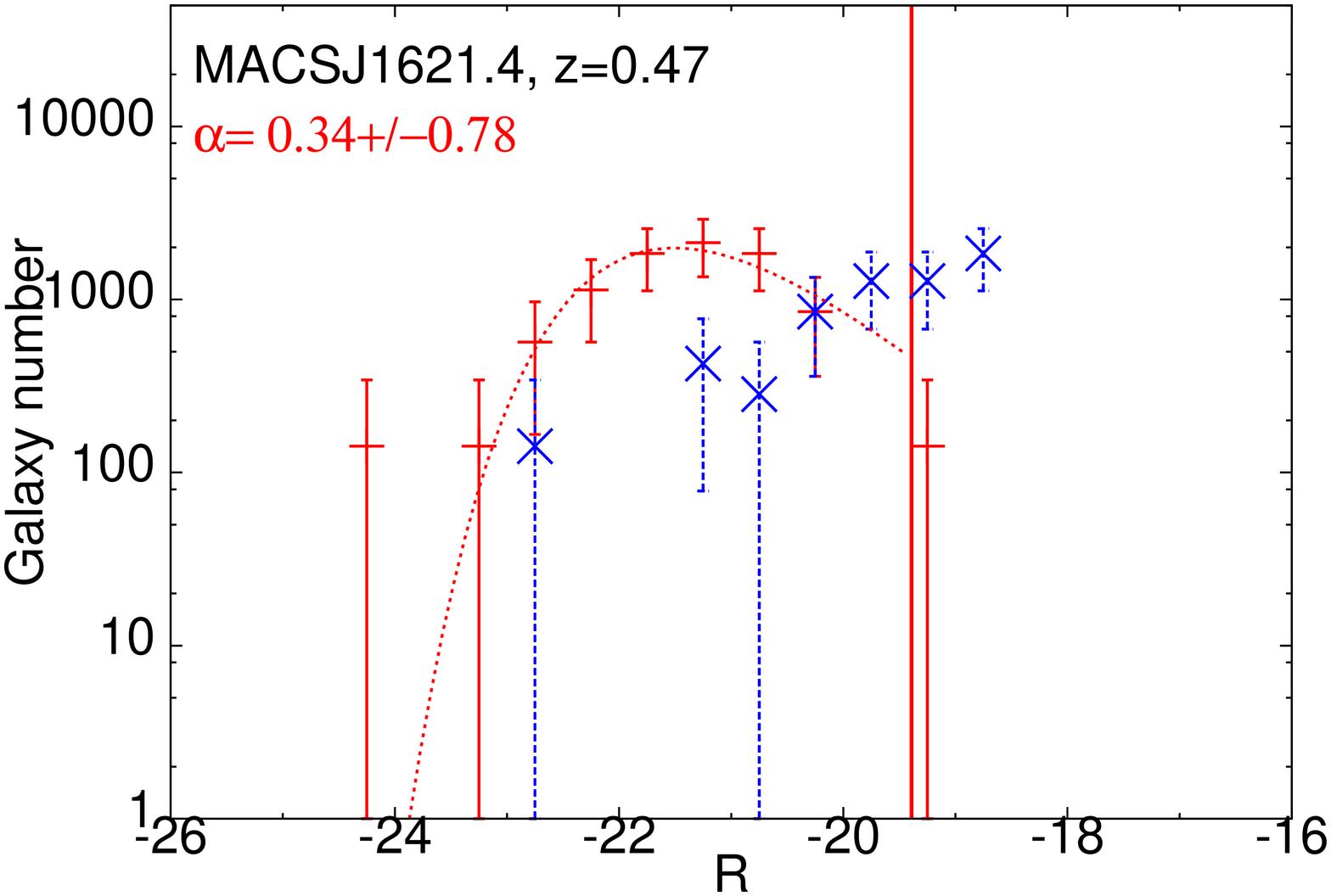}
\includegraphics[width=0.17\textwidth,clip,angle=270]{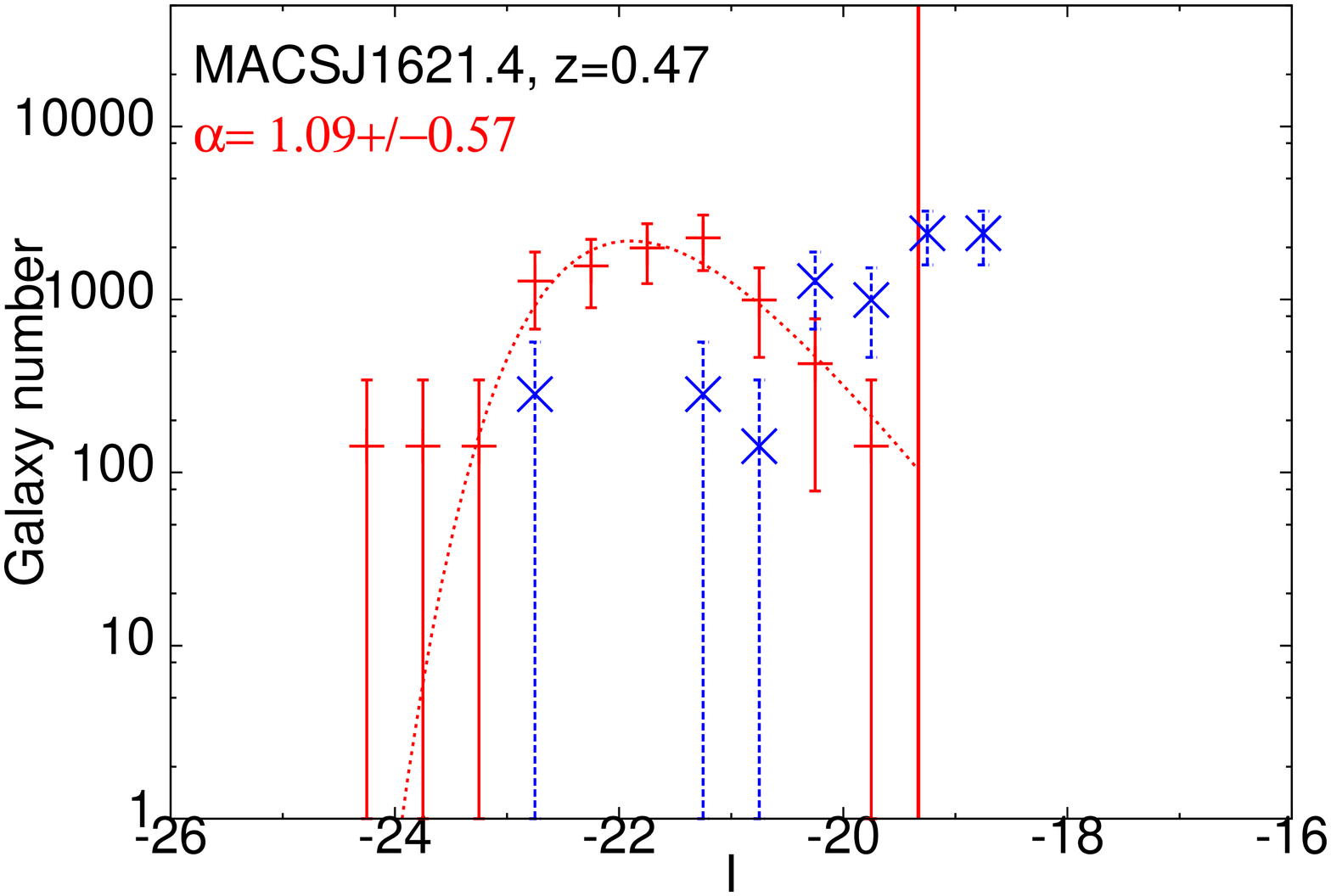} \\
\includegraphics[width=0.17\textwidth,clip,angle=270]{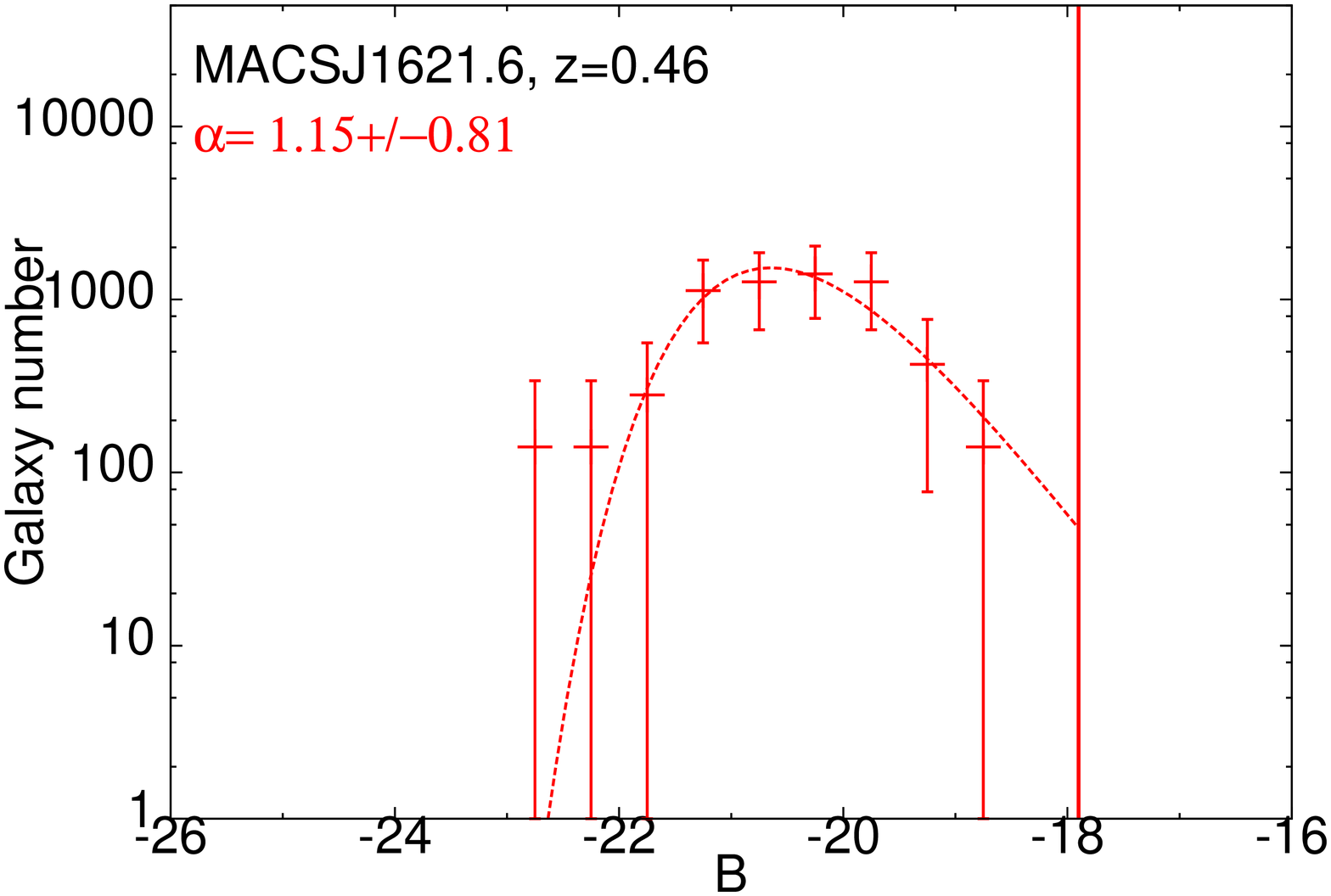}
\includegraphics[width=0.17\textwidth,clip,angle=270]{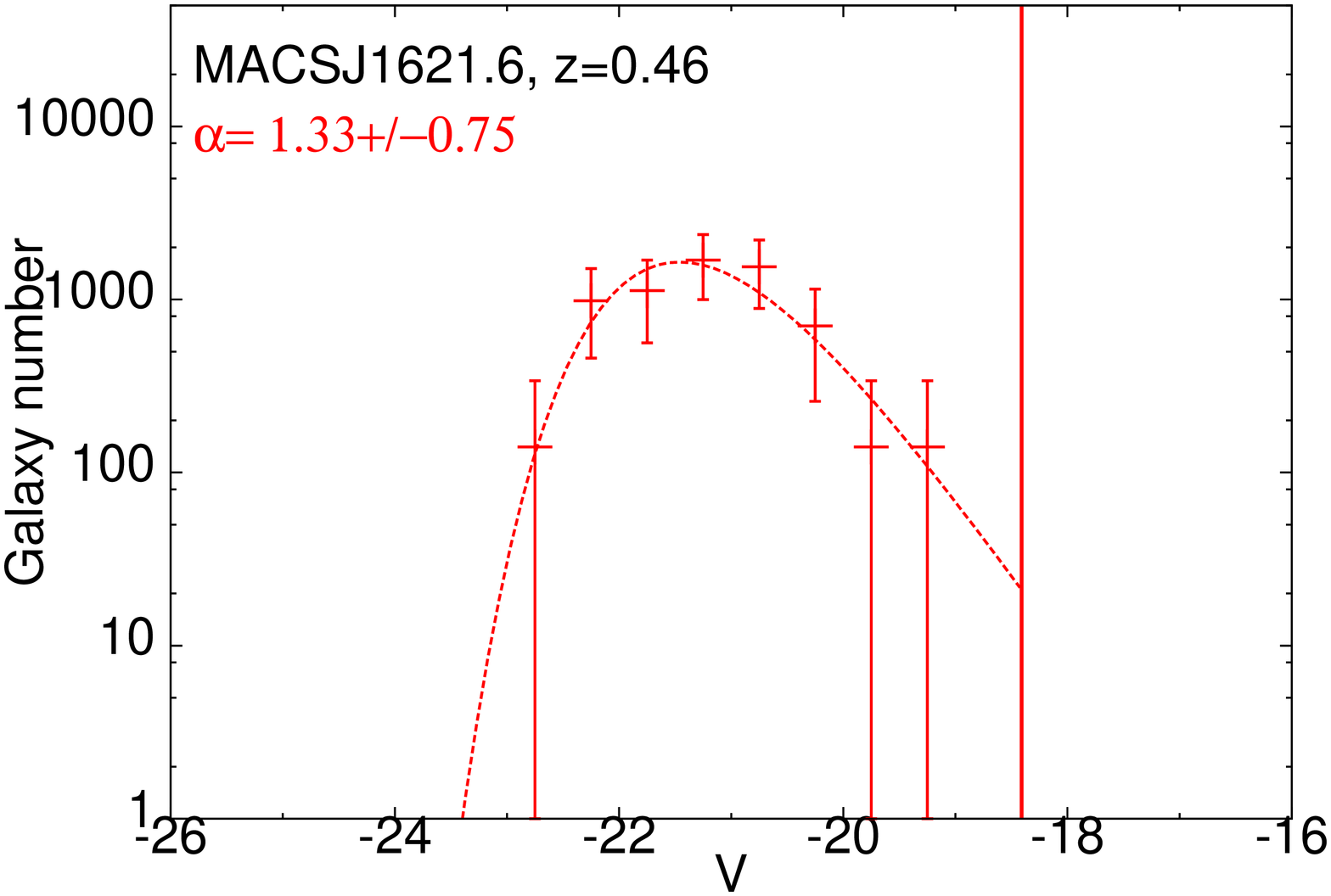}
\includegraphics[width=0.17\textwidth,clip,angle=270]{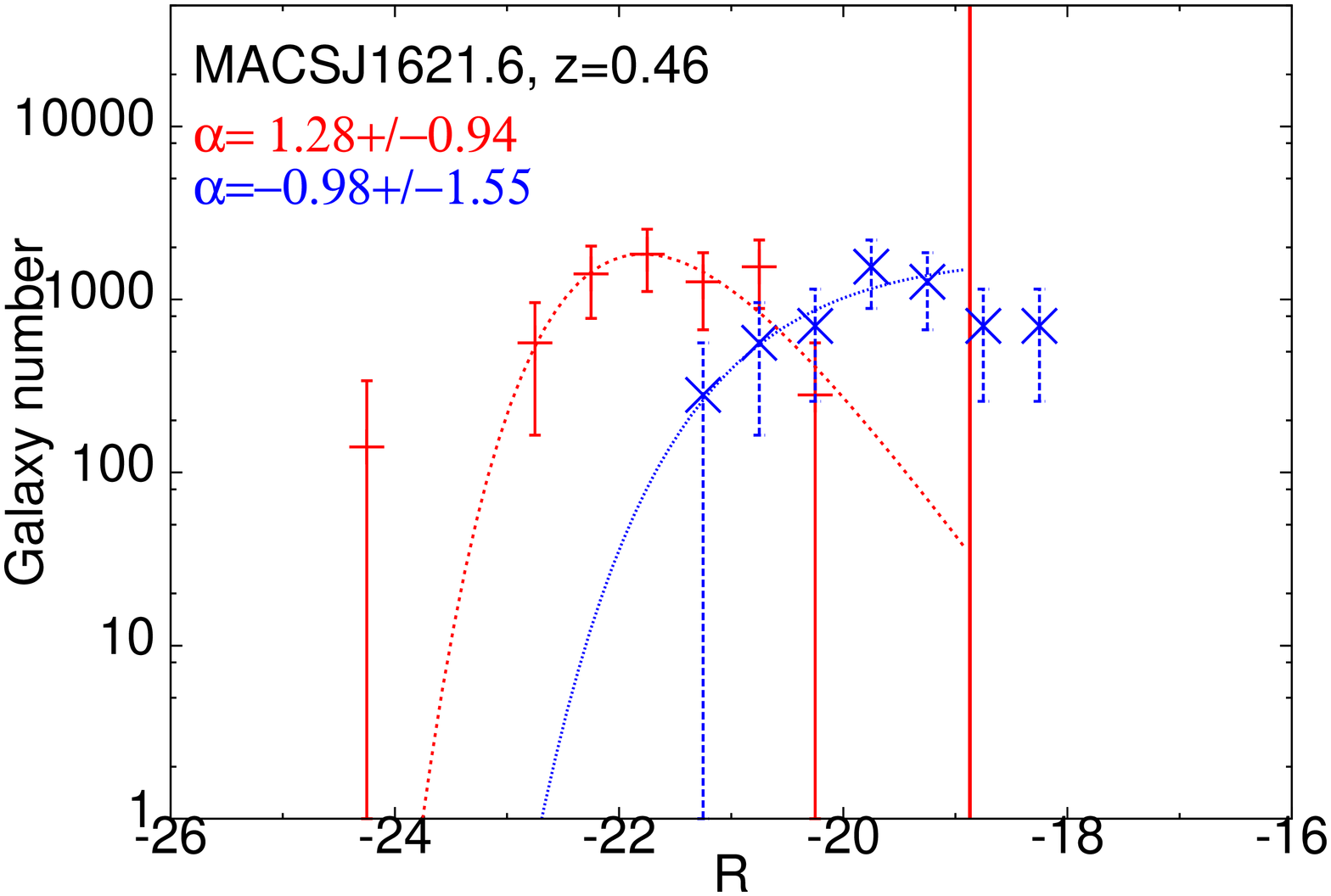}
\includegraphics[width=0.17\textwidth,clip,angle=270]{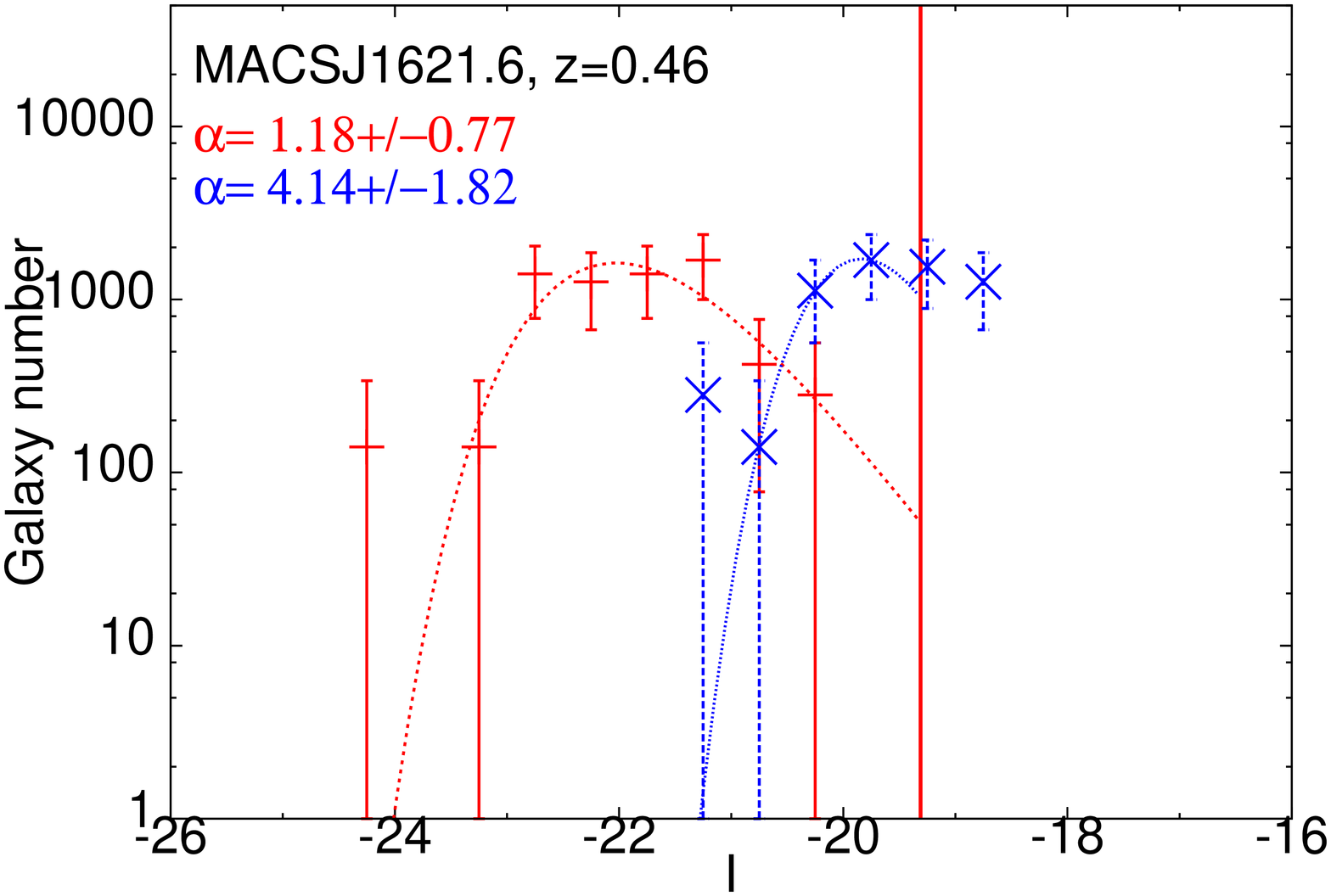} \\
\includegraphics[width=0.17\textwidth,clip,angle=270]{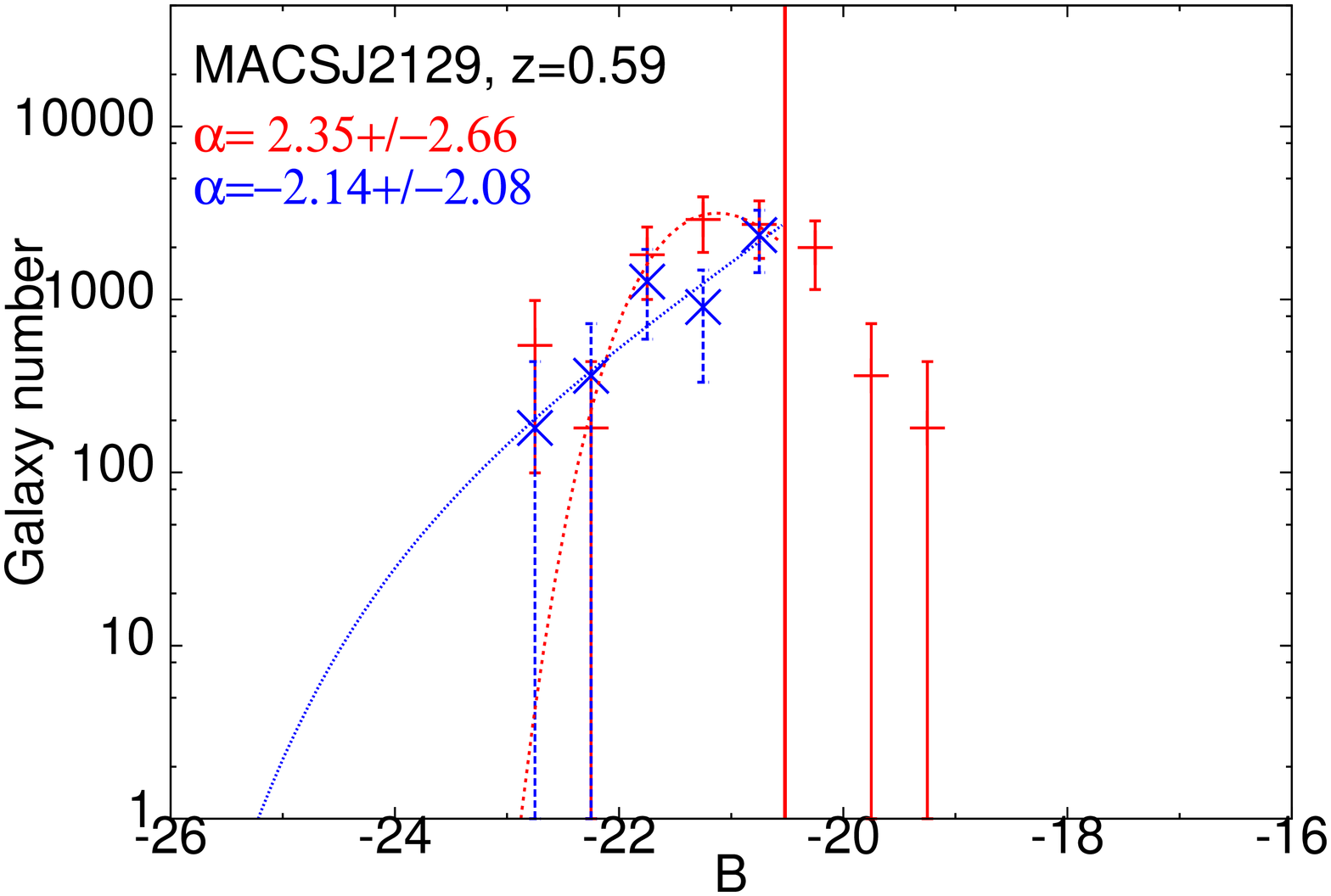}
\includegraphics[width=0.17\textwidth,clip,angle=270]{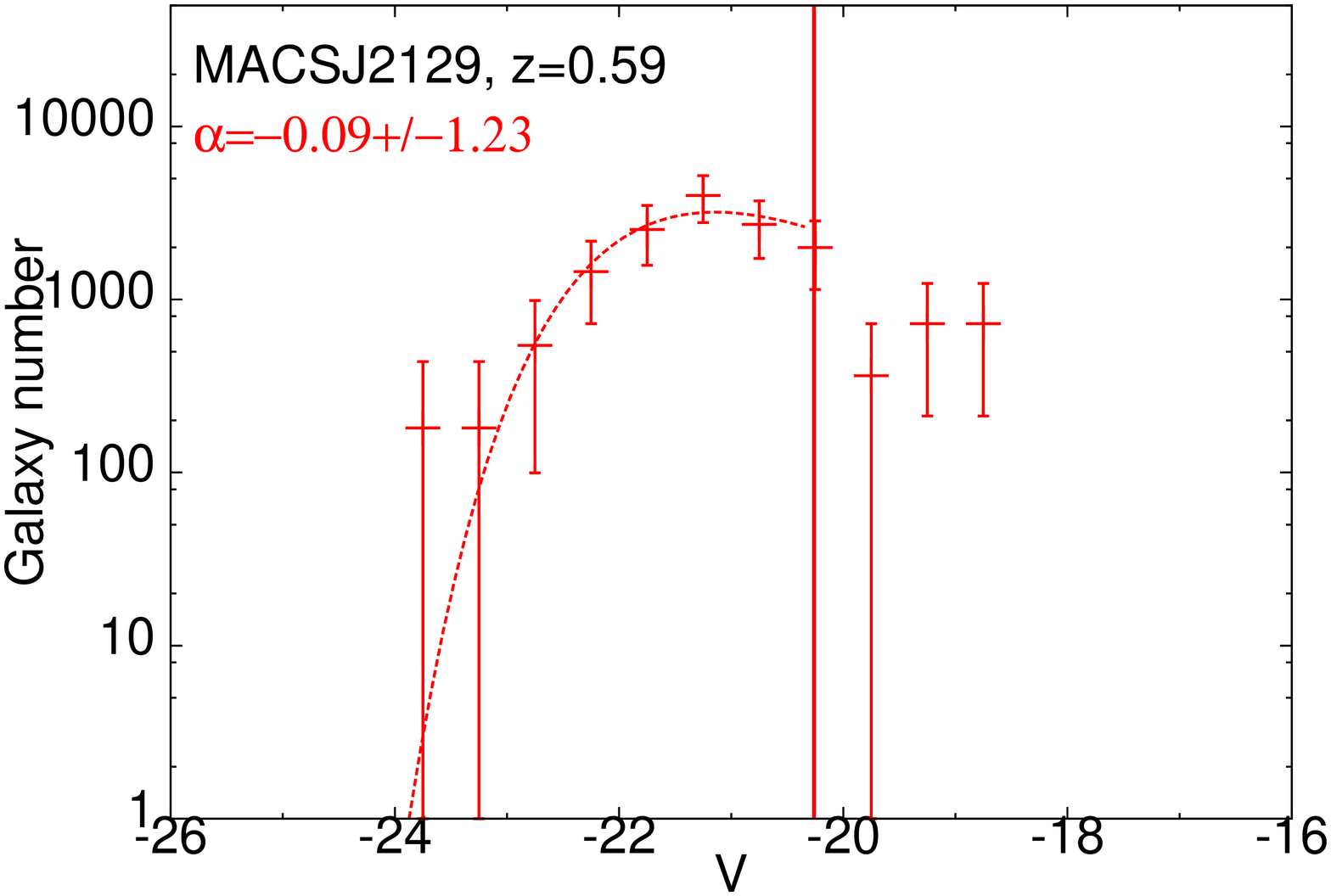}
\includegraphics[width=0.17\textwidth,clip,angle=270]{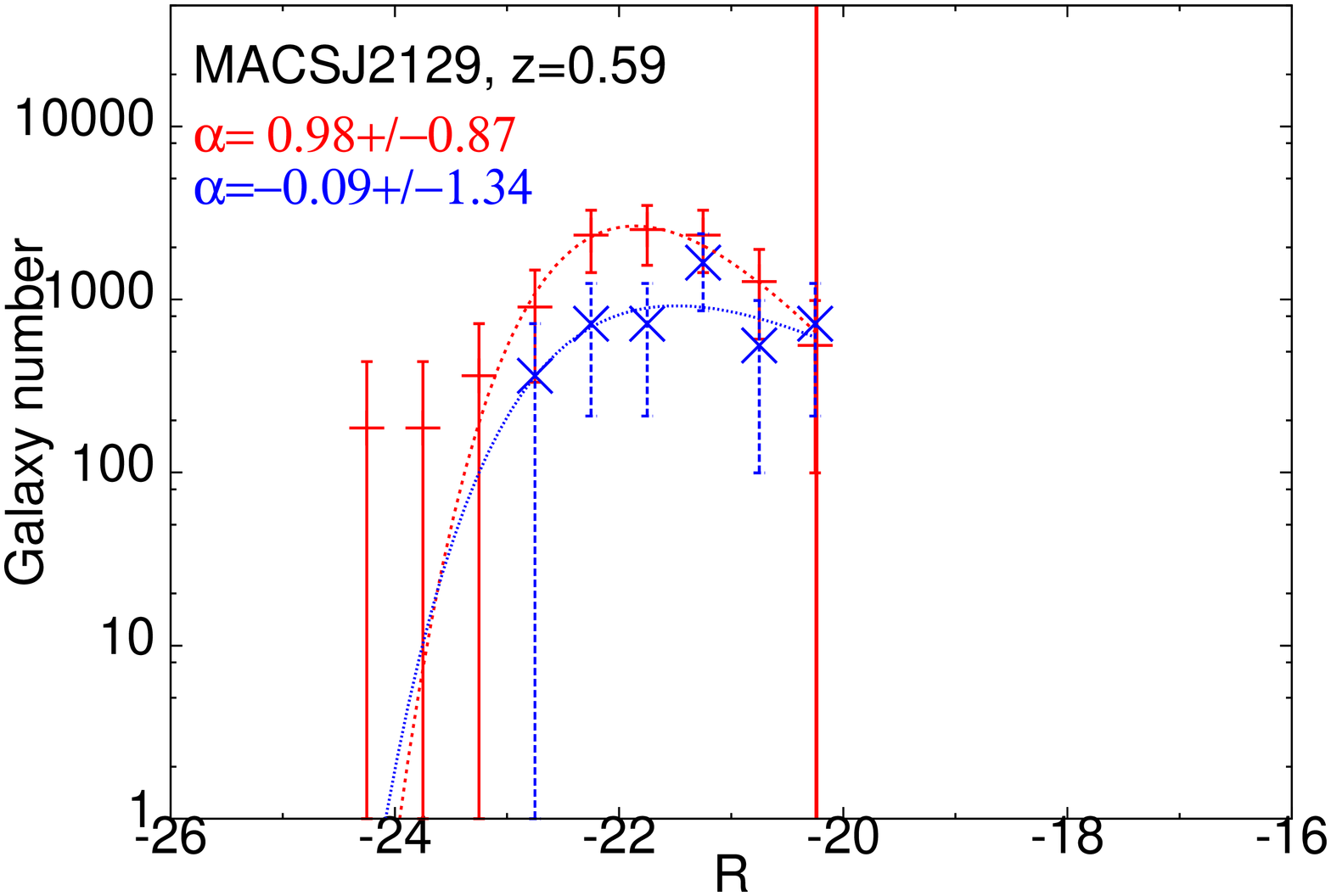}
\includegraphics[width=0.17\textwidth,clip,angle=270]{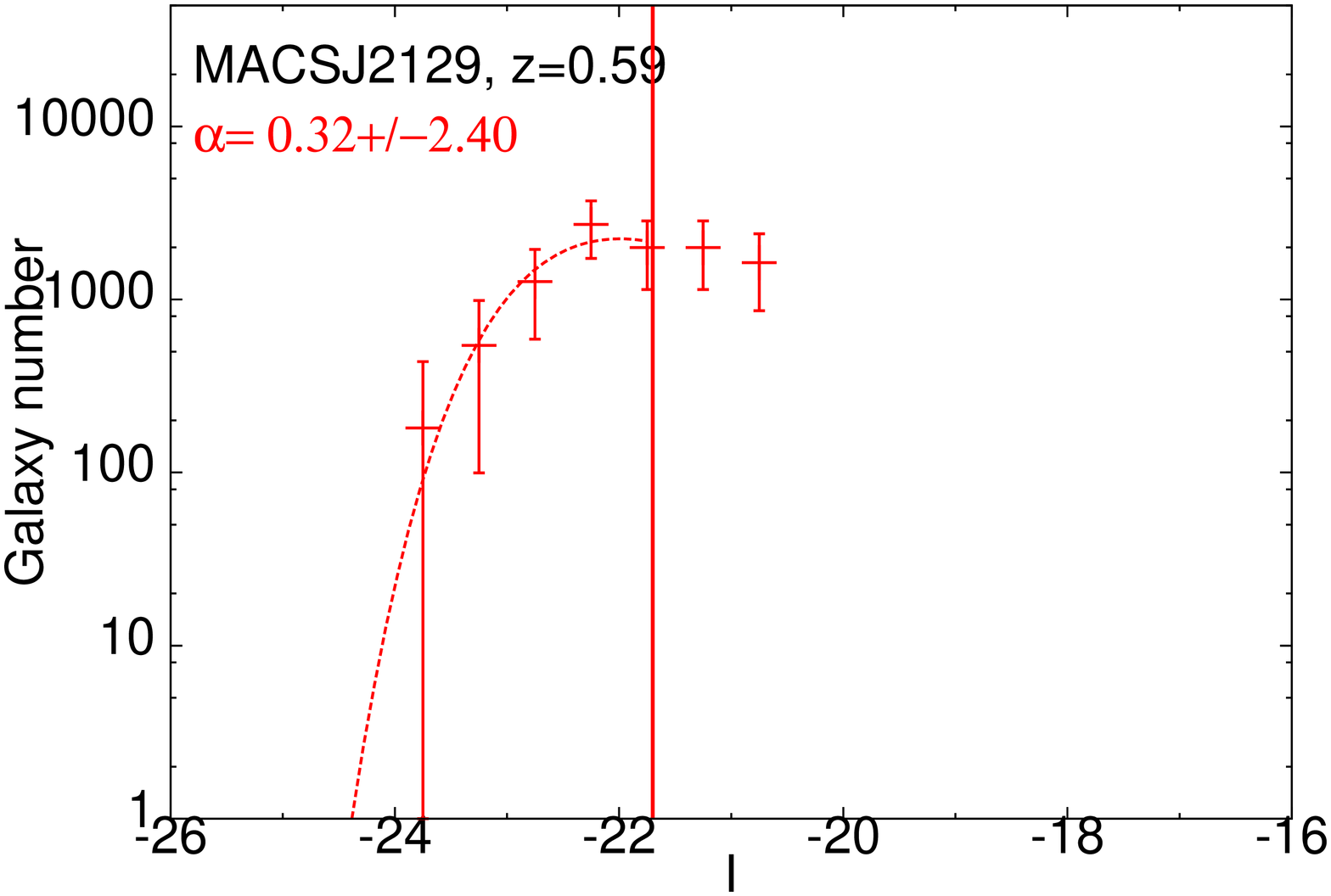} \\

\end{tabular}
\caption{Continued}
\end{figure*}

\begin{landscape}
\begin{table*}
\centering
\setcounter{table}{0}
\caption{Parameters of the best Schechter function fit for every
  cluster normalized to 1 deg$^2$. Fits are done to the
  cluster 90\% completeness limit in the selected band. A '$-$'
  means the fit has not converged due to a too bright completeness
  limit or because the selected cluster population is poorer than 20. Clusters with too few galaxy counts in every band (see section
  ~\ref{sec:indiv} for details) are not displayed. As noted above,
  clusters for which the fit with a Schechter function does not
  converge are kept in the table when they are taken into account in
  the stacked GLFs. There can be large differences between our
  parameters and the literature for some clusters, as we do not
  adapt the completeness limit nor any step of our method to
  individual clusters. This choice is made not to bias the study of
  stacked GLFs.} 
\begin{tabular}{lcc|ccc|ccc}

\hline\hline 
&&& & red-sequence GLFs  && & blue GLFs& \\
\hline
&$z$&comp& $\alpha$ & M*  &$\phi$* ($deg^{-2}$)&          $\alpha$ & M*  &$\phi$* ($deg^{-2}$) \\
\hline
CL0016 & 0.55& & & & & & & \\
B & &  -18.3 & -0.97$\pm$    0.62 & -20.4$\pm$  0.8 &  6052$\pm$  5477 & - & - & - \\
V & &  -18.5 & -0.49$\pm$    0.34 & -20.9$\pm$  0.4 &  8154$\pm$  2893 & - & - & - \\
R & &  -18.5 & -0.54$\pm$    0.27 & -21.3$\pm$  0.4 &  8890$\pm$  3012 & - & - & - \\
I & &  -19.9 & -0.52$\pm$    0.57 & -21.8$\pm$  0.5 &  9801$\pm$  4369 & - & - & - \\
\hline
PDCS18 & 0.40& & & & & & & \\
B & & -17.5 & - & - & - & - & - & - \\
V & &  -17.4 & -0.94$\pm$    0.57 & -23.1$\pm$  3.3 &   848$\pm$  1401 & -0.51$\pm$    0.35 & -22.8$\pm$  2.7 &  2708$\pm$   976 \\
R & &  -18.4 &  1.44$\pm$    1.58 & -20.3$\pm$  0.6 &  1264$\pm$  1478 & -0.24$\pm$    1.32 & -19.8$\pm$  1.2 &  2487$\pm$  1518 \\
I & &  -18.3 &  2.49$\pm$    1.92 & -20.2$\pm$  0.6 &   474$\pm$  1020 & -0.46$\pm$    0.83 & -20.9$\pm$  1.0 &  2451$\pm$  2304 \\
\hline
XDCS & 0.41& & & & & & & \\
B & &  -18.6 &  1.70$\pm$    2.38 & -19.5$\pm$  1.1 &  1074$\pm$  2218 & - & - & - \\
V & &  -18.5 & -0.88$\pm$    0.59 & -23.1$\pm$  2.0 &  1184$\pm$  1611 & -0.87$\pm$    1.13 & -22.0$\pm$  2.6 &  1412$\pm$  3337 \\
R & &  -19.5 &  0.05$\pm$    2.58 & -21.5$\pm$  2.5 &  2165$\pm$  1106 & - & - & - \\
I & &  -18.9 &  0.64$\pm$    1.26 & -21.1$\pm$  0.8 &  1416$\pm$   643 & - & - & - \\
\hline
MACS0454 & 0.54& & & & & & & \\
B & & -17.8 & - & - & - & - & - & - \\
V & &  -19.4 & -0.52$\pm$    0.77 & -21.9$\pm$  1.3 &  2114$\pm$  1841 &  1.80$\pm$    3.53 & -19.0$\pm$  1.0 &  1364$\pm$  4734 \\
R & &  -19.4 &  1.73$\pm$    2.27 & -20.6$\pm$  0.8 &  1402$\pm$  2655 & -1.40$\pm$    2.12 & -22.0$\pm$  7.0 &   925$\pm$  6462 \\
I & &  -20.3 & -0.08$\pm$    1.56 & -21.9$\pm$  1.4 &  3687$\pm$  1988 & -0.15$\pm$    1.93 & -21.1$\pm$  1.3 &  3700$\pm$  1604 \\
\hline
RXJ0848 & 0.54& & & & & & & \\
B & &  -19.3 & -2.18$\pm$    0.90 & -22.9$\pm$  6.5 &   173$\pm$  1607 & - & - & - \\
V & & -19.0 & - & - & - & - & - & - \\
R & & -20.9 & - & - & - & - & - & - \\
I & & -21.4 & - & - & - & - & - & - \\
\hline
A851 & 0.41& & & & & & & \\
B & & -19.1 & - & - & - & - & - & - \\
V & & -19.5 & - & - & - & - & - & - \\
R & & -18.9 & - & - & - & - & - & - \\
I & &  -18.9 & -0.98$\pm$    0.70 & -23.4$\pm$  6.2 &   691$\pm$  1625 &  1.88$\pm$    1.58 & -19.2$\pm$  0.5 &  2047$\pm$  3034 \\
\hline
LCDCS0130 & 0.70& & & & & & & \\
B & & -18.5 & - & - & - & - & - & - \\
V & & -18.0 & - & - & - & - & - & - \\
R & & -19.5 & - & - & - & - & - & - \\
I & & -19.5 & - & - & - & - & - & - \\
\hline
MS1054 & 0.82& & & & & & & \\
B & &  -19.3 &  1.97$\pm$    0.53 & -20.0$\pm$  0.2 & 12219$\pm$  6210 & - & - & - \\
V & &  -20.4 & -0.13$\pm$    0.74 & -22.3$\pm$  0.9 &  8736$\pm$  3317 & - & - & - \\
R & &  -20.9 &  1.07$\pm$    0.98 & -21.7$\pm$  0.5 &  7890$\pm$  4505 & - & - & - \\
I & & -22.8 & - & - & - & - & - & - \\
\hline
\hline
\end{tabular}
\label{tab:glf_indiv}
\end{table*}
\end{landscape}

\begin{landscape}
\begin{table*}[ht!]
\centering
\setcounter{table}{0}
  \caption{Continued.}
\begin{tabular}{lcc|ccc|ccc}
\hline
\hline
&&& & red-sequence GLFs  && & blue GLFs& \\
\hline
&$z$&comp& $\alpha$ & M*  &$\phi$* ($deg^{-2}$)&          $\alpha$ & M*  &$\phi$* ($deg^{-2}$) \\
\hline
RXCJ1206 & 0.44& & & & & & & \\
B & &  -18.8 & -1.53$\pm$    0.36 & -22.6$\pm$  1.9 &  1108$\pm$  2244 & - & - & - \\
V & &  -18.8 & -0.68$\pm$    0.20 & -22.0$\pm$  0.5 &  4719$\pm$  1692 & - & - & - \\
R & &  -19.7 & -0.63$\pm$    0.27 & -22.3$\pm$  0.4 &  4830$\pm$  1798 & - & - & - \\
I & &  -20.1 & -1.35$\pm$    0.32 & -23.3$\pm$  0.9 &  1355$\pm$  1463 & - & - & - \\
\hline
LCDCS0504 & 0.79& & & & & & & \\
B & &  -19.0 & - & - & - & -0.29$\pm$    0.44 & -21.7$\pm$  0.6 & 13856$\pm$  4540 \\
V & &  -19.7 &  0.26$\pm$    0.89 & -22.1$\pm$  0.8 &  2945$\pm$   928 & -0.66$\pm$    0.40 & -21.9$\pm$  0.6 &  6768$\pm$  3392 \\
R & &  -20.7 &  2.90$\pm$    3.22 & -21.0$\pm$  1.0 &   427$\pm$  1710 & -0.64$\pm$    0.55 & -22.4$\pm$  0.7 &  6929$\pm$  3919 \\
I & &  -20.6 &  0.96$\pm$    1.47 & -22.3$\pm$  0.7 &  3216$\pm$  2523 & -0.86$\pm$    0.64 & -22.9$\pm$  1.2 &  5148$\pm$  5522 \\
\hline
BMW HRI 2265 & 0.89& & & & & & & \\
B & &  -19.3 & - & - & - & -2.49$\pm$    0.51 & -21.4$\pm$  0.4 &   760$\pm$   280 \\
V & & -20.4 & - & - & - & - & - & - \\
R & & -20.9 & - & - & - & - & - & - \\
I & & -22.3 & - & - & - & - & - & - \\
\hline
LCDCS0531 & 0.64& & & & & & & \\
B & &  -19.2 &  3.02$\pm$    6.97 & -18.6$\pm$  1.7 &   298$\pm$  2694 & - & - & - \\
V & &  -18.6 &  0.97$\pm$    1.12 & -19.7$\pm$  0.9 &  4316$\pm$  2356 & - & - & - \\
R & &  -19.6 & -1.15$\pm$    0.79 & -22.1$\pm$  1.6 &  1661$\pm$  2945 & - & - & - \\
I & & -21.5 & - & - & - & - & - & - \\
\hline
HDF & 0.85& & & & & & & \\
B & &  -19.5 &  2.48$\pm$    1.39 & -20.3$\pm$  0.5 &   253$\pm$   397 & -0.20$\pm$    0.59 & -20.5$\pm$  0.5 &  6700$\pm$  1326 \\
V & &  -21.1 & -2.26$\pm$    1.22 & -23.3$\pm$  3.3 &   101$\pm$   705 & - & - & - \\
R & &  -21.6 & -0.11$\pm$    1.22 & -22.5$\pm$  0.9 &   898$\pm$   298 & - & - & - \\
I & & -23.0 & - & - & - & - & - & - \\
\hline
MJM & 0.60& & & & & & & \\
B & & -19.0 & - & - & - & - & - & - \\
V & & -19.3 & - & - & - & - & - & - \\
R & & -20.3 & - & - & - & - & - & - \\
I & &  -20.8 & - & - & - & -1.08$\pm$    0.76 & -23.2$\pm$  1.7 &  1237$\pm$  2135 \\
\hline
LCDCS0829 & 0.45& & & & & & & \\
B & &  -15.9 &  0.53$\pm$    0.53 & -18.5$\pm$  0.4 & 10607$\pm$  1628 & - & - & - \\
V & &  -17.3 & -0.32$\pm$    0.44 & -19.9$\pm$  0.5 &  9134$\pm$  2822 & -1.28$\pm$    0.69 & -25.1$\pm$  5.3 &   169$\pm$  1395 \\
R & &  -17.8 & -0.79$\pm$    0.38 & -20.8$\pm$  0.7 &  4557$\pm$  2803 & - & - & - \\
I & &  -18.2 & -0.20$\pm$    0.35 & -20.3$\pm$  0.4 &  7954$\pm$  1915 & - & - & - \\
\hline
MACS1423 & 0.55& & & & & & & \\
B & &  -20.3 &  3.50$\pm$    6.42 & -19.2$\pm$  1.1 &   289$\pm$  2779 & - & - & - \\
V & &  -19.0 &  1.49$\pm$    0.67 & -20.3$\pm$  0.3 &  4896$\pm$  2472 & - & - & - \\
R & &  -19.0 &  1.42$\pm$    0.77 & -20.8$\pm$  0.4 &  5075$\pm$  2773 & - & - & - \\
I & &  -19.9 &  0.76$\pm$    0.60 & -21.3$\pm$  0.4 &  7855$\pm$  2017 & - & - & - \\
\hline
GHO1603 & 0.54& & & & & & & \\
B & &  -17.8 &  0.94$\pm$    0.60 & -19.5$\pm$  0.4 &  5867$\pm$  1806 & - & - & - \\
V & &  -17.9 &  0.05$\pm$    0.42 & -21.5$\pm$  0.6 &  4176$\pm$  1046 & - & - & - \\
R & &  -19.9 & -0.88$\pm$    0.40 & -24.1$\pm$  2.2 &  1798$\pm$  1822 & - & - & - \\
I & &  -20.9 & -0.20$\pm$    1.02 & -22.2$\pm$  0.8 &  4730$\pm$  1836 & - & - & - \\
\hline
\hline
\end{tabular}
\end{table*}
\end{landscape}

\begin{landscape}
\begin{table*}[ht!]
\centering
\setcounter{table}{0}
  \caption{Continued.}
\begin{tabular}{lcc|ccc|ccc}
\hline
\hline
&&& & red-sequence GLFs  && & blue GLFs& \\
\hline
&$z$&comp& $\alpha$ & M*  &$\phi$* ($deg^{-2}$)&          $\alpha$ & M*  &$\phi$* ($deg^{-2}$) \\
\hline
MACSJ1621.4 & 0.47& & & & & & & \\
B & &  -17.4 &  1.10$\pm$    0.87 & -19.8$\pm$  0.6 &  3178$\pm$  1590 & - & - & - \\
V & &  -18.9 &  0.33$\pm$    0.82 & -21.2$\pm$  0.9 &  4839$\pm$   977 & - & - & - \\
R & &  -19.4 &  0.34$\pm$    0.78 & -21.2$\pm$  0.6 &  5552$\pm$  1098 & - & - & - \\
I & &  -19.3 &  1.09$\pm$    0.57 & -21.1$\pm$  0.3 &  4098$\pm$  1396 & - & - & - \\
\hline
MACSJ1621.6 & 0.46& & & & & & & \\
B & &  -17.9 &  1.15$\pm$    0.81 & -19.8$\pm$  0.5 &  2735$\pm$  1358 & - & - & - \\
V & &  -18.4 &  1.33$\pm$    0.75 & -20.5$\pm$  0.4 &  2554$\pm$  1344 & - & - & - \\
R & &  -18.9 &  1.28$\pm$    0.94 & -20.9$\pm$  0.5 &  2974$\pm$  1916 & -0.98$\pm$    1.55 & -20.5$\pm$  2.1 &  2120$\pm$  4595 \\
I & &  -19.3 &  1.18$\pm$    0.77 & -21.2$\pm$  0.4 &  2855$\pm$  1412 &  4.14$\pm$    1.82 & -18.1$\pm$  0.3 &    71$\pm$   210 \\
\hline
MACSJ2129 & 0.59& & & & & & & \\
B & &  -20.5 &  2.35$\pm$    2.66 & -19.8$\pm$  0.6 &  1702$\pm$  5290 & -2.14$\pm$    2.08 & -24.0$\pm$  9.0 &    81$\pm$  1959 \\
V & &  -20.3 & -0.09$\pm$    1.23 & -21.2$\pm$  0.9 &  9395$\pm$  2567 & - & - & - \\
R & &  -20.2 &  0.98$\pm$    0.87 & -21.1$\pm$  0.5 &  5394$\pm$  2441 & -0.09$\pm$    1.34 & -21.6$\pm$  1.3 &  2702$\pm$  1393 \\
I & &  -21.7 &  0.32$\pm$    2.40 & -21.7$\pm$  1.2 &  6318$\pm$  4242 & - & - & - \\
\hline
\hline

\end{tabular}
\end{table*}
\end{landscape}

\end{appendix}

\end{document}